\documentclass[12pt]{article}
\usepackage{amsmath,amsfonts, graphicx, float, caption}
\usepackage{lifecon} % Symbols for life contingencies
\usepackage[titletoc,title]{appendix}
\usepackage{color}
\usepackage{xcolor}
\usepackage{algorithm}
\usepackage{algpseudocode}
\usepackage{pifont}
\usepackage[dvips]{epsfig}
\usepackage{subfigure}
\usepackage{caption}
\usepackage{natbib}
\usepackage{eurosym}
\usepackage{natbib}
\usepackage{textcomp}
\usepackage{endnote}
\usepackage{lscape,verbatim}
\usepackage{setspace}
\usepackage{rotating}
\usepackage{rotating,ctable,bm}
\usepackage[latin1]{inputenc}
\usepackage[dvips]{epsfig}
\usepackage{subfigure}
\usepackage{pdflscape}
\usepackage{bm}
\usepackage{calc}
\usepackage{placeins}

\newtheorem{remark}{Remark}[section]
\newtheorem{lemma}{Lemma}[section]

\allowdisplaybreaks

\begin{document}

%\begin{frontmatter}

\title{A unified approach to mortality modelling using state-space framework: characterisation, identification, estimation and forecasting}

\author{Man Chung Fung$\dag$, Gareth W. Peters$\ddag$ $\star$ $\ast$ \and Pavel V. Shevchenko$\dag$ $\ddag$\\
{\small{
$\dag$ Risk Analytics Group, Data61, CSIRO, Sydney, Australia (simon.fung@csiro.au)}}\\
{\small{
$\ddag$ Department of Statistical Science, University College London}} \\
{\small{
$\star$ Associate Fellow, Oxford Mann Institute, Oxford University}}\\
{\small{
$\ast$ Associate Fellow, Systemic Risk Center, London School of Economics}}
}

%\begin{document}
\maketitle

\begin{abstract}
\begin{centering}
This paper explores and develops alternative statistical representations and estimation approaches for dynamic mortality models. The framework we adopt is to reinterpret popular mortality models such as the Lee-Carter class of models in a general state-space modelling methodology, which allows modelling, estimation and forecasting of mortality under a unified framework. Furthermore, we propose an alternative class of model identification constraints which is more suited to statistical inference in filtering and parameter estimation settings based on maximization of the marginalized likelihood or in Bayesian inference. We then develop a novel class of Bayesian state-space models which incorporate \textsl{apriori} beliefs about the mortality model characteristics as well as for more flexible and appropriate assumptions relating to heteroscedasticity that present in observed mortality data. We show that multiple period and cohort effect can be cast under a state-space structure. To study long term mortality dynamics, we introduce stochastic volatility to the period effect.  The estimation of the resulting stochastic volatility model of mortality is performed using a recent class of Monte Carlo procedure specifically designed for state and parameter estimation in Bayesian state-space models, known as the class of particle Markov chain Monte Carlo methods. We illustrate the framework we have developed using Danish male mortality data, and show that incorporating heteroscedasticity and stochastic volatility markedly improves model fit despite an increase of model complexity. Forecasting properties of the enhanced models are examined with long term and short term calibration periods on the reconstruction of life tables.
\end{centering}
\end{abstract}

%\begin{keywords}
\begin{centering}
\end{centering}
%\end{keywords}
%JEL Classification: G22, G23, G13

\maketitle\thispagestyle{empty}

%\end{frontmatter}

%%%%%%%%%%%%%%%%%%%%%%%%%%%%% Introduction %%%%%%%%%%%%%%%%%%%%%%%%%%%%%%%%%%%%%%%%%%%%%%%%%%%%
\newpage
%%%%%%%%%%%%%%%%%%%%%%%%%%%%%%%%%%%%%%%%%%%%%%%%%%%%%%%%%%%%%%%%%%%%%%%%%%%%%%%%%%%%%%%%%%%%%%%%%%%
%%%%%%%%%%%%%%%%%%%%%%%%%%%%%%%%%%%%%%%%%%%%%%%%%%%%%%%%%%%%%%%%%%%%%%%%%%%%%%%%%%%%%%%%%%%%%%%%%%%
\section{Introduction}\label{sec:intro}
%%%%%%%%%%%%%%%%%%%%%%%%%%%%%%%%%%%%%%%%%%%%%%%%%%%%%%%%%%%%%%%%%%%%%%%%%%%%%%%%%%%%%%%%%%%%%%%%%%%
%%%%%%%%%%%%%%%%%%%%%%%%%%%%%%%%%%%%%%%%%%%%%%%%%%%%%%%%%%%%%%%%%%%%%%%%%%%%%%%%%%%%%%%%%%%%%%%%%%%

An ageing population is a major challenge that many countries are facing today. The problem arises from the fact that fertility rates are declining while life expectancy has been increasing in the past several decades without any sign of slowing down. The adverse financial outcome of people living longer than expected, and hence the possibility of outliving their retirement savings, is known as longevity risk. This long term demographic risk has significant implications for societies and manifests as a systematic risk for pension plans and annuity providers. Policymakers rely on mortality projection to determine appropriate pension benefits and to understand the costing of different economic assumptions and regulations regarding the age of retirement of a given population. For instance, in the UK and Australia defined-benefit pension plans prior to 2000's had limited exposure to effects of longevity risk since high equity returns on pension fund wealth management portfolios were masking the impact of longevity risk, however post 2000 declining equity returns coupled with record low interest rate financial environments has demonstrated the significance of decades of longevity improvements, posing a very real problem for pension schemes. Furthermore, by regulation, insurers who offer retirement income products are required to hold additional reserving capital to cover longevity risk. A key input to address longevity risk is the development of advanced mortality modelling methodology, so that human longevity can be predicted with better accuracy and any uncertainties can be accounted for in mortality forecasting.

Since the introduction of the Lee-Carter model (\cite{LeeCa92}), a range of stochastic mortality models have been proposed in the literature. \cite{RenshawHa03} and \cite{RenshawHa06} introduce multiple period effects and cohort effect to capture the change of mortality with respect to year and year-of-birth, respectively, to the Lee-Carter model. \cite{CairnsBlDo06b} proposed a two-factor period effect mortality model, known as the Cairns-Blake-Dowd (CBD) model, for pensioner ages. A cohort extension of the CBD model was studied in \cite{Cairnsetal09}. \cite{Plat09} draws on the strengths of the Lee-Carter model and the CBD model to produce an age-period-cohort model that includes a term to capture young mortality dynamics. In these well known cases it is common practice in actuarial settings to estimate stochastic mortality models based on a singular value decomposition approach (\cite{LeeCa92}, \cite{KoissiShHo06}) or via a maximum likelihood based approach if a discrete Poisson regression setting is considered (\cite{BrouhnsDeVe02}, \cite{Cairnsetal09}).

A common feature of the estimation methods adopted in the frameworks mentioned above is that the dynamics of the period effect, the stochastic latent processes, are not directly incorporated into joint parameter and state estimation, and instead form a component of a second stage of estimation. Typically this involves specifying a model for the period effect for forecasting purpose only after an estimation is performed. Such approaches often suffer from a statistical lack of efficiency compared to methods that perform joint static model parameter estimation and latent process filtering. \textsl{Hence, the first argument we make is that recasting different classes of mortality models in a state-space formulation can better facilitate state-space based inference under either frequentist or Bayesian estimation. This is especially true in the case that the inference is performed jointly on the latent process and static model parameters, rather than in a less statistically efficient two-stage procedure}.

Typically the studies carried out in practice and in the literature have the feature that only mortality data from the past several decades is considered. For many countries, age-specific death rates are evolving rather smoothly except for some potential change of trend in the last 50 years or so in some developed countries. Besides ARIMA models, structural change model have been proposed to take into account the trend-changing behaviour of the period effect (\cite{LiChCh11}, \cite{vanBerkumAnVe14}). Despite this, the implication of including earlier periods that exhibit significant volatility of mortality, which can be attributed to some life-critical events such as wars and epidemics, is still not yet being investigated. The ability to incorporate such structural information into a mortality model is greatly facilitated when recasting the model in a state-space formulation. Furthermore, extensions to mortality models that can also be facilitated in a state-space formulation are increasingly able to be considered and may better explain the stochastic dynamics of such processes. These include features such as: time varying volatility; cross-sectional volatility between different age groups; extremal dependence features; cohort effects; structure breaks in regimes; long memory or persistence in mortality features in different age-groups; cointegration and non-stationarity features; as well as regression based structures that decompose mortality according to categorical features such as official death causes, regional categories etc.

Moreover, additional stochastic factor models such as two and three factor models can be easily considered. This can be particularly relevant when modelling features such as trends in excess mortality in particular age-groups resulting from disease epidemics (\cite{zucs2005influenza}, \cite{dawood2012estimated}), cold and heat-waves (\cite{fouillet2006excess}, \cite{analitis2008effects}) and other effects such as medical impairements, occupational hazards, hazardous persuites, geographical location of residence and ethnic origin, see \cite{eloranta2012partitioning} and \cite{england1993new}. \textsl{ Hence, the second argument we make is that all these different model structures can readily be encoded in state-space model structures. Furthermore, they can be consistently combined in joint estimation procedures in such state-space model structures in either frequentist and Bayesian formualtions, whilst also admitting consistent joint forecasting models for predictive purposes}.

A variety of state-space model approaches exist in a range of different literatures, in this paper we propose to begin with the widely adopted frameworks typically introduced in state-space modelling settings in \cite{Harvey89} or \cite{WestHa}, which we develop to address some of the aforementioned issues. In contrast to the singular value decomposition and Poisson regression estimation approaches where the period effect is treated as parameter without any temporal structure in the first-stage estimation, period effect is regarded as a latent process with a Markovian structure under the state-space approach. In other words, a state-space formulation permits modelling, estimation and forecasting of mortality under a unified framework. Recent progress in sampling-based techniques has allowed statistical inference to be conducted on sophisticated state-space models that can incorporate multiple latent driving factors which may exhibit non-linear and non-Gaussian stochastic dynamics. We take advantage of this development and utilise realistic model to capture the long term volatility structure of mortality time series.

\cite{Pedroza06} and \cite{KogureKu10} consider Bayesian estimation of the Lee-Carter model in state-space form. A maximum likelihood approach is studied in \cite{deJongTi06}. Here, we extend such frameworks to show how to adopt a combination of filtering procedures with Rao-Blackwellization to obtain gradient based Fisher score equation recursions to accurately and efficiently perform optimal filtering of the latent state process, in the sense of mean square error minimization, and recursive least squares estimation for the static model parameters jointly in a recursive manner. Furthermore, we extend such state-space models to incorporate non-linear and non-Gaussian features in the state-space structure that no longer admit simple Kalman filter forward backward algorithm recursions, leading us to more cutting edge filtering techniques based on Sequential Monte Carlo methods. In this regard, we estimate and examine the Lee-Carter model with heteroscedasticity using both gradient-based maximum likelihood and Bayesian analysis. Alternative models that have tried to include such features include, for example, the Poisson regression in \cite{BrouhnsDeVe02} and \cite{CzadoDeDe05}, who aimed to replace the homogeneous additive error term in the Lee-Carter model by a Poisson error structure. Also we note a recently developed framework for modelling death counts with common latent risk factors via credit risk plus methodology with model estimation via Markov chain Monte Carlo (MCMC) in \cite{HirzScSh15}. However, we argue that the state-space formulation allows heteroscedasticity to be accounted for in a more straightforward manner.

Through reformulation and extensions of the Lee-Carter type mortality models in a state-space model structure, we investigate several key properties observed in mortality data. First, the cross sectional variance-covariance matrix between age-group structures is non-homogeneous. Second, examination of mortality data over a long period indicates that volatility of the evolution of death rates is not constant, i.e. homoskedascity is present. We show that the incorporation of a second stochastic volatility latent factor will allow us to identify the periods in which mortality demonstrates heightened volatility. This will aid in interpretation and forecasting from such models. Specifically, we introduce a stochastic volatility model for the period effect, aiming to capture long term mortality dynamics. The state-space framework provides a natural platform to analyse stochastic volatility models (\cite{KimShCh98}, \cite{ChibNaSh02}). In this paper we develop a particle Markov chain Monte Carlo (PMCMC) (\cite{AndrieuDoHo10}) Bayesian model formulation in order to estimate the resulting stochastic volatility model of mortality jointly with the other latent stochastic factors and the static model parameters.

We introduce to mortality modelling the estimation framework based around the PMCMC algorithm which utilises sequential Monte Carlo (SMC) (\cite{DoucetFrGo01}, \cite{peters2012sequential}) to obtain required quantities in Metropolis-Hastings algorithms that has found many applications in a variety of areas, for example finance (\cite{PetersBrShDo11}), economics (\cite{FlurySh11}), non-life insurance (\cite{peters2010chain}), risk management (\cite{targino2015sequential}) and computational biology (\cite{GolightlyWi11}). We apply this powerful tool in mortality modelling and it allows us to develop efficient algorithms to estimate a stochastic volatility extension of the Lee-Carter model.

The paper is organised as follows. In Section 2 we give an overview of the conventional mortality modelling and estimation methodology in the literature. A state-space approach for mortality modelling is formulated and discussed in Section 3.  Section 4 is devoted to state-space inference for stochastic mortality models in a frequentist approach. Section 5 focuses on Bayesian inference for dynamic mortality models in state-space framework. In Section 6 we analyse Danish mortality data based on the enhanced models and methodologies proposed in the paper. Section 7 provides concluding remarks.

%%%%%%%%%%%%%%%%%%%%%%%%%%%%% Lee-Carter Model %%%%%%%%%%%%%%%%%%%%%%%%%%%%%%%%%%%%%%%%%%%%%%%%%%%%
%%%%%%%%%%%%%%%%%%%%%%%%%%%%%%%%%%%%%%%%%%%%%%%%%%%%%%%%%%%%%%%%%%%%%%%%%%%%%%%%%%%%%%%%%%%%%%%%%%%
\section{Classical Bayesian and Frequentist Approaches}\label{sec:mortalitySS}
%%%%%%%%%%%%%%%%%%%%%%%%%%%%%%%%%%%%%%%%%%%%%%%%%%%%%%%%%%%%%%%%%%%%%%%%%%%%%%%%%%%%%%%%%%%%%%%%%%%
%%%%%%%%%%%%%%%%%%%%%%%%%%%%%%%%%%%%%%%%%%%%%%%%%%%%%%%%%%%%%%%%%%%%%%%%%%%%%%%%%%%%%%%%%%%%%%%%%%%

In this section we first briefly recall some important definitions on mortality modelling. We then review stochastic mortality models that are commonly found in the literature. Standard estimation procedures under frequentist and Bayesian approaches are discussed.

%%%%%%%%%%%%%%%%%%%%%%%%%%%%%%%%%%%%%%%%%%%%%%%%%%%%%%%%%%%%%%%%%%%%%%%%%%%%%%%%%%%%%%%%%%%%%%%%%%%
\subsection{Definitions and Notation}\label{subsec:NotationsDefinitions}
%%%%%%%%%%%%%%%%%%%%%%%%%%%%%%%%%%%%%%%%%%%%%%%%%%%%%%%%%%%%%%%%%%%%%%%%%%%%%%%%%%%%%%%%%%%%%%%%%%%
We use the following standard definitions from actuarial literature on mortality modelling (\cite{DicksonHaWa09}, \cite{PitaccoDeHa}). Let $T_x$ be a random variable representing the remaining lifetime of a person aged $x$. The cumulative distribution function and survival function of $T_x$ are written as ${}_{\tau}q_x = P(T_x \leq \tau)$ and ${}_{\tau}p_x = P(T_x > \tau)$ respectively. For a person aged $x$, the force of mortality at age $x+\tau$  is defined as
\begin{equation}\label{eqn:forcemortality}
    \mu_{x+\tau} := \lim_{h\rightarrow 0} \frac{1}{h}P(T_x<\tau+h|T_x>\tau)
    = -\frac{d}{d\tau}\ln{{}_{\tau}p_x}.
\end{equation}
Let $f_x(t)$ be the density function of $T_x$, then from \eqref{eqn:forcemortality} we have
${}_{\tau}q_x = \int^\tau_0 f_x(s)\,ds =\int^\tau_0 {}_{s}p_x\,\mu_{x+s}\,ds$.
The central death rate for a $x$-year-old, where $x \in \mathbb{N}$, is defined as
\begin{equation}\label{eqn:centralmortality}
m_x := \frac{q_x}{\int^1_0 {}_{s}p_x\,ds} = \frac{\int^1_0
{}_{s}p_x\,\mu_{x+s}\,ds}{\int^1_0 {}_{s}p_x\,ds},
\end{equation}
which is a weighted-average of the force of mortality (here $q_x := {}_1q_x$). Under the so-called piecewise constant force of mortality assumption, that is $\mu_{x+s} = \mu_x$ where $0 \leq s<1$ and $x \in \mathbb{N}$, we have, from \eqref{eqn:centralmortality}, $m_x = \mu_x$. Moreover, if a Poisson assumption is made for the actual number of deaths, then the resulting maximum likelihood estimate of the force of mortality $\hat{\mu}_x$ (and hence $\hat{m}_x$) is given by $\hat{\mu}_x=D_x/E_x=\hat{m}_x$ where $D_x$ is the number of deaths recorded at age $x$ last birthday and the exposure-to-risk $E_x$ is the average number of people aged $x$ last birthday, during the observation year. Note that $E_x$ is approximated by an estimate of the population aged $x$ last birthday in the middle of the observation year. We refer to $\hat{m}_{x}$ as the crude death rate.

In the above setup it is assumed that the force of mortality $\mu$ is deterministic. The stochastic case can be handled by the intensity-based framework where death time is modeled as the first jump time of a doubly stochastic process (\cite{Biffis05}). Hereafter we treat the force of mortality $\mu_{x+t}(t)$, the central death rate $m_{x,t}$ and the crude death rate $\hat{m}_{x,t}$ as stochastic processes. For a detailed discussion of the background of stochastic mortality modelling in discrete-time and continuous-time, see \cite{CairnsBlDo08}.  %The survival probability, conditional on the path $s \rightarrow \mu_{x+s}(s)$, becomes
%\begin{equation}
%    P(T_x>\tau|\mu_{x+s}(s),0 \leq s\leq \tau) = e^{-\int^\tau_0 \mu_{x+s}(s)ds}.
%\end{equation}

%%%%%%%%%%%%%%%%%%%%%%%%%%%%%%%%%%%%%%%%%%%%%%%%%%%%%%%%%%%%%%%%%%%%%%%%%%%%%%%%%%%%%%%%%%%%%%%%%%%
\subsection{Stochastic Mortality Models}\label{subsec:mortalitymodels}
%%%%%%%%%%%%%%%%%%%%%%%%%%%%%%%%%%%%%%%%%%%%%%%%%%%%%%%%%%%%%%%%%%%%%%%%%%%%%%%%%%%%%%%%%%%%%%%%%%%
One of the most widely considered examples of stochastic factor model in the context of mortality modelling is the approach first presented in \citet{LeeCa92} who proposed a stochastic mortality model for the age-specific crude death rate $\hat{m}_{x,t}$, where $x=x_1,\dots,x_p$ and $t=1,\dots,T$ represent age (or age-group) and year (time) respectively. Under the model, the dynamics of the log crude death rates, $y_{x,t}=\ln{\hat{m}_{x,t}}$, is given by\footnote{Alternatively, one may treat the Lee-Carter model as a model for the log \textit{central} death rate $\ln{m_{x,t}}=\alpha_x+\beta_x \kappa_t$. The distinction of the crude and central death rate is of particular importance when one considers a Poisson regression setup of death counts (discussed in Section~\ref{sec:regressionBased}) where the dynamics of the central death rate is being modeled (\cite{Cairnsetal09} and \cite{Dowdetal10}).}
\begin{equation}
    y_{x,t} = \alpha_x+\beta_x\kappa_t+ \varepsilon_{x,t},
    \quad \varepsilon_{x,t} \overset{iid}{\sim} \text{N}(0,\sigma^2_\varepsilon), \label{eqn:LCy} \\
\end{equation}
where $\text{N}(0,\sigma^2_\varepsilon)$ denotes a Gaussian distribution with zero mean and variance $\sigma^2_\varepsilon$. The vector $\boldsymbol{\alpha}=\alpha_{x_1:x_p}:=[\alpha_{x_1},\dots,\alpha_{x_p}]$ represents the age-profile of the log death rates and $\boldsymbol{\beta}=\beta_{x_1:x_p}$ measures the sensitivity of of death rates for different age group to a change of the time series $\kappa_t$. The period effect, $\kappa_t$, for forecasting purpose, is assumed to satisfy the equation
\begin{equation}
    \kappa_t = \kappa_{t-1}+\theta+\omega_{t}, \quad \omega_t \overset{iid}{\sim}
    \text{N}(0,\sigma^2_\omega), \label{eqn:LCk}
\end{equation}
where $\varepsilon_{x,t}$ and $\omega_t$ are independent.

Under this specification, it is clear that the Lee-Carter model is not identifiable, since \eqref{eqn:LCy} is invariant up to some linear transformations of the parameters:
\begin{equation}\label{eqn:LCtransform}
\boldsymbol{y}_t = \boldsymbol{\alpha}+\boldsymbol{\beta}\kappa_t+
\boldsymbol{\varepsilon}_t  = \boldsymbol{\alpha}+
\boldsymbol{\beta}c +
\frac{\boldsymbol{\beta}}{d}\left((\kappa_t-c)d\right)+\boldsymbol{\varepsilon}_t
=
\tilde{\boldsymbol{\alpha}}+\tilde{\boldsymbol{\beta}}\tilde{\kappa}_t+
\boldsymbol{\varepsilon}_t,
\end{equation}
where $\tilde{\boldsymbol{\alpha}}=\boldsymbol{\alpha}+
\boldsymbol{\beta}c$,
$\tilde{\boldsymbol{\beta}}=\boldsymbol{\beta}/ d$ and
$\tilde{\kappa}_t=(\kappa_t-c)d$.

To overcome this identification issue when estimating the Lee-Carter model, one has to impose a non-unique choice of constraints to restrict the model to an identifiable class. It is standard practice in actuarial literature to consider the following two constraints:
\begin{equation}\label{eqn:constraints}
    \sum^{x_p}_{x=x_1}\beta_x=1, \quad \sum^T_{t=1}\kappa_t=0,
\end{equation}
as suggested in \cite{LeeCa92} to remedy the identifiability issue. This choice of constraints is equivalent to fixing $c = (1/T)\sum^T_{t=1} \kappa_t$ and $d = \sum^{x_p}_{x=x_1} \beta_x$. Consequently we have $\sum^T_{t=1}\tilde{\kappa}_t=0$ and $\sum^{x_p}_{x=x_1}\tilde{\beta}_x=1$. The reason for these particular form of identification constraints relates to the fact that the constraint on the path space of the stochastic factor $\kappa_1,\ldots,\kappa_T$ is intended to have the effect of centering the $\kappa_t$ values over the range $t \in \left\{1,\ldots,T\right\}$, such that the structure is designed to capture age-period effects with the $\alpha_x$ terms incorporating the main age effects, averaged over time, and the bilinear terms $\beta_x \kappa_t$ incorporating the age specific period trends (relative to the main age effects).

Since the introduction of the Lee-Carter model it has found a widespread uptake of this class of factor model in both practice, where the Lee-Carter model is now used as a benchmark methodology by the US Bureau of the Census, and in academia where a range of stochastic mortality model extensions have been proposed in the literature, see Table~\ref{table:modelsliterature}.
\begin{table}[h]
\center \setlength{\tabcolsep}{1em}
\renewcommand{\arraystretch}{1.1}
\scalebox{0.85}{\begin{tabular}{ll}
\hline \hline
Model &  Dynamics \\
\hline
\cite{LeeCa92}   & $\ln(m_{x,t}) = \alpha_x+\beta_x\kappa_t$     \\
\cite{RenshawHa03}   & $\ln(m_{x,t}) = \alpha_x+\sum_{i=1}^k\beta^{(i)}_x\kappa^{(i)}_t$     \\
\cite{RenshawHa06}   & $\ln(m_{x,t}) = \alpha_x+\beta^{(1)}_x\kappa_t+ \beta^{(2)}_x\,\zeta_{t-x}$     \\
\cite{Currie06}   & $\ln(m_{x,t}) = \alpha_x+ \kappa_t+ \zeta_{t-x}$     \\
\cite{CairnsBlDo06b}   & $\text{logit}(q_{x,t}) = \kappa^{(1)}_t+ \kappa^{(2)}_t(x-\bar{x})$     \\
\cite{Cairnsetal09}   & $\text{logit}(q_{x,t}) = \kappa^{(1)}_t+ \kappa^{(2)}_t(x-\bar{x})+ \zeta_{t-x}$     \\
\cite{Plat09}   & $\ln(m_{x,t}) = \alpha_x+ \kappa^{(1)}_t+ \kappa^{(2)}_t(\bar{x}-x)+\kappa^{(3)}_t(\bar{x}-x)^+ + \zeta_{t-x}$     \\
\hline \hline
\end{tabular}}
\center\caption{\label{table:modelsliterature} \small{Several popular stochastic mortality models.}}
\end{table}
We note here that \cite{RenshawHa03} and \cite{RenshawHa06} introduces multi-period ($\sum_{i=1}^k\beta^{(i)}_x\kappa^{(i)}_t$) and cohort factor ($\zeta_{t-x}$), respectively, to the Lee-Carter method. \cite{Currie06} considers a simplified version of the model in \cite{RenshawHa06}. \cite{CairnsBlDo06b} propose to model $\text{logit}(q_{x,t}):=\ln\left(q_{x,t}/(1-q_{x,t})\right)$ instead of log death rates and $\bar{x}$ is the average age in the sample range. An addition of cohort factor is studied in \cite{Cairnsetal09}. \cite{Plat09} introduces a model which combines the desirable features of the previous models and include a term $(\bar{x}-x)^+ := \text{max}(\bar{x}-x,0)$ to capture better young age mortality. The specification of identification constraints for the Lee-Carter type models, that is for those where the log death rate is being modeled in Table~\ref{table:modelsliterature}, is discussed in \cite{HuntVi15}.

%%%%%%%%%%%%%%%%%%%%%%%%%%%%%%%%%%%%%%%%%%%%%%%%%%%%%%%%%%%%%%%%%%%%%%%%%%%%%%%%%%%%%%%%%%%%%%%%%%%
\subsection{Two-Stage Estimation Approaches: Frequentist View}
%%%%%%%%%%%%%%%%%%%%%%%%%%%%%%%%%%%%%%%%%%%%%%%%%%%%%%%%%%%%%%%%%%%%%%%%%%%%%%%%%%%%%%%%%%%%%%%%%%%
Several ``classical'' approaches to Lee-Carter model estimation have been proposed in the literature, though they typically involve a two-stage procedure looking first at the observation equation as a regression (ignoring the latent factor structure explicitly) and then in the second stage they fit time series models to the latent factor structures. A good overview of such methods is obtained in \cite{PitaccoDeHa}. This two-stage procedure is at odds with modern state-space modelling procedures which have been progressively moving towards joint parameter estimation and latent state estimation in frequentist and Bayesian formulations, which will be discussed in subsequent sections. This is reflected in the first attempt to improve the calibration approaches as reflected in the comment in \cite{Cairnsetal11} where they highlight that the ``..\textsl{key element of the proposed framework is our single-stage approach to model fitting and process parameter estimation.}'' Such sentiments, relating to consistent single stage joint estimation are also echoed in the work of \cite{CzadoDeDe05}.

\subsubsection{Multi-factor Lee-Carter SVD-based two-stage calibration}
One of the most commonly adopted approaches to estimate stochastic mortality models is via singular value decomposition (SVD). We use the multi-period ($k$-factor) Lee-Carter model (\cite{RenshawHa03}) with identification constraints given by
\begin{equation}
    \sum^T_{t=1}\kappa^{(i)}_t=0, \quad \sum^{x_p}_{x=x_1}\beta^{(i)}_x=1,
\end{equation}
where $i=1,\dots,k$, as an example to illustrate the methodology below (\cite{KoissiShHo06}).
\begin{description}
\item[Stage 1a - Observation Equation Estimation Stage:]  We first notice that the constraint $\sum^T_{t=1} \kappa^{(i)}_t = 0$ will lead to an estimator for the level $\bm{\alpha}$ given by
    \begin{equation}
        \hat{\alpha}_{x}=\frac{1}{T}\sum_{y=1}^Ty_{x,t}.
    \end{equation}
\item[Stage 1b - Observation Equation Estimation Stage:] The next stage is to de-trend the observations $\left\{\bm{y}_{1:T}\right\}$ by the level estimate $\hat{\bm{\alpha}}$ and then to perform a SVD on the resulting $(p\times T)$ matrix of residual observations to obtain the decomposition
    \begin{equation}
        \text{SVD}[\bm{y}_{1:T}-\hat{\bm{\alpha}}] = \sum^h_{i=1}\rho_i \bm{u}_{i}\bm{v}^\top_{i},
    \end{equation}
    where $\top$ denotes transposition and $\rho_i$, for $i\in\{1,\dots,h\}$, are the descending singular values where $h$ is the rank of the data matrix. Here $\bm{u}_{i}$ and $\bm{v}_{i}$ are the corresponding left and right singular vectors of the singular value $\rho_i$ with dimension $p$ and $T$ respectively. For a $k$-rank, where $k \leq h$, approximation of the matrix, we have
    \begin{equation}
        \bm{y}_{1:T}-\hat{\bm{\alpha}} = \sum^k_{i=1} \rho_i \bm{u}_{i}\bm{v}^\top_{i} + \bm{\varphi}_{1:T},
    \end{equation}
    where $\bm{\varphi}_{1:T}=\sum^h_{i=k+1}\rho_i \bm{u}_{i}\bm{v}^\top_{i}$ is the $k$-rank residuals. We then identify $\tilde{\bm{\beta}}^{(i)}=\bm{u}_{i}$ and $\tilde{\bm{\kappa}}^{(i)} = \rho_i\bm{v}_{i}$, for $i=1,\dots,k$. One then performs the transformation
    \begin{equation}
        \kappa^{(i)}_t = \tilde{\kappa}^{(i)}_t\sum_x\tilde{\beta}^{(i)}_x, \quad \beta^{(i)}_x = \frac{\tilde{\beta}^{(i)}_x}{\sum_x\tilde{\beta}^{(i)}_x},
    \end{equation}
    to ensure the constraints $\sum_x\beta^{(i)}_x=1$, for $i=1,\dots,k$, are satisfied.
\item[Stage 2 - Latent Process Factor Estimation Stage:] At this stage\footnote{We omit here the refitting procedure for $\kappa$ suggested in \cite{LeeCa92}.}, the estimation of the latent factors can be performed by specifying a time series model structure such as ARIMA model for each of the factors:
    \begin{equation}
        \kappa_t^{(j)} = \theta^{(j)} + \sum_{r=1}^p\kappa_{t-r}^{(j)} + \sum_{s=1}^q \epsilon_{t-s}^{(j)} + \epsilon_t,
    \end{equation}
    or alternatively one could fit the equivalent Vector Auto-Regressive (VAR) model structure not treating each factor as independent in the time series specification. One would typically perform this stage of estimation via the Yule-Walker equations, see for instance discussions in \cite{tsay1984consistent}. Under such specifications, one then obtain closed form distributions and estimators for period effect latent factor forecasts that can be substituted into the observation model for forecasts of the mortality by age in future forecast horizons and used to construct life tables.
\end{description}

\subsubsection{Regression-based approaches}\label{sec:regressionBased}
It is important to note that the SVD approach assumes homoscedasticity in the error structure. Therefore, to  account for heteroscedasticity in mortality data for different ages, \cite{BrouhnsDeVe02} propose to model death counts, instead of death rates, via Poisson regression where the addition error term in the Lee-Carter approach is replaced by Poison random variation. Specifically, the number of death $D_{x,t}$ is modeled as
\begin{equation}
    D_{x,t} \sim \text{Poisson}(E_{x,t}\,m_{x,t}(\Phi)),
\end{equation}
where $E_{x,t}$ is the death exposure, $m_{x,t}(\Phi)$ is a model of the central death rate and $\Phi$ is the parameter vector according to the model being used, including time dynamic factors such as period and cohort effect, see for example Table~\ref{table:modelsliterature}. The parameter vector is then estimated by maximising the log-likelihood function, which is given by
\begin{equation}
    l(\Phi;D,E) = \sum_t \sum_x \left(D_{x,t}\ln(E_{x,t}\,m_{x,t}(\Phi)) - E_{x,t}\,m_{x,t}(\Phi) - \ln(D_{x,t}!)\right),
\end{equation}
where $D_{x,t}!$ indicates the factorial of $D_{x,t}$. Times series models are then used to model the time dynamic factors forming a second stage estimation procedure for forecasting purpose. Note that the CBD type models can be estimated under this approach since we have $q_{x,t}=1-\exp\{-m_{x,t}\}$ (\cite{Cairnsetal09}).

%Such models also allow one to incorporate additional dispersion features where appropriate. Basically, such models are variations of the estimation of Genearlized Linear Model (GLM) regression structures. In such contexts, there have been several developments such as the two-factor model with multiplicative terms given by a Gompertz-Makeham graduation term and an age specific trend adjustment term studied in \cite{renshaw1996modelling} and variations in \cite{sithole2000investigation}. Spline based variations of the GLM model structures have also been considered in works such as \cite{currie2004smoothing}, where bivariate penalized B-splines were considered to smooth over both age and time within a penalized GLM framework. In this case, it is documented that when comparing forecasts to the classical Lee-Carter stochastic factor model, there was a significantly slower mortality decline under the GLM smoothing spline forecasts. For a detailed discussion on when such GLM structures and Lee-Carter model is identical in model specification, see the works of \cite{renshaw2000modelling}.

\begin{remark}In all the discussed cases above, there is the general idea that the two-stage estimation approaches (SVD and regression) treat the unobserved factors corresponding to for instance a period effect $\kappa_t$ and a cohort effect $\zeta_{t-x}$ as parameters. For forecasting purpose, these dynamics factors are then modeled as time series, typically under the ARIMA framework. In this paper we argue that a more consistent approach involves embedding the specification of the model formally within a state-space model structure and to perform the estimation via a joint combination of filtering and static-parameter estimation, which can be achieved either in Bayesian (posterior-based) or frequentist (likelihood-based) settings. We will demonstrate both in this paper.
\end{remark}

%%%%%%%%%%%%%%%%%%%%%%%%%%%%%%%%%%%%%%%%%%%%%%%%%%%%%%%%%%%%%%%%%%%%%%%%%%%%%%%%%%%%%%%%%%%%%%%%%%%
\subsection{Estimation Approaches: Bayesian View}
\label{ClassicalBayes}
%%%%%%%%%%%%%%%%%%%%%%%%%%%%%%%%%%%%%%%%%%%%%%%%%%%%%%%%%%%%%%%%%%%%%%%%%%%%%%%%%%%%%%%%%%%%%%%%%%%
From the Bayesian modelling perspective there are few papers that study stochastic mortality models, the main papers in this area involve the works of \cite{CzadoDeDe05}, \cite{kogure2009bayesian} and \cite{Cairnsetal11}. As observed in these studies, there are many possible advantages to adopting a Bayesian approach for mortality modelling, especially in the context of small populations which may also have substantial quantities of missing data.

An important point to note is that all Bayesian model formulations to date in the mortality modelling literature, that we are aware of, have utilised what would, in modern statistical approaches be considered rudimentary sampling based approaches to performing Bayesian estimation of the Lee-Carter type models. The criticism here can be leveled in two ways.
\begin{enumerate}
\item The first relates to the fact that in these Bayesian formulations the latent dynamic process states are still treated in the MCMC sampling procedures as if they were a set of static model parameters. The issues with doing this have been mentioned in numerous places, see for example \cite{CarterKo94}. Recently new approaches to such inference in Bayesian models have been developed to avoid having to make univariate conjugate Gibbs or Metropolis-within-Gibbs steps for the latent processes. The reason for this is that it is known in general to be very inefficient in performing inference and can be prone to misleading posterior inference results due to poor mixing performance of the Markov chain for a finite computational budget. Detailed discussions have been provided on such problems in \cite{AndrieuDoHo10} and subsequently in work such as \cite{chopin2013smc2} and the specific case to population based state-space models in ecology in \cite{peters2010ecological}.
\item Secondly, all existing MCMC sampling-based approaches we are aware of for Bayesian inference in the mortality modelling literature tends to neglect the issue of model identification in the likelihood which can cause issues in the Bayesian formulation. In fact, some approaches implement identification constraint in the Bayesian model and develop an MCMC sampler that tries to impose the identification constraint in such a manner that the resulting Markov chain may not be consistent with preserving the correct invariant stationary distribution if one applies the constraints inappropriately. We investigate this issue in a separate paper (\cite{PetersFuSc16}).
\end{enumerate}

These two considerations need to be resolved to update the approaches to more efficient sampling approaches with enhanced specifications of the model formulation to deal with such issues directly. In particular, modern approaches to such model estimations are to treat the latent unobserved process not as static parameters but as a state-space model in which filtering based methods (Kalman Filter variants, SMC) can be utilised for the latent process estimations jointly with consistent estimation of the `static' model parameters. We will detail such estimation procedures which are also consistent with imposing specific identification constraints of relevance to the Lee-Carter model formulations, that are developed to ensure the correct invariant Bayesian posterior model is preserved by the Markov chain sampler and filters developed.

\begin{remark}[Likelihood Identification Issues and Bayesian Modelling]
We note the fact that model parameters that are not identified in the likelihood pose no formal problem in a Bayesian analysis. Identification is a property of the likelihood function, whereas Bayesian inference simply uses the likelihood function to map through the data from prior beliefs to posterior beliefs. However, it is often the case that working with unidentified likelihood functions is usually unsatisfactory from a practical perspective as it may lead to partial identification issues in the posterior or problematic multimodality in the posterior. In general if one utilises a proper prior distribution it may act to provide a ``near-identification'' in the sense that one considers parameter restrictions as limiting forms of prior densities, then there is at least a functional equivalence between introducing prior information about parameters, and imposing identifying restrictions.
\end{remark}

%%%%%%%%%%%%%%%%%%%%%%%%%%%%%%%%%%%%%%%%%%%%%%%%%%%%%%%%%%%%%%%%%%%%%%%%%%%%%%%%%%%%%%%%%%%%%%%%%%%
%%%%%%%%%%%%%%%%%%%%%%%%%%%%%%%%%%%%%%%%%%%%%%%%%%%%%%%%%%%%%%%%%%%%%%%%%%%%%%%%%%%%%%%%%%%%%%%%%%%
\section{State-Space Formulations of Mortality Models}\label{subsec:SSM}
%%%%%%%%%%%%%%%%%%%%%%%%%%%%%%%%%%%%%%%%%%%%%%%%%%%%%%%%%%%%%%%%%%%%%%%%%%%%%%%%%%%%%%%%%%%%%%%%%%%
%%%%%%%%%%%%%%%%%%%%%%%%%%%%%%%%%%%%%%%%%%%%%%%%%%%%%%%%%%%%%%%%%%%%%%%%%%%%%%%%%%%%%%%%%%%%%%%%%%%
We are now in a position to present an alternative representations of stochastic mortality modelling based on state-space methodology (\cite{Harvey89}, \cite{WestHa}).  A key advantage of this approach is that the two-stage estimation and forecasting procedure under the SVD or Poisson regression maximum likelihood approaches can be combined in a single setting. The improved statistical consistency of a single stage approach is recognised in \cite{Cairnsetal11}. Another key advantage comes from the recent progress in sampling-based techniques in the estimation of state-space models.  The advancement allows statistical inference to be conducted on sophisticated state-space models. We take advantage of this development and utilise realistic model aiming to capture long term mortality dynamics.

A general state-space model consists of a state equation
\begin{equation}\label{eqn:generalSSMstate}
    \boldsymbol{\phi}_t = a(\boldsymbol{\phi}_{t-1},\boldsymbol{u}_t),
\end{equation}
and an observation equation
\begin{equation}\label{eqn:generalSSMmeasurement}
    \boldsymbol{z}_t = b(\boldsymbol{\phi}_t,\boldsymbol{v}_t),
\end{equation}
where the states $\boldsymbol{\phi}_t$ form a hidden/latent Markov process with disturbance $\boldsymbol{u}_t$, and the observed time series data $\boldsymbol{z}_t$ depends only on $\boldsymbol{\phi}_t$ and disturbance $\boldsymbol{v}_t$. Here $a(.)$ and $b(.)$ are possibly nonlinear functions, and the states $\boldsymbol{\phi}_t$ and observations $\boldsymbol{z}_t$ can be multi-dimensional.

It is clear that the models in Table~\ref{table:modelsliterature} specify the observation equation of different state-space models that can be considered. For example, for the multi-period Lee-Carter model (\cite{RenshawHa03}), the observed data is $z_{x,t} = \ln(\hat{m}_{x,t})$ for different age $x$ and the latent states are the period effects $\boldsymbol{\phi}_t = \left(\kappa^{(1)}_t,\dots,\kappa^{(k)}_t\right)$. We also note that multi-population (i.e. multi-curve) structures can be incorporated in the following state-space models in a number of different ways and the approaches we will develop for estimation will accommodate such settings. In the following sub-sections we will discuss a few different classes of mortality models that are difficult to deal with in the approaches mentioned in Section 2, but can be handled straightforwardly in state-space framework.

%%%%%%%%%%%%%%%%%%%%%%%%%%%%%%%%%%%%%%%%%%%%%%%%%%%%%%%%%%%%%%%%%%%%%%%%%%%%%%%%%%%%%%%%%%%%%%%%%%%
\subsection{Lee-Carter Model with Heteroscedasticity: LC-H model}
\label{sec:LChetero}
%%%%%%%%%%%%%%%%%%%%%%%%%%%%%%%%%%%%%%%%%%%%%%%%%%%%%%%%%%%%%%%%%%%%%%%%%%%%%%%%%%%%%%%%%%%%%%%%%%%
We present here a state-space formulation of the Lee-Carter model with heteroscedasticity structure. In this context, the hetroscedasticity refers to a relaxation of the constant single degree of freedom diagonal covariance assumption typically made on the observation vector for each year $t$ across the panel of age group stratefications $\bm{y}_{t} = \left(y_{x_1,t},y_{x_2,t},\ldots,y_{x_p,t}\right)$. Within this state-space model structure we propose an alternative identification constraint which is tailored for the estimation under the state-space approach.

The Lee-Carter model with heteroscedasticity structure can be written in state-space form by combining the processes $\boldsymbol{y}_t=(y_{x_1,t},\dots, y_{x_p,t})$ and $\kappa_t$ into one dynamical system
\begin{subequations}\label{eqn:LCH}
\begin{align}
    \boldsymbol{y}_t &= \boldsymbol{\alpha}+\boldsymbol{\beta}\kappa_t+ \boldsymbol{\varepsilon}_t, \quad \boldsymbol{\varepsilon}_t \overset{iid}{\sim} \text{N}(0,\Sigma), \\
    \kappa_t &= \kappa_{t-1}+\theta+\omega_{t}, \quad \omega_t \overset{iid}{\sim} \text{N}(0,\sigma^2_\omega),
\end{align}
\end{subequations}
where $\boldsymbol{\alpha}=\alpha_{x_1:x_p}$, $\boldsymbol{\beta}=\beta_{x_1:x_p}$ and $\kappa_t$ is the latent state of the resulting linear Gaussian state-space model. Here $\Sigma$ is a $p$ by $p$ diagonal matrix with $\sigma^2_{\varepsilon,x_1:x_p}$ on the diagonal. We refer to this model as LC-H model, and the special case with $\sigma^2_{\varepsilon,x_i} = \sigma^2_\varepsilon$, $i \in \{1,\dots,p\}$, as LC model.

Instead of the identification constraint \eqref{eqn:constraints}, we suggest an alternative constraint which is simpler and more readily applicable to Monte Carlo based procedures such as MCMC and SMC. Our formulation of the identification constraints are given by setting
\begin{equation}\label{eqn:newConstraints}
    \alpha_{x_1} = \text{constant}, \quad
    \beta_{x_1} = \text{constant}.
\end{equation} Such a choice is a valid identification constraint since if one of the elements of each $\boldsymbol{\alpha}$ and $\boldsymbol{\beta}$ are known (here we have arbitrarily chosen $\alpha_{x_1}$ and $\beta_{x_1}$), then a non-trivial linear transformation in \eqref{eqn:LCtransform} is not allowed anymore; that is, we must have $c=0$ and $d=1$. Note that implementing the proposed constraint is straightforward in both maximum likelihood and Bayesian setting compared to the constraint \eqref{eqn:constraints}. %The choice of the constants in \eqref{eqn:newConstraints} will be discussed in Section~\ref{sec:Empirical}.

%%%%%%%%%%%%%%%%%%%%%%%%%%%%%%%%%%%%%%%%%%%%%%%%%%%%%%%%%%%%%%%%%%%%%%%%%%%%%%%%%%%%%%%%%%%%%%%%%%%
\subsection{Two Factor Lee-Carter Model with Age Based Heteroscedasticity: LC2-H model}
\label{sec:LC2hetero}
%%%%%%%%%%%%%%%%%%%%%%%%%%%%%%%%%%%%%%%%%%%%%%%%%%%%%%%%%%%%%%%%%%%%%%%%%%%%%%%%%%%%%%%%%%%%%%%%%%%
A natural extension of the LC-H model is to include a second stochastic factor for the cohort effect. We denote this model by LC2-H model. The cohort effect (\cite{RenshawHa06}) can be modeled under the state-space framework as follows
\begin{equation}\label{eqn:LCcohortObs}
    \begin{bmatrix}
        y_{x_1,t} \\
        y_{x_2,t} \\
        \vdots \\
        y_{x_p,t}
    \end{bmatrix}
    =
    \begin{bmatrix}
        \alpha_{x_1} \\
        \alpha_{x_2} \\
        \vdots \\
        \alpha_{x_p}
    \end{bmatrix}
    +
    \begin{bmatrix}
        \beta^{(1)}_{x_1} & \beta^{(2)}_{x_1} & 0 & \cdots & 0  \\
        \beta^{(1)}_{x_2} & 0 & \beta^{(2)}_{x_2} & \cdots & 0  \\
        \vdots & \vdots & \vdots & \ddots & \vdots \\
        \beta^{(1)}_{x_p} & 0 & 0 & \cdots & \beta^{(2)}_{x_p}
    \end{bmatrix}
    \begin{bmatrix}
        \kappa_t \\
        \zeta^{x_1}_t \\
        \zeta^{x_2}_t \\
        \vdots \\
        \zeta^{x_p}_t
    \end{bmatrix}
    +
    \begin{bmatrix}
        \varepsilon_{x_1,t} \\
        \varepsilon_{x_2,t} \\
        \vdots \\
        \varepsilon_{x_p,t}
    \end{bmatrix},
\end{equation}
where $\zeta^x_t := \zeta_{t-x}$. The state equation can be expressed as
\begin{equation}\label{eqn:LCcohortState}
    \begin{bmatrix}
        \kappa_{t} \\
        \zeta^{x_1}_{t} \\
        \zeta^{x_2}_{t} \\
        \vdots \\
        \zeta^{x_{p-1}}_{t}\\
        \zeta^{x_p}_{t}
    \end{bmatrix}
    =
    \begin{bmatrix}
        1 & 0 & 0 & \cdots & 0 & 0  \\
        0 & \vartheta & 0 & \cdots & 0 & 0  \\
        0 & 1 & 0 &  \cdots & 0 & 0 \\
        0 & 0 & 1 &\cdots & 0 & 0 \\
        \vdots & \vdots & \vdots & \ddots & \vdots & \vdots \\
        0 & 0 & 0 & \cdots & 1 & 0
    \end{bmatrix}
    \begin{bmatrix}
        \kappa_{t-1} \\
        \zeta^{x_1}_{t-1} \\
        \zeta^{x_2}_{t-1} \\
        \vdots \\
        \zeta^{x_{p-1}}_{t-1}\\
        \zeta^{x_p}_{t-1}
    \end{bmatrix}
    +
    \begin{bmatrix}
        \theta \\
        0 \\
        0 \\
        \vdots \\
        0 \\
        0
    \end{bmatrix}
    +
    \begin{bmatrix}
        \omega^\kappa_t \\
        \omega^{\zeta_1}_t \\
        0 \\
        \vdots \\
        0 \\
        0
    \end{bmatrix}.
\end{equation}
Here we assume $\kappa_t$ is a random walk with drift process and an AR(1) process is assumed for the cohort effect, that is $\zeta^{x_1}_t = \vartheta \zeta^{x_1}_{t-1}+\omega^{\zeta_1}_t$, where $|\vartheta|<1$. Note that, from \eqref{eqn:LCcohortState}, we have $\zeta^{x_i}_t = \zeta^{x_{i-1}}_{t-1}$ for $i=2,\dots,p$, which is the defining property of the cohort effect and consequently we are only required to model the dynamics of $\zeta^{x_1}_t$. We can write the model \eqref{eqn:LCcohortObs} - \eqref{eqn:LCcohortState} in the following form
\begin{subequations}\label{eqn:LCH}
\begin{align}
    \boldsymbol{y}_t &= \boldsymbol{\alpha}+ B\left[\kappa_t,\bm{\zeta}_t\right]^\top+ \boldsymbol{\varepsilon}_t, \quad \boldsymbol{\varepsilon}_t \overset{iid}{\sim} \text{N}(0,\Sigma), \\
    \kappa_t &= \kappa_{t-1}+\theta+\omega^{\kappa}_{t}, \quad \omega^\kappa_t \overset{iid}{\sim} \text{N}(0,\sigma^2_{\omega^\kappa}),\\
\bm{\zeta}_t &= C\bm{\zeta}_{t-1} + D\bm{\omega}^\zeta_t, \quad \omega^{\zeta_1}_t  \overset{iid}{\sim} \text{N}(0,\sigma^2_{\omega^\zeta}),
\end{align}
\end{subequations}
where $B$ is the $p$ by $p+1$ matrix in \eqref{eqn:LCcohortObs}, $C$ is the corresponding $p$ by $p$ sub-matrix in \eqref{eqn:LCcohortState} and $D$ is a zero $p$ by $p$ matrix except for the $(1,1)$ element with value $1$.
%%%%%%%%%%%%%%%%%%%%%%%%%%%%%%%%%%%%%%%%%%%%%%%%%%%%%%%%%%%%%%%%%%%%%%%%%%%%%%%%%%%%%%%%%%%%%%%%%%%
\subsection{Three Factor Lee-Carter Model with Dynamic Time Based Heteroscedasticity: LC3-H2 model}
\label{sec:LC2hetero}
%%%%%%%%%%%%%%%%%%%%%%%%%%%%%%%%%%%%%%%%%%%%%%%%%%%%%%%%%%%%%%%%%%%%%%%%%%%%%%%%%%%%%%%%%%%%%%%%%%%
We can further extend the LC2-H model by adding a third factor for the dynamic of volatility in the observation vector over time. The state-space dynamics is given by:
\begin{subequations}\label{eqn:LCH}
\begin{align}
    \boldsymbol{y}_t &= \boldsymbol{\alpha}+ B\left[\kappa_t,\bm{\zeta}_t\right]^\top+ \sqrt{\gamma^y_t}\boldsymbol{\varepsilon}_t, \quad \boldsymbol{\varepsilon}_t \overset{iid}{\sim} \text{N}(0,\Sigma), \\
    \kappa_t &= \kappa_{t-1}+\theta+\omega^\kappa_{t}, \quad \omega^\kappa_t \overset{iid}{\sim} \text{N}(0,\sigma^2_{\omega^\kappa}),\\
\bm{\zeta}_t &= C\bm{\zeta}_{t-1} + D\bm{\omega}^\zeta_t, \quad \omega^{\zeta_1}_t  \overset{iid}{\sim} \text{N}(0,\sigma^2_{\omega^\zeta}),\\
\gamma^y_{t} &= a(b-\gamma^y_{t-1}) + \gamma^y_{t-1} + \sigma \sqrt{\gamma^y_{t-1}} \epsilon^{\gamma^y}_t , \quad \epsilon^{\gamma^y}_t \overset{iid}{\sim} \text{N}(0,1),
\end{align}
\end{subequations}
where $\gamma^y_t$ is a process obtained via an Euler discretization of a square Bessel process corresponding to the Cox-Ingersoll-Ross process given by
$$
d\gamma^y_t = a(b-\gamma^y_t)\, dt + \sigma\sqrt{\gamma^y_t}\, dW_t,
$$
where $2ab \geq \sigma^2$ to ensure $\gamma^y_t$ is strictly positive. Such a dynamic volatility factor can be used to explain time varying periods of heightened observation variance, which potentially occur in some populations over time. These may be attributed to disease, war, famine, environmental factors or shocks as well as changes in migration and immigration patterns that could influnce the volatility of the observed death counts in different age groups.

%%%%%%%%%%%%%%%%%%%%%%%%%%%%%%%%%%%%%%%%%%%%%%%%%%%%%%%%%%%%%%%%%%%%%%%%%%%%%%%%%%%%%%%%%%%%%%%%%%%
\subsection{Multi-Factor Model with Stochastic Volatility in the Latent Process: LCSV model}
\label{sec:SV}
%%%%%%%%%%%%%%%%%%%%%%%%%%%%%%%%%%%%%%%%%%%%%%%%%%%%%%%%%%%%%%%%%%%%%%%%%%%%%%%%%%%%%%%%%%%%%%%%%%%
A common assumption in mortality modelling is that the period effect is derived from a discretization of a random walk with drift process. Such a process may be sufficient for modelling simple dynamics, but can be insufficient if time varying periods of volatility are present in the time series. In fact, much of the literature focuses mainly on capturing the trend of the period effect $\kappa_t$ for the past several decades where mortality time series for many countries are reasonably smooth.

%As Figure~\ref{fig:timeSeriesDEN} reveals, Danish mortality rates exhibit significant volatility in some periods, mainly before year 1950. While mortality jumps can be included in the Lee-Carter methodology, see for example \cite{ChenCo09} and \cite{CoxLiPe10}, the implication of introducing stochastic volatility to the model is still yet to be investigated in the literature.

Here we extend the Lee-Carter framework to incorporate stochastic volatility in the latent process. As a result, the impact of epidemics, natural disasters, medical breakthrough or wars on the evolution of mortality can be taken into account. This will produce different structural effects on the calibration and importantly on the forecasting when compared to the previously developed model of LC3-H2.  We refer \eqref{eqn:LCSVa}-\eqref{eqn:LCSVc} as Lee-Carter stochastic volatility model which we denote as LCSV model:
\begin{subequations}\label{eqn:LCSV}
\begin{align}
    \boldsymbol{y}_t &= \boldsymbol{\alpha} + \boldsymbol{\beta} \kappa_t + \boldsymbol{\varepsilon}_t, \quad \boldsymbol{\varepsilon}_t
    \overset{iid}{\sim} \text{N}(\boldsymbol{0},\sigma^2_\varepsilon\boldsymbol{1}_p), \label{eqn:LCSVa} \\
    \kappa_t &= \kappa_{t-1} + \theta + \omega_t, \quad \omega_t|\gamma_t \sim
    \text{N}(0,\exp\{\gamma_t\}), \label{eqn:LCSVb} \\
    \gamma_t &= \lambda_1 \gamma_{t-1} + \lambda_2 + \eta_t, \quad \eta_t \overset{iid}{\sim}
    \text{N}(0,\sigma^2_\gamma). \label{eqn:LCSVc}
\end{align}
\end{subequations}
The log-volatility process $\gamma_t$ is introduced in the state equation for $\kappa_t$ via the error term $\omega_t$. The process $\gamma_t$ is an autoregressive model of order 1 (AR(1)) with $|\lambda_1|<1$ and the mean reverting level is given by $\lambda_2/(1-\lambda_1)$. A heteroscedasticity structure can be introduced in \eqref{eqn:LCSVa} and will be referred to as LCSV-H model. Cohort effect can also be incorporated as follows:
\begin{subequations}\label{eqn:FPS3_SV}
\begin{align}
    \boldsymbol{y}_t &= \boldsymbol{\alpha} + B\left[\kappa_t,\bm{\zeta}_t\right]^\top + \boldsymbol{\varepsilon}_t, \quad \boldsymbol{\varepsilon}_t
    \overset{iid}{\sim} \text{N}(\boldsymbol{0},\sigma^2_\varepsilon\boldsymbol{1}_p),  \\
    \kappa_t &= \kappa_{t-1} + \theta + \omega_t, \quad \omega_t|\gamma_t \sim
    \text{N}(0,\exp\{\gamma_t\}), \\
    \gamma_t &= \lambda_1 \gamma_{t-1} + \lambda_2 + \eta_t, \quad \eta_t \overset{iid}{\sim}
    \text{N}(0,\sigma^2_\gamma), \\
\bm{\zeta}_t &= C\bm{\zeta}_{t-1} + D\bm{\omega}^\zeta_t, \quad \omega^{\zeta_1}_t  \overset{iid}{\sim} \text{N}(0,\sigma^2_{\omega^\zeta}).
\end{align}
\end{subequations}
Compared to the LC3-H2 model where stochastic volatility is included in the observation noise term, introducing stochastic volatility in the latent period process has the advantage, in terms of simplicity and ease of interpretation, that the variability of mortality data in the time dimension is captured purely by the latent process.
%We assume a homogenous error structure in the observation equation ~\eqref{eqn:LCSVa} for simplicity in this section.\footnote{A Lee-Carter stochastic volatility model with heteroscedasticity structure is considered in Sec.~\ref{sec:Empirical}} Since a stochastic process is only introduced in the variance term, no more additional identification constraint is required. %Similarly as before, we fix $\alpha_{x_1}$ and $\beta_{x_1}$ to be known constants as an identification constraint.

%Considering the dynamics of $\kappa_t$ and $\gamma_t$, \eqref{eqn:LCSVb}-\eqref{eqn:LCSVc}, our task is to filter out $\gamma_t$ given $\kappa_{1:t}$. Since the unconditional distribution of $\omega_t$ is not Gaussian (it is Gaussian if $\gamma_t$ is given). The proposed model is thus a non-Gaussian state-space model and numerical approximation is required for the filtering of the log-volatility process $\gamma_t$.

%%%%%%%%%%%%%%%%%%%%%%%%%%%%%%%%%%%%%%%%%%%%%%%%%%%%%%%%%%%%%%%%%%%%%%%%%%%%%%%%%%%%%%%%%%%%%%%%%%%
%%%%%%%%%%%%%%%%%%%%%%%%%%%%%%%%%%%%%%%%%%%%%%%%%%%%%%%%%%%%%%%%%%%%%%%%%%%%%%%%%%%%%%%%%%%%%%%%%%%
\section{Frequentist State-Space Inference}
%%%%%%%%%%%%%%%%%%%%%%%%%%%%%%%%%%%%%%%%%%%%%%%%%%%%%%%%%%%%%%%%%%%%%%%%%%%%%%%%%%%%%%%%%%%%%%%%%%%
%%%%%%%%%%%%%%%%%%%%%%%%%%%%%%%%%%%%%%%%%%%%%%%%%%%%%%%%%%%%%%%%%%%%%%%%%%%%%%%%%%%%%%%%%%%%%%%%%%%

Given these different state-space model structures, the next task to consider is the inference for the joint single stage state and parameter estimation. In this section we consider full likelihood based joint inference procedures based on filtering and gradient estimation. To achieve this we must describe both filtering in linear Gaussian and non-linear / non-Gaussian filtering via SMC method (particle filters) and their application to gradient based estimation in the marginal likelihood, having integrated out the latent state processes. We will do this in a general way and then present particular examples of relevance to this paper.

Under the classical maximum likelihood approach, parameters are estimated by maximizing a model's log-likelihood function. In the case of state-space models in the form of \eqref{eqn:generalSSMstate}-\eqref{eqn:generalSSMmeasurement}, the likelihood is in two forms: the complete data likelihood, assuming $\phi_0$ fixed, is given by
\begin{equation}
p_{\bm{\psi}}\left(\bm{\phi}_{1:T},\bm{z}_{1:T}\right) = \prod_{t=1}^T p_{\bm{\psi}}\left(\bm{z}_{t}|\phi_{t}\right) p_{\bm{\psi}}\left(\bm{\phi}_{t}|\bm{\phi}_{t-1}\right),
\end{equation}
and the marginal likelihood, typically used for the static model based inference, is given by
\begin{equation}\label{eqn:marginalLikelihood}
p_{\bm{\psi}}\left(\bm{z}_{1:T}\right) = \int \prod_{t=1}^T p_{\bm{\psi}}\left(\bm{z}_{t}|\bm{\phi}_{t}\right) p_{\bm{\psi}}\left(\bm{\phi}_{t}|\bm{\phi}_{t-1}\right)\, d\bm{\phi}_t,
\end{equation}
where $\bm{\psi}$ denotes the $d$-dimensional parameter vector of the model. Two challenges now arise. The first is that typically the integral in \eqref{eqn:marginalLikelihood} cannot be evaluated in closed form, except for linear Gaussian state-space model systems. The second issue is that the gradient equations for such a state-space model marginal likelihood, even if they can be calculated in closed form, requires a non-linear multiple equation solver. For this reason it is common to adopt a solution based on a recursive estimation using gradient and Hessian information from the marginal likelihood. Under the gradient-based approach, the optimal parameter vector can be found by iterations, where the $(m+1)$-th estimate is obtained by:
\begin{equation}
    \boldsymbol{\psi}^{(m+1)} = \boldsymbol{\psi}^{(m)} - \left[\nabla^2_{\bm{\psi}}\,\ell(\boldsymbol{\psi}^{(m)})\right]^{-1}\nabla_{\bm{\psi}}\,\ell\left(\boldsymbol{\psi}^{(m)}\right),
\end{equation}
where $\ell(\bm{\psi})$, $\nabla_{\bm{\psi}}\ell(\boldsymbol{\psi})$ and $-\nabla^2_{\bm{\psi}}\,\ell(\boldsymbol{\psi})$ denote the log-likelihood function, the gradient (or score) vector and the Hessian information matrix of the log-likelihood function respectively, defined with respect to grad and Laplacian differential operators given by:
\begin{equation}
\begin{split}
\left[\nabla_{\bm{\psi}}\right]_{i} &:= {\partial \over \partial \psi_i}, \;\; \forall i \in \{1,\ldots,n\} \\
\left[\nabla^2_{\bm{\psi}}\right]_{i,j} &:= \frac{\partial^2}{\partial \psi_i \partial \psi_j}, \;\; \forall i,j \in \left\{1,\ldots,n\right\}.
\end{split}
\end{equation}
The iterating scheme will stop once certain criterion is met, for example when the magnitude of the score vector is small enough. This will be illustrated using the LC-H model as an example in Section \ref{subsec:MLE}.

The result developed are based on the marginal likelihood of the state-space model, with generic static model parameters $\bm{\psi}$ for observations $\bm{z}_{1:T} = \bm{z}_{1},\ldots,\bm{z}_T$ having integrated out latent states $\boldsymbol{\phi}_1,\ldots,\boldsymbol{\phi}_T$, denoted by $p_{\bm{\psi}}\left(\bm{z}_{1:T}\right)$. We are then interested in forming recursive filtering to integrate the complete data likelihood to find the marginalized likelihood and then working with recursive gradient based estimation to update static model parameters in Newton-Descent type algorithm, or for linear Gaussian systems a recursive least squares based approach.

As observed in \cite{poyiadjis2005maximum} and \cite{poyiadjis2011particle}, it is useful to consider two classes of identities for the gradient and Hessian of the marginalized likelihood, given by the Fisher's identity and the Louis' identity, respectively according to
\begin{equation}
\begin{split}
\nabla_{\bm{\psi}}p_{\bm{\psi}}\left(\bm{z}_{1:T}\right)  &= \int \nabla_{\bm{\psi}} \ln p_{\bm{\psi}}\left(\boldsymbol{\phi}_{1:T},\bm{z}_{1:T}\right) p_{\bm{\psi}}\left(\left. \boldsymbol{\phi}_{1:T}\right|\bm{z}_{1:T}\right)\, d\boldsymbol{\phi}_{1:T},\\
-\nabla^2_{\bm{\psi}}p_{\bm{\psi}}\left(\bm{z}_{1:T}\right) &= \nabla_{\bm{\psi}} \ln p_{\bm{\psi}}\left(\bm{z}_{1:T}\right) \nabla_{\bm{\psi}} \ln p_{\bm{\psi}}\left(\bm{z}_{1:T}\right)^\top - \frac{\nabla_{\bm{\psi}}^2\ln p_{\bm{\psi}}\left(\bm{z}_{1:T}\right)}{p_{\bm{\psi}}\left(\bm{z}_{1:T}\right)},  \\
\end{split}
\end{equation}
where
\begin{equation}
\begin{split}
\frac{\nabla_{\bm{\psi}}^2 p_{\bm{\psi}}\left(\bm{z}_{1:T}\right)}{p_{\bm{\psi}}\left(\bm{z}_{1:T}\right)} &= \int \nabla_{\bm{\psi}} \ln p_{\bm{\psi}}\left(\bm{\phi}_{1:T},\bm{z}_{1:T}\right)\nabla_{\bm{\psi}} \ln p_{\bm{\psi}}\left(\bm{\phi}_{1:T},\bm{z}_{1:T}\right)^\top p_{\bm{\psi}}\left(\bm{\phi}_{1:T}|\bm{z}_{1:T}\right)d\bm{\phi}_{1:T}\\
& \;\; + \int \nabla^2_{\bm{\psi}} \ln p_{\bm{\psi}}\left(\bm{\phi}_{1:T},\bm{z}_{1:T}\right)  p_{\bm{\psi}}\left(\bm{\phi}_{1:T}|\bm{z}_{1:T}\right)d\bm{\phi}_{1:T}.
\end{split}
\end{equation}
An important point about these recursions is that the integrals for the gradient vector and Hessian matrix are expressed in terms of the path-space distribution $p_{\bm{\psi}}\left(\bm{\phi}_{1:T}|\bm{z}_{1:T}\right)$. In the case of the linear Gaussian dynamics this distribution can be obtained based on variations of the Kalman filter recursion, however when the state-space model is non-linear or non-Gaussian this distribution must be estimated via Monte Carlo methods. The most efficient of these methods for state-space modelling purposes is known as the class of SMC methods (particle filters). In this case it will be more accurate from the perspective of the variance of the estimated gradient and Hessian matrices to utilise the filter distribution based estimators in a recursive fashion based on the local estimates of the distributions $p_{\bm{\psi}}\left(\bm{\phi}_{t}|\bm{z}_{1:t}\right)$, for each $t\in \left\{1,\ldots,T\right\}$ rather than the path space estimator which is based on distribution $p_{\bm{\psi}}\left(\bm{\phi}_{1:T}|\bm{z}_{1:T}\right)$ at the final time $T$. The result explaining the difference in estimation precision for the gradient and Hessian, from the perspective of variance of the solution on the path space distribution versus filter distributions is provided in Theorem 1 of \cite{poyiadjis2011particle}. This motivates the need to work with the filter recursions.

To achieve this one can replace in the Fisher and Louis' identities the path-space quantities $p_{\bm{\psi}}\left(\bm{\phi}_{1:T},\bm{z}_{1:T}\right)$ and $p_{\bm{\psi}}\left(\bm{\phi}_{1:T}|\bm{z}_{1:T}\right)$ by the filter quantities given by $p_{\bm{\psi}}\left(\bm{\phi}_{t},\bm{z}_{1:t}\right)$ and $p_{\bm{\psi}}\left(\bm{\phi}_{t}|\bm{z}_{1:t}\right)$. After this substitution, one may utilise the following recursive formulations to evaluate the gradient and Hessian, see \cite{poyiadjis2005maximum} and \cite{poyiadjis2011particle}. In this case the Fisher identity is recursively given by:
\begin{equation*}
\begin{split}
\nabla_{\bm{\psi}}\ln p_{\bm{\psi}}\left(\bm{z}_{1:t}\right) &= \int \nabla_{\bm{\psi}}\ln p_{\bm{\psi}}\left(\boldsymbol{\phi}_{t}, \bm{z}_{1:t}\right) p_{\bm{\psi}}\left(\boldsymbol{\phi}_{t}| \bm{z}_{1:t}\right) \,d \boldsymbol{\phi}_{t},\\
\nabla_{\bm{\psi}} \ln p_{\bm{\psi}}\left(\bm{\phi}_t,\bm{z}_{1:t}\right)  &= \frac{p_{\bm{\psi}}\left(\bm{z}_{1:t-1}\right)p_{\bm{\psi}}\left(\bm{z}_{t}|\bm{\phi}_t\right)}{p_{\bm{\psi}}\left(\bm{\phi}_t,\bm{z}_{1:t}\right)} \int p_{\bm{\psi}}\left(\bm{\phi}_t|\bm{\phi}_{t-1}\right)p_{\bm{\psi}}\left(\bm{\phi}_{t-1}|\bm{z}_{1:t-1}\right) \\
&\times \left[\nabla_{\bm{\psi}}\ln p_{\bm{\psi}}\left(\bm{z}_{t}|\bm{\phi}_t\right) + \nabla_{\bm{\psi}}\ln p_{\bm{\psi}}\left(\bm{\phi}_t|\bm{\phi}_{t-1}\right) + \nabla_{\bm{\psi}}\ln p_{\bm{\psi}}\left(\bm{\phi}_{t-1},\bm{z}_{1:t-1}\right)\right]d\bm{\phi}_{t-1},\\
p_{\bm{\psi}}\left(\bm{\phi}_t,\bm{z}_{1:t}\right) &= p_{\bm{\psi}}\left(\bm{z}_{1:t-1}\right)p_{\bm{\psi}}\left(\bm{z}_{t}|\bm{\phi}_t\right)\int p_{\bm{\psi}}\left(\bm{\phi}_t|\bm{\phi}_{t-1}\right) p_{\bm{\psi}}\left(\bm{\phi}_{t-1}|\bm{z}_{1:t-1}\right) \, d\bm{\phi}_{t-1}.
\end{split}
\end{equation*}

The recursive form of Luis' identity is given by:
\begin{equation*}
\begin{split}
\frac{\nabla^2_{\bm{\psi}} p_{\bm{\psi}}\left(\bm{z}_{1:t}\right)}{p_{\bm{\psi}}\left(\bm{\phi}_t,\bm{z}_{1:t}\right)} &= \int \nabla_{\bm{\psi}} \ln p_{\bm{\psi}}\left(\bm{\phi}_t,\bm{z}_{1:t}\right) \nabla_{\bm{\psi}} \ln p_{\bm{\psi}}\left(\bm{\phi}_t,\bm{z}_{1:t}\right)^\top p_{\bm{\psi}}\left(\bm{\phi}_t|\bm{z}_{1:t}\right)\, d\bm{\phi}_t\\
& + \int \nabla^2_{\bm{\psi}} \ln p_{\bm{\psi}}\left(\bm{\phi}_t,\bm{z}_{1:t}\right) p_{\bm{\psi}}\left(\bm{\phi}_t|\bm{z}_{1:t}\right)\, d\bm{\phi}_t,\\
\nabla^2_{\bm{\psi}} \ln p_{\bm{\psi}}\left(\bm{\phi}_t,\bm{z}_{1:t}\right) &= \frac{\nabla^2_{\bm{\psi}} p_{\bm{\psi}}\left(\bm{\phi}_t,\bm{z}_{1:t}\right)}{p_{\bm{\psi}}\left(\bm{\phi}_t,\bm{z}_{1:t}\right)} - \nabla_{\bm{\psi}} \ln p_{\bm{\psi}}\left(\bm{\phi}_t,\bm{z}_{1:t}\right) \nabla_{\bm{\psi}} \ln p_{\bm{\psi}}\left(\bm{\phi}_t,\bm{z}_{1:t}\right)^\top,\\
\nabla^2_{\bm{\psi}} p_{\bm{\psi}}\left(\bm{\phi}_t,\bm{z}_{1:t}\right) &= p_{\bm{\psi}}\left(\bm{z}_{1:t-1}\right)p_{\bm{\psi}}\left(\bm{z}_{t}|\bm{\phi}_t\right) \int p_{\bm{\psi}}\left(\bm{\phi}_t|\bm{\phi}_{t-1}\right)p_{\bm{\psi}}\left(\bm{\phi}_{t-1}|\bm{z}_{1:t-1}\right) \\
&\times \left\{\left[\nabla_{\bm{\psi}}\ln p_{\bm{\psi}}\left(\bm{z}_{t}|\bm{\phi}_t\right) + \nabla_{\bm{\psi}}\ln p_{\bm{\psi}}\left(\bm{\phi}_t|\bm{\phi}_{t-1}\right) + \nabla_{\bm{\psi}}\ln p_{\bm{\psi}}\left(\bm{\phi}_{t-1},\bm{z}_{1:t-1}\right)\right]\right.\\
&\times \left[\nabla_{\bm{\psi}}\ln p_{\bm{\psi}}\left(\bm{z}_{t}|\bm{\phi}_t\right) + \nabla_{\bm{\psi}}\ln p_{\bm{\psi}}\left(\bm{\phi}_t|\bm{\phi}_{t-1}\right) + \nabla_{\bm{\psi}}\ln p_{\bm{\psi}}\left(\bm{\phi}_{t-1},\bm{z}_{1:t-1}\right)\right]^\top\\
&\left.  + \left[\nabla^2_{\bm{\psi}}\ln p_{\bm{\psi}}\left(\bm{z}_{t}|\bm{\phi}_t\right) + \nabla^2_{\bm{\psi}}\ln p_{\bm{\psi}}\left(\bm{\phi}_t|\bm{\phi}_{t-1}\right) + \nabla^2_{\bm{\psi}}\ln p_{\bm{\psi}}\left(\bm{\phi}_{t-1},\bm{z}_{1:t-1}\right)\right]
\right\} d\bm{\phi}_{t-1}.
\end{split}
\end{equation*}
In general, the solution to these recursions can be achieved via SMC as detailed in \cite{poyiadjis2005maximum} and \cite{poyiadjis2011particle}.

In the following sections we will illustrate the use of these recursive identities for the special case of the LC-H model where the state-space takes a linear Gaussian form. In this case the integrals and recursive evaluation of the gradient and Hessian can be written in closed form. To proceed we first introduce the optimal filter recursion, with respect to minimization of mean squared error, in the case of linear Gaussian state-space models, known as the Kalman filter (\cite{kalman1960Paper} and \cite{Harvey89}).

%%%%%%%%%%%%%%%%%%%%%%%%%%%%%%%%%%%%%%%%%%%%%%%%%%%%%%%%%%
\subsection{Closed Form Filter Recursions for LC-H Model} \label{subsec:KF}
%%%%%%%%%%%%%%%%%%%%%%%%%%%%%%%%%%%%%%%%%%%%%%%%%%%%%%%%%%
The aim of filtering is to obtain the distribution of the latest state given observations. For a general state-space model, \eqref{eqn:generalSSMstate}-\eqref{eqn:generalSSMmeasurement}, the filtering density $\pi(\bm{\phi}_t|\bm{z}_{1:t})$ at time $t$ can be calculated sequentially by first assuming the filtering density $\pi(\bm{\phi}_{t-1}|\bm{z}_{1:t-1})$ at time $t-1$ is known. Then the one-step ahead predictive density for the state is given by
\begin{equation}\label{eqn:predictState}
    \pi(\boldsymbol{\phi}_t|\boldsymbol{z}_{1:t-1}) = \int\pi(\boldsymbol{\phi}_t|\boldsymbol{\phi}_{t-1})\pi(\boldsymbol{\phi}_{t-1}|\boldsymbol{z}_{1:t-1})d\boldsymbol{\phi}_{t-1}.
\end{equation}
From Bayes' Formula and the structure of the conditional dependency of the state-space model, one can obtain the filtering density as
\begin{equation}\label{eqn:filteringKF}
    \pi(\boldsymbol{\phi}_t|\boldsymbol{z}_{1:t})=\frac{\pi(\boldsymbol{\phi}_t|\boldsymbol{z}_{1:t-1})
    \pi(\boldsymbol{z}_t|\boldsymbol{\phi}_t)}{\int\pi(\boldsymbol{\phi}_t|\boldsymbol{z}_{1:t-1})\pi(\boldsymbol{z}_t|\boldsymbol{\phi}_t)d\boldsymbol{\phi}_t}.
\end{equation}
For nonlinear and non-Gaussian state-space models, numerical techniques such as SMC methods are required to estimate the filtering density, see \cite{DoucetFrGo01}, \cite{DoucetGoAn00} and \cite{Liu08}.

In the case of the LC-H model, since it is a linear and Gaussian state-space model, the filtering distribution can be obtained analytically via Kalman filtering. In particular we can find the conditional distributions of the key quantities in the filtering recursions are all Gaussian distributions as follows:
\begin{subequations}
    \begin{align}
        \kappa_{t-1}|\boldsymbol{y}_{1:t-1} &\sim \text{N}(m_{t-1},C_{t-1}), \label{filterdisttminus1} \\
        \kappa_t|\boldsymbol{y}_{1:t-1} &\sim \text{N}(a_t, R_t), \\
        \boldsymbol{y}_t|\boldsymbol{y}_{1:t-1} &\sim \text{N}(\boldsymbol{f}_t,\boldsymbol{Q}_t), \label{eqn:KFpredictY} \\
        \kappa_t|\boldsymbol{y}_{1:t} &\sim \text{N}(m_t,C_t) \label{eqn:filterdistt}
    \end{align}
\end{subequations}
where the recursive nature of these distributions arises from the recursions of the sufficient statistics:
\begin{subequations}
    \begin{align}
        &a_t=m_{t-1}+\theta, \quad R_t=C_{t-1}+\sigma^2_\omega, \label{eqn:kf1} \\
        &\boldsymbol{f}_t=\boldsymbol{\alpha}+\boldsymbol{\beta} a_t, \quad \boldsymbol{Q}_t=\boldsymbol{\beta}\boldsymbol{\beta}^\top R_t+\Sigma, \\
        &m_t=a_t+R_t\boldsymbol{\beta}^\top\boldsymbol{Q}_t^{-1}(\boldsymbol{y}_t-\boldsymbol{f}_t), \quad
        C_t=R_t-R_t\boldsymbol{\beta}^\top\boldsymbol{Q}_t^{-1}\boldsymbol{\beta}R_t. \label{eqn:kf4}
    \end{align}
\end{subequations}
That is, given the filtering distribution at $t-1$, \eqref{filterdisttminus1}, the filtering distribution at $t$ is given by \eqref{eqn:filterdistt} using \eqref{eqn:kf1}-\eqref{eqn:kf4}.

%%%%%%%%%%%%%%%%%%%%%%%%%%%%%%%%%%%%%%%%%%%%%%%%%%%%%%%%%%
\subsection{Closed form Gradient-Based Estimation via Score and Hessian Recursions for LC-H Model} \label{subsec:MLE}
%%%%%%%%%%%%%%%%%%%%%%%%%%%%%%%%%%%%%%%%%%%%%%%%%%%%%%%%%%

For the LC-H model, the log-likelihood function $\ell(\boldsymbol{\psi}) := \ln \pi(\boldsymbol{y}_{1:T}|\boldsymbol{\psi})$ is given by
\begin{equation}\label{eqn:logLikelihood}
    \ell(\boldsymbol{\psi}) =\ln\left(\prod^T_{t=1} \pi(\boldsymbol{y}_t|\boldsymbol{y}_{1:t-1},\boldsymbol{\psi})\right) =
    -\frac{pT}{2}\ln{2\pi}-\frac{1}{2}\sum^T_{t=1}\left(\ln{|\boldsymbol{Q}_t|}+\boldsymbol{v}^\top_t\boldsymbol{Q}^{-1}_t\boldsymbol{v}_t\right),
\end{equation}
where $\boldsymbol{v}_t := \boldsymbol{y}_t-\boldsymbol{f}_t$, and $\boldsymbol{\psi} =(\alpha_{x_2:x_p},\beta_{x_2:x_p},\theta,\sigma^2_{\varepsilon,x_1:x_p},\sigma^2_\omega)$ is an $n$-dimensional parameter vector. The log-likelihood function \eqref{eqn:logLikelihood} can be derived directly from \eqref{eqn:KFpredictY}.

It can be shown that (\cite{Harvey89}) the elements of the score vector and the information matrix are given in closed form for the LC-H model according to the expressions:
\begin{equation}\label{eqn:score}
    \frac{\partial \ell}{\partial \psi_i} =
    \frac{1}{2}\sum^T_{t=1}\left\{\text{tr}\left[\left(\boldsymbol{Q}^{-1}_t\frac{\partial
    \boldsymbol{Q}_t}{\partial
    \psi_i}\right)(\boldsymbol{1}_p-\boldsymbol{Q}^{-1}_t\boldsymbol{v}_t\boldsymbol{v}^\top_t)\right]+2\frac{\partial
    \boldsymbol{v}^\top_t}{\partial \psi_i}\boldsymbol{Q}^{-1}_t\boldsymbol{v}_t\right\}, \quad i=1,\dots,n
\end{equation}
where $\text{tr}[\cdot]$ denotes the trace operator and
\begin{equation}\label{eqn:infomatrix}
    -\text{E}\left[\frac{\partial^2\ell}{\partial \psi_i \partial \psi_j}\right]=
    \frac{1}{2}\sum^T_{t=1}\left[\text{tr}\left(\boldsymbol{Q}^{-1}_t\frac{\partial
    \boldsymbol{Q}_t}{\partial \psi_i}\boldsymbol{Q}^{-1}_t\frac{\partial \boldsymbol{Q}_t}{\partial
    \psi_j}\right)\right] + \text{E}\left[\sum^T_{t=1}\frac{\partial
    \boldsymbol{v}^\top_t}{\partial \psi_i}\boldsymbol{Q}^{-1}_t\frac{\partial \boldsymbol{v}_t}{\partial
    \psi_j}\right], \quad i,j=1,\dots,n
\end{equation}
and the expectation operator $\text{E}[\cdot]$ on the
second term in \eqref{eqn:infomatrix} can be dropped (since the expressions are asymptotically equivalent). In order to evaluate the score vector and the information matrix, we need
\begin{equation}\label{eqn:dv}
    \frac{\partial \boldsymbol{v}_t}{\partial \psi_i} =
    -\frac{\partial \boldsymbol{\alpha}}{\partial \psi_i}-\frac{\partial \boldsymbol{\beta}}{\partial
    \psi_i}a_t-\boldsymbol{\beta} \frac{\partial a_t}{\partial
    \psi_i}
\end{equation}
and
\begin{equation}\label{eqn:dF}
    \frac{\partial \boldsymbol{Q}_t}{\partial \psi_i}=\frac{\partial
    \boldsymbol{\beta}}{\psi_i}R_t\boldsymbol{\beta}^\top+\boldsymbol{\beta}\frac{\partial
    R_t}{\partial \psi_i}\boldsymbol{\beta}^\top+\boldsymbol{\beta} R_t\frac{\partial
    \boldsymbol{\beta}^\top}{\partial \psi_i}+\frac{\partial
    \Sigma}{\partial \psi_i}.
\end{equation}
The expressions \eqref{eqn:dv} and \eqref{eqn:dF} require, for $t=1,\dots,T$ and
$i=1,\dots,n$,
\begin{equation}\label{eqn:dap}
    \frac{\partial a_t}{\partial \psi_i}=\frac{\partial
    m_{t-1}}{\partial \psi_i}+\frac{\partial \theta}{\partial
    \psi_i} \quad \text{and} \quad     \frac{\partial R_t}{\partial \psi_i}=\frac{\partial
    C_{t-1}}{\partial \psi_i}+\frac{\partial \sigma^2_\omega}{\partial
    \psi_i}.
\end{equation}
%and
%\begin{equation}\label{eqn:dPp}
%    \frac{\partial R_t}{\partial \psi_i}=\frac{\partial
%    C_{t-1}}{\partial \psi_i}+\frac{\partial \sigma^2_\omega}{\partial
%    \psi_i}.
%\end{equation}
The expressions in \eqref{eqn:dap} in turn require, for
$t=1,\dots,T-1$ and $i=1,\dots,n$,
\begin{align}
    \frac{\partial m_t}{\partial \psi_i}&=\frac{\partial
    a_t}{\partial \psi_i}+\frac{\partial R_t}{\partial
    \psi_i}\boldsymbol{\beta}^\top \boldsymbol{Q}^{-1}_t \boldsymbol{v}_t + R_t\frac{\partial
    \boldsymbol{\beta}}{\partial \psi_i} \boldsymbol{Q}^{-1}_t \boldsymbol{v}_t - R_t\boldsymbol{\beta}^\top
    \boldsymbol{Q}^{-1}_t\frac{\partial \boldsymbol{Q}_t}{\partial \psi_i}\boldsymbol{Q}^{-1}_t \boldsymbol{v}_t +
    R_t\boldsymbol{\beta}^\top \boldsymbol{Q}^{-1}_t\frac{\partial \boldsymbol{v}_t}{\partial \psi_i}
\end{align}
and
\begin{align}
    \frac{\partial C_t}{\partial \psi_i}&=\frac{\partial
    R_t}{\partial \psi_i}-\frac{\partial
    R_t}{\partial \psi_i}\boldsymbol{\beta}^\top \boldsymbol{Q}^{-1}_t\boldsymbol{\beta} R_t-
    R_t\frac{\partial \boldsymbol{\beta}}{\partial \psi_i}\boldsymbol{Q}^{-1}_t\boldsymbol{\beta}
    R_t \notag \\
    &+ R_t\boldsymbol{\beta}^\top \boldsymbol{Q}^{-1}_t \frac{\partial \boldsymbol{Q}_t}{\partial
    \psi_i} \boldsymbol{Q}^{-1}_t\boldsymbol{\beta} R_t-R_t\boldsymbol{\beta}^\top
    \boldsymbol{Q}^{-1}_t\frac{\partial \boldsymbol{\beta}}{\partial \psi_i}
    R_t - R_t \boldsymbol{\beta}^\top \boldsymbol{Q}^{-1}_t \boldsymbol{\beta} \frac{\partial
    R_t}{\partial \psi_i}.
\end{align}
Note that $\frac{\partial m_0}{\partial \psi_i} = \frac{\partial
C_0}{\partial \psi_i} = 0$ for $i=1,\dots,n$ and the required differentiation matrices $\frac{\partial \boldsymbol{\alpha}}{\partial \boldsymbol{\psi}_i}$, $\frac{\partial \boldsymbol{\beta}}{\partial \boldsymbol{\psi}_i}$, $\frac{\partial \boldsymbol{\Sigma}}{\partial \boldsymbol{\psi}_i}$, $\frac{\partial \boldsymbol{\theta}}{\partial \boldsymbol{\psi}_i}$ and $\frac{\partial \boldsymbol{\sigma^2_\omega}}{\partial \boldsymbol{\psi}_i}$ are displayed in Appendix~\ref{appA:Matrices}. The gradient-based estimation for the LC-H model is described in Algorithm \ref{MLAlgorithm}.

\begin{algorithm}[H]
\caption{Gradient-based approach for estimating parameters $\boldsymbol{\psi}$}
\label{MLAlgorithm}
\begin{algorithmic}[1]
\State{Initialise $\boldsymbol{\psi}=\boldsymbol{\psi}^{(0)}$}; specify $m_0$ and $C_0$.
\While{stopping criterion is not met}
    \State Count the number of iteration performed as $m$;
    \State Run Kalman filter using $\boldsymbol{\psi}^{(m)}$; obtain \eqref{eqn:dv} and \eqref{eqn:dF} for $t=1,\dots, T$;
    \State Evaluate the score vector $\nabla_{\boldsymbol{\psi}}\,\ell(\bm{\psi}^{(m)})$ using \eqref{eqn:score};
    \State Evaluate the information matrix $-\text{E}\left[\nabla^2_{\bm{\psi}} \, \ell(\bm{\psi}^{(m)})\right]$ given by \eqref{eqn:infomatrix};
    \State Set $\boldsymbol{\psi}^{(m+1)} = \boldsymbol{\psi}^{(m)} + \left[\text{E}[\nabla^2_{\bm{\psi}} \, \ell(\bm{\psi}^{(m)})]\right]^{-1}\nabla_{\boldsymbol{\psi}}\,\ell(\bm{\psi}^{(m)})$.
\EndWhile
\end{algorithmic}
\end{algorithm}

%Note that once the parameters are sufficiently converged so that the log-likelihood function reaches a (local) maximum, the mean and variance $m_t$ and $C_t$ of the filtering densities $\pi(\kappa_t|\boldsymbol{y}_{1:t})_{t=1,\dots,T}$ obtained by Kalman filtering at the last iteration can be used to form the filtering distribution.\footnote{If the last iteration is $\tilde{m}$, then the quantities $m_t$ and $C_t$ obtained are evaluated using parameters $\boldsymbol{\psi}^{(\tilde{m}-1)}$. We can allow $\tilde{m}$ to be sufficiently large so that $\boldsymbol{\psi}^{(\tilde{m}-1)}$ can also be considered as optimal parameters.}

%%%%%%%%%%%%%%%%%%%%%%%%%%%%%%%%%%%%%%%%%%%%%%%%%%%%%%%%%%%%%%%%%%%%%%%%%%%%%%%%%%%%%%%%%%%%%%%%%%%
%%%%%%%%%%%%%%%%%%%%%%%%%%%%%%%%%%%%%%%%%%%%%%%%%%%%%%%%%%%%%%%%%%%%%%%%%%%%%%%%%%%%%%%%%%%%%%%%%%%
\section{Bayesian State-Space Inference}
\label{subsec:Bayesian}
%%%%%%%%%%%%%%%%%%%%%%%%%%%%%%%%%%%%%%%%%%%%%%%%%%%%%%%%%%%%%%%%%%%%%%%%%%%%%%%%%%%%%%%%%%%%%%%%%%%
%%%%%%%%%%%%%%%%%%%%%%%%%%%%%%%%%%%%%%%%%%%%%%%%%%%%%%%%%%%%%%%%%%%%%%%%%%%%%%%%%%%%%%%%%%%%%%%%%%%
In contrast to the classical maximum likelihood approach where parameters are deterministic but unknown, in a Bayesian view the parameters are treated as random variables. In this way the Bayesian paradigm can take parameter uncertainty into account and incorporate in a consistent manner apriori beliefs on the important model parameters, as encoded through the prior.

In this section we aim to develop modern approaches to Bayesian inference for state-space modelling that do not rely on potentially inefficient sampling approaches based on Gibbs or Metropolis-within-Gibbs for the latent state process. Instead we will introduce to stochastic mortality modelling state-space models of two classes of Bayesian inference:
\begin{itemize}
\item \textbf{Linear Gaussian Stochastic Mortality Models:} a Rao-Blackwellised Gibbs sampler approach which is based on a combination of Metropolis-within-Gibbs and Gibbs sampling steps for the static model parameters, combined with a Forward-Backward Kalman filter recursion for the state process. We assume the proposed constraints \eqref{eqn:newConstraints} throughout, avoiding the constraint issue when performing MCMC as discussed in Section \ref{ClassicalBayes}.
\item \textbf{Non-Linear / Non-Gaussian Stochastic Mortality Models:} in the case of non-linear and or non-Gaussian state-space model dynamics such as the stochastic volatility models of LC3-H2 and the LCSV models, the sampler we develop is based on novel developments of the Particle Metropolis Hastings samplers of \cite{AndrieuDoHo10} adapted to the stochastic mortality models. In particular we consider a combination of Rao-Blackwellized Kalman filter and particle filter for the latent state process full posterior conditionals, combined with a combination of Metropolis-within-Gibbs and Gibbs sampling steps for the static model parameters.
\end{itemize}
In general under all the Bayesian approaches we consider, we aim to obtain the joint posterior density
\begin{equation}
    \pi(\kappa_{0:T},\boldsymbol{\psi}|\boldsymbol{y}_{1:T})
\end{equation}
of the states $\kappa_{0:T}$ as well as the parameters, $\boldsymbol{\psi}$, given the observations $\boldsymbol{y}_{1:T}$. We begin with the first case of the linear Gaussian state-space stochastic mortality models and we use the LC-H model as an example where the parameter vector is $\boldsymbol{\psi} :=(\alpha_{x_2:x_p},\beta_{x_2:x_p},\theta,\sigma^2_{\varepsilon,x_1:x_p},\sigma^2_\omega)$ as we use the constraint proposed in \eqref{eqn:newConstraints}.

%%%%%%%%%%%%%%%%%%%%%%%%%%%%%%%%%%%%%%%%%%%%%%%%%%%%%%%%%%%%%%%%%%%%%%%%%%%%%%%%%%%%%%%%%%%%%%%%%%%
\subsection{Linear Gaussian State-Space Inference}
%%%%%%%%%%%%%%%%%%%%%%%%%%%%%%%%%%%%%%%%%%%%%%%%%%%%%%%%%%%%%%%%%%%%%%%%%%%%%%%%%%%%%%%%%%%%%%%%%%%
We develop an efficient approach involving a combined Gibbs sampling conjugate model sampler for the marginal target distributions of the static model parameters along with a forward backward Kalman filter sampler for the latent process $\kappa_{1:T}$. A sample of the targeted density is obtained via Gibbs sampling where $M$ is the number of MCMC iterations (Algorithm \ref{GibbsAlgorithm}).

\begin{algorithm}[H]
\caption{Rao-Blackwellized Forward-Backward Kalman Filter and Gibbs sampling for $\pi(\kappa_{0:T},\boldsymbol{\psi}|\boldsymbol{y}_{1:T})$}
\label{GibbsAlgorithm}
\begin{algorithmic}[1]
\State{Initialise: $\boldsymbol{\psi}=\boldsymbol{\psi}^{(0)}$.}
\For{$i=1,\dots,M$}
    \State Sample $\kappa^{(i)}_{0:T}$ from
        $\pi(\kappa_{0:T}|\boldsymbol{\psi}^{(i-1)},\boldsymbol{y}_{1:T})$ via FFBS (Section \ref{sec:EstimationFirstDensity}).
    \For{$h=1,\dots,n$}
        \State Sample $\psi^{(i)}_h$ from $\pi(\psi_h|\kappa^{(i)}_{0:T},\boldsymbol{\psi}_{-h}^{(i)},\boldsymbol{y}_{1:T})$,
        \State where $\boldsymbol{\psi}_{-h}^{(i)}=(\psi^{(i)}_1,\dots,\psi^{(i)}_{h-1},\psi^{(i-1)}_{h+1},\dots,\psi^{(i-1)}_n)$.
    \EndFor
\EndFor
\end{algorithmic}
\end{algorithm}

The general block Gibbs sampling algorithm steps require to sample from the full conditional densities $\pi(\kappa_{0:T}|\boldsymbol{\psi},\boldsymbol{y}_{1:T})$ and $\pi(\boldsymbol{\psi}|\kappa_{0:T},\boldsymbol{y}_{1:T})$, which are shown in the following.

%%%%%%%%%%%%%%%%%%%%%%%%%%%%%%%%%%%%%%%%%%%%%%%%%%%%%%%%%%%%%%%%%%%%%%%%%%%%%%%%%%%%%%%%%%%%%%%%%%%
\subsubsection{Sampling from the full conditional density $\pi(\kappa_{0:T}|\boldsymbol{\psi},\boldsymbol{y}_{1:T})$} \label{sec:EstimationFirstDensity}
%%%%%%%%%%%%%%%%%%%%%%%%%%%%%%%%%%%%%%%%%%%%%%%%%%%%%%%%%%%%%%%%%%%%%%%%%%%%%%%%%%%%%%%%%%%%%%%%%%%
Samples of the full conditional density $\pi(\kappa_{0:T}|\boldsymbol{\psi},\boldsymbol{y}_{1:T})$ can be obtained via the so-called forward-filtering-backward sampling (FFBS) procedure (\cite{CarterKo94}). We can write
\begin{equation}\label{eqn:ffbs}
    \pi(\kappa_{0:T}|\boldsymbol{\psi},\boldsymbol{y}_{1:T})=\prod^T_{t=0}\pi(\kappa_t|\kappa_{t+1:T},\boldsymbol{\psi},\boldsymbol{y}_{1:T})=
    \prod^T_{t=0}\pi(\kappa_t|\kappa_{t+1},\boldsymbol{\psi},\boldsymbol{y}_{1:t}),
\end{equation}
where the last term in the product, $\pi(\kappa_T|\boldsymbol{\psi},\boldsymbol{y}_{1:T})$, is distributed as $\text{N}(m_T,C_T)$ which is obtained from the last iteration of the Kalman filtering procedure.

Once we draw a sample $\kappa_T$ from $\text{N}(m_T,C_T)$, then \eqref{eqn:ffbs} suggests that we can draw recursively and backwardly $\kappa_t$ from $\pi(\kappa_t|\kappa_{t+1},\boldsymbol{\psi},\boldsymbol{y}_{1:t})$ where $t=T-1,T-2,\dots,1,0$. Moreover, we have
\begin{equation}
    \kappa_t|\kappa_{t+1},\boldsymbol{\psi},\boldsymbol{y}_{1:t} \sim \text{N}(h_t,H_t),
\end{equation}
where
\begin{subequations}
\begin{align}
    h_t &= m_t+C_tR^{-1}_{t+1}(\kappa_{t+1}-a_{t+1}), \\
    H_t &= C_t-C_tR^{-1}_{t+1}C_t,
\end{align}
\end{subequations}
which can be derived based on Kalman smoother (\cite{PetrisPeCa}).

The FFBS procedure is displayed in Algorithm~\ref{FFBSalgorithm}.
Note that the prior distribution for $\kappa_0$ can be set to be vague to run the Kalman filter; the output of the algorithm includes the posterior distribution of $\kappa_0$.

\begin{algorithm}
\caption{FFBS Algorithm: Forward Filtering Backward Sampling}
\label{FFBSalgorithm}
\begin{algorithmic}[1]
\State{Run Kalman filter to obtain $m_T$ and $C_T$.}
\State{Sample $\kappa_T$ from $\text{N}(m_T,C_T)$.}
\For{$t=T-1,\dots,0$}
\State Sample $\kappa_t$ from $\text{N}(h_t,H_t)$ using the sample $\kappa_{t+1}$ obtained in the previous step.
\EndFor
\end{algorithmic}
\end{algorithm}

\FloatBarrier

%%%%%%%%%%%%%%%%%%%%%%%%%%%%%%%%%%%%%%%%%%%%%%%%%%%%%%%%%%%%%%%%%%%%%%%%%%%%%%%%%%%%%%%%%%%%%%%%%%%
\subsubsection{Sampling from the full conditional density $\pi(\boldsymbol{\psi}|\kappa_{0:T},\boldsymbol{y}_{1:T})$}\label{subsubsec:paraPostDistLCH}
%%%%%%%%%%%%%%%%%%%%%%%%%%%%%%%%%%%%%%%%%%%%%%%%%%%%%%%%%%%%%%%%%%%%%%%%%%%%%%%%%%%%%%%%%%%%%%%%%%%
The first thing to observe is that under the reparameterization of the identification constraints \eqref{eqn:newConstraints}, the following Gibbs sampling stages can be performed exactly.

We assume that the prior for $(\alpha_{x_2:x_p}, \beta_{x_2:x_p}, \theta, \sigma^2_{\varepsilon,x_1:x_p}, \sigma^2_\omega)$ are given by
\begin{subequations}
    \begin{align}
        & \alpha_x \sim \text{N}(\tilde{\mu}_\alpha,\tilde{\sigma}^2_\alpha), \quad
        \beta_x \sim \text{N}(\tilde{\mu}_\beta,\tilde{\sigma}^2_\beta), \quad
        \theta \sim \text{N}(\tilde{\mu}_\theta,\tilde{\sigma}^2_\theta), \\
        & \sigma^2_{\varepsilon,x} \sim \text{IG}(\tilde{a}_\varepsilon, \tilde{b}_\varepsilon), \quad
        \sigma^2_\omega \sim \text{IG}(\tilde{a}_\omega, \tilde{b}_\omega),
    \end{align}
\end{subequations}
where $\text{IG}(\tilde{a}_\omega,\tilde{b}_\omega)$ denotes an inverse-gamma
distribution with mean $\tilde{b}_\omega/(\tilde{a}_\omega-1)$ and variance $\tilde{b}^2_\omega/((\tilde{a}_\omega-1)^2(\tilde{a}_\omega-2))$ for $\tilde{a}_\omega > 2$. We assume that the priors for all parameters are independent. In this case the posterior densities of parameters are of the same type as the prior densities, a so-called conjugate prior. The posterior distribution for each parameter is given by (we write, for ease of notation,  $\boldsymbol{y}=\boldsymbol{y}_{1:T}$, $\boldsymbol{\kappa}=\kappa_{0:T}$,
family $\boldsymbol{\psi}_{-\lambda}$ means ``parameter vector $\boldsymbol{\psi}$ without the parameter $\lambda$"):
\begin{align}
\alpha_x|\boldsymbol{y}_{1:T},\boldsymbol{\kappa}, \boldsymbol{\psi}_{-\alpha_x} &\sim
        \text{N}\left(\frac{\tilde{\mu}_\alpha\sigma^2_{\varepsilon,x}+\tilde{\sigma}^2_\alpha\sum_t
           (y_{xt}-\beta_x\kappa_t)}{\tilde{\sigma}^2_\alpha T+\sigma^2_{\varepsilon,x}},
            \frac{\tilde{\sigma}^2_\alpha\sigma^2_{\varepsilon,x}}{\tilde{\sigma}^2_\alpha T+\sigma^2_{\varepsilon,x}}\right),\\
\beta_x|\boldsymbol{y}_{1:T},\boldsymbol{\kappa}, \boldsymbol{\psi}_{-\beta_x} &\sim
            \text{N} \left(\frac{\tilde{\sigma}^2_\beta\sum_t(y_{xt}-\alpha_x)\kappa_t+\tilde{\mu}_\beta\sigma^2_{\varepsilon,x}}
            {\tilde{\sigma}^2_\beta\sum_t\kappa^2_t+\sigma^2_{\varepsilon,x}},
            \frac{\tilde{\sigma}^2_\beta\sigma^2_{\varepsilon,x}}{\tilde{\sigma}^2_\beta\sum_t\kappa^2_t+\sigma^2_{\varepsilon,x}}\right),\\
\theta|\boldsymbol{y}_{1:T},\boldsymbol{\kappa},\boldsymbol{\psi}_{-\theta} &\sim
           \text{N} \left(\frac{\tilde{\sigma}^2_\theta\sum^T_{t=1}(\kappa_t-\kappa_{t-1})+\tilde{\mu}_\theta\sigma^2_\omega}
           {\tilde{\sigma}^2_\theta T+\sigma^2_\omega},
           \frac{\tilde{\sigma}^2_\theta \sigma^2_\omega}{\tilde{\sigma}^2_\theta T+\sigma^2_\omega}\right),\\
\sigma^2_{\varepsilon,x}|\boldsymbol{y}_{1:T},\boldsymbol{\kappa},\boldsymbol{\psi}_{-\sigma^2_{\varepsilon,x}} &\sim
            \text{IG} \left(\tilde{a}_\varepsilon+\frac{pT}{2},\,
            \tilde{b}_\varepsilon+\frac{1}{2}
           \sum^T_{t=1}\left(y_{xt}-(\alpha_x+\beta_x\kappa_t)\right)^2\right),\\
\sigma^2_\omega|\boldsymbol{y}_{1:T},\boldsymbol{\kappa}, \boldsymbol{\psi}_{-\sigma^2_\omega} &\sim
            \text{IG} \left(\tilde{a}_\omega+\frac{T}{2},\,
           \tilde{b}_\omega+\frac{1}{2}
           \sum^T_{t=1}\left(\kappa_t-(\kappa_{t-1}+\theta)\right)^2\right).
\end{align}

%%%%%%%%%%%%%%%%%%%%%%%%%%%%%%%%%%%%%%%%%%%%%%%%%%%%%%%%%%%%%%%%%%%%%%%%%%%%%%%%%%%%%%%%%%%%%%%%%%%
\subsection{Non-Linear / Non-Gaussian State-Space Inference}\label{subsec:estimationNLNGModels}
%%%%%%%%%%%%%%%%%%%%%%%%%%%%%%%%%%%%%%%%%%%%%%%%%%%%%%%%%%%%%%%%%%%%%%%%%%%%%%%%%%%%%%%%%%%%%%%%%%%
In the case of non-linear / non-Gaussian state-space model dynamics such as the stochastic volatility models of LC3-H2 and the LCSV models, the sampler we develop is based on novel developments of the Particle Metropolis Hastings samplers of \cite{AndrieuDoHo10} adapted to the stochastic mortality models. In particular we consider a combination of Rao-Blackwellized Kalman filter and particle filter for the latent state process full posterior conditionals, combined with a combination of Metropolis-within-Gibbs and Gibbs sampling steps for the static model parameters, both embedded within a PMCMC framework. We will illustrate the idea of this methodology for the LCSV model where a stochastic volatility dynamics is included in the latent process for the period effect.

%%%%%%%%%%%%%%%%%%%%%%%%%%%%%%%%%%%%%%%%%%%%%%%%%%%%%%%%%%%%%%%%%%%%%%%%%%%%%%%%%%%%%%%%%%%%%%%%%%%
\subsubsection{Estimation for the LCSV Mortality Model}
%%%%%%%%%%%%%%%%%%%%%%%%%%%%%%%%%%%%%%%%%%%%%%%%%%%%%%%%%%%%%%%%%%%%%%%%%%%%%%%%%%%%%%%%%%%%%%%%%%%
The static parameter vector is denoted as $\boldsymbol{\psi}=(\alpha_{x_2:x_p},\beta_{x_2:x_p},\theta,\sigma^2_\varepsilon,\sigma^2_\gamma,\lambda_1,\lambda_2,\gamma_0)$. Note that we treat $\gamma_0$ as a static parameter and our task is to obtain samples from the joint posterior distribution:
\begin{equation} \label{eqnFullPostFPS}
    \pi(\kappa_{0:T},\gamma_{1:T}, \boldsymbol{\psi}|\boldsymbol{y}_{1:T}).
\end{equation}
In this setting one can try a number of different approaches, the first would be to sample jointly from the full posterior distribution \eqref{eqnFullPostFPS} via PMCMC methods to be described below. A second approach would be to combine PMCMC methods within a block-Gibbs based sampler such as the following sampling scheme, where we apply Gibbs sampling to sample from the full conditional densities
\begin{subequations}
\begin{align}
    &\pi(\kappa_{0:T}|\boldsymbol{\psi},\gamma_{1:T},\boldsymbol{y}_{1:T}), \label{eqn:FCDkappa} \\
    &\pi(\boldsymbol{\psi}|\kappa_{0:T},\gamma_{1:T},\boldsymbol{y}_{1:T}), \label{eqn:FCDparameters} \\
    &\pi(\gamma_{1:T}|\boldsymbol{\psi},\kappa_{0:T},\boldsymbol{y}_{1:T}). \label{eqn:FCDgamma}
\end{align}
\end{subequations}
Note that sampling from \eqref{eqn:FCDkappa} can be achieved by the FFBS procedure described in Algorithm \ref{FFBSalgorithm}, as one can apply Kalman filtering since $\gamma_{1:T}$ is assumed to be given. The only difference compared to Section \ref{sec:EstimationFirstDensity} is that the term $\sigma^2_\omega$ is replaced by $\exp\{\gamma_t\}$ in Kalman filtering.

In the following we provide details on how to sample from either the full posterior \eqref{eqnFullPostFPS} or from full conditionals such as the density in \eqref{eqn:FCDgamma}, via PMCMC method. Sampling from the posteriors of the static parameters \eqref{eqn:FCDparameters} is detailed in Section~\ref{sec:LCSVstaticPosteriors}.

%%%%%%%%%%%%%%%%%%%%%%%%%%%%%%%%%%%%%%%%%%%%%%%%%%%%%%%%%%%%%%%%%%%%%%%%%%%%%%%%%%%%%%%%%%%%%%%%%%%
\subsubsection{Particle Markov chain Monte Carlo (PMCMC) for  Mortality Models}
%%%%%%%%%%%%%%%%%%%%%%%%%%%%%%%%%%%%%%%%%%%%%%%%%%%%%%%%%%%%%%%%%%%%%%%%%%%%%%%%%%%%%%%%%%%%%%%%%%%
In this section we explain the generic form of the PMCMC methodology that can be applied in a range of approaches for state-space stochastic mortality models. In general a PMCMC sampling method is a class of MCMC method where SMC algorithm is used as a proposal distribution within a MCMC algorithm. Though this seems trivial, it is actually based on a key observation that by using such a filter within the MCMC, the dimension of the acceptance probability in the Metropolis-Hastings acceptance-rejection stage is significantly reduced and can therefore facilitate much better mixing performance of the resulting Markov chain, reducing variance in estimation, see discussion in detail in \cite{AndrieuDoHo10}.

To bring out the essence of PMCMC, we first discuss a generic approach to sample from a target distribution
\begin{equation}
    \pi(\bm{\phi}_{1:T},\boldsymbol{\psi}|\boldsymbol{z}_{1:T}),
\end{equation}
where $\bm{\phi}_{1:T}$ and $\boldsymbol{\psi}$ are the latent state and static parameters of a general state-space model. Note, the state processes in this context are generally non-linear and potentially non-Gaussian.

From the perspective of obtaining the most efficiently mixing Markov chain to sample from this posterior, the ideal proposal distribution for constructing the Markov chain for $(\bm{\phi}'_{1:T},\boldsymbol{\psi}')$ is easily seen to be given by
\begin{equation}
    q(\boldsymbol{\psi}'|\boldsymbol{\psi})p_{\boldsymbol{\psi}'}(\bm{\phi}'_{1:T}|\boldsymbol{z}_{1:T}),
\end{equation}
where $q(\boldsymbol{\psi}'|\boldsymbol{\psi})$ is a proposal for the parameters and the proposal for the latent state, $p_{\boldsymbol{\psi}'}(\bm{\phi}'_{1:T}|\boldsymbol{z}_{1:T})$, is from the state equation (given $\boldsymbol{\psi}'$). Here $(\bm{\phi}_{1:T},\boldsymbol{\psi})$ is the current state at MCMC iteration $j-1$ and $(\bm{\phi}'_{1:T},\boldsymbol{\psi}')$ is the proposed next move at MCMC iteration $j$.

In this ideal case, the acceptance probability of this ideal proposal is given by:
\begin{align}
    \alpha((\bm{\phi}'_{1:T},\boldsymbol{\psi}'),(\bm{\phi}_{1:T},\boldsymbol{\psi})) &= 1 \wedge \frac{p(\bm{\phi}'_{1:T},\boldsymbol{\psi}'|\boldsymbol{z}_{1:T})q(\boldsymbol{\psi}|\boldsymbol{\psi}')
       p_{\boldsymbol{\psi}}(\bm{\phi}_{1:T}|\boldsymbol{z}_{1:T})}
       {p(\bm{\phi}_{1:T},\boldsymbol{\psi}|\boldsymbol{z}_{1:T})q(\boldsymbol{\psi}'|\boldsymbol{\psi})
       p_{\boldsymbol{\psi}'}(\bm{\phi}'_{1:T}|\boldsymbol{z}_{1:T})} \\
    &= 1 \wedge \frac{p_{\boldsymbol{\psi}'}(\bm{\phi}'_{1:T}|\boldsymbol{z}_{1:T})p(\boldsymbol{\psi}'|\boldsymbol{z}_{1:T})
        q(\boldsymbol{\psi}|\boldsymbol{\psi}')p_{\boldsymbol{\psi}}(\bm{\phi}_{1:T}|\boldsymbol{z}_{1:T})}
        {p_{\boldsymbol{\psi}}(\bm{\phi}_{1:T}|\boldsymbol{z}_{1:T})p(\boldsymbol{\psi}|\boldsymbol{z}_{1:T})
        q(\boldsymbol{\psi}'|\boldsymbol{\psi})p_{\boldsymbol{\psi}'}(\bm{\phi}'_{1:T}|\boldsymbol{z}_{1:T})} \\
    &= 1 \wedge \frac{p(\boldsymbol{\psi}'|\boldsymbol{z}_{1:T})
        q(\boldsymbol{\psi}|\boldsymbol{\psi}')}{p(\boldsymbol{\psi}|\boldsymbol{z}_{1:T})
        q(\boldsymbol{\psi}'|\boldsymbol{\psi})} \\
    &= 1 \wedge \frac{p_{\boldsymbol{\psi}'}(\boldsymbol{z}_{1:T})p(\boldsymbol{\psi}')q(\boldsymbol{\psi}|\boldsymbol{\psi}')}
    {p_{\boldsymbol{\psi}}(\boldsymbol{z}_{1:T})p(\boldsymbol{\psi})q(\boldsymbol{\psi}'|\boldsymbol{\psi})}, \label{eqn:marginallh}
\end{align}
where $r_1 \wedge r_2 := \text{min}(r_1,r_2)$. A desirable property of the ideal proposal is that the acceptance probability depends only on the marginal likelihood, together with the prior and proposal for the static parameters. This is optimal in the sense that the dimension of the numerator and denominator is reduced significantly to the static model parameter dimensions, and not including explicitly the path-space latent process dimensions, a reduction of $d\times T$ dimensions for a d-dimensional state vector $\bm{\phi}_t$. However, clearly one can never achieve this goal as it requires perfect knowledge of $p_{\boldsymbol{\psi}'}(\bm{\phi}'_{1:T}|\boldsymbol{z}_{1:T})$ as well as the ability to sample this distribution, both of which are unachievable except in the special case of the Linear-Gaussian case explained in Section \ref{sec:EstimationFirstDensity}.

To circumvent this problem, the particle marginal Metropolis-Hastings sampler (PMMH; \cite{AndrieuDoHo10}) applies SMC method to obtain an approximate of the state transition density (which is also the state proposal)
\begin{equation}
    \hat{p}_{\boldsymbol{\psi}'}(\bm{\phi}_{1:T}|\boldsymbol{z}_{1:T})=\sum^N_{i=1}w^{(i)}_{T}\delta_{\bm{\phi}^{(i)}_{1:T}}(\bm{\phi}_{1:T}),
\end{equation}
where $w^{(i)}_T$ is the importance weight, $\delta_{x}(X)$ denotes a Dirac mass function centered at $X$ and a proposed next move of the latent state is drawn from this discrete approximate distribution. Moreover, a by-product of a SMC algorithm is the marginal likelihood, $\hat{p}_{\boldsymbol{\psi}}(\boldsymbol{z}_{1:T})$, which has the following important property:
\begin{lemma}\label{lem:unbiasedMarginal}
A SMC proposal admits as a by-product an unbiased estimator of the marginal likelihood $p_{\boldsymbol{\psi}}(\boldsymbol{z}_{1:T})$ given by
\begin{equation}
    \hat{p}_{\boldsymbol{\psi}}(\boldsymbol{z}_{1:T}) := \prod_{t=2}^T \hat{p}_{\boldsymbol{\psi}}(\boldsymbol{z}_{t}|\boldsymbol{z}_{1:t-1}),
\end{equation}
where a SMC approximation with $N$-particles produces, for all $t$,
\begin{equation}
    \hat{p}_{\boldsymbol{\psi}}(\boldsymbol{z}_{t}|\boldsymbol{z}_{1:t-1}) = \frac{1}{N}\sum_{i=1}^N w^{(i)}_t,
\end{equation}
which is an unbiased particle estimate of $p_{\boldsymbol{\psi}}(\boldsymbol{z}_{t}|\boldsymbol{z}_{1:t-1})$.
This non-trivial unbiasedness was first presented in \cite{Del2004} and has since been utilised to great advantage as explained in \cite{chopin2013smc2}. In addition the variance of this estimator typically only grows linealy with $T$.
\end{lemma}
The unbiased approximate marginal likelihoods are then used in the acceptance probability \eqref{eqn:marginallh}:
\begin{equation}
    \alpha((\bm{\phi}'_{1:T},\boldsymbol{\psi}'),(\bm{\phi}_{1:T},\boldsymbol{\psi})) = 1 \wedge \frac{\hat{p}_{\boldsymbol{\psi}'}(\boldsymbol{z}_{1:T})p(\boldsymbol{\psi}')q(\boldsymbol{\psi}|\boldsymbol{\psi}')}
    {\hat{p}_{\boldsymbol{\psi}}(\boldsymbol{z}_{1:T})p(\boldsymbol{\psi})q(\boldsymbol{\psi}'|\boldsymbol{\psi})}.
\end{equation}
Due to the unbiasedness of the estimated marginal likelihood, \cite{AndrieuDoHo10} show that, even though only SMC approximates are used (with finite number of particles $N$), the invariant distribution of PMMH is the target distribution $\pi(\bm{\phi}_{1:T},\boldsymbol{\psi}|\boldsymbol{z}_{1:T})$.

To apply PMCMC for an efficient estimation of the LCSV model, we first notice that we can obtain explicitly the posteriors of static parameters via conjugate priors. As a result we are only required to sample from the density $\pi(\gamma_{1:T}|\boldsymbol{\psi},\kappa_{0:n},\boldsymbol{y}_{1:T})$, instead of the joint density
$\pi(\gamma_{1:T},\boldsymbol{\psi}|\kappa_{0:n},\boldsymbol{y}_{1:T})$. It turns out that there is a class of PMCMC algorithm, called Particle Independent Metropolis-Hastings sampler (PIMH), which provide a mechanism to sample exactly from $\pi(\gamma_{1:T}|\boldsymbol{\psi},\kappa_{0:n},\boldsymbol{y}_{1:T})$.

Our approach to sampling from the joint posterior distribution, $\pi(\kappa_{0:T},\gamma_{1:T},\boldsymbol{\psi}|\boldsymbol{y}_{1:T})$, of the LCSV model
is summarised in Algorithm~\ref{GibbsAlgorithmLCSV}.
\begin{algorithm}[H]
\caption{Sampling from $\pi(\kappa_{0:T},\gamma_{1:T},\boldsymbol{\psi}|\boldsymbol{y}_{1:T})$}
\label{GibbsAlgorithmLCSV}
\begin{algorithmic}[1]
\State{Initialise: $\boldsymbol{\psi}=\boldsymbol{\psi}^{(0)}$, $\gamma_{1:T}=\gamma_{1:T}^{(0)}$.}
\For{$i=1,\dots,M$}
    \State Sample $\kappa^{(i)}_{0:T}$ from
        $\pi(\kappa_{0:T}|\gamma_{1:T}^{(i-1)},\boldsymbol{\psi}^{(i-1)},\boldsymbol{y}_{1:T})$ via FFBS;
    \State Sample $\gamma^{(i)}_{1:T}$ from
        $\pi(\gamma_{1:T}|\kappa_{0:T}^{(i)},\boldsymbol{\psi}^{(i-1)},\boldsymbol{y}_{1:T})$ via PIMH (Section \ref{sec:PIMH});
    \For{$h=1,\dots,n$}
        \State Sample $\psi^{(i)}_h$ from $\pi(\psi_h|\kappa^{(i)}_{0:T},\gamma_{1:T}^{(i)},\psi_{-h}^{(i)},\boldsymbol{y}_{1:T})$,
        \State where $\psi_{-h}^{(i)}=(\psi^{(i)}_1,\dots,\psi^{(i)}_{h-1},\psi^{(i-1)}_{h+1},\dots,\psi^{(i-1)}_n)$ via conjugate prior.
    \EndFor
\EndFor
\end{algorithmic}
\end{algorithm}

%%%%%%%%%%%%%%%%%%%%%%%%%%%%%%%%%%%%%%%%%%%%%%%%%%%%%%%%%%%%%%%%%%%%%%%%%%%%%%%%%%%%%%%%%%%%%%%%%%%
\subsubsection{PIMH: Sampling from $\pi(\gamma_{1:T}|\boldsymbol{\psi},\kappa_{0:n},\boldsymbol{y}_{1:T})$}
\label{sec:PIMH}
%%%%%%%%%%%%%%%%%%%%%%%%%%%%%%%%%%%%%%%%%%%%%%%%%%%%%%%%%%%%%%%%%%%%%%%%%%%%%%%%%%%%%%%%%%%%%%%%%%%
We first note that $\pi(\gamma_{1:T}|\boldsymbol{\psi},\kappa_{0:n},\boldsymbol{y}_{1:T}) =
\pi_{\boldsymbol{\psi}}(\gamma_{1:T}|\kappa_{0:n})$ given the structure of the LCSV model. Using an independent proposal density, $q_{\boldsymbol{\psi}}(\gamma_{1:T}|\kappa_{0:n})$, in the Metropolis-Hastings algorithm, the acceptance probability is given by
\begin{equation}
    \alpha(\gamma'_{1:T},\gamma_{1:T}) = 1 \wedge \frac{\pi_{\boldsymbol{\psi}}(\gamma'_{1:T}|\kappa_{0:n})q_{\boldsymbol{\psi}}(\gamma_{1:T}|\kappa_{0:n})}
    {\pi_{\boldsymbol{\psi}}(\gamma_{1:T}|\kappa_{0:n})q_{\boldsymbol{\psi}}(\gamma'_{1:T}|\kappa_{0:n})}.
\end{equation}
Ideally, one may take $q_{\boldsymbol{\psi}}(\gamma_{1:T}|\kappa_{0:n}) = \pi_{\boldsymbol{\psi}}(\gamma_{1:T}|\kappa_{0:n})$. However, in most cases such an ideal choice is impossible to sample from and to evaluate. The PIMH sampler proposes instead to use the SMC approximation $\hat{\pi}_{\boldsymbol{\psi}}(\gamma_{1:T}|\kappa_{0:n})$ as the proposal density and calculate the acceptance probability as
\begin{equation}\label{eqn:PIMHacceptance}
    \alpha(\gamma'_{1:T},\gamma_{1:T}) = 1 \wedge \frac{\hat{\pi}_{\boldsymbol{\psi}}(\kappa_{0:n})'}{\hat{\pi}_{\boldsymbol{\psi}}(\kappa_{0:n})[j-1]},
\end{equation}
where $\hat{\pi}_{\boldsymbol{\psi}}(\kappa_{0:n})'$ and $\hat{\pi}_{\boldsymbol{\psi}}(\kappa_{0:n})[j-1]$ are unbiased marginal likelihoods estimated by SMC (see Lemma~\ref{lem:unbiasedMarginal}) in the current MCMC iteration $j$ and the previous iteration $j-1$ respectively. It can be shown that the invariant distribution of the PIMH sampler is the target distribution $\pi_{\boldsymbol{\psi}}(\gamma_{1:T}|\kappa_{0:n})$ (\cite{AndrieuDoHo10}).

It remains to specify an SMC approximation $\hat{\pi}_{\boldsymbol{\psi}}(\gamma_{1:T}|\kappa_{0:n})$ (Appendix~\ref{app:SMC}). We use the so-called bootstrap filter, that is, the proposal distribution in the SMC algorithm to draw $\gamma_t$ is given by the state equation \eqref{eqn:LCSVc}:
\begin{equation}
    g_t(\gamma_t|\gamma_{1:t-1},\kappa_{0:t}) :=
    \pi_{\boldsymbol{\psi}}(\gamma_t|\gamma_{t-1}).
\end{equation}
Consequently,
the importance weight is evaluated as
\begin{equation}
    \tilde{w}_t \propto \tilde{w}_{t-1}\pi_{\boldsymbol{\psi}}(\kappa_t|\gamma_t,\kappa_{t-1}),
\end{equation}
where $\pi(\kappa_t|\gamma_t,\kappa_{t-1})$ is the incremental importance weight. Our approach for sampling from $\pi_{\boldsymbol{\psi}}(\gamma_{1:T}|\kappa_{0:T})$ is summarised in Algorithm~\ref{PIMHAlgorithm} (together with Algorithm~\ref{BootstrapAlgorithm}).

\begin{algorithm}
\caption{PIMH: sampling from $\pi_{\boldsymbol{\psi}}(\gamma_{1:T}|\kappa_{0:T})$}
\label{PIMHAlgorithm}
\begin{algorithmic}[1]
\State Iteration $j=0$: obtain an SMC approximation $\hat{\pi}_{\boldsymbol{\psi}}(\gamma_{1:T}|\kappa_{0:T})$ via Algorithm~\ref{BootstrapAlgorithm}. Draw $\gamma_{1:T}[0] \sim \hat{\pi}_{\boldsymbol{\psi}}(\gamma_{1:T}|\kappa_{0:T})$ and obtain the corresponding marginal likelihood estimate $\hat{\pi}_{\boldsymbol{\psi}}(\kappa_{0:n})[0]$.
\For{$j=1,\dots,N_{PIMH}$}
\State Obtain an SMC approximation $\hat{\pi}_{\boldsymbol{\psi}}(\gamma_{1:T}|\kappa_{0:T})$ via Algorithm~\ref{BootstrapAlgorithm}. Draw $\gamma'_{1:T} \sim \hat{\pi}_{\boldsymbol{\psi}}(\gamma_{1:T}|\kappa_{0:T})$ and obtain the corresponding marginal likelihood estimate $\hat{\pi}_{\boldsymbol{\psi}}(\kappa_{0:n})'$.
\State Draw $u \sim U(0,1)$. If
    \begin{equation}
        u < \frac{\hat{\pi}_{\boldsymbol{\psi}}(\kappa_{0:n})'}{\hat{\pi}_{\boldsymbol{\psi}}(\kappa_{0:n})[j-1]},
    \end{equation}
    set $\gamma_{1:T}[j] = \gamma'_{1:T}$ and $\hat{\pi}_{\boldsymbol{\psi}}(\kappa_{0:n})[j]=\hat{\pi}_{\boldsymbol{\psi}}(\kappa_{0:n})'$; otherwise set
    $\gamma_{1:T}[j] = \gamma_{1:T}[j-1]$ and $\hat{\pi}_{\boldsymbol{\psi}}(\kappa_{0:n})[j]=\hat{\pi}_{\boldsymbol{\psi}}(\kappa_{0:n})[j-1]$.
\EndFor
\State Obtain $\gamma_{1:T}[N_{PIMH}]$ as a sample of $\pi_{\boldsymbol{\psi}}(\gamma_{1:T}|\kappa_{0:T})$.
\end{algorithmic}
\end{algorithm}

\begin{algorithm}
\caption{Bootstrap filter of $\pi_{\boldsymbol{\psi}}(\gamma_{1:T}|\kappa_{0:T})$; see Appendix~\ref{app:SMC}}
\label{BootstrapAlgorithm}
\begin{algorithmic}[1]
\State{At $t=1$: draw $\gamma_1^{(i)}$ from $\pi_{\boldsymbol{\psi}}(\gamma_1|\gamma_0)$. Set $\tilde{w}^{(i)}_1=\pi_{\boldsymbol{\psi}}(\kappa_1|\gamma_1,\kappa_0)$ and $w^{(i)}_1=\tilde{w}^{(i)}_1/\sum^N_{j=1}\tilde{w}^{(j)}_1$.}
\For{$t=2,\dots,T$}
\State Draw $\gamma^{(i)}_t$ from $\pi_{\boldsymbol{\psi}}(\gamma_t|\gamma^{(i)}_{t-1})$ and set
    \begin{equation}
        \gamma^{(i)}_{1:t} = (\gamma^{(i)}_{1:t-1},\gamma^{(i)}_t);
    \end{equation}
\State Evaluate
    \begin{equation}
        \tilde{w}^{(i)}_t = \tilde{w}^{(i)}_{t-1}\cdot \pi_{\boldsymbol{\psi}}(\kappa_t|\gamma^{(i)}_t,\kappa_{t-1});
    \end{equation}
\State Normalise:
    \begin{equation}
        w^{(i)}_t = \frac{\tilde{w}^{(i)}}{\sum^N_{j=1}\tilde{w}^{(j)}_t};
    \end{equation}
\State Evaluate
    \begin{equation}
        N_{eff} = \left(\sum^N_{i=1}(w^{(i)}_t)^2\right)^{-1};
    \end{equation}
\State{If $N_{eff} < 0.8N$, resample $\gamma^{(i)}_{1:t}$
    from $\left(w^{(j)}_t, \gamma^{(j)}_{1:t}\right)^N_{j=1}$ and set $w^{(i)}_t =\frac{1}{N}$.}
\EndFor
\State Obtain $\hat{\pi}_{\boldsymbol{\psi}}(\gamma_{1:T}|\kappa_{0:T})=
    \sum^N_{i=1}w^{(i)}_T\delta_{\gamma^{(i)}_{1:T}}(\gamma_{1:T})$.
\end{algorithmic}
\end{algorithm}

\subsubsection{Sampling from $\pi(\boldsymbol{\psi}|\kappa_{0:T},\gamma_{1:T},\boldsymbol{y}_{1:T})$}
\label{sec:LCSVstaticPosteriors}
We assume the prior for $(\alpha_{x_2:x_p}, \beta_{x_2:x_p}, \theta, \sigma^2_\varepsilon,
\sigma^2_\gamma, \lambda_1, \lambda_2, \gamma_0)$ is given by
\begin{subequations}
    \begin{align}
      & \alpha_x \sim
        \text{N}(\tilde{\mu}_\alpha, \tilde{\sigma}^2_\alpha), \quad
        \beta_x \sim
        \text{N}(\tilde{\mu}_\beta, \tilde{\sigma}^2_\beta), \quad
        \theta \sim
        \text{N}(\tilde{\mu}_\theta, \tilde{\sigma}^2_\theta), \quad
        \sigma^2_\varepsilon \sim
        \text{IG}(\tilde{a}_\varepsilon, \tilde{b}_\varepsilon), \\
      & \sigma^2_\gamma \sim
        \text{IG}(\tilde{a}_\gamma, \tilde{b}_\gamma),\,
        \lambda_1 \sim
        \text{N}_{[-1,1]}(\tilde{\mu}_{\lambda_1}, \tilde{\sigma}^2_{\lambda_1}),\,
        \lambda_2 \sim
        \text{N}(\tilde{\mu}_{\lambda_2}, \tilde{\sigma}^2_{\lambda_2}),\,
        \gamma_0 \sim
        \text{N}(\tilde{\mu}_{\gamma_0}, \tilde{\sigma}^2_{\gamma_0}),
    \end{align}
\end{subequations}
where $x = x_2,\dots,x_p$ and $\text{N}_{[-1,1]}$ denotes a truncated Gaussian with support $[-1,1]$. It is assumed that the priors for all parameters are independent.

Samples from the density
$\pi(\boldsymbol{\psi}|\kappa_{0:T},\gamma_{1:T},\boldsymbol{y}_{1:T})$ are
obtained by sampling from the following posteriors:
\begin{align}
    \alpha_x|\boldsymbol{y}_{1:T},\boldsymbol{\kappa}, \bm{\gamma}, \boldsymbol{\psi}_{-\alpha_x} &\sim \text{N}\left(\frac{\tilde{\mu}_\alpha \sigma^2_\varepsilon +\tilde{\sigma}^2_\alpha\sum_t(y_{xt}-\beta_x\kappa_t)}{T\tilde{\sigma}^2_\alpha+\sigma^2_\varepsilon},
        \frac{\tilde{\sigma}^2_\alpha\sigma^2_\varepsilon}{T\tilde{\sigma}^2_\alpha+\sigma^2_\varepsilon}\right), \\
    \beta_x|\boldsymbol{y}_{1:T},\boldsymbol{\kappa}, \bm{\gamma}, \boldsymbol{\psi}_{-\beta_x}  &\sim \text{N} \left(\frac{\tilde{\sigma}^2_\beta \sum_t(y_{xt}-\alpha_x)\kappa_t+\tilde{\mu}_\beta \sigma^2_\varepsilon}{\tilde{\sigma}^2_\beta\sum_t\kappa^2_t+\sigma^2_\varepsilon},
        \frac{\tilde{\sigma}^2_\beta\sigma^2_\varepsilon}{\tilde{\sigma}^2_\beta\sum_t\kappa^2_t+\sigma^2_\varepsilon}\right), \\
    %\theta_x|(\kappa_{0:T},\gamma_{1:T},\boldsymbol{y}_{1:T}) &\sim  \text{N} \left(\left(\frac{\tilde{\mu}_\theta}{\tilde{\sigma}^2_\theta}+\sum_t\frac{1}{e^{\gamma_t}}(\kappa_t-\kappa_{t-1})\right)
     %   \left(\frac{1}{\tilde{\sigma}^2_\theta}+\sum_t\frac{1}{e^{\gamma_t}}\right)^{-1},
    %    \left(\frac{1}{\tilde{\sigma}^2_\theta}+\sum_t\frac{1}{e^{\gamma_t}}\right)^{-1}\right) \\
    \theta_x|\boldsymbol{y}_{1:T},\boldsymbol{\kappa}, \bm{\gamma}, \boldsymbol{\psi}_{-\theta} &\sim  \text{N} \left(\frac{\tilde{\mu}_\theta/\tilde{\sigma}^2_\theta+\sum_t(\kappa_t-\kappa_{t-1})/e^{\gamma_t}}
        {1/\tilde{\sigma}^2_\theta+\sum_t 1/e^{\gamma_t}},
        \frac{1}{1/\tilde{\sigma}^2_\theta+\sum_t 1/e^{\gamma_t}}\right), \\
    \sigma^2_\varepsilon|\boldsymbol{y}_{1:T},\boldsymbol{\kappa}, \bm{\gamma}, \boldsymbol{\psi}_{-\sigma^2_\varepsilon} &\sim  \text{IG} \left(\tilde{a}_\varepsilon+\frac{pT}{2},\,
            \tilde{b}_\varepsilon+\frac{1}{2}\sum^T_{t=1}\sum^{p}_{x=1}\left(y_{xt}-(\alpha_x+\beta_x\kappa_t)\right)^2\right), \\
    \sigma^2_\gamma|\boldsymbol{y}_{1:T},\boldsymbol{\kappa}, \bm{\gamma}, \boldsymbol{\psi}_{-\sigma^2_\gamma} &\sim  \text{IG} \left(\tilde{a}_\gamma+\frac{T}{2},\,
        \tilde{b}_\gamma+\frac{1}{2} \sum^T_{t=1}\left(\gamma_t-\lambda\gamma_{t-1}\right)^2\right), \\
    \lambda_1|\boldsymbol{y}_{1:T},\boldsymbol{\kappa}, \bm{\gamma}, \boldsymbol{\psi}_{-\lambda_1} &\sim    \text{N}_{[-1,1]} \left(\frac{\sigma^2_\gamma\tilde{\mu}_{\lambda_1}+\tilde{\sigma}^2_{\lambda_1}\sum_t\gamma_{t-1}\gamma_t}
       {\sigma^2_\gamma+\tilde{\sigma}^2_{\lambda_1}\sum_t\gamma^2_{t-1}},
      \frac{\tilde{\sigma}^2_{\lambda_1}\sigma^2_\gamma}{\sigma^2_\gamma+\tilde{\sigma}^2_{\lambda_1}\sum_t\gamma^2_{t-1}}\right), \\
    \lambda_2|\boldsymbol{y}_{1:T},\boldsymbol{\kappa}, \bm{\gamma}, \boldsymbol{\psi}_{-\lambda_2} &\sim   \text{N} \left(\frac{\sigma^2_\gamma\tilde{\mu}_{\lambda_2}+\tilde{\sigma}^2_{\lambda_2}\sum_t(\gamma_t-\lambda_1\gamma_{t-1})}
        {\sigma^2_\gamma+T\tilde{\sigma}^2_{\lambda_2}}, \frac{\tilde{\sigma}^2_{\lambda_2}\sigma^2_\gamma}{\sigma^2_\gamma+T\tilde{\sigma}^2_{\lambda_2}}\right), \\
    \gamma_0|\boldsymbol{y}_{1:T},\boldsymbol{\kappa}, \bm{\gamma}, \boldsymbol{\psi}_{-\gamma_0} &\sim \text{N} \left(\frac{\sigma^2_\gamma\tilde{\mu}_{\gamma_0}+\tilde{\sigma}^2_{\gamma_0}\lambda\gamma_1}
        {\sigma^2_\gamma+\tilde{\sigma}^2_{\gamma_0}\lambda^2},
        \frac{\tilde{\sigma}^2_{\gamma_0}\sigma^2_\gamma}{\sigma^2_\gamma+\tilde{\sigma}^2_{\gamma_0}\lambda^2}\right).
\end{align}
where the posterior distributions are obtained similarly as in Section \ref{subsubsec:paraPostDistLCH}.

\section{Empirical Analysis: Danish Male Population}
\label{sec:Empirical}

In this section a real data empirical study\footnote{We have also performed numerous simulation studies using synthetic data to confirm the effectiveness of our estimation approaches but these are omitted here for space considerations. They are available upon request.} is conducted on Danish mortality data using the models summarised in Table \ref{table:ModelSummary}. The LC, LC-H and LCSV models are described in Section \ref{subsec:SSM}. While the LC-H model addresses heteroscedasticity in the observation equation, the LCSV model attempts to incorporate stochastic volatility in the state dynamics. The LCSV-H model includes both features of the LC-H model and the LCSV model, thus allowing for a full consideration of variability in long term mortality dynamics.

The Human Mortality Database\footnote{http://www.mortality.org/ (accessed on September 2015)} provides a particularly long time series of mortality data from year 1835 to 2011 for the Danish population, supplemented with a detailed document analysing the data (\cite{Andreev02}). The provision of a long time series is important to our analysis concerning stochastic volatility. In the past several decades, mortality trend for developed countries generally exhibit a rather smooth pattern. The inclusion of periods that involve wars, epidemics or other life-critical events are crucial factors in witnessing significant volatility in mortality time series. In the following we analyse the population mortality from Denmark based on the models in Table~\ref{table:ModelSummary} and Bayesian methodologies studied in this paper. We then examine the models in terms of the forecasting properties of death rates and life expectancies. We also comment on the linear trend assumption and jump-off bias in mortality forecasting.

\begin{table}[h]
\center \setlength{\tabcolsep}{1em}
\renewcommand{\arraystretch}{1.1}
\scalebox{0.9}{\begin{tabular}{lll}
\hline \hline
Model & Name & Dynamics \\
\hline
Lee-Carter (LC) model               & LC     & \eqref{eqn:LCy} - \eqref{eqn:LCk} \\
LC model with heteroscedasticity    & LC-H   & \eqref{eqn:LCH} \\
LC stochastic volatility (SV) model & LCSV   & \eqref{eqn:LCSV} \\
LC SV model with heteroscedasticity & LCSV-H & Combination of LC-H and LCSV  \\
\hline \hline
\end{tabular}}
\center\caption{\label{table:ModelSummary} \small{A summary of state-space mortality models considered in our empirical study.}}
\end{table}

\subsection{Data description}
\begin{figure}[h]
\begin{center}
\includegraphics[width=8cm, height=6cm]{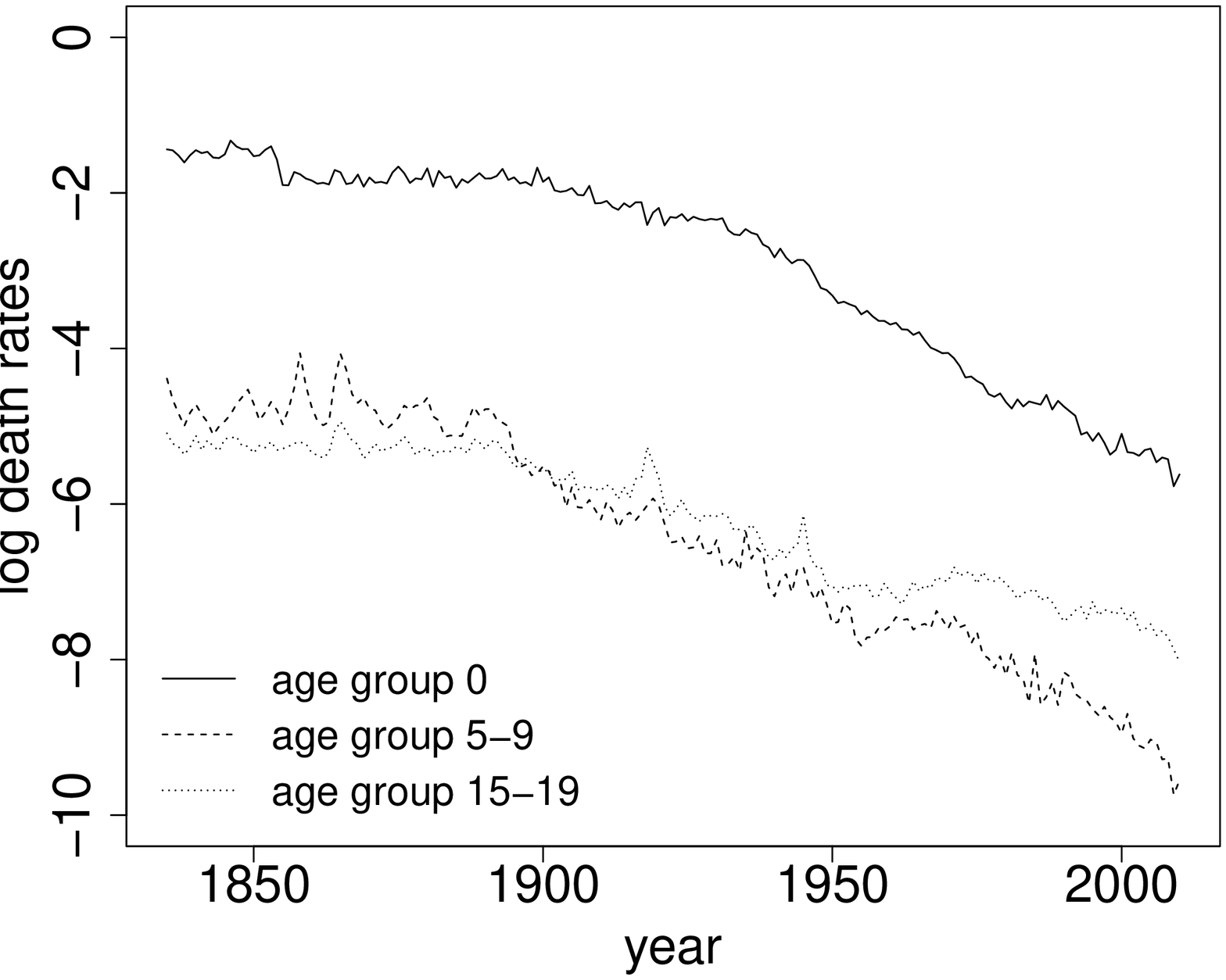}\includegraphics[width=8cm, height=6cm]{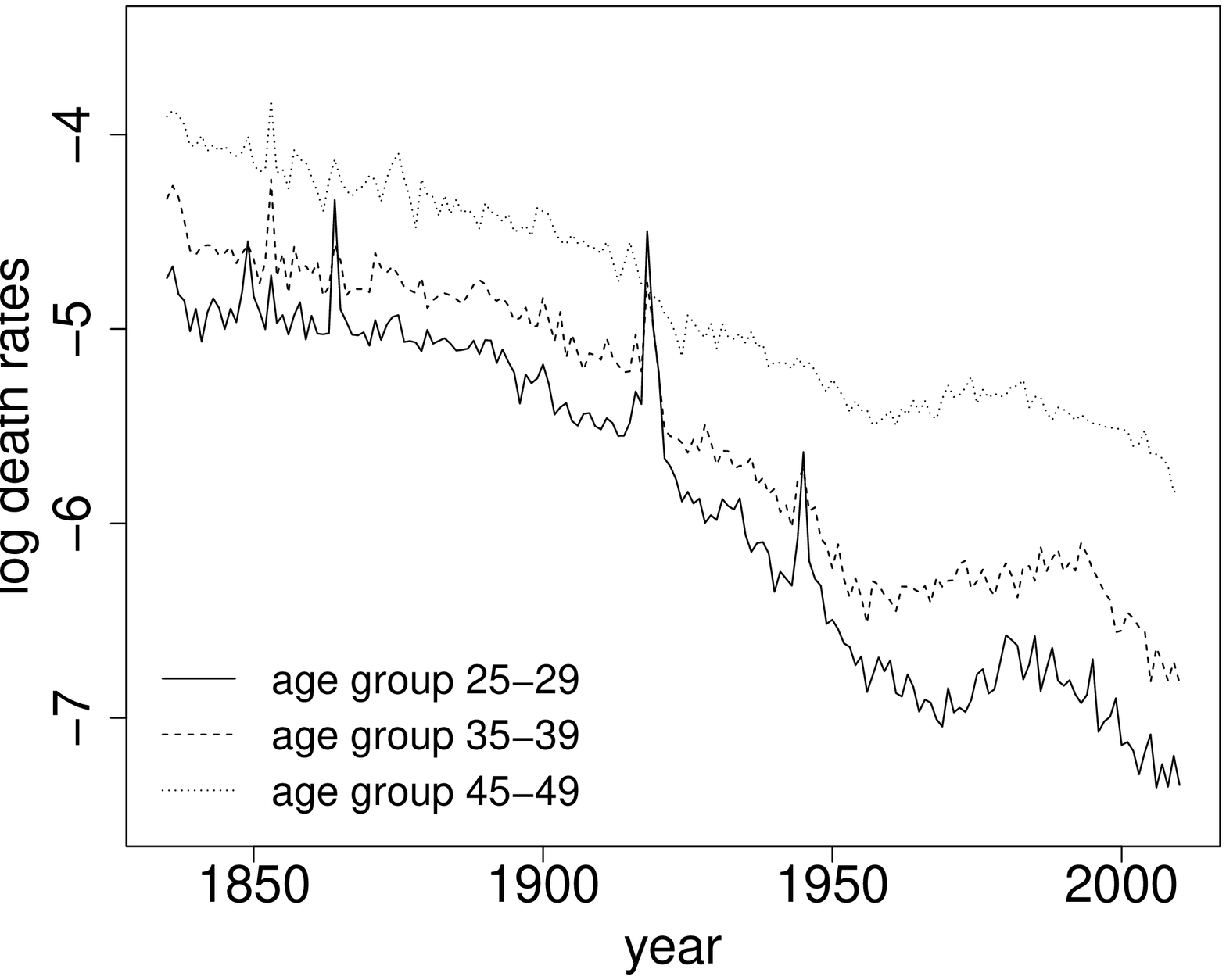}
\includegraphics[width=8cm, height=6cm]{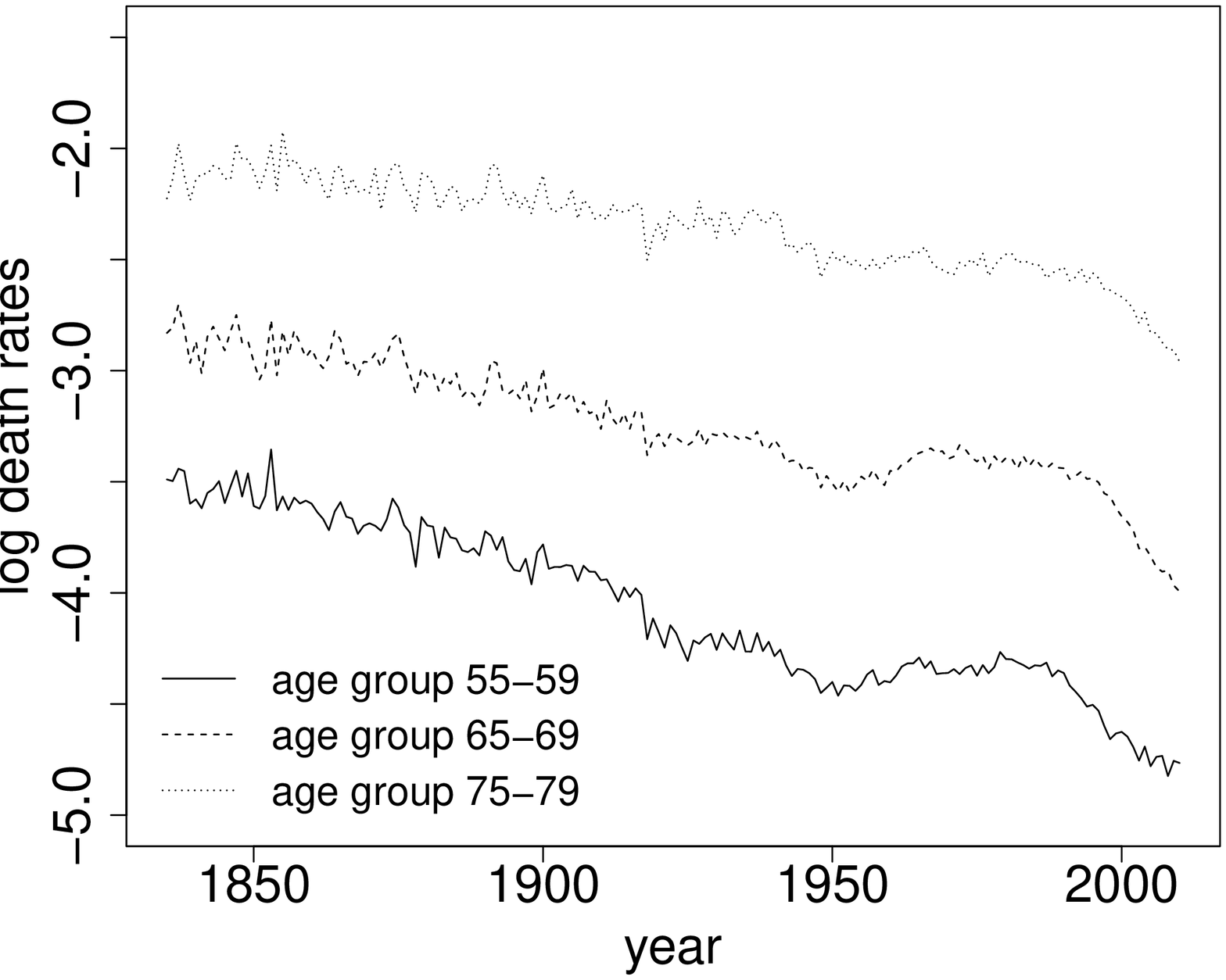}\includegraphics[width=8cm, height=6cm]{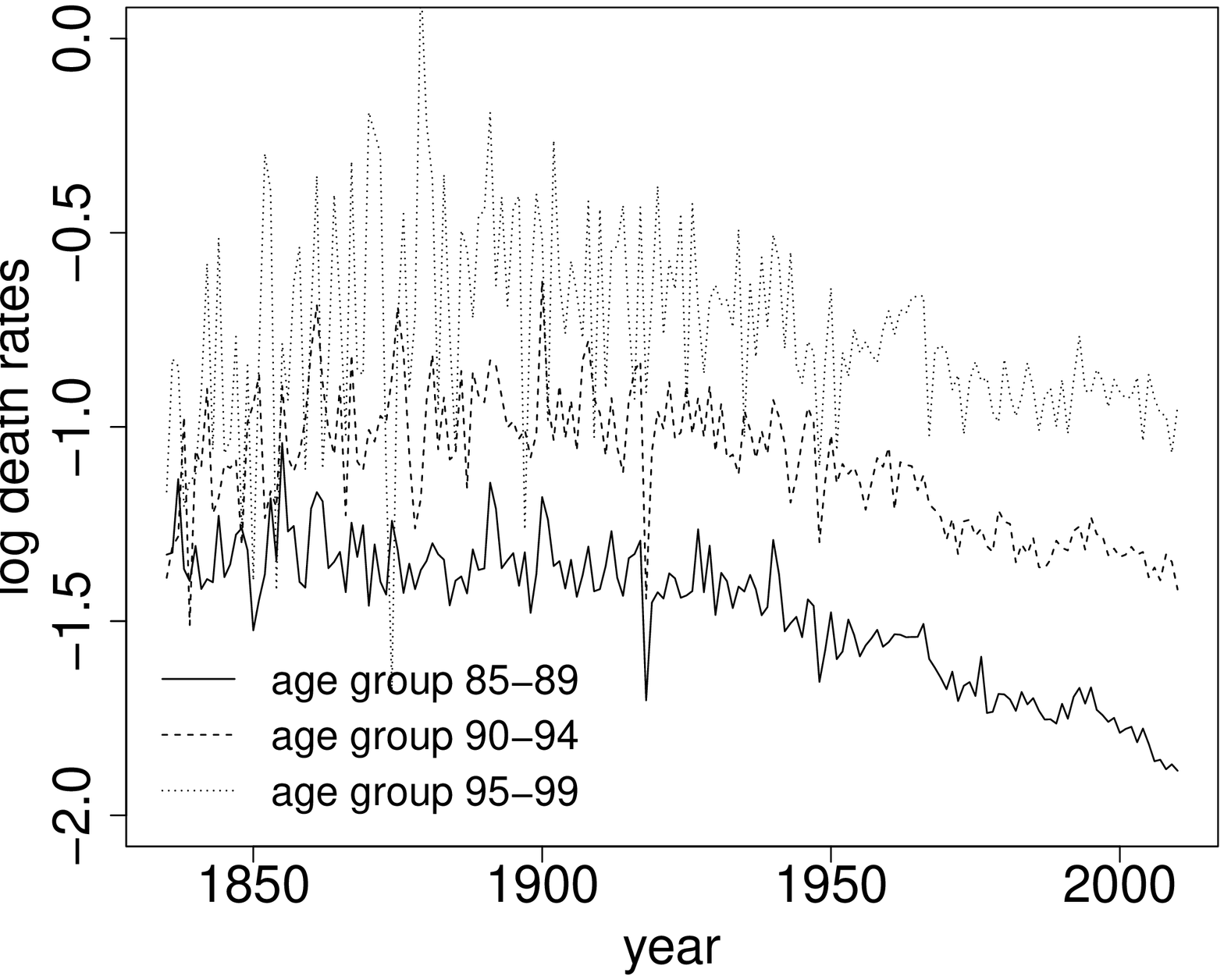}
\caption{\small{Time Series of log death rates for Danish male population from year 1835-2010.}}
\label{fig:timeSeriesDEN}
\end{center}
\end{figure}

The data set consists of Danish male population death rates for 21 age groups (0, 1-4, 5-9, $\dots$, 95-99) from year 1835-2010 where we fix year 2010 as the end year. Figure~\ref{fig:timeSeriesDEN} displays some of the time series of the log death rates for the Danish male population. It is clear that the multi-dimensional time series exhibit different volatility for different age groups, which justify the introduction of heteroscedasticity into the observation equation as discussed in Section~\ref{sec:LChetero}. We also observe that, mainly before 1950, there are periods that the volatility of death rates for some age groups are markedly different. Such a change of volatility in the temporal dimension suggests that stochastic volatility may be present in the underlying time preiod effect.

\subsection{Estimation Results}
In our empirical study we focus on Bayesian inference and forecasting. We assume vague priors so that all inferences are mainly based on the data and the impact of the prior is not material. Taking the LCSV model as an example, we assume $\kappa_0 \sim \text{N}(0,10)$, $\alpha_x \sim \text{N}(0,10)$, $\beta_x \sim \text{N}(0,10)$, $\theta \sim \text{N}(0,10)$, $\sigma^2_{\varepsilon} \sim \text{IG}(2.001, 0.001)$, $\sigma^2_\gamma \sim \text{IG}(2.001, 0.001)$, $\lambda_1 \sim \text{N}(0, 10)$, $\lambda_2 \sim \text{N}(0, 10)$ and $\gamma_0 \sim \text{N}(0, 10)$, where $x \in \{x_2,\dots,x_p\}$. The number of iterations of the Markov chain is $15000$ with $5000$ burn-in. We fix $\alpha_{x_1}= \frac{1}{T}\sum_{t=1}^T y_{x_1,t}$ and $\beta_{x_1}=0.2$ as an identification constraint.

\begin{figure}[h]
\begin{center}
\includegraphics[width=5.5cm, height=5cm]{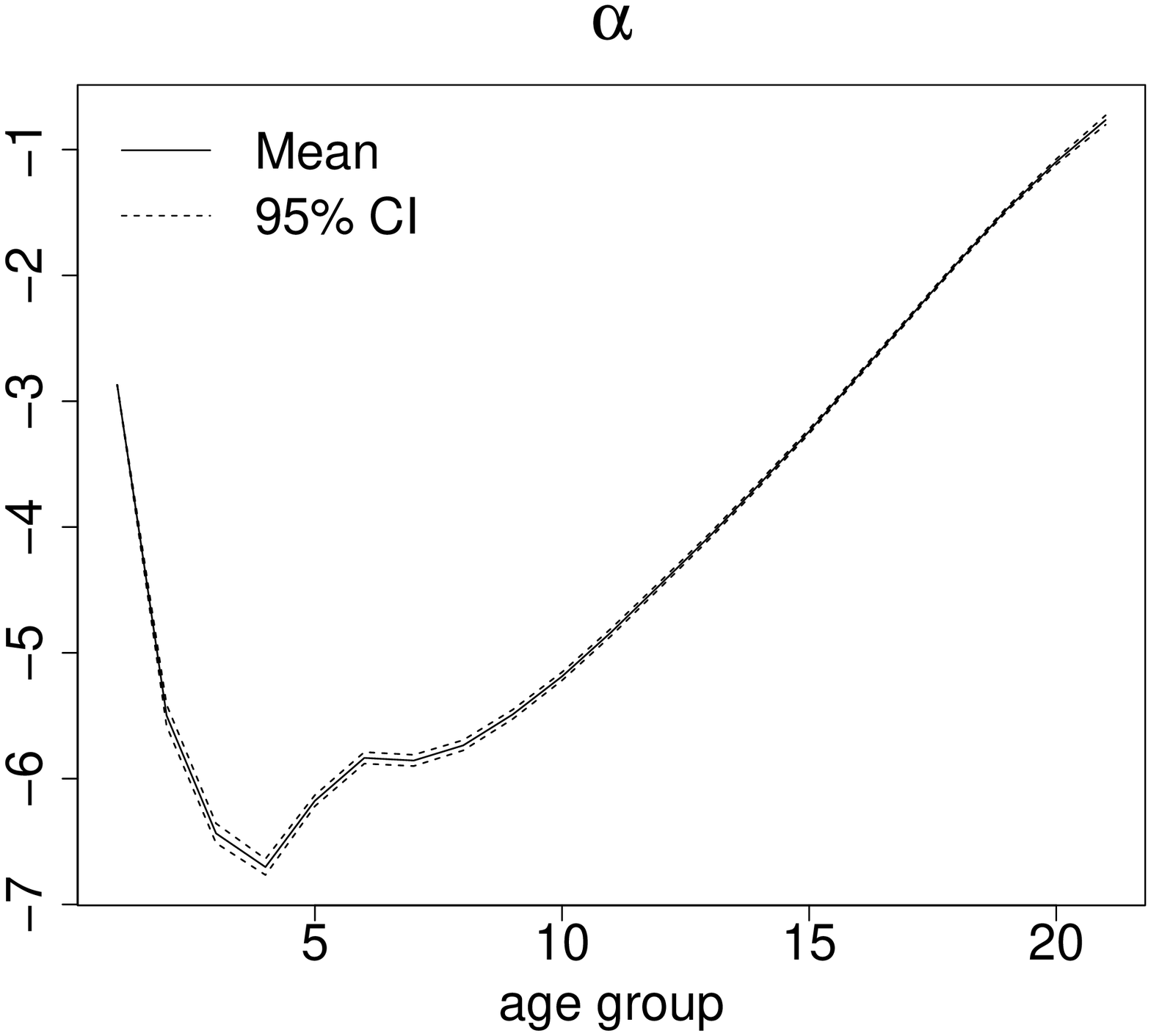}\includegraphics[width=5.5cm, height=5cm]{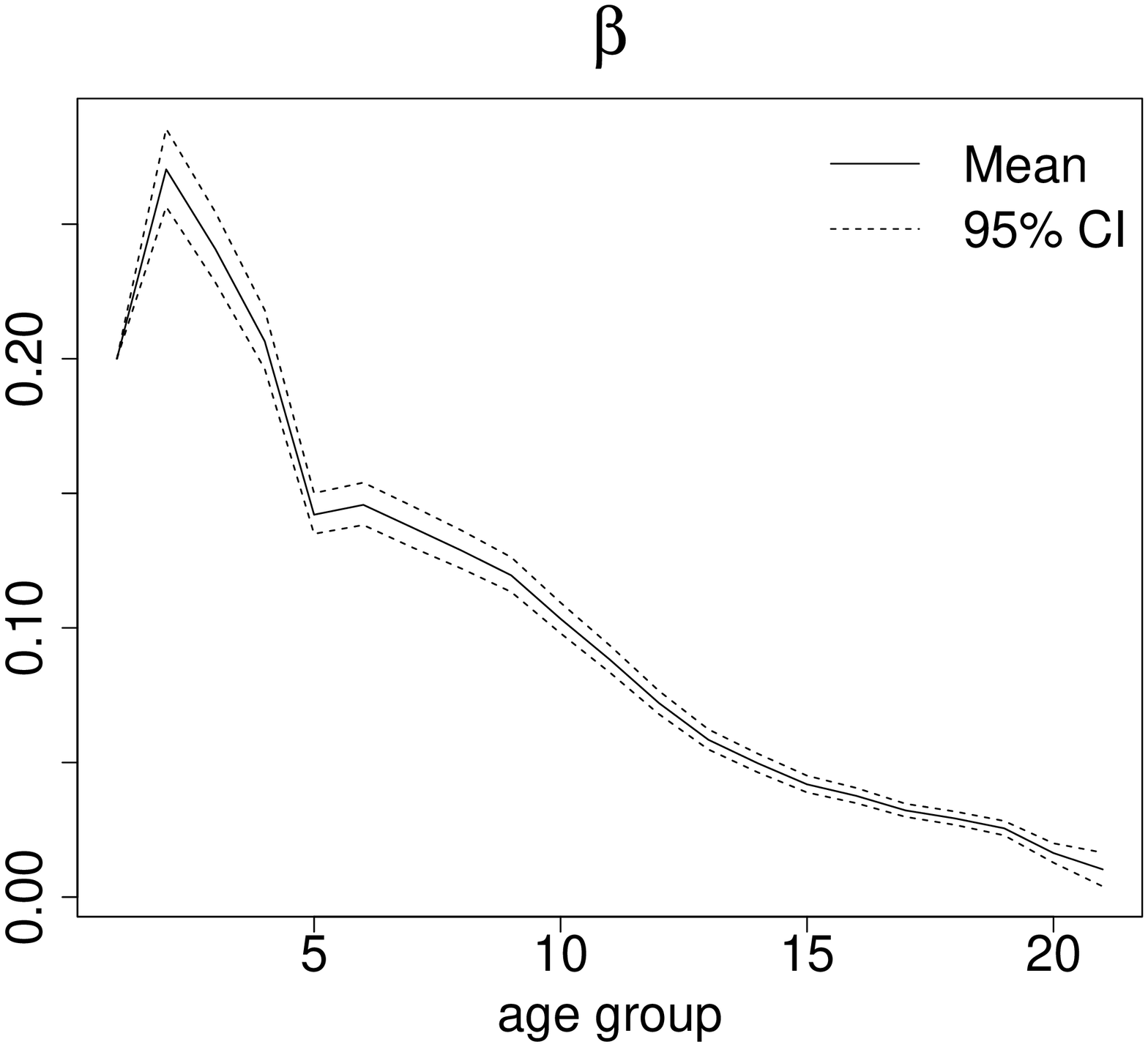}\includegraphics[width=5.5cm, height=5cm]{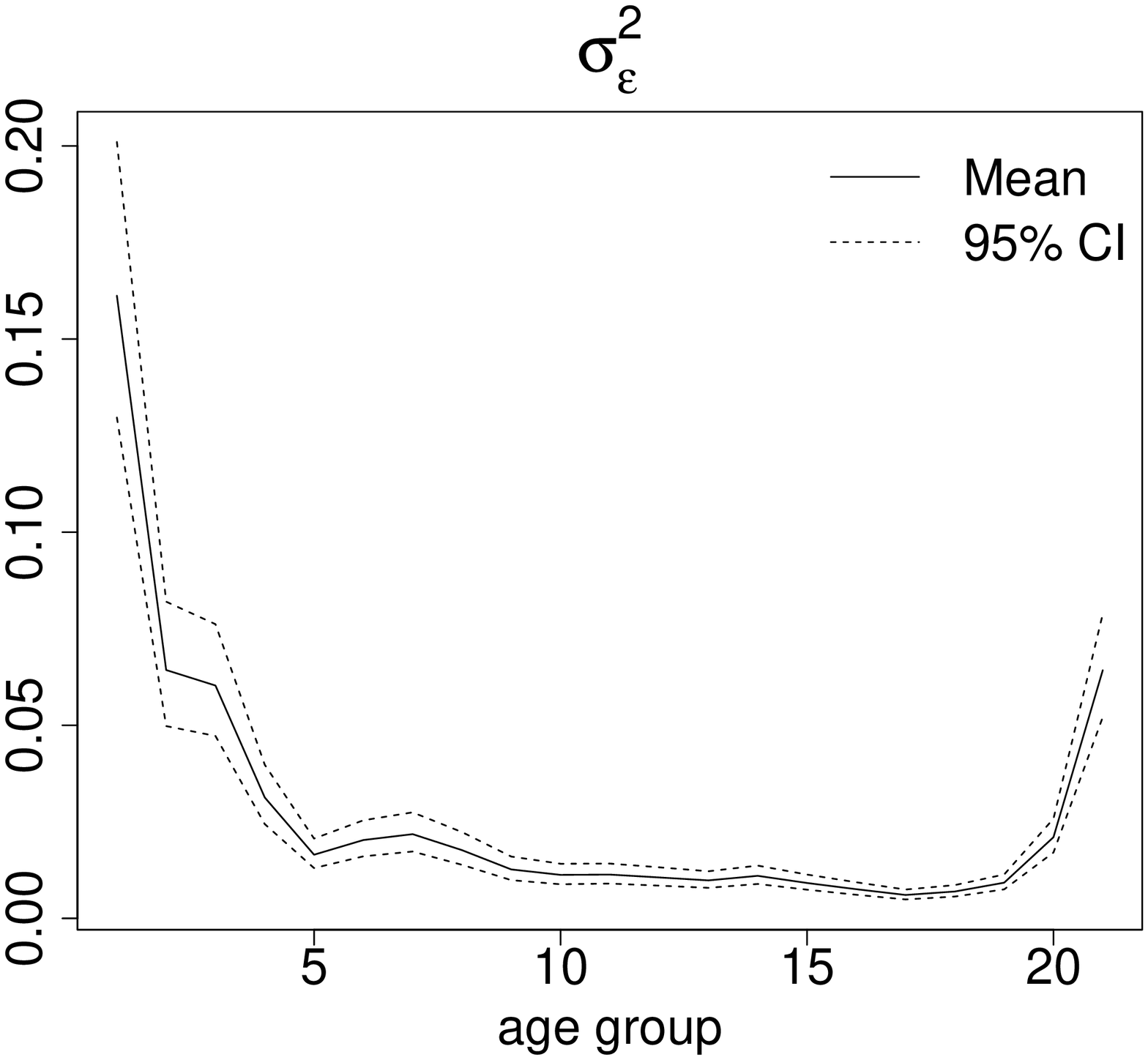}
\includegraphics[width=5.5cm, height=5cm]{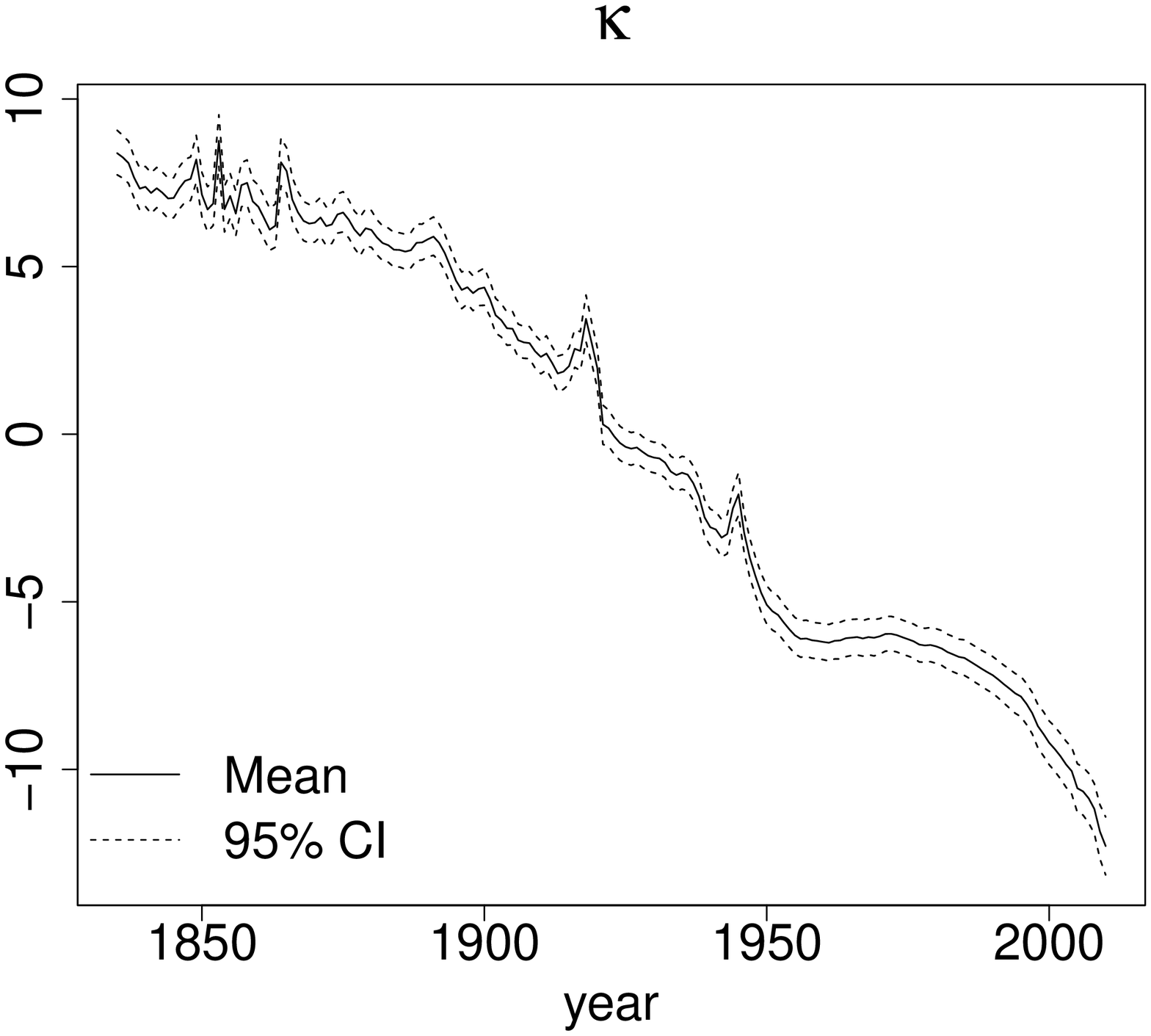}\includegraphics[width=5.5cm, height=5cm]{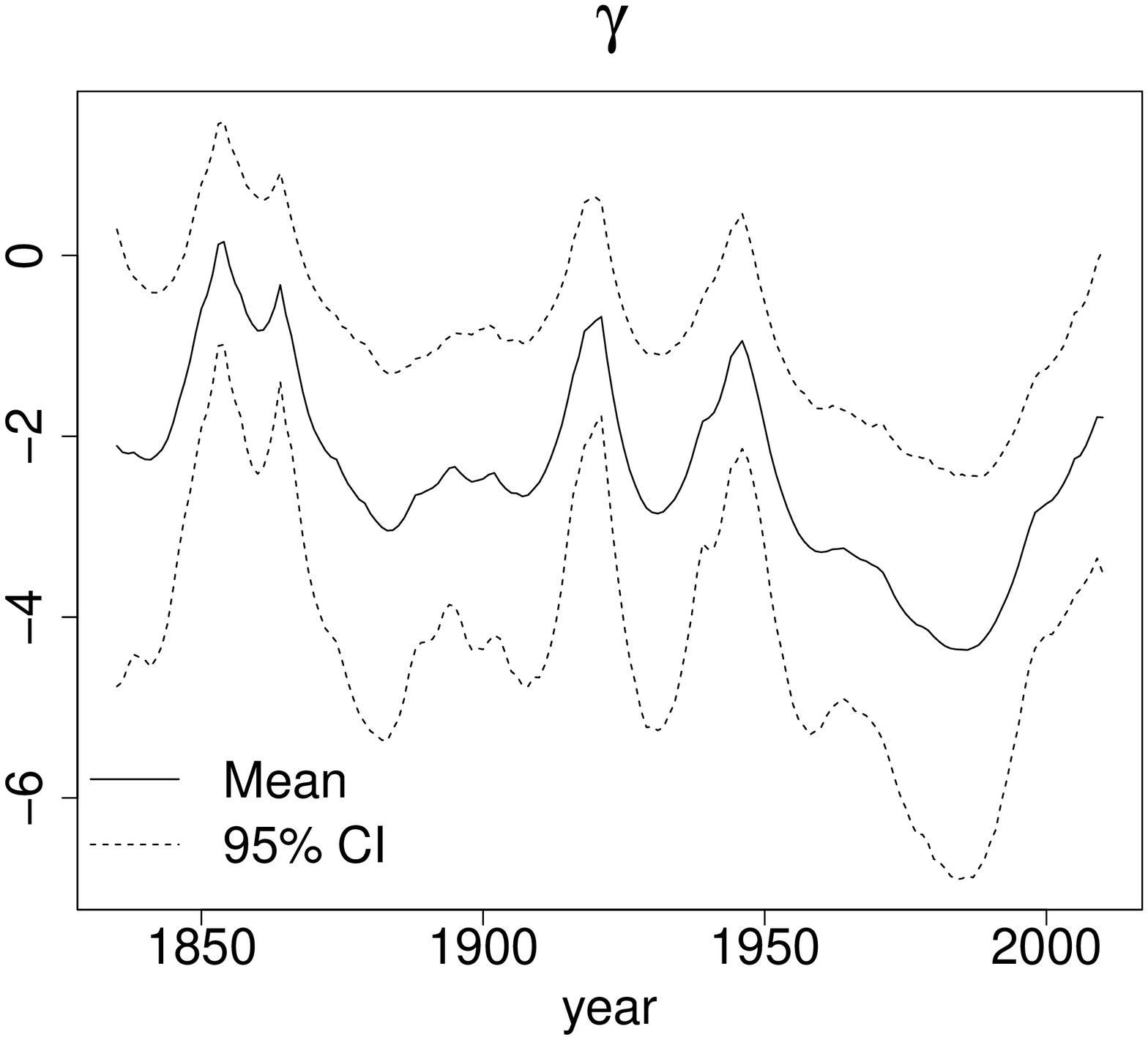}\includegraphics[width=5.5cm, height=5cm]{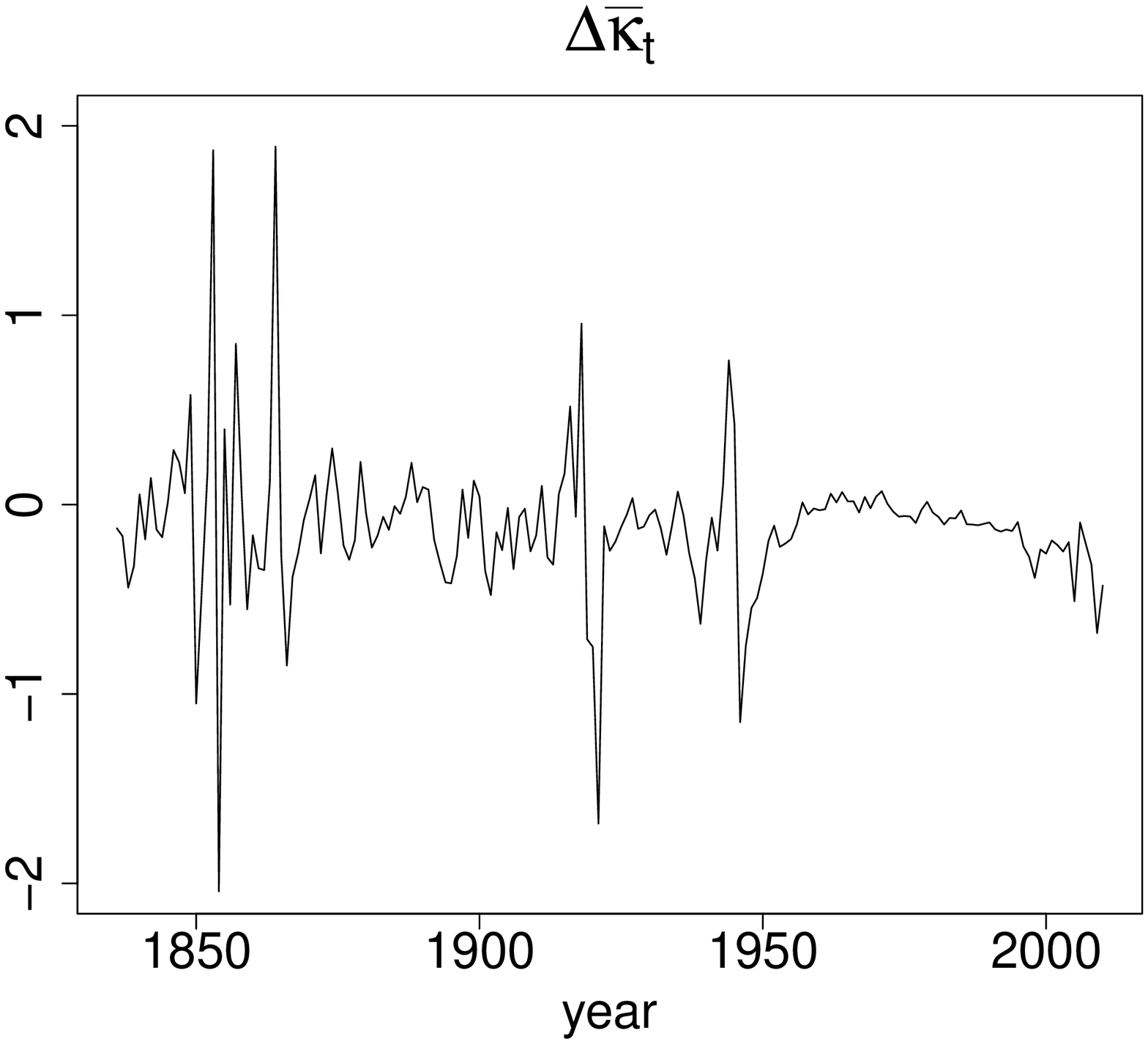}
\caption{\small{Estimation of (upper panels) $\boldsymbol{\alpha}$, $\boldsymbol{\beta}$ and $\sigma^2_{x_1:x_{21},\varepsilon}$; (lower panels) time effect $\kappa_{1834:2010}$, log-volatility $\gamma_{1835:2010}$ and first difference $\Delta \bar{\kappa}_t$, for Danish male mortality data (1835-2010) using the LCSV-H model.}}
\label{fig:DENLCSVH18352010}
\end{center}
\end{figure}

Estimated values of the static parameters (except $\boldsymbol{\alpha}$ and $\boldsymbol{\beta}$) for the Danish mortality data (1835-2010) are shown in Table~\ref{table:DENStaticPara}. The rest of the estimated parameters and states are displayed in Figure~\ref{fig:DENLCSVH18352010}. Here we only show the plots for the LCSV-H model since the corresponding figures obtained from the LC, LC-H and LCSV model are visually similar to the case of the LCSV-H model.

It is evident from Figure~\ref{fig:DENLCSVH18352010} that there are periods, namely 1850-1870, 1910-20, 1930-1950, that the time effect $\kappa$ exhibits higher volatility compared with other periods. We also observe that $\kappa$ accelerates markedly downward after 1990 and is relatively smooth in the recent period 1950-2010. The filtering of the log-volatility process $\gamma_{1835:2010}$ (Figure~\ref{fig:DENLCSVH18352010}) quantifies the volatility level ($e^{\gamma_t}$) of the time effect and gives further evidence on the stochastic volatility nature of mortality. To see more clearly the phenomenon of changing volatility, we plot the first difference $\Delta \bar{\kappa}_t = \bar{\kappa}_t-\bar{\kappa}_{t-1}$  in Figure~\ref{fig:DENLCSVH18352010} for the LCSV-H model, where $\bar{\kappa}_t$ denotes the posterior mean of $\kappa_t$, $t=1836,\dots,2010$. It shows evidently the change of volatility level in the latent process $\kappa_t$. The patterns of the estimated log-volatility $\gamma_{1835:2010}$ and the first difference $\Delta \bar{\kappa}_t$ clearly suggest that it is not appropriate to assume constant volatility ($\sigma^2_\omega$) for the time effect.

The state-space modelling approach is able to uncover the age-specific heteroscedasticity structure hidden in the Danish mortality time series. Figure~\ref{fig:DENLCSVH18352010} reveals that variability is particularly high for the very young and very old age group. Implications of the heteroscedastic structure on forecasting will be discussed in Section~\ref{sec:forecast}.

\begin{figure}[h]
\begin{center}
\includegraphics[width=5.5cm, height=5cm]{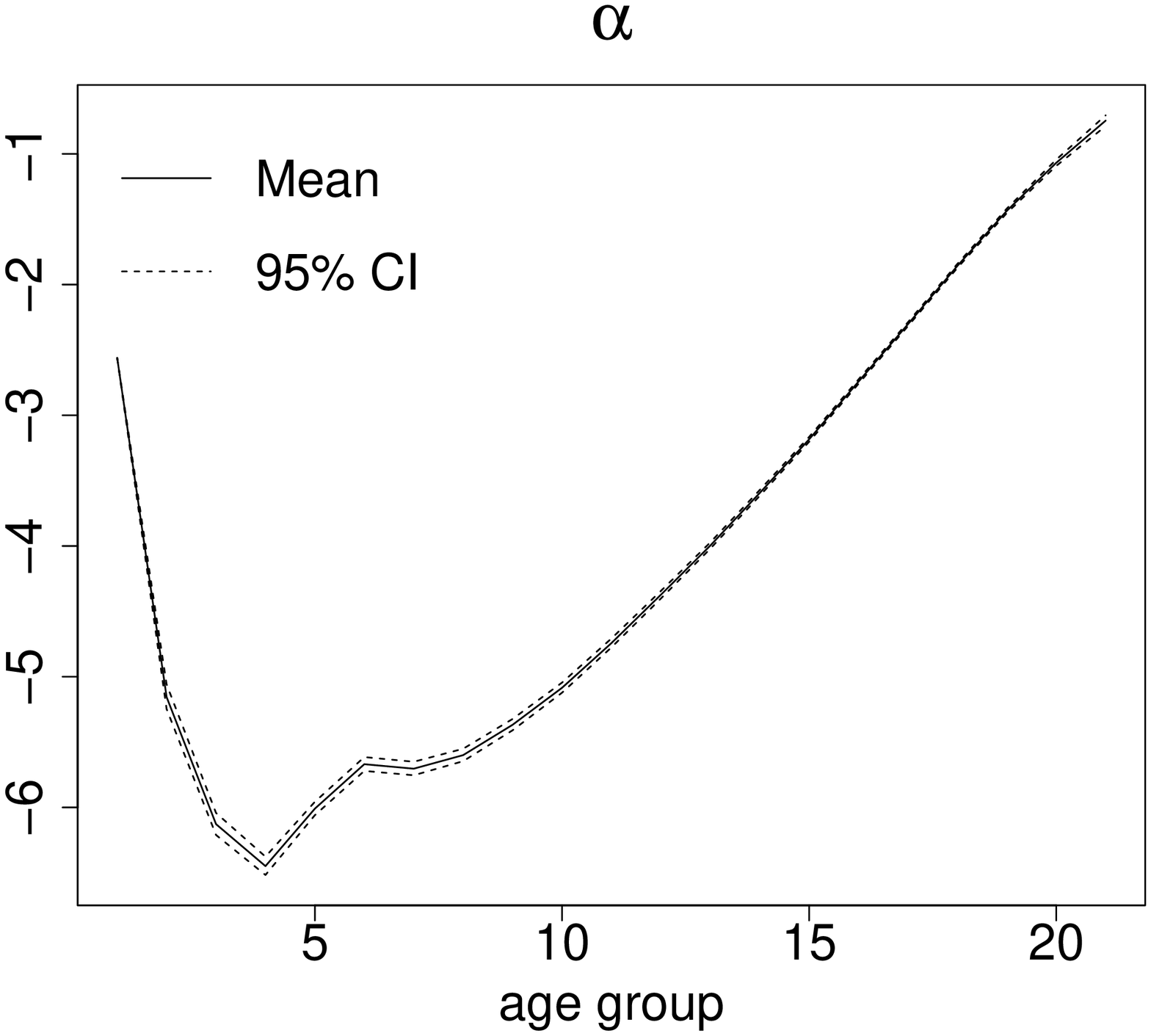}\includegraphics[width=5.5cm, height=5cm]{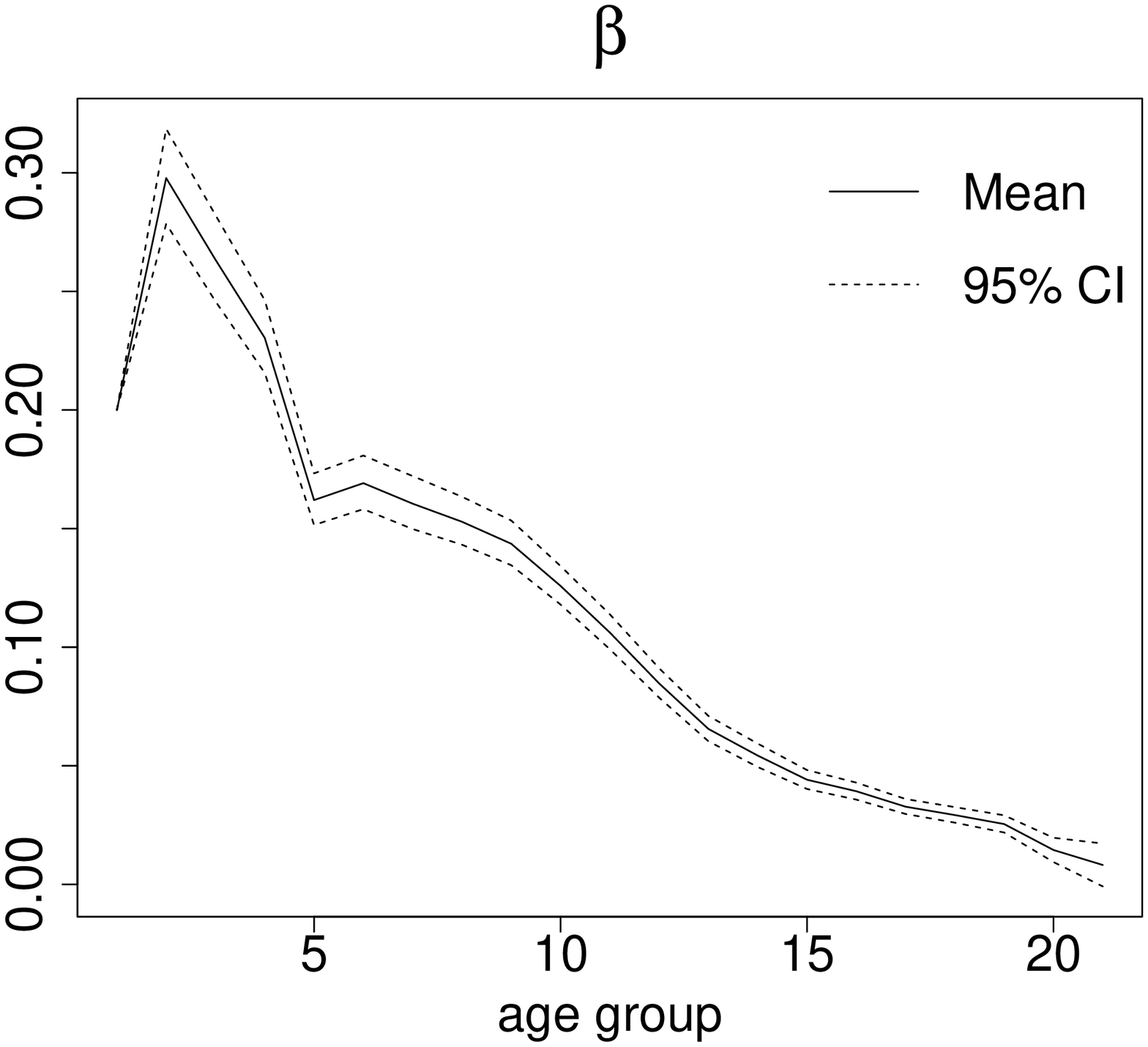}\includegraphics[width=5.5cm, height=5cm]{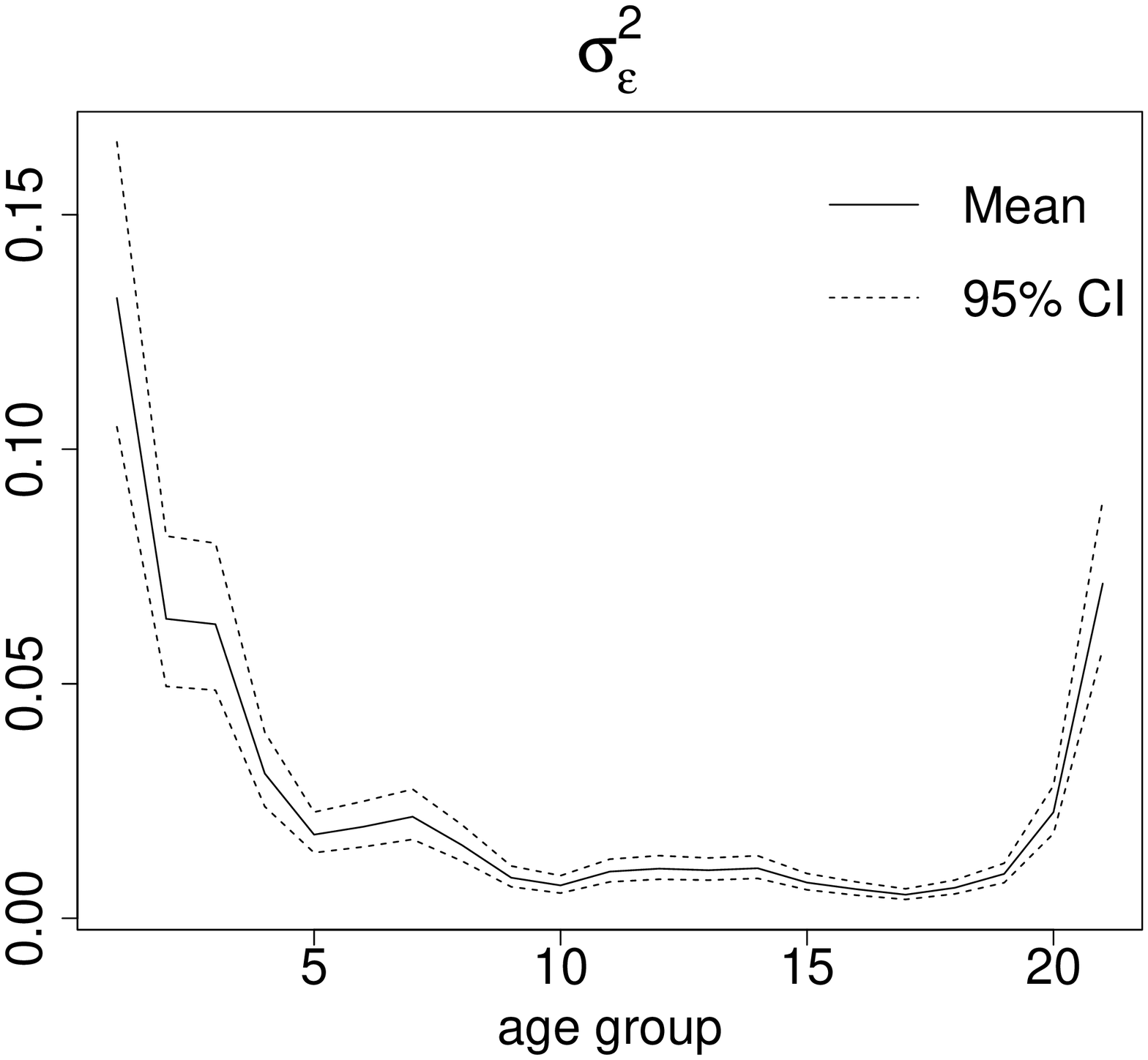}
\includegraphics[width=5.5cm, height=5cm]{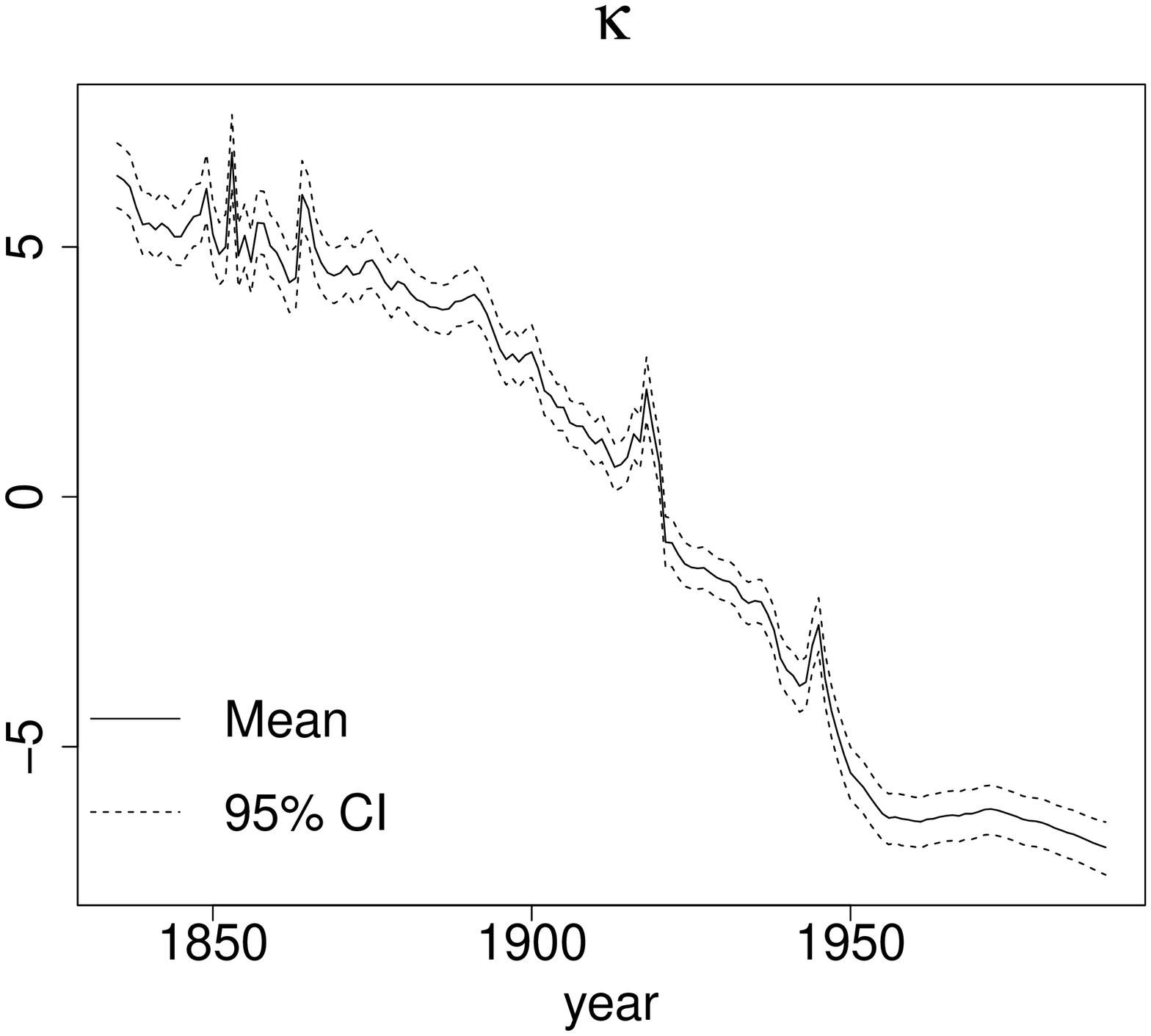}\includegraphics[width=5.5cm, height=5cm]{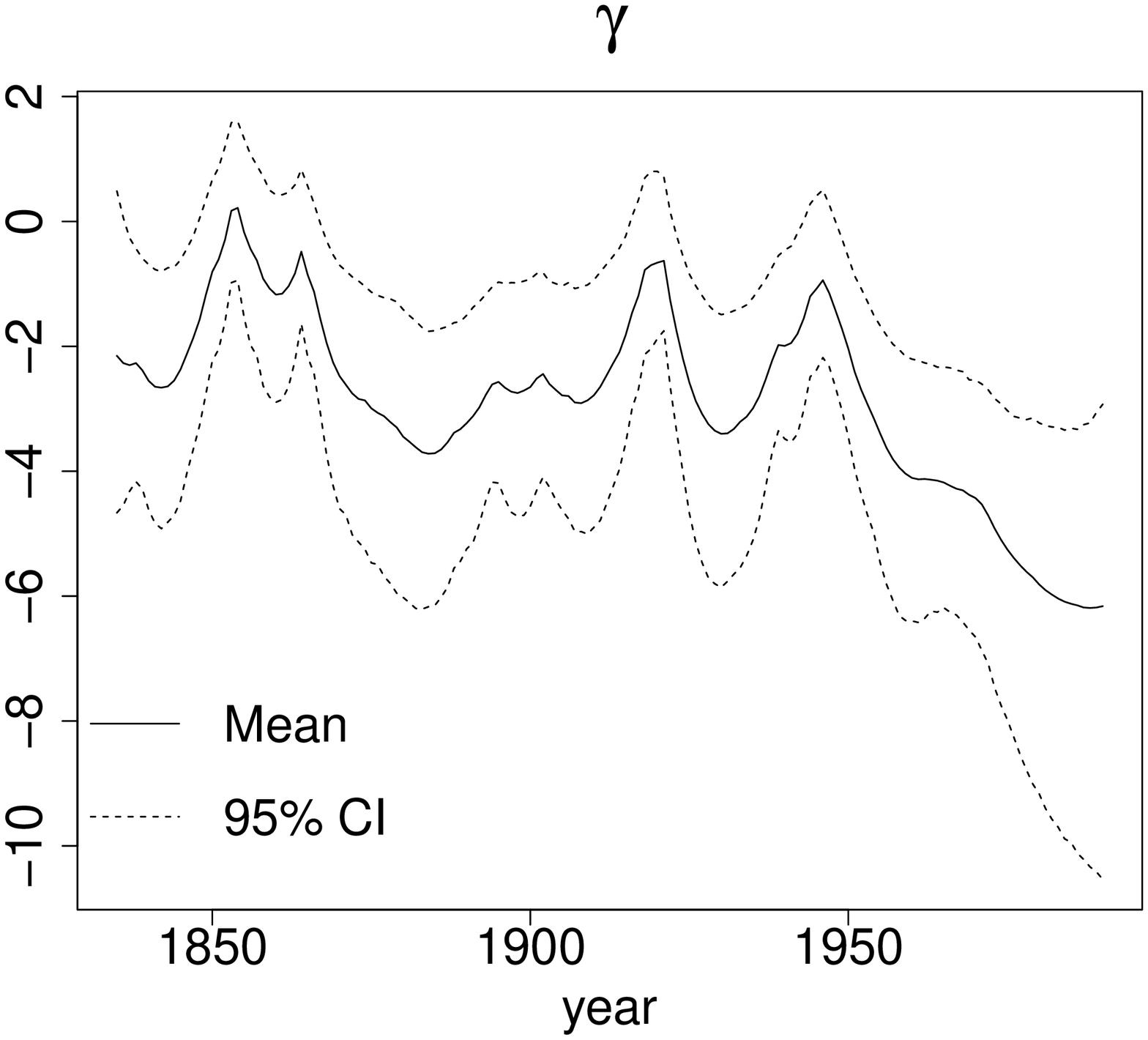}\includegraphics[width=5.5cm, height=5cm]{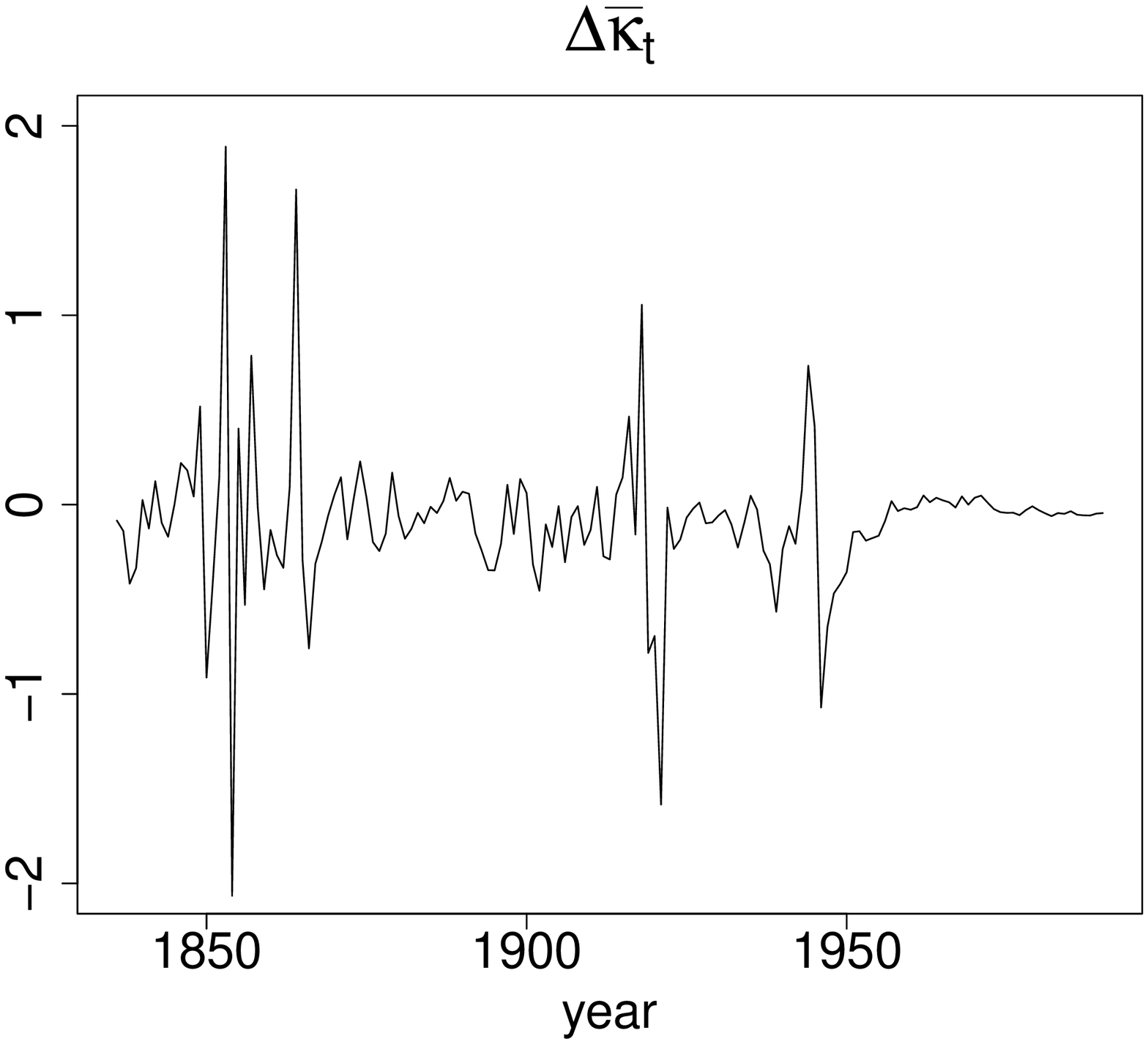}
\caption{\small{Estimation of (upper panels) $\boldsymbol{\alpha}$, $\boldsymbol{\beta}$ and $\sigma^2_{x_1:x_{21},\varepsilon}$; (lower panels) time effect $\kappa_{1834:1990}$, log-volatility $\gamma_{1835:1990}$ and first difference $\Delta \bar{\kappa}_t$, for Danish male mortality data (1835-1990) using the LCSV-H model.}}
\label{fig:DENLCSVH18351990}
\end{center}
\end{figure}

\begin{figure}[h]
\begin{center}
\includegraphics[width=5.5cm, height=5cm]{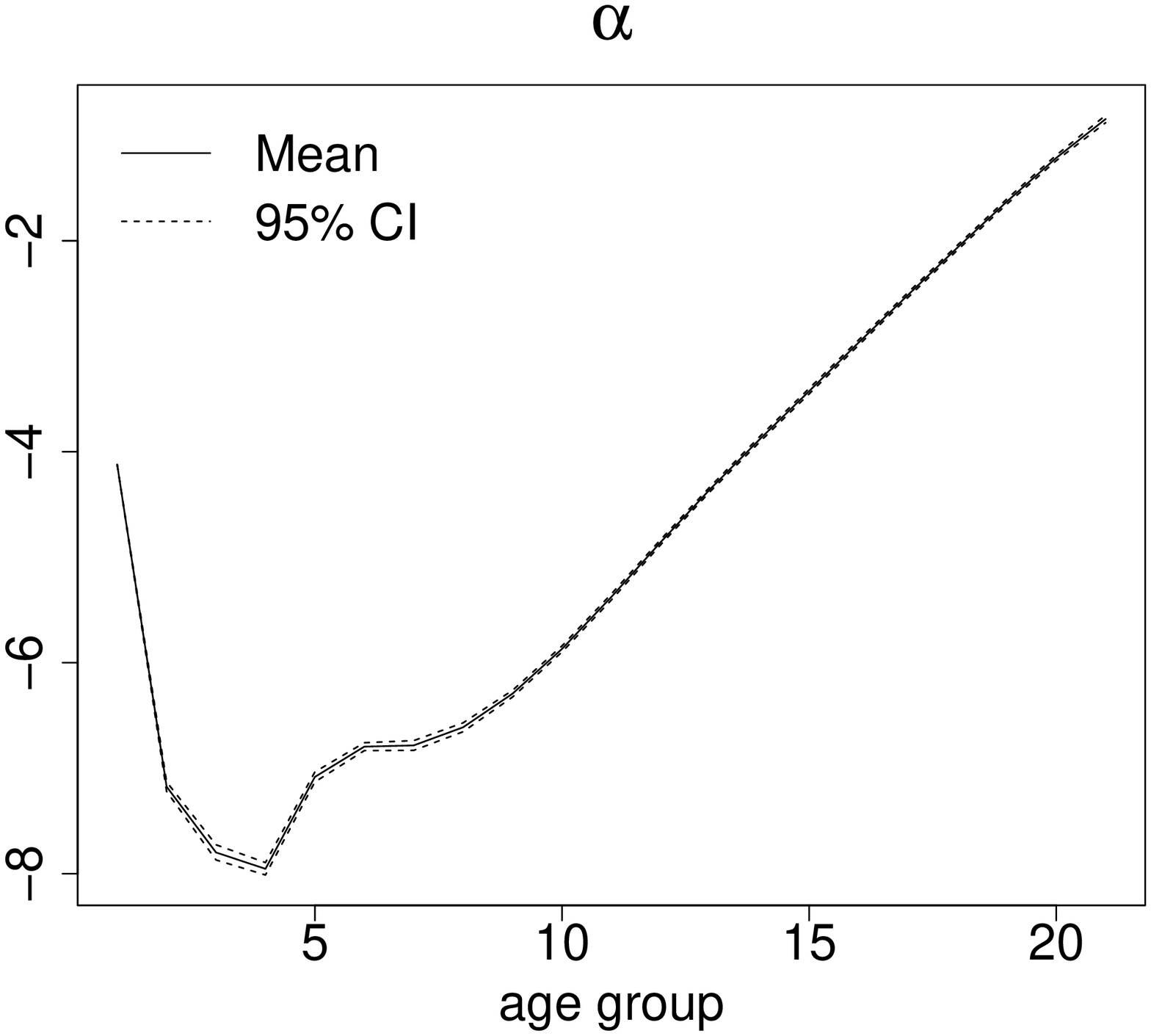}\includegraphics[width=5.5cm, height=5cm]{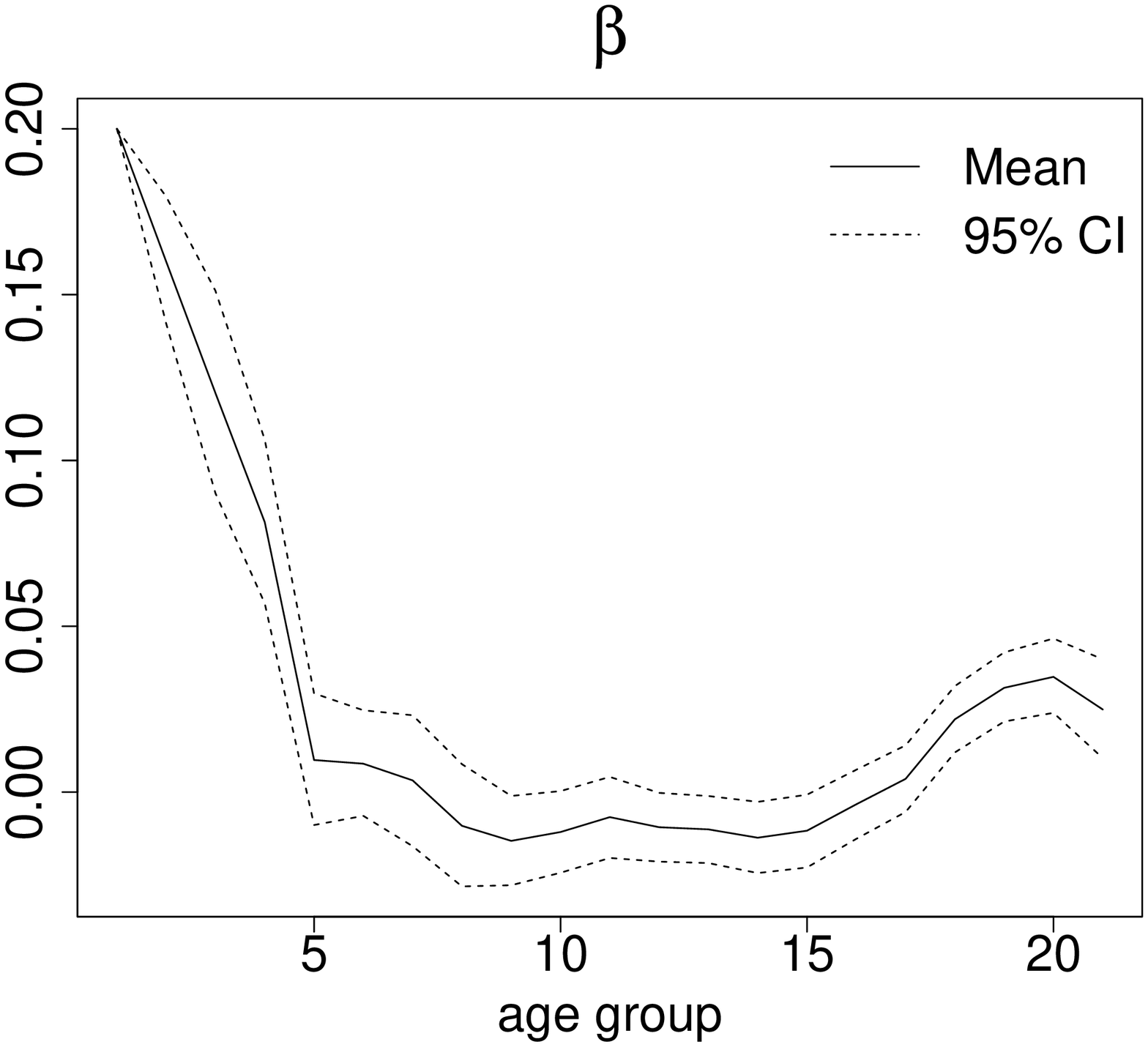}\includegraphics[width=5.5cm, height=5cm]{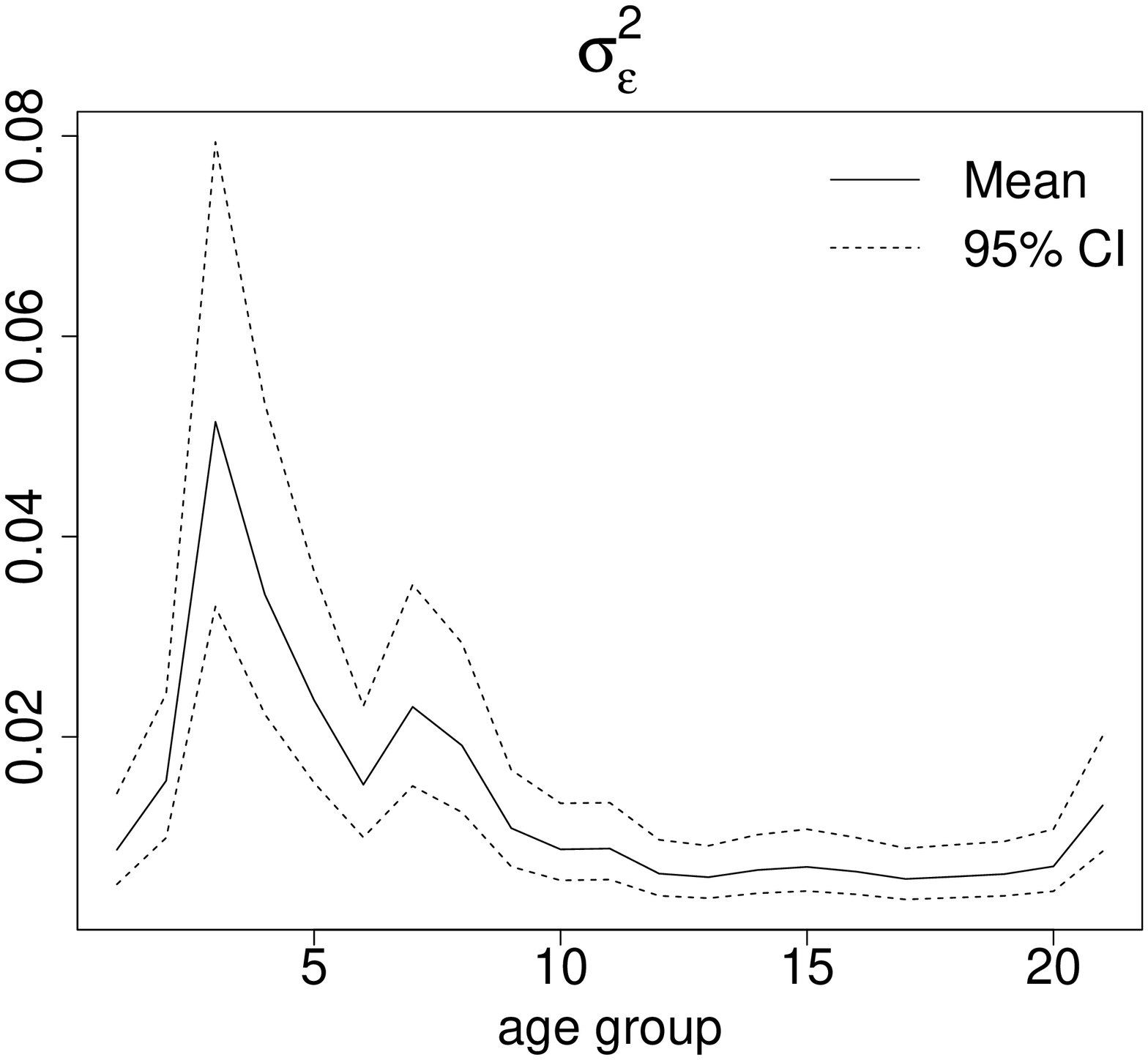}
\includegraphics[width=5.5cm, height=5cm]{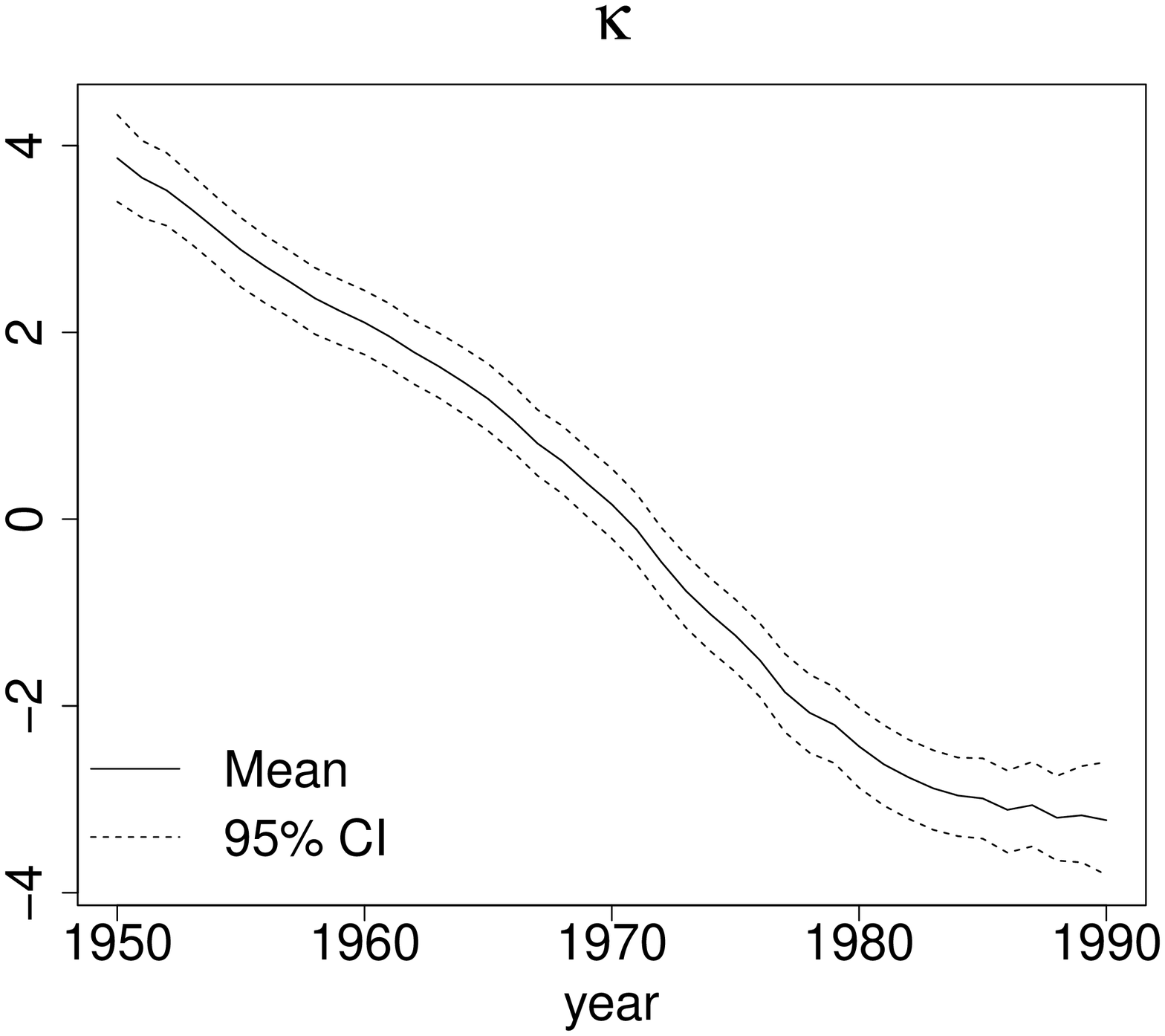}\includegraphics[width=5.5cm, height=5cm]{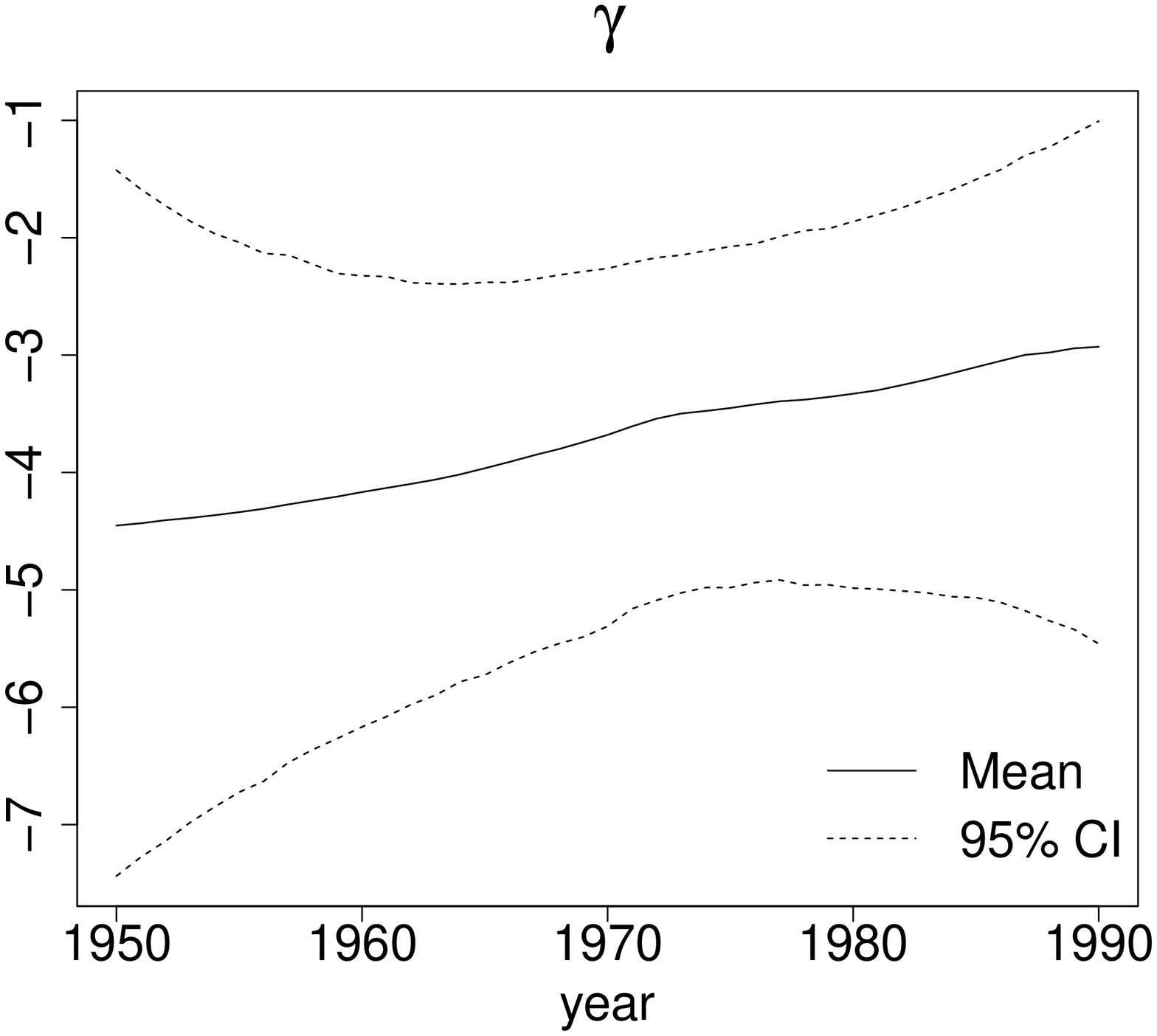}\includegraphics[width=5.5cm, height=5cm]{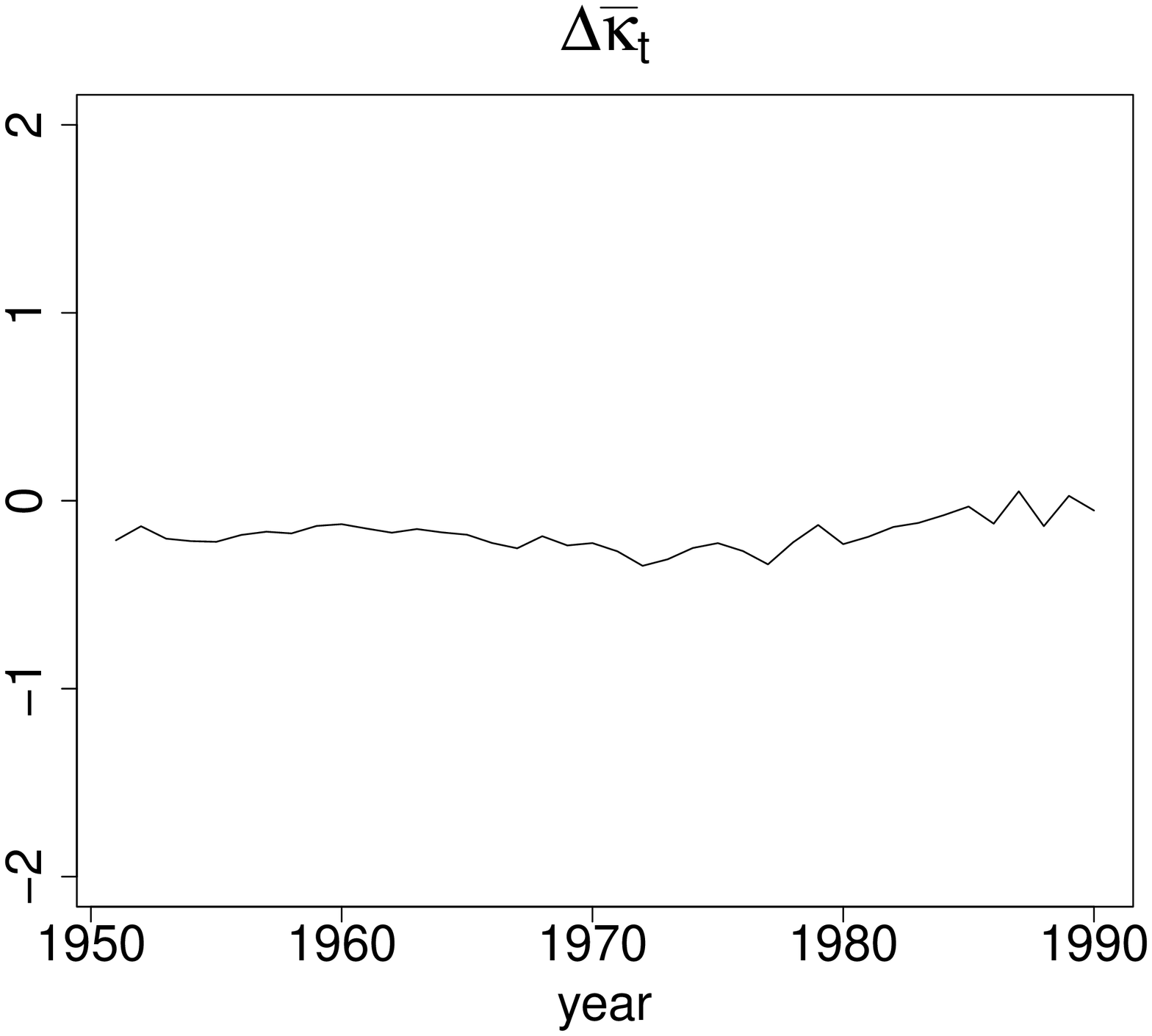}
\caption{\small{Estimation of (upper panels) $\boldsymbol{\alpha}$, $\boldsymbol{\beta}$ and $\sigma^2_{x_1:x_{21},\varepsilon}$; (lower panels) time effect $\kappa_{1949:1990}$, log-volatility $\gamma_{1950:1990}$ and first difference $\Delta \bar{\kappa}_t$, for Danish male mortality data (1950-1990) using the LCSV-H model.}}
\label{fig:DENLCSVH19501990}
\end{center}
\end{figure}

To investigate the forecasting properties of the stochastic volatility model, we also estimate the models based on calibration periods 1835-1990 and 1950-1990. Figure~\ref{fig:DENLCSVH18351990} and \ref{fig:DENLCSVH19501990} show the estimated parameters and states for the LCSV-H model in those periods.

\begin{table}[h]
\center \setlength{\tabcolsep}{0.7em}
\renewcommand{\arraystretch}{1.0}
\scalebox{0.9}{\begin{tabular}{c|c|c|c|c}
\hline \hline
 & LC & LC-H & LCSV & LCSV-H  \\
\hline
$\theta$               & -0.11 (-0.17, -0.06) & -0.11 (-0.17, -0.06)            & -0.11 (-0.15, -0.07)  & -0.09 (-0.14, -0.04) \\
$\sigma^2_\varepsilon$ & 0.023 (0.022, 0.024) & Similar to Fig.~\ref{fig:DENLCSVH18352010} & 0.023 (0.022, 0.024)  & Fig.~\ref{fig:DENLCSVH18352010} \\
$\sigma^2_\omega$      & 0.13 (0.09, 0.18)    & 0.15 (0.10, 0.21)               & N.A.                  & N.A.                    \\
$\lambda_1$            & N.A.                 & N.A.                            & 0.989 (0.962, 0.999)  & 0.984 (0.949, 0.999)    \\
$\lambda_2$            & N.A.                 & N.A.                            & -0.025 (-0.11, 0.042) & -0.03 (-0.15, 0.05)   \\
$\sigma^2_\gamma$      & N.A.                 & N.A.                            & 0.15 (0.03, 0.48)     & 0.25 (0.06, 0.67)       \\
$\gamma_0$             & N.A.                 & N.A.                            & -2.09 (-4.52, 0.23)   & -2.11 (-5.04, 0.47)     \\
\hline \hline
\end{tabular}}
\center\caption{\label{table:DENStaticPara}\small{Estimated values of the static parameters (except $\boldsymbol{\alpha}$ and $\boldsymbol{\beta}$) for the Danish male mortality data (1835-2010). The range in $(,)$ represents $95\%$ credible interval. (N.A.: Not Applicable)}}
\end{table}

\subsection{Model Assessment}

To compare the fit of the models to the data, we apply deviance information criterion (DIC) as a Bayesian measures of model complexity and fit (\cite{Spiegelhalteretal02}). It is common to assess and compare models with latent variables using conditional DIC (\cite{BergMeYu04}, \cite{CeleuxFoRoTi06}). Specifically, we use the so-called conditional log-likelihood which is calculated as
\begin{equation}
    \ln f(\boldsymbol{y}_{1:T}|\boldsymbol{\psi},\kappa_{1:T}) = \sum_{x=x_1}^{x_p} \sum_{t=1}^T \left(-\frac{1}{2}\ln 2\pi -\ln \sigma_{\varepsilon,x}
        -\frac{1}{2} \left(\frac{y_{x,t}-(\alpha_x+\beta_x \kappa_t)}{\sigma_{\varepsilon,x}}\right)^2 \right).
\end{equation}
Note that the likelihood is conditional on parameters that include both static parameters and the latent process $\kappa$. Using the conditional log-likelihood function, the deviance is defined as
\begin{equation}
    D(\boldsymbol{\Psi}) = -2\ln f(\boldsymbol{y}_{1:T}|\boldsymbol{\Psi}) + 2\ln h(\boldsymbol{y}_{1:T}),
\end{equation}
where $\boldsymbol{\Psi} = (\boldsymbol{\psi},\kappa_{1:T})$ and we assume $h(\boldsymbol{y}_{1:T})=1$ since in the models we consider it plays the role of a constant which is the same for competing models. The effective dimension, $p_{D}$, is evaluated as
\begin{equation}
    p_{D} = \bar{D}(\boldsymbol{\Psi})- D(\bar{\boldsymbol{\Psi}}),
\end{equation}
where $\bar{D}(\boldsymbol{\Psi})$ and $\bar{\boldsymbol{\Psi}}$ denote, respectively, the mean of $D(\boldsymbol{\Psi})$ and the mean of the posterior distribution of $\boldsymbol{\Psi}$. The conditional DIC is then given by
\begin{equation}
    \text{DIC} := \bar{D}(\boldsymbol{\Psi}) + p_{D} = 2\bar{D}(\boldsymbol{\Psi}) - D(\bar{\boldsymbol{\Psi}}),
\end{equation}
which can be evaluated straightforwardly using MCMC samples.

\begin{table}[h]
\center \setlength{\tabcolsep}{0.7em}
\renewcommand{\arraystretch}{1.0}
\scalebox{1.0}{\begin{tabular}{c|c|c|c}
\hline \hline
Calibration period: & 1835 - 2010 & 1835 - 1990 & 1950 - 1990  \\
\hline
LC     & -3218.6 & -3087.5  & -1567.3 \\
LC-H   & -4469.1 & -4269.7  & -1793.6 \\
LCSV   & -3250.8 & -3109.7  & -1559.7 \\
LCSV-H & -4518.3 & -4326.8  & -1794.1 \\
\hline \hline
\end{tabular}}
\center\caption{\label{table:DIC}\small{DIC of models with different calibration periods.}}
\end{table}

The DIC values for the models with different calibration periods are shown in Table~\ref{table:DIC}.\footnote{The lower the DIC value, the better the model in terms of a trade-off of fit and complexity.} The inclusion of heteroscedasticity structure has markedly improved the LC and LCSV model. For long calibration period, the LCSV model has outperformed the LC model. It indicates that the better fit of the LCSV model has more than compensated for its increased complexity. For short calibration period (1950 - 1990), the LC model performed better than the LCSV model which is expected, since over the short period the evolution of mortality rates is rather smooth and there is no clear advantage in introducing stochastic volatility to the LC model.

\subsection{Forecasting}
\label{sec:forecast}
In this section, we investigate the forecasting properties of the mortality models summarised in Table~\ref{table:ModelSummary} where heteroscedasticity as well as stochastic volatility structures are incorporated. Our analysis is based on the forecasting distributions of (log) death rates and life expectancy. The Bayesian state-space framework allows us to obtain the forecasting distributions using MCMC samples which is shown below.

\subsubsection{Death rates}\label{sec:deathrates}

For the LC (LC-H) model, the $k$-step ahead forecasting distribution of $\boldsymbol{y}_{T+k}$, given $\boldsymbol{y}_{1:T}$, is given by
\begin{equation}\label{eqn:forecastLC}
    \pi(\boldsymbol{y}_{T+k}|\boldsymbol{y}_{1:T})=\int \pi(\boldsymbol{y}_{T+k}|\kappa_{T+k},\boldsymbol{\psi})\pi(\kappa_{T+k}|\kappa_{T+k-1},\boldsymbol{\psi})
    \dots \pi(\kappa_T,\boldsymbol{\psi}|\boldsymbol{y}_{1:T})\,
    d\boldsymbol{\psi}d\kappa_{T:T+k},
\end{equation}
where $\boldsymbol{\psi}$ is the parameter vector for the LC (LC-H) model. \eqref{eqn:forecastLC} suggests that we can sample recursively to obtain the forecasting distribution, for $k \geq 1$, as follows
\begin{subequations}\label{eqn:forecastLCsample}
\begin{align}
    \kappa_{T+k}^{(\ell)} &\sim \text{N}\left(\kappa^{(\ell)}_{T+k-1}+\theta^{(\ell)},\left(\sigma^2_\omega\right)^{(\ell)}\right), \\
    \boldsymbol{y}^{(\ell)}_{T+k} &\sim \text{N}\left(\boldsymbol{\alpha}^{(\ell)}+\boldsymbol{\beta}^{(\ell)}\kappa^{(\ell)}_{T+k},
    \Sigma^{(\ell)}\right),
\end{align}
\end{subequations}
where $\ell=1,\dots,L$ and $L$ is the number of MCMC iterations after burn-in. Here $\Sigma$ is a diagonal matrix with $\sigma^2_{\varepsilon,x}$ on the diagonal for the LC-H model and $\sigma^2_\varepsilon$ for the LC model. This procedure generates an estimate of the forecasting distribution.

Similarly, the forecasting distribution of $\boldsymbol{y}_{T+k}$, given $\boldsymbol{y}_{T}$,
for the LCSV (LCSV-H) model is given by
\begin{align}
    \pi(\boldsymbol{y}_{T+k}|\boldsymbol{y}_{1:T})=&\int \pi(\boldsymbol{y}_{T+k}|\kappa_{T+k},\boldsymbol{\psi})\pi(\kappa_{T+k}|\kappa_{T+k-1},\gamma_{T+k},
    \boldsymbol{\psi})\dots \notag \\
    & \pi(\gamma_{T+1}|\gamma_T,\boldsymbol{\psi})\pi(\kappa_T,\gamma_T,\boldsymbol{\psi}|\boldsymbol{y}_{1:T})\,
    d\boldsymbol{\psi}d\kappa_{T:T+k}d\gamma_{T:T+k}.
\end{align}
For $k \geq 1$, the forecasting distribution can be obtained by sampling recursively
\begin{subequations}\label{eqn:forecastLCSVsample}
\begin{align}
    \gamma_{T+k}^{(\ell)} &\sim \text{N}\left(\lambda^{(\ell)}_1 \gamma^{(\ell)}_{T+k-1}+\lambda^{(\ell)}_2, \left(\sigma^2_\gamma\right)^{(\ell)}\right), \\
    \kappa_{T+k}^{(\ell)} &\sim \text{N}\left(\kappa^{(\ell)}_{T+k-1}+\theta^{(\ell)},\exp\{\gamma_{T+k}^{(\ell)}\}\right), \\
    \boldsymbol{y}^{(\ell)}_{T+k} &\sim \text{N}\left(\boldsymbol{\alpha}^{(\ell)}+\boldsymbol{\beta}^{(\ell)}\kappa^{(\ell)}_{T+k},
        \Sigma^{(\ell)}\right),
\end{align}
\end{subequations}
where $\ell=1,\dots,L$, and $\Sigma$ is a diagonal matrix with $\sigma^2_{\varepsilon,x}$ on the diagonal for the LCSV-H model and $\sigma^2_\varepsilon$ for the LCSV model.

\begin{figure}[h]
\begin{center}
\includegraphics[width=5.5cm, height=5cm]{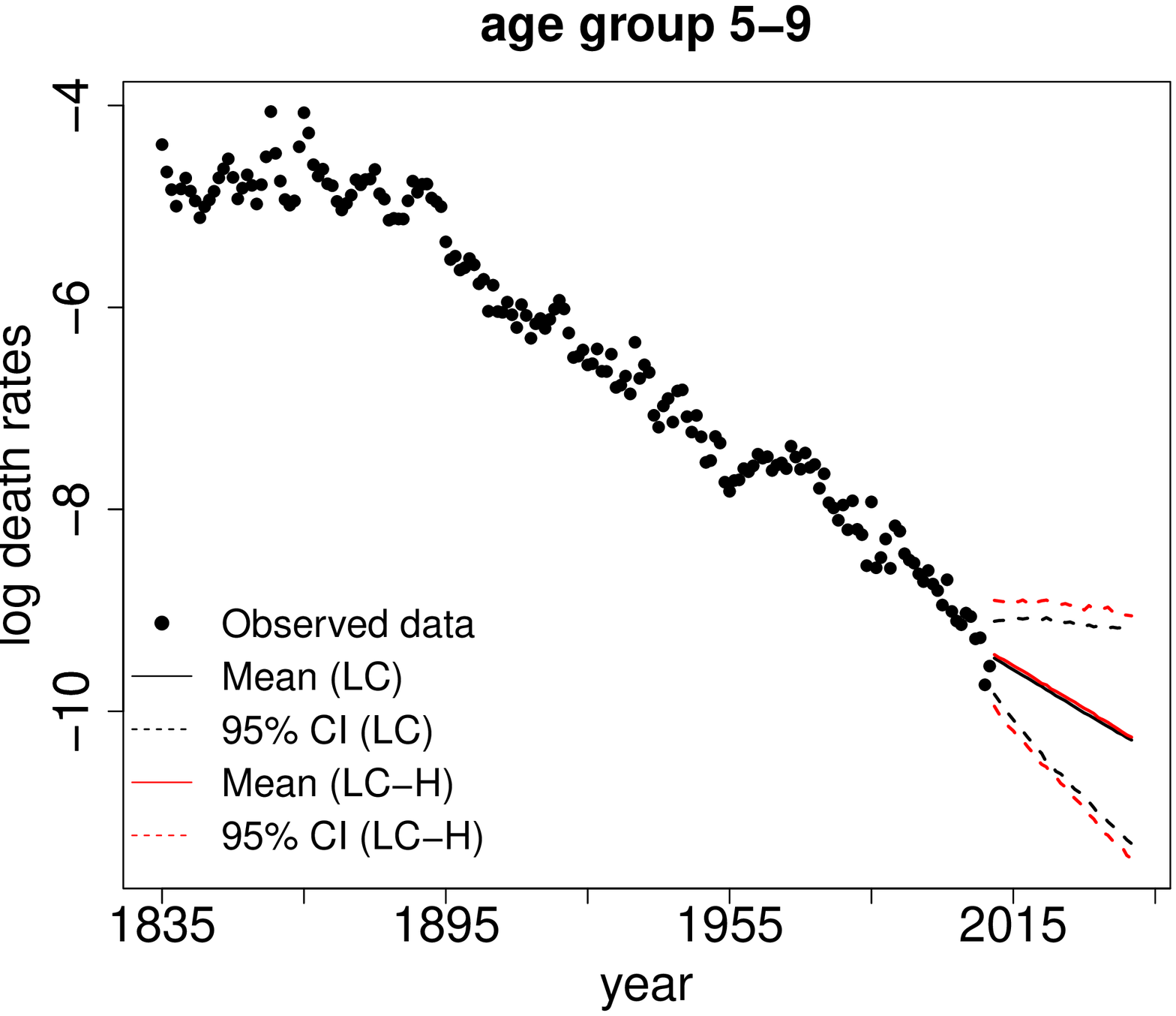}\includegraphics[width=5.5cm, height=5cm]{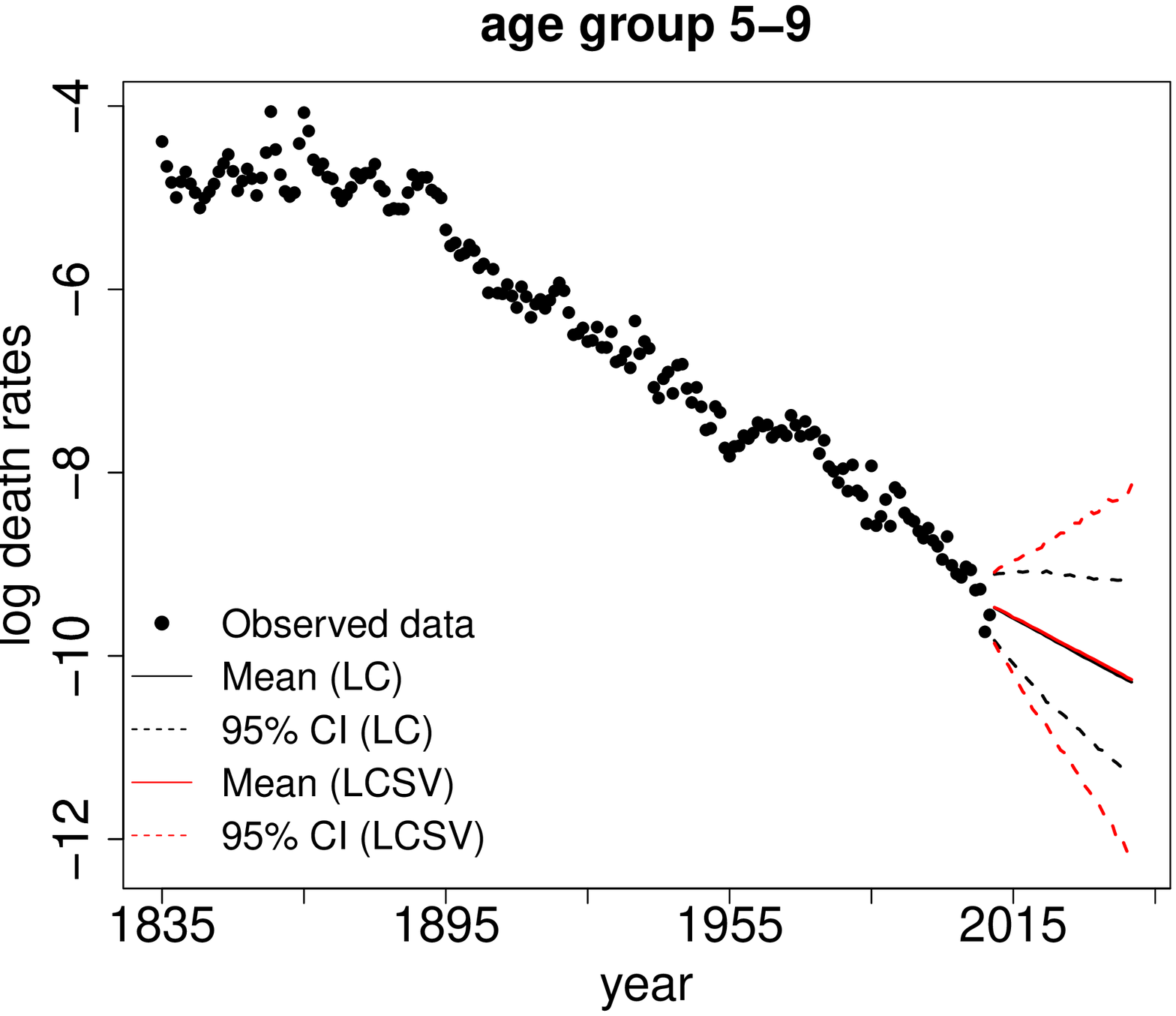}\includegraphics[width=5.5cm, height=5cm]{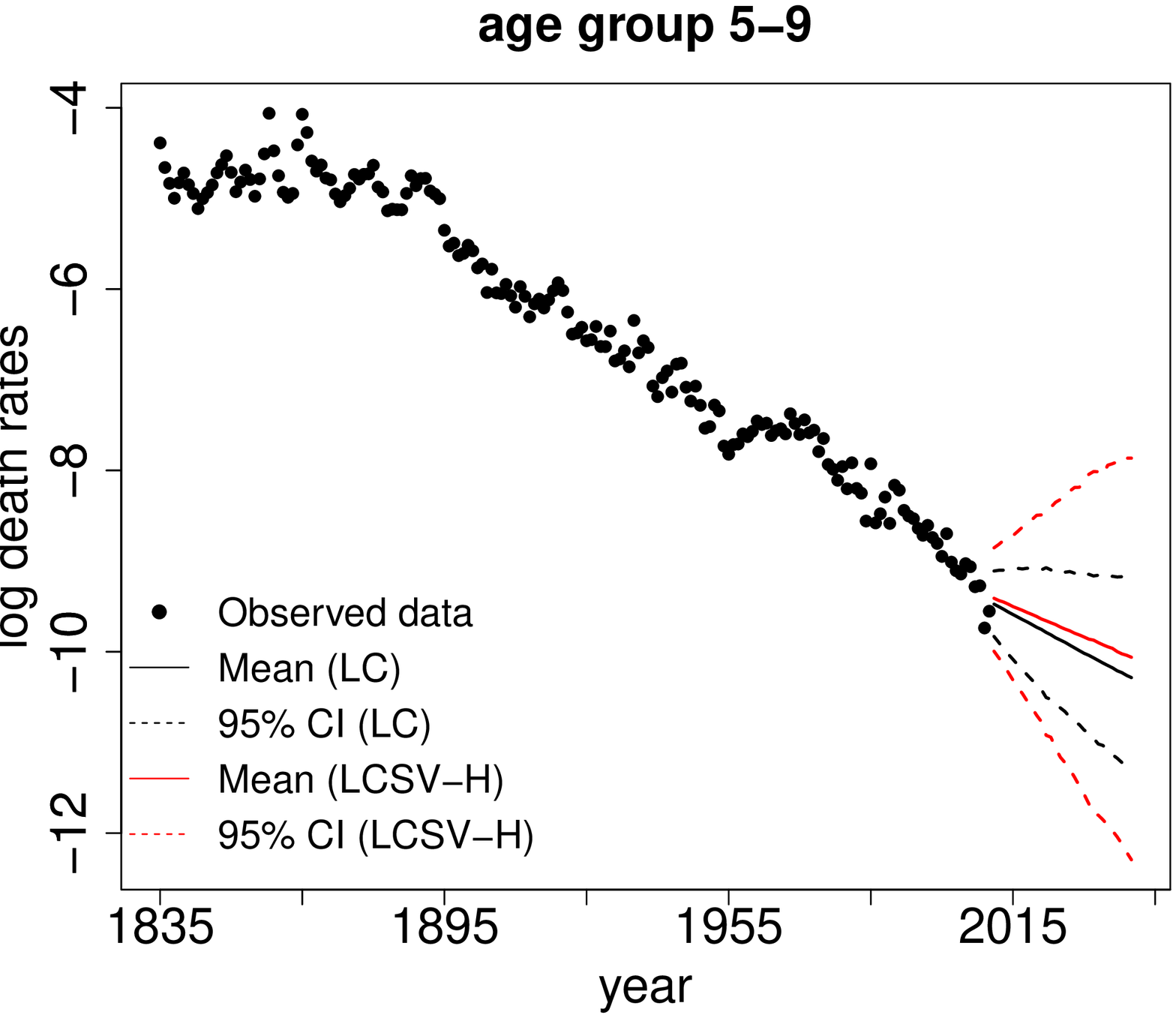}
\includegraphics[width=5.5cm, height=5cm]{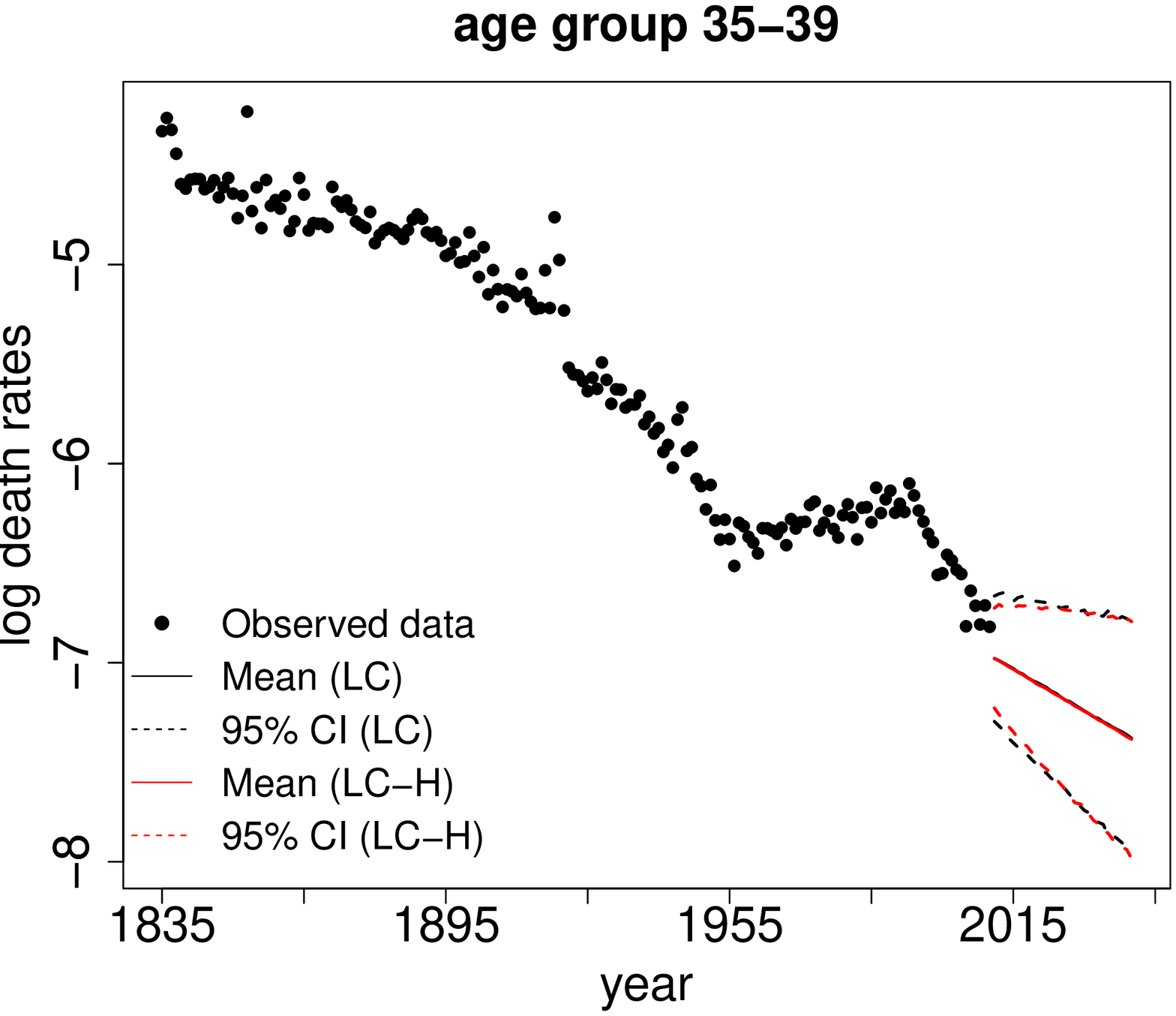}\includegraphics[width=5.5cm, height=5cm]{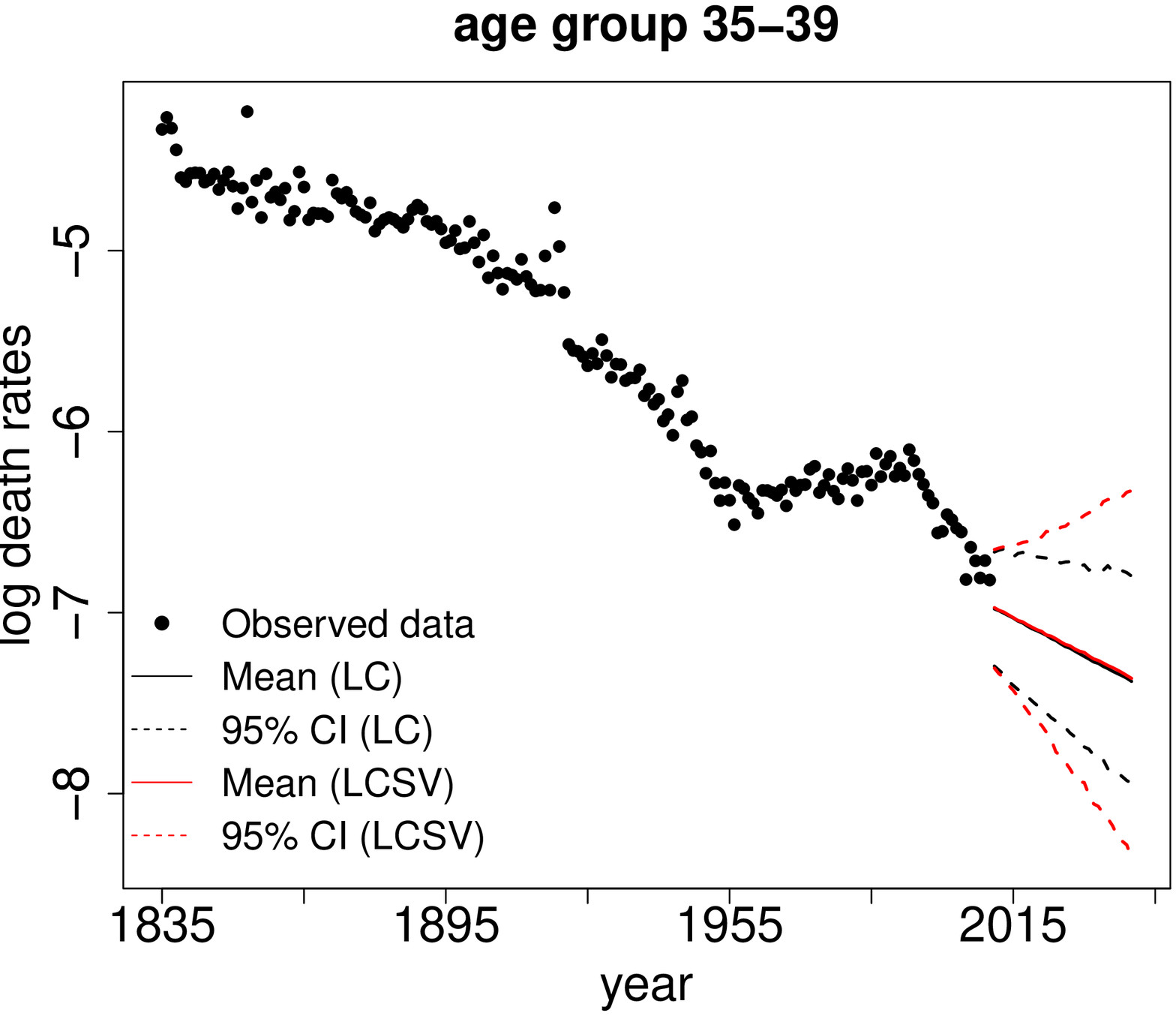}\includegraphics[width=5.5cm, height=5cm]{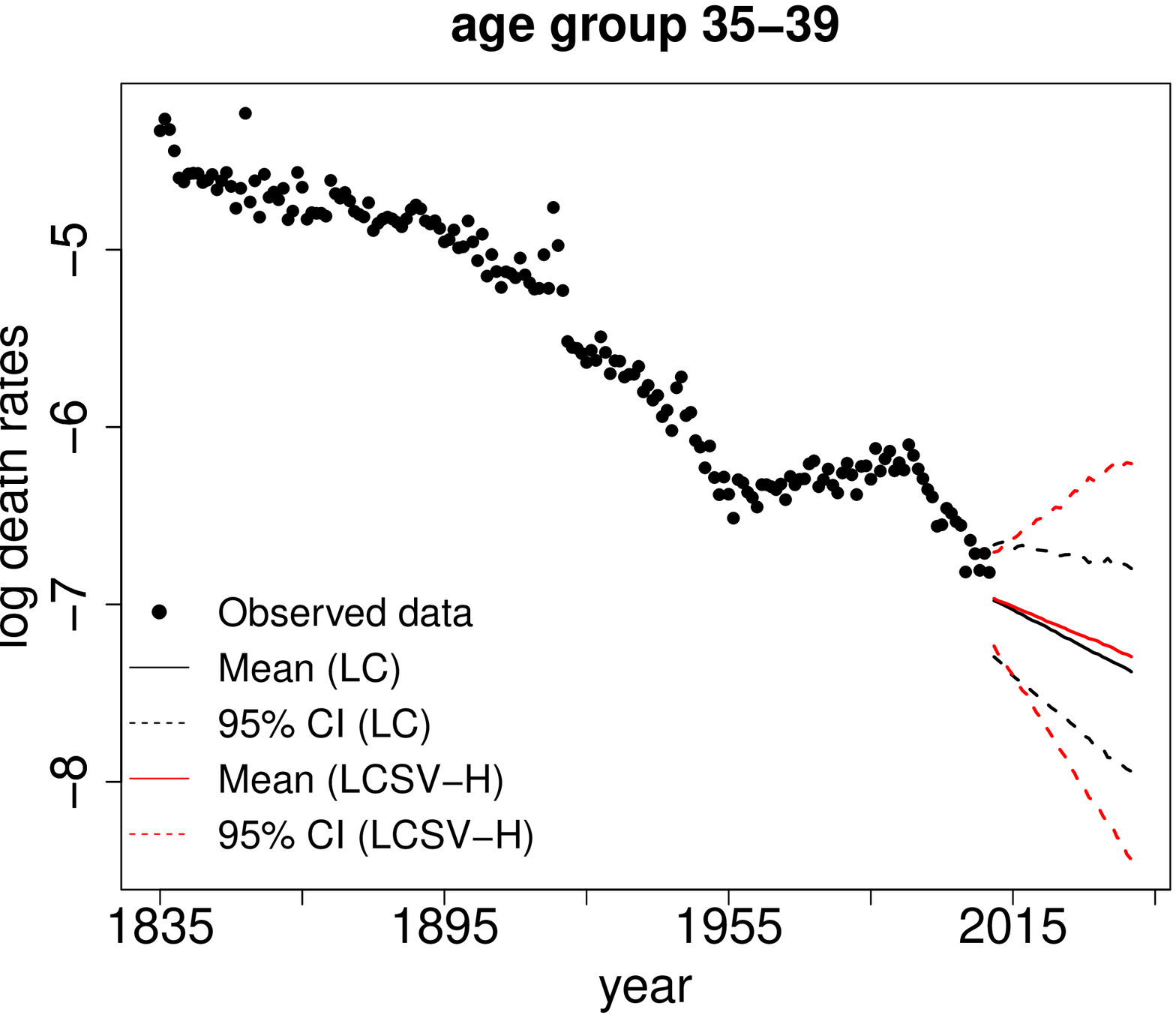}
\includegraphics[width=5.5cm, height=5cm]{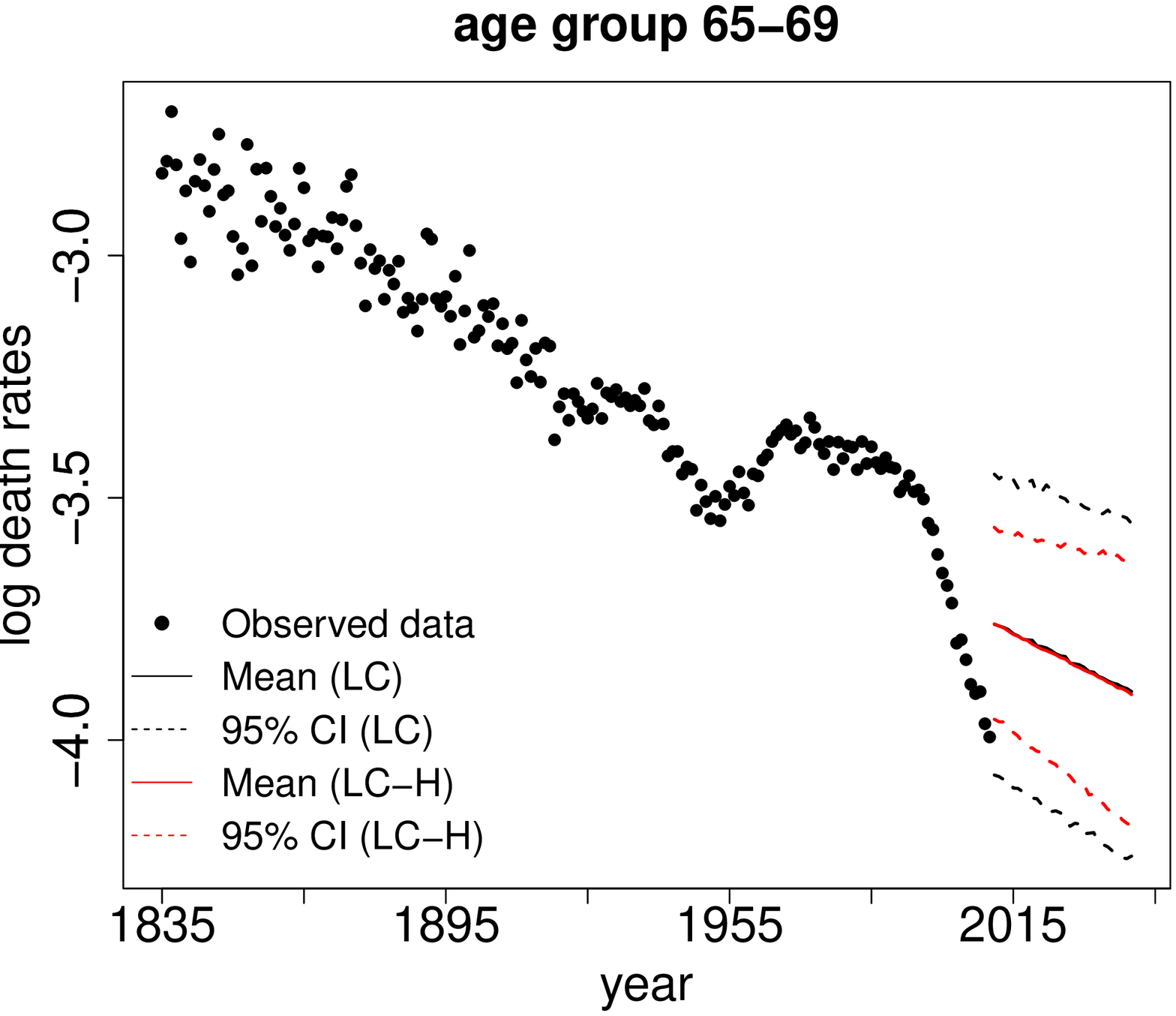}\includegraphics[width=5.5cm, height=5cm]{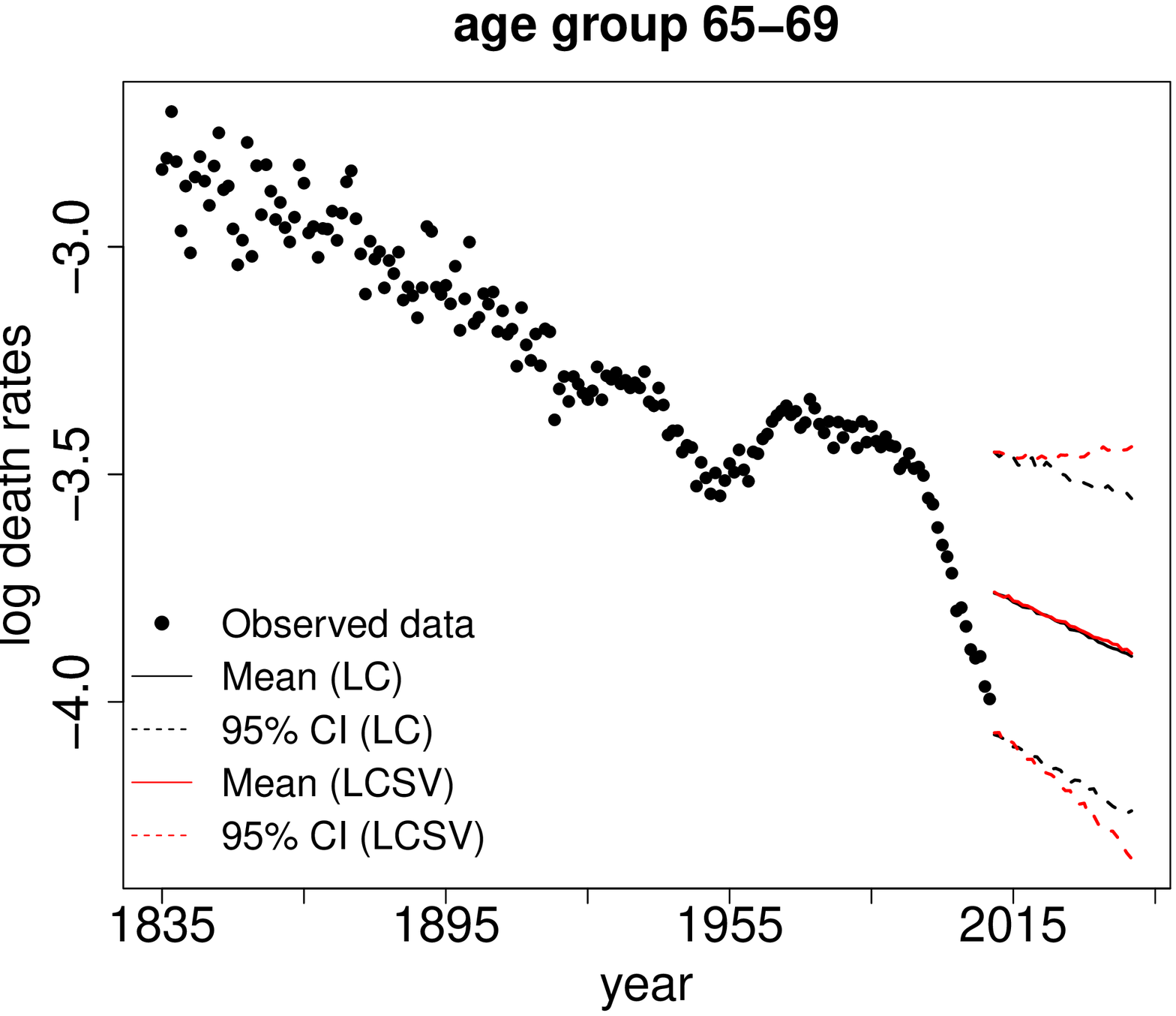}\includegraphics[width=5.5cm, height=5cm]{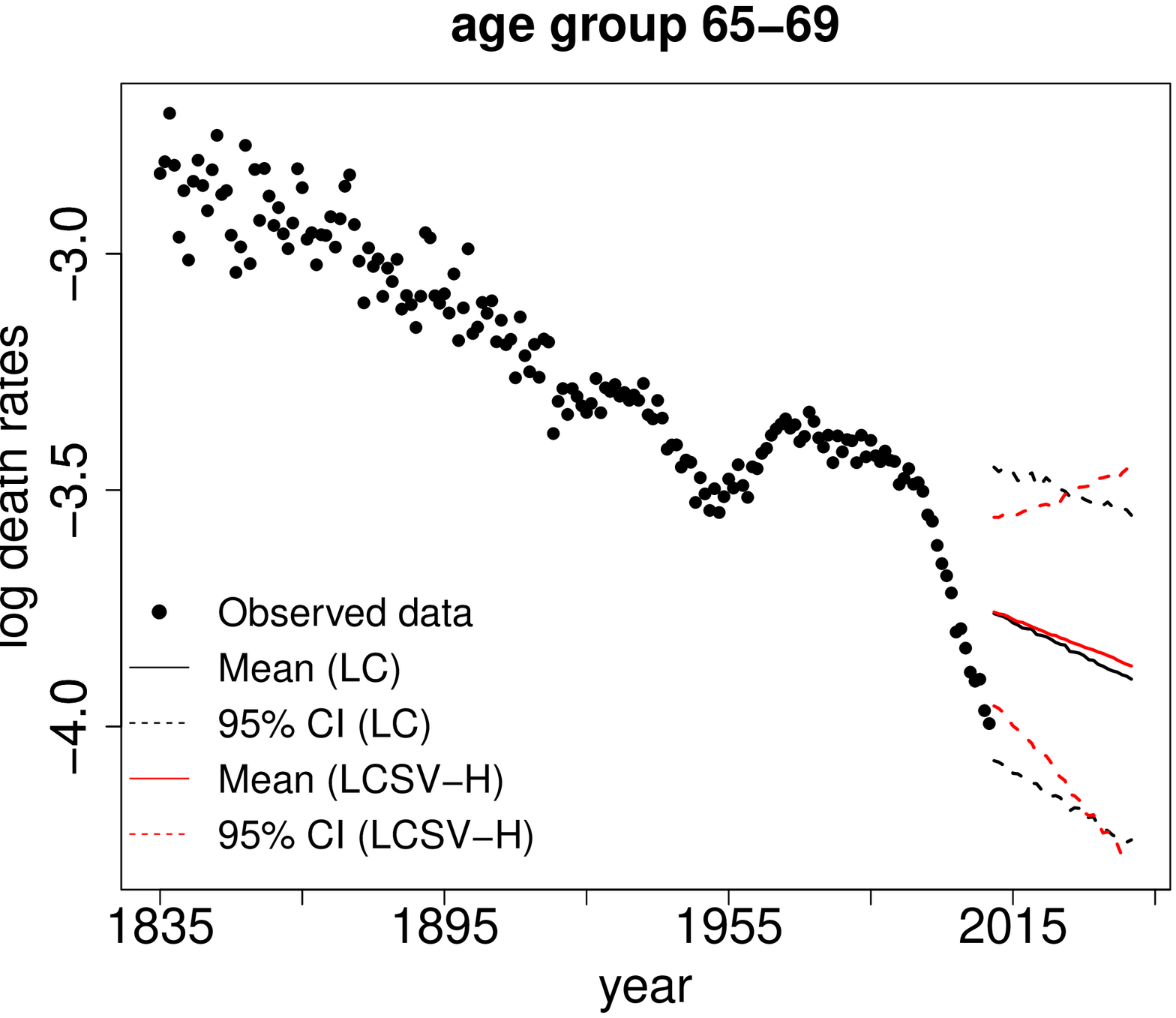}
\includegraphics[width=5.5cm, height=5cm]{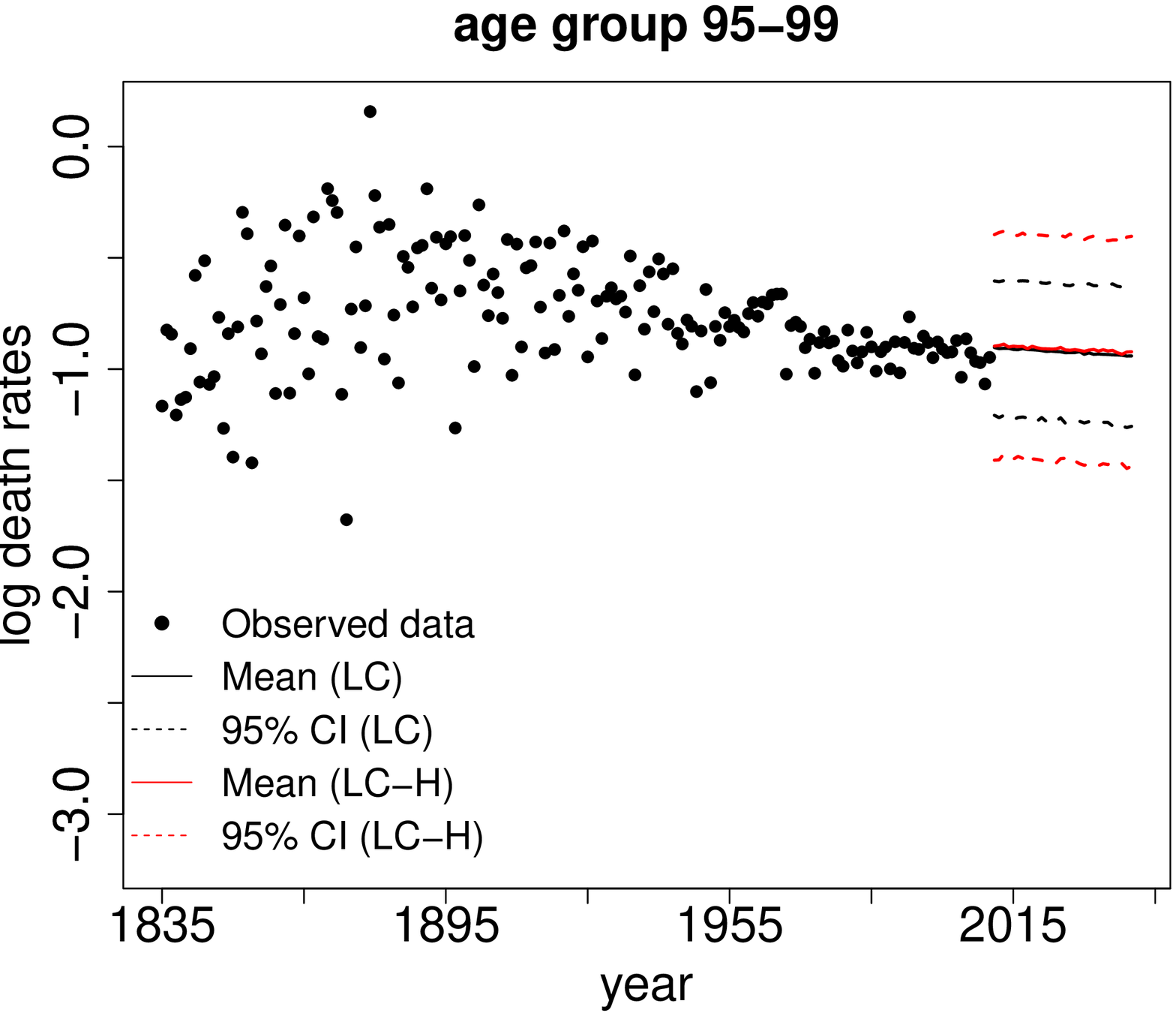}\includegraphics[width=5.5cm, height=5cm]{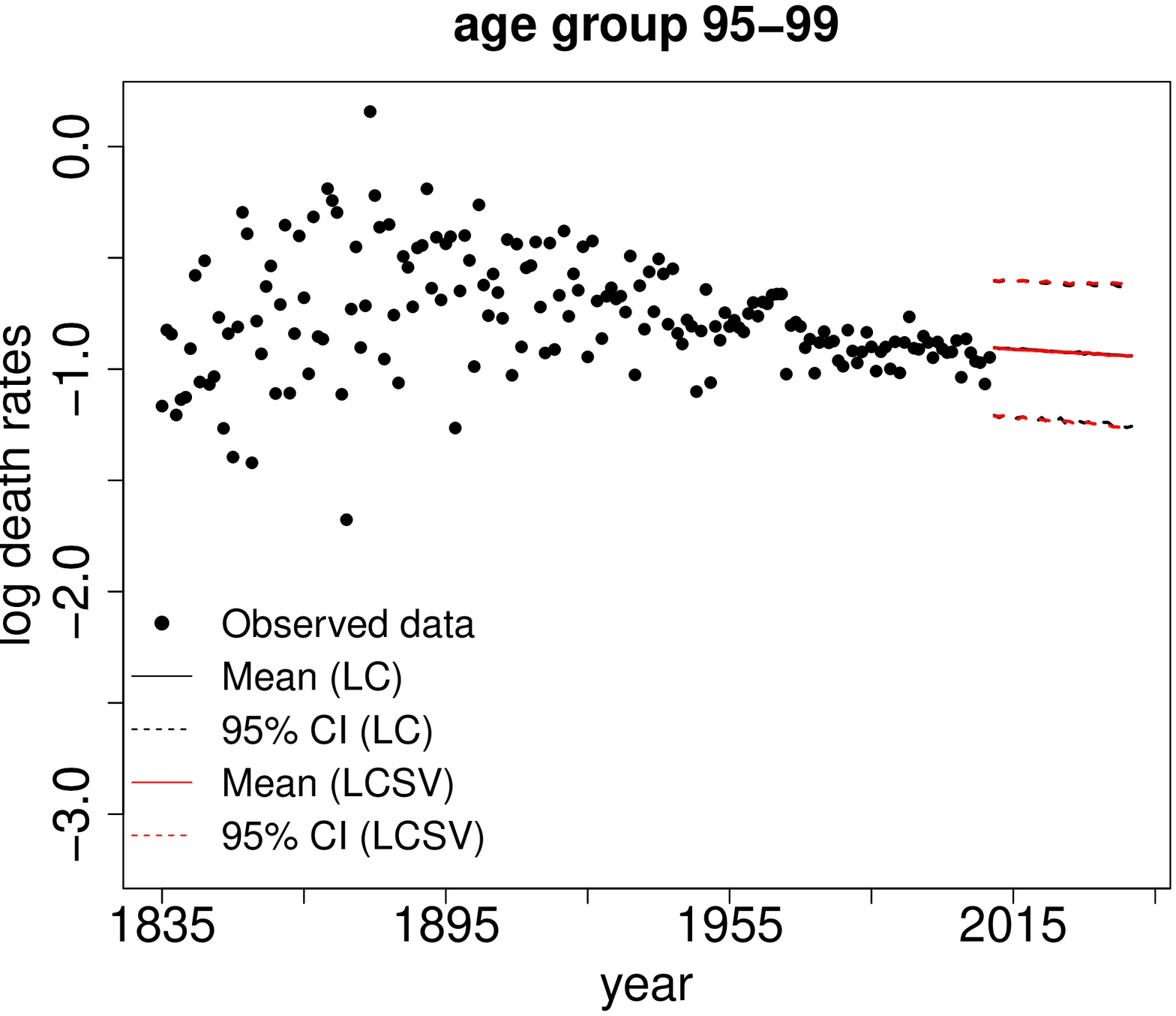}\includegraphics[width=5.5cm, height=5cm]{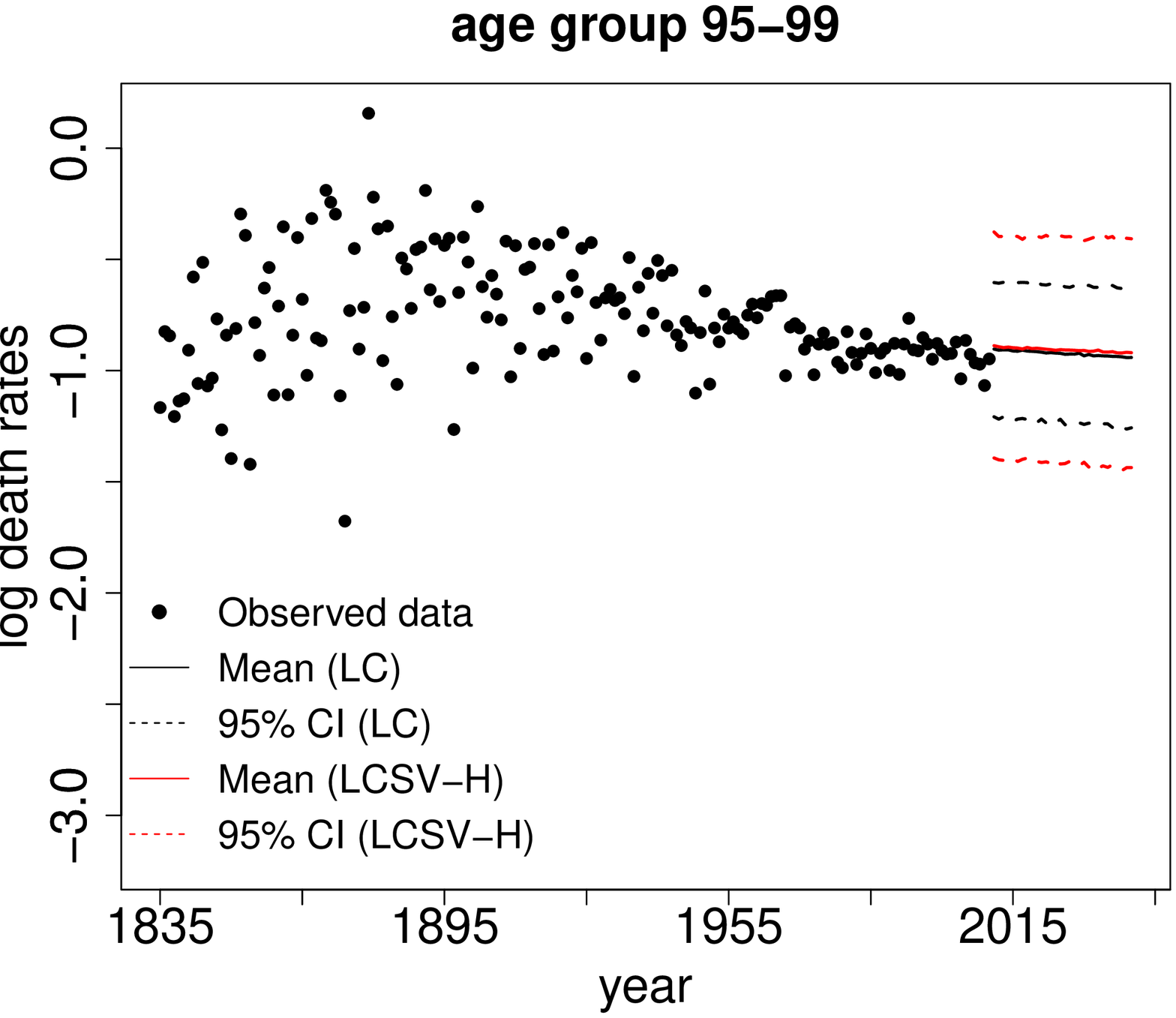}
\caption{\small{30-year forecasted log death rates (2011-2041) for Danish male population under (left column) LC-H model, (middle column) LCSV model and (right column) LCSV-H model in comparison with LC model. Calibration period: 1835-2010.}}
\label{fig:forecastDeathRates18352010}
\end{center}
\end{figure}

Figure~\ref{fig:forecastDeathRates18352010} shows the forecasted log death rates based on the LC-H, LCSV and LCSV-H model, using the LC model as a benchmark. We show age groups 5-9, 35-39, 65-69 and 95-99 as representatives of young, adult, old and very old age. The models are estimated using data for the period 1835-2010 and forecast for 30 years.

The heteroscedasticity structure, from the LC-H model, gives rise to materially larger forecasting intervals for the young and very old age group, while the forecasting interval for the age group 35-39 is narrower than predicted by the LC model. The LCSV model, on the other hand, produces a wider forecasting interval compared to the LC model except for the very old age group. The observed wider forecasting interval is due to the fact that the volatility level is increasing in the last estimation periods and is larger than $\sigma^2_\omega$ estimated in the LC model. Moreover, as the estimated $\beta_x$ is close to zero at older ages (Figure~\ref{fig:DENLCSVH18352010}), the impact of the forecasted $\kappa$ on the prediction of death rates diminished significantly as older ages are considered. The LCSV-H model exhibits similar features of the LC-H and the LCSV model. It is interesting to note that the forecasted means obtained from the different models are very similar and their differences mainly lie in the forecasting interval.

\begin{figure}[h]
\begin{center}
\includegraphics[width=5.5cm, height=5cm]{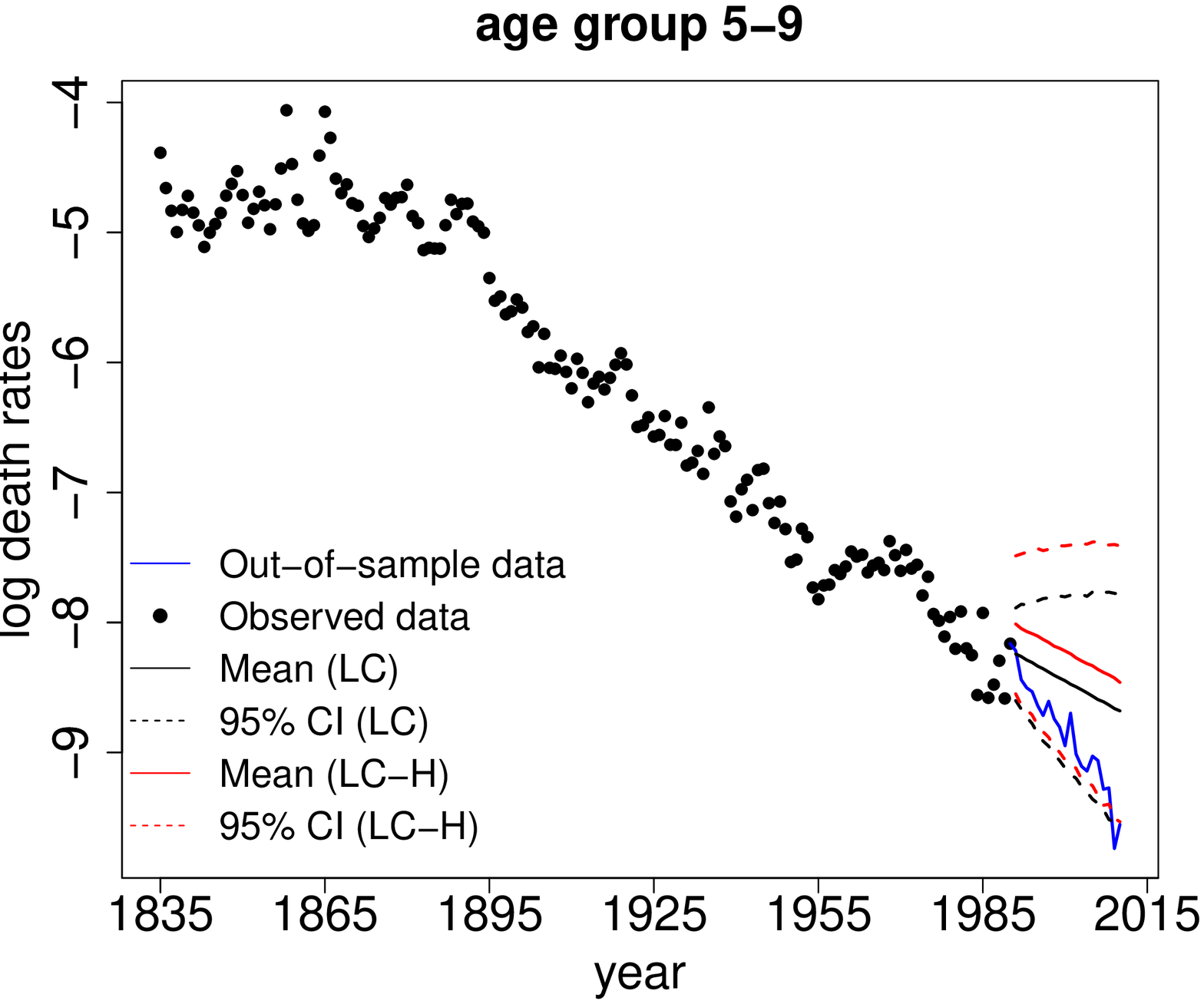}\includegraphics[width=5.5cm, height=5cm]{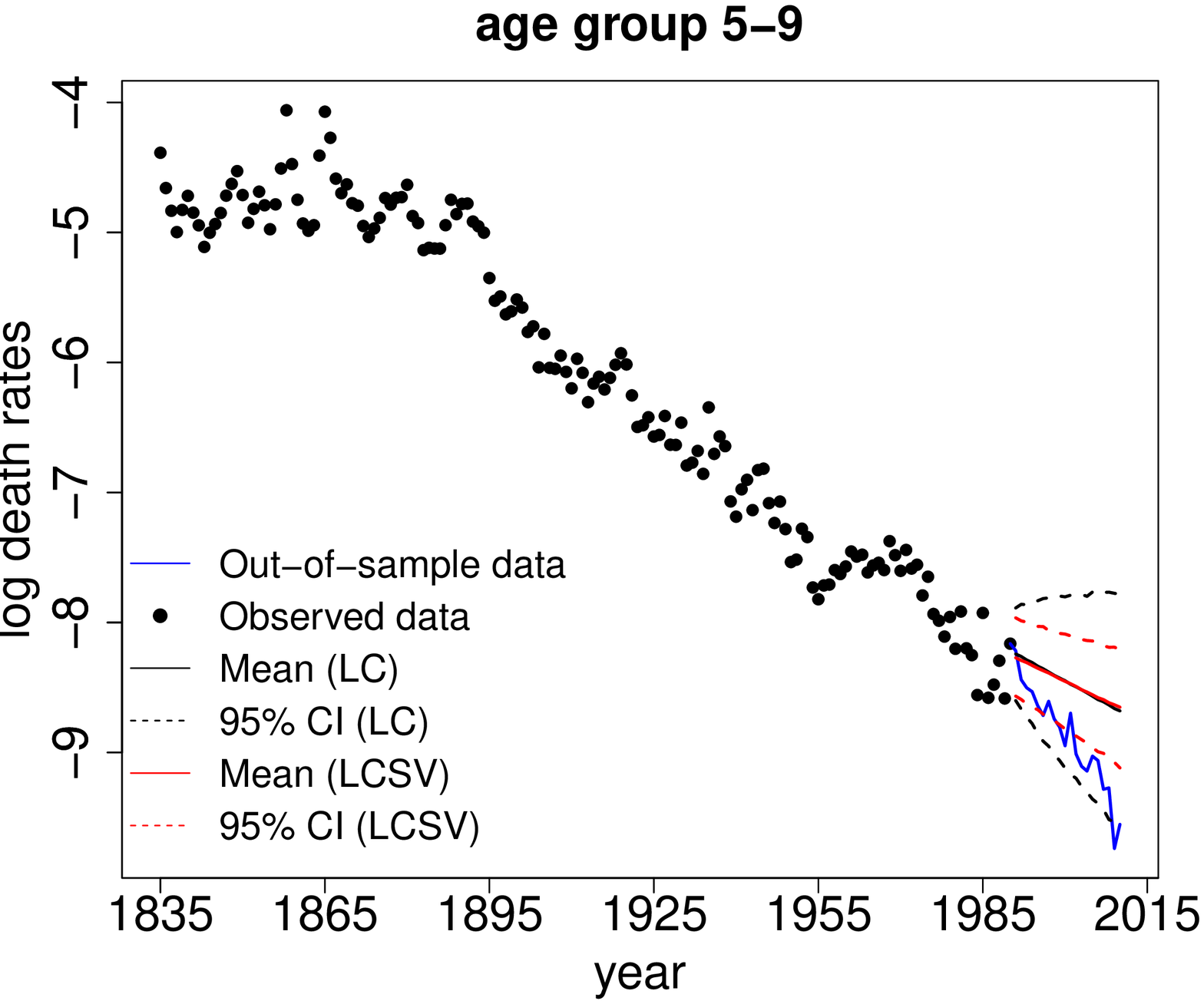}\includegraphics[width=5.5cm, height=5cm]{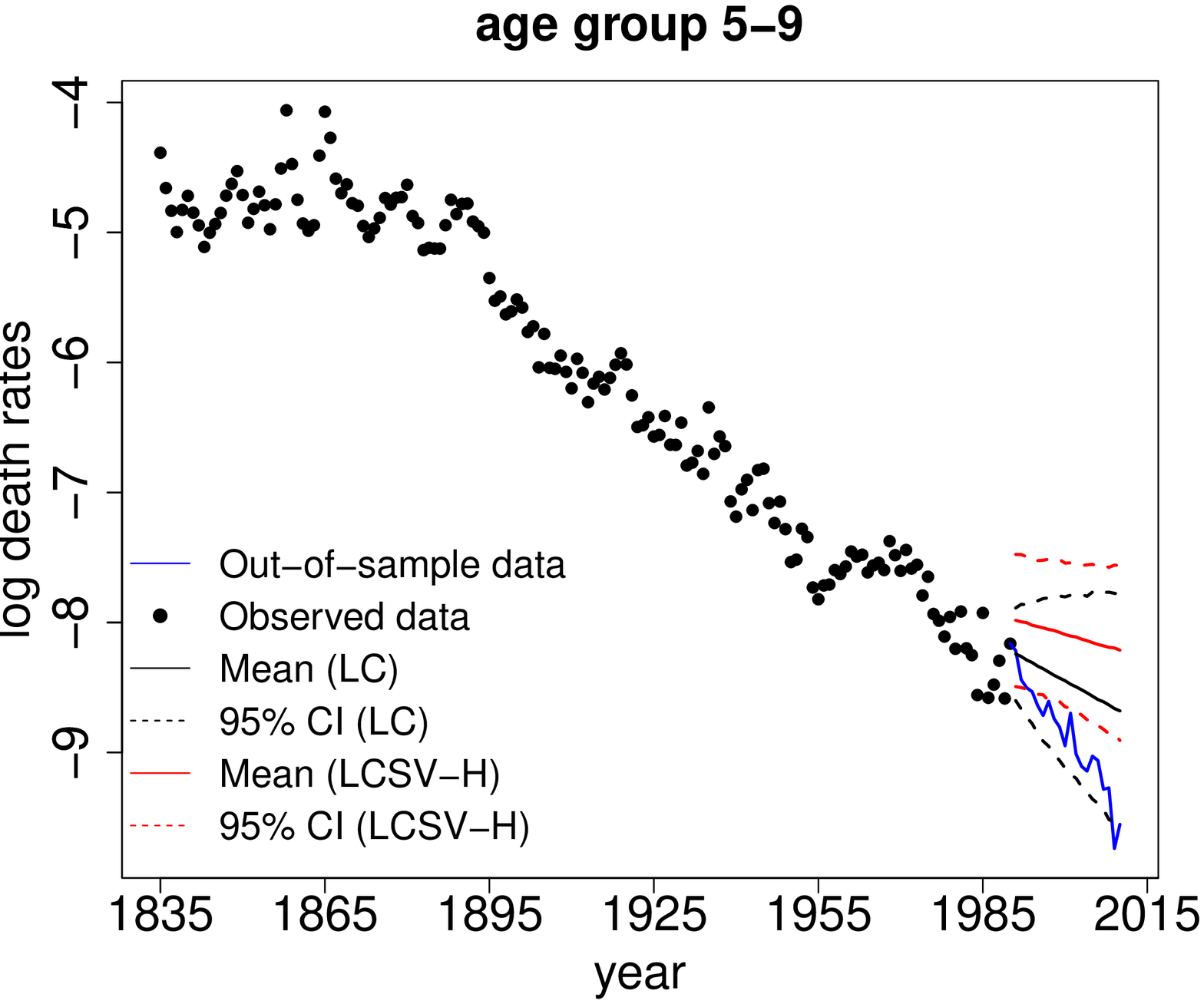}
\includegraphics[width=5.5cm, height=5cm]{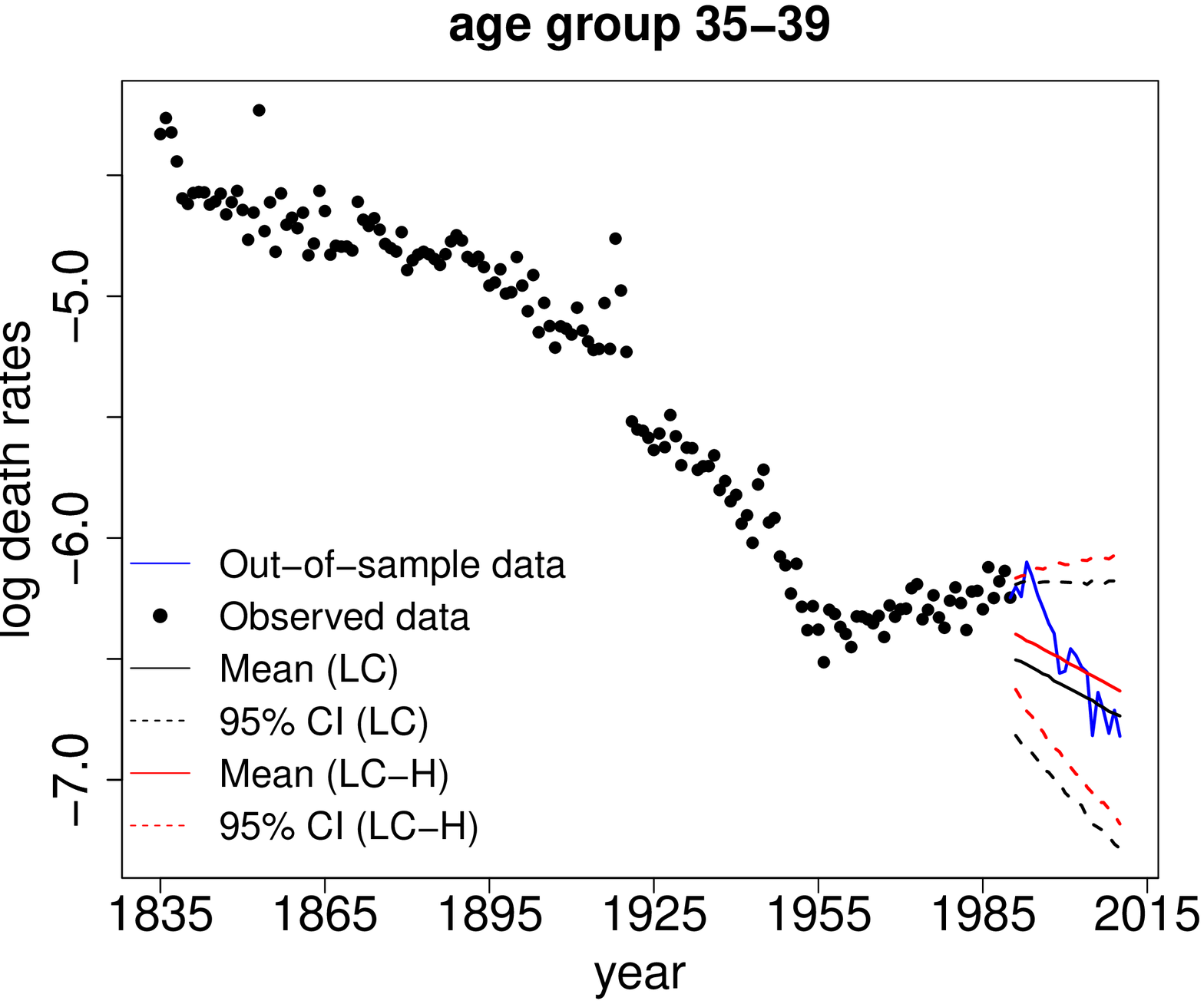}\includegraphics[width=5.5cm, height=5cm]{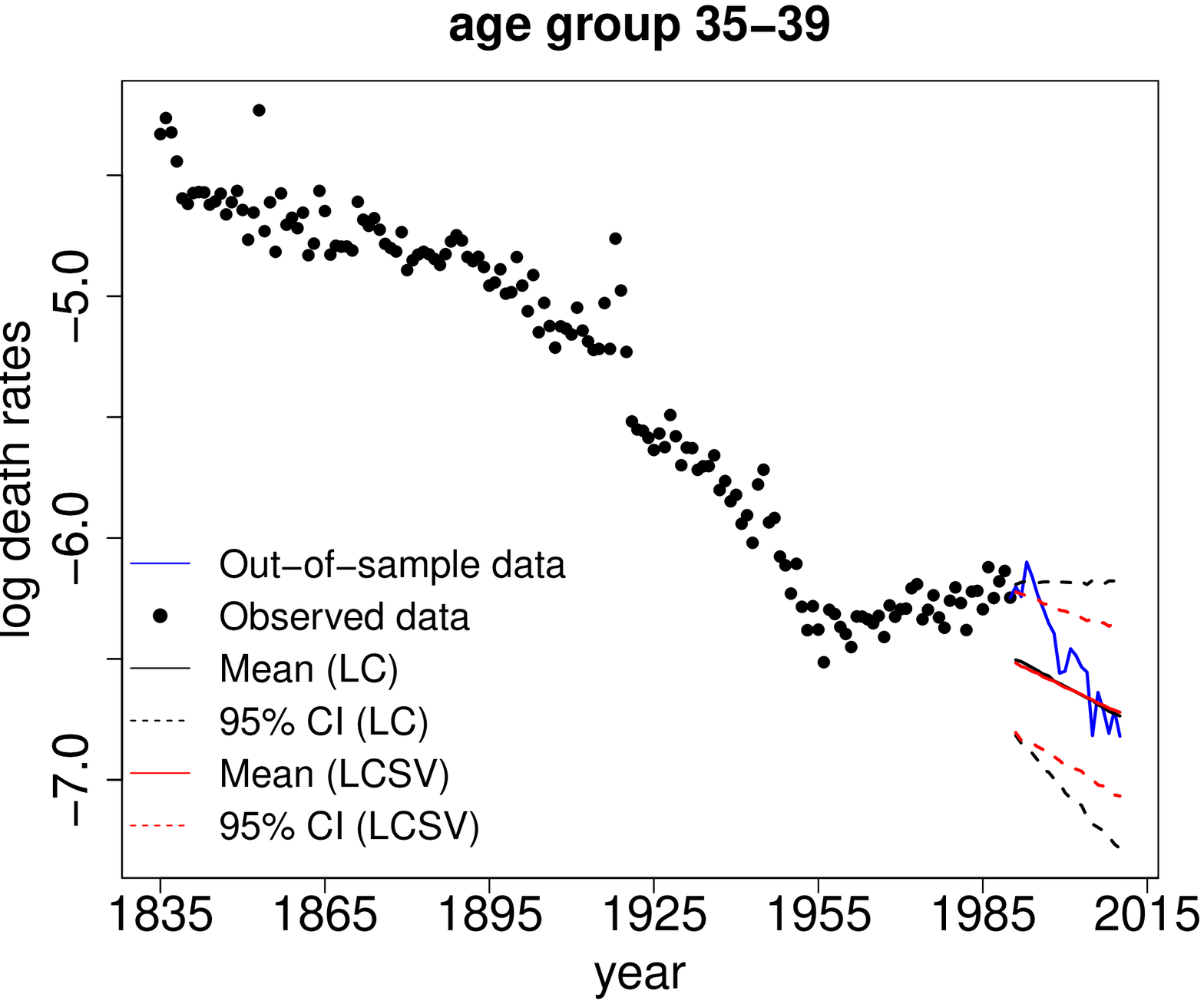}\includegraphics[width=5.5cm, height=5cm]{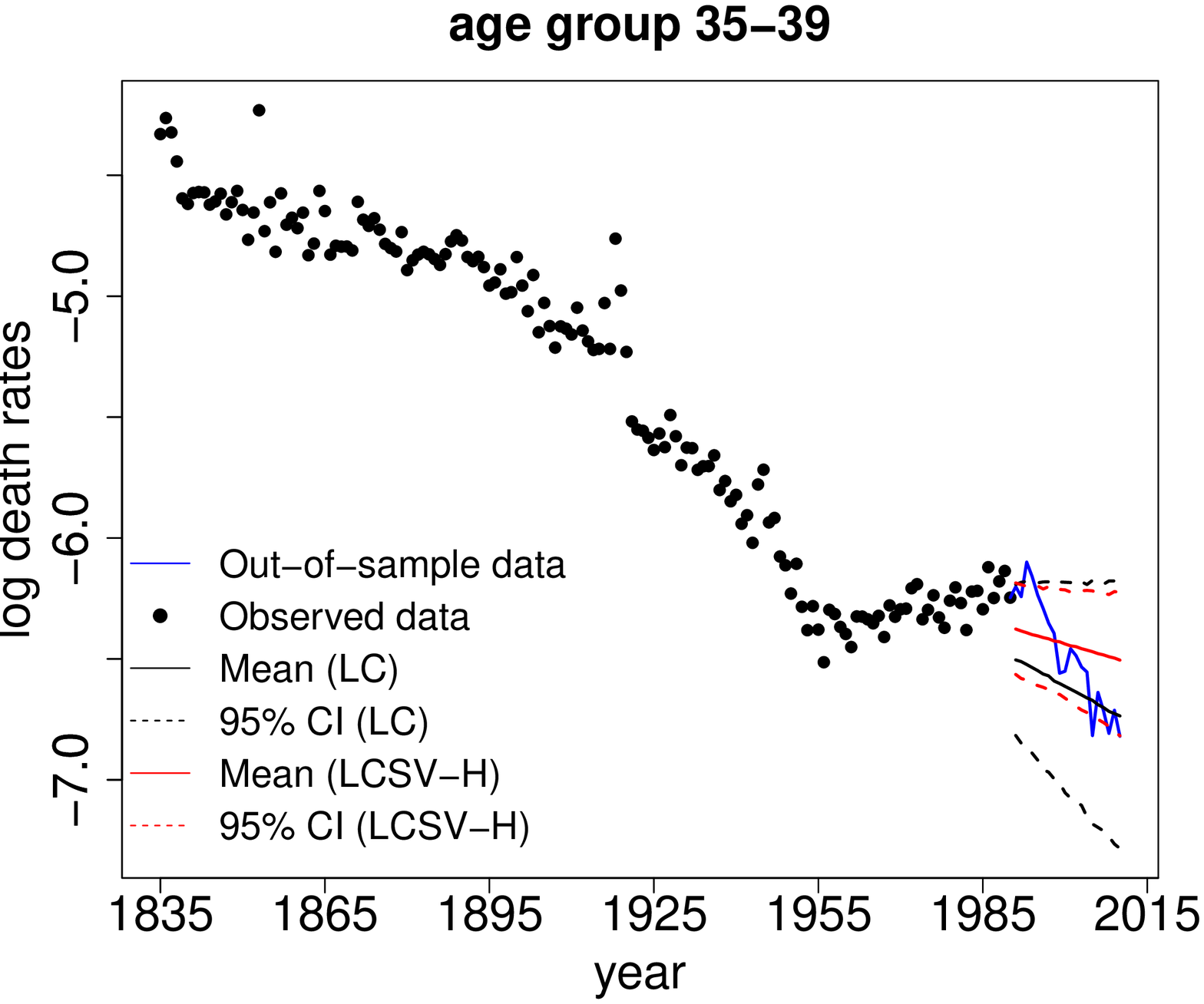}
\includegraphics[width=5.5cm, height=5cm]{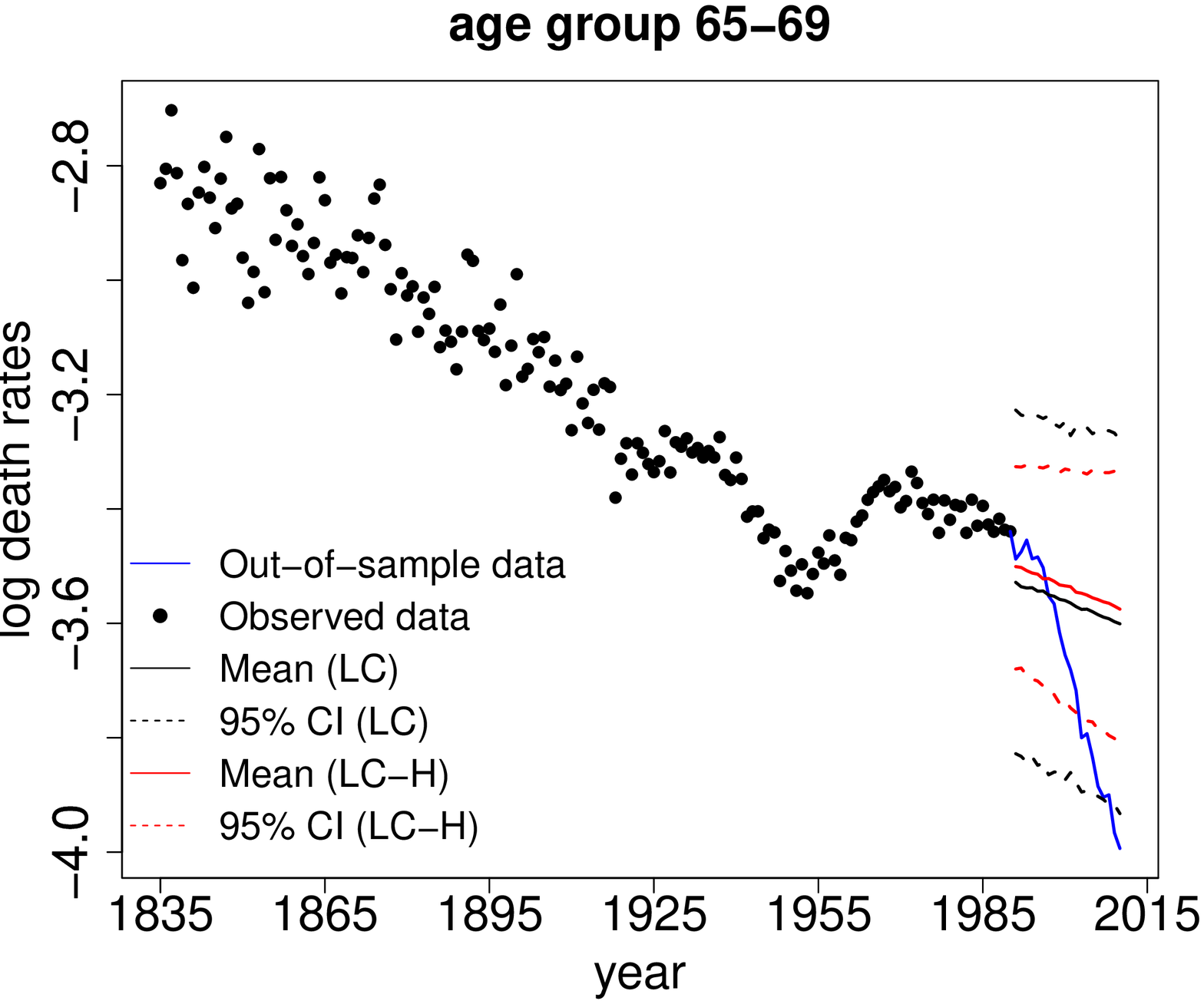}\includegraphics[width=5.5cm, height=5cm]{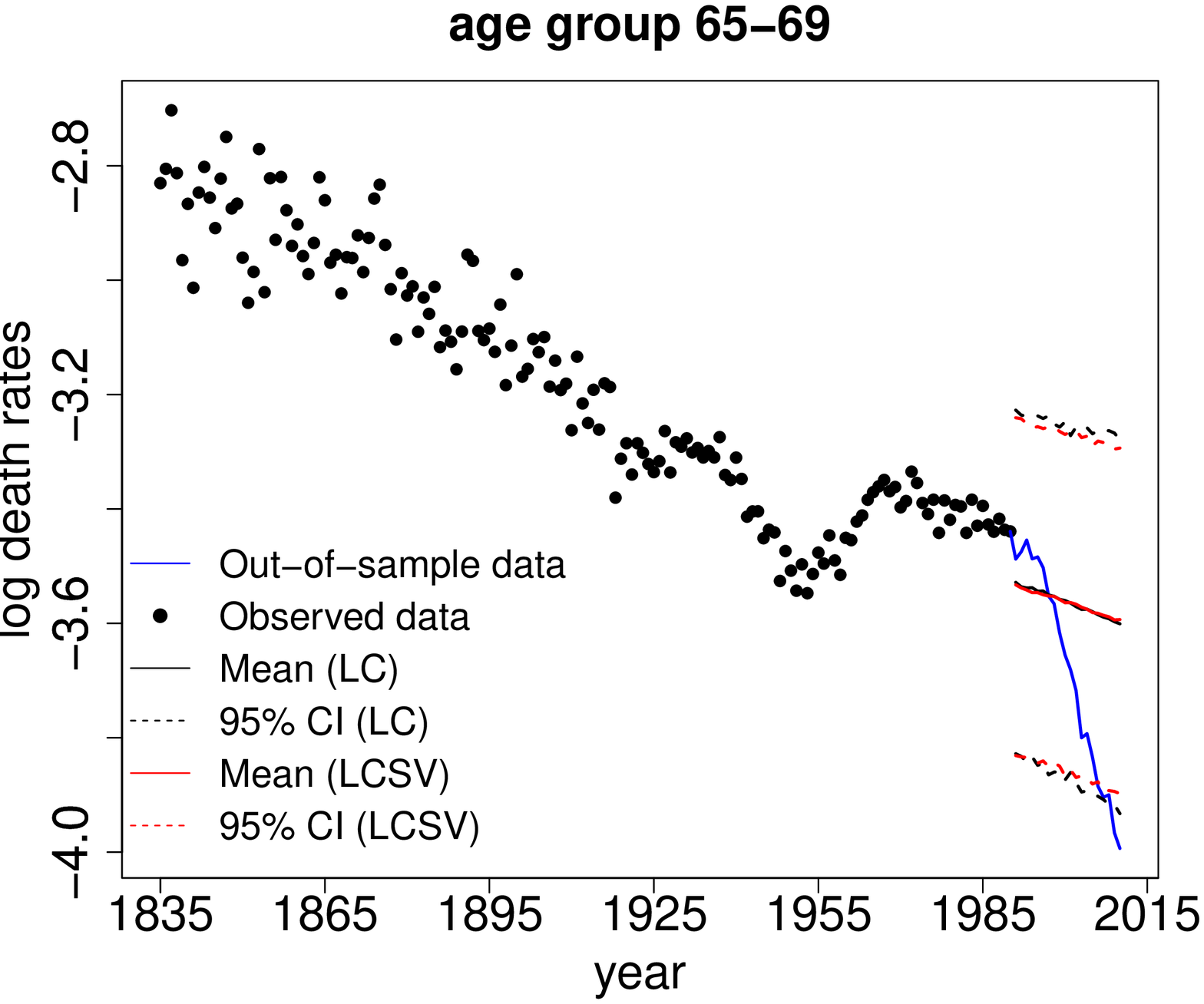}\includegraphics[width=5.5cm, height=5cm]{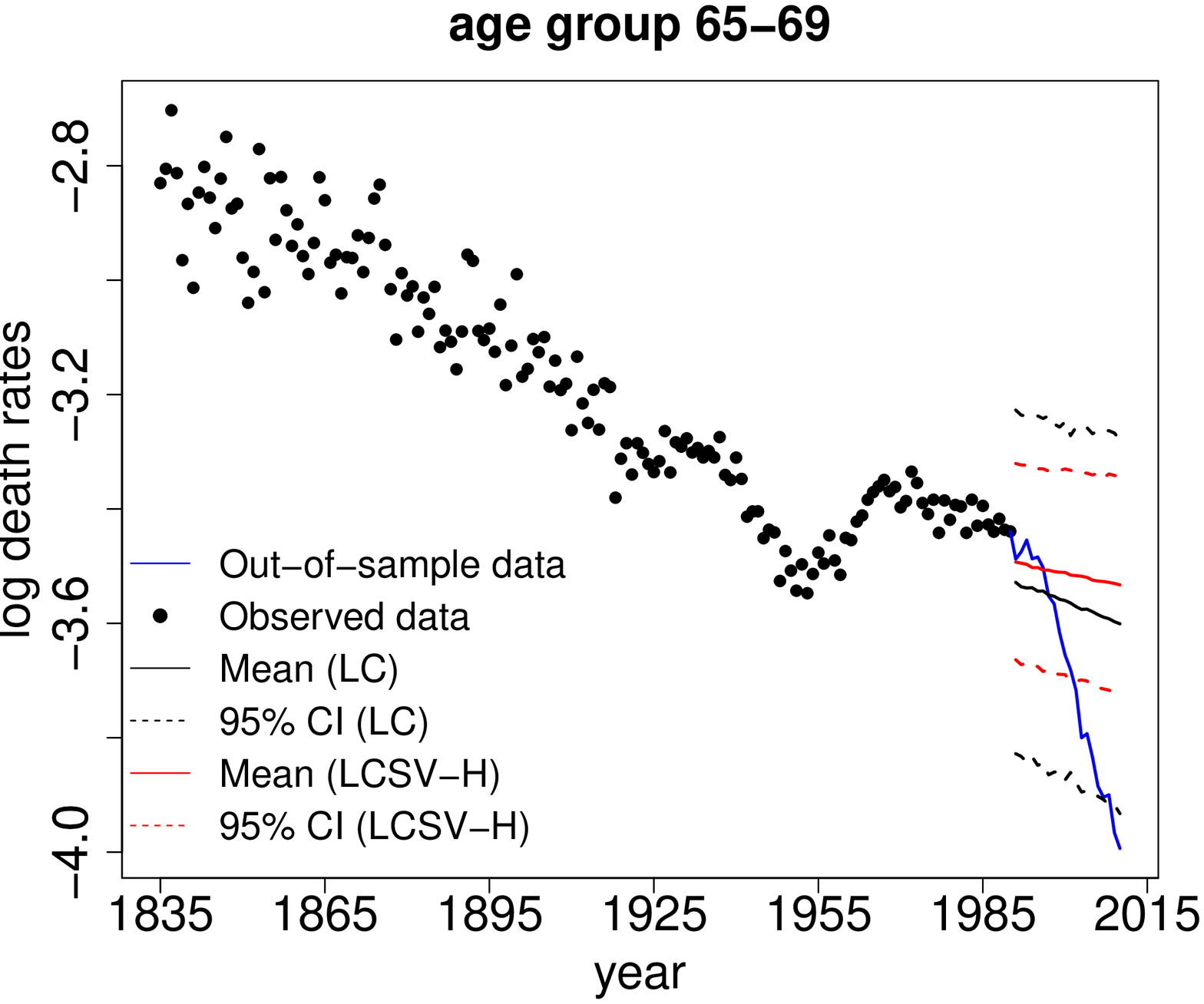}
\includegraphics[width=5.5cm, height=5cm]{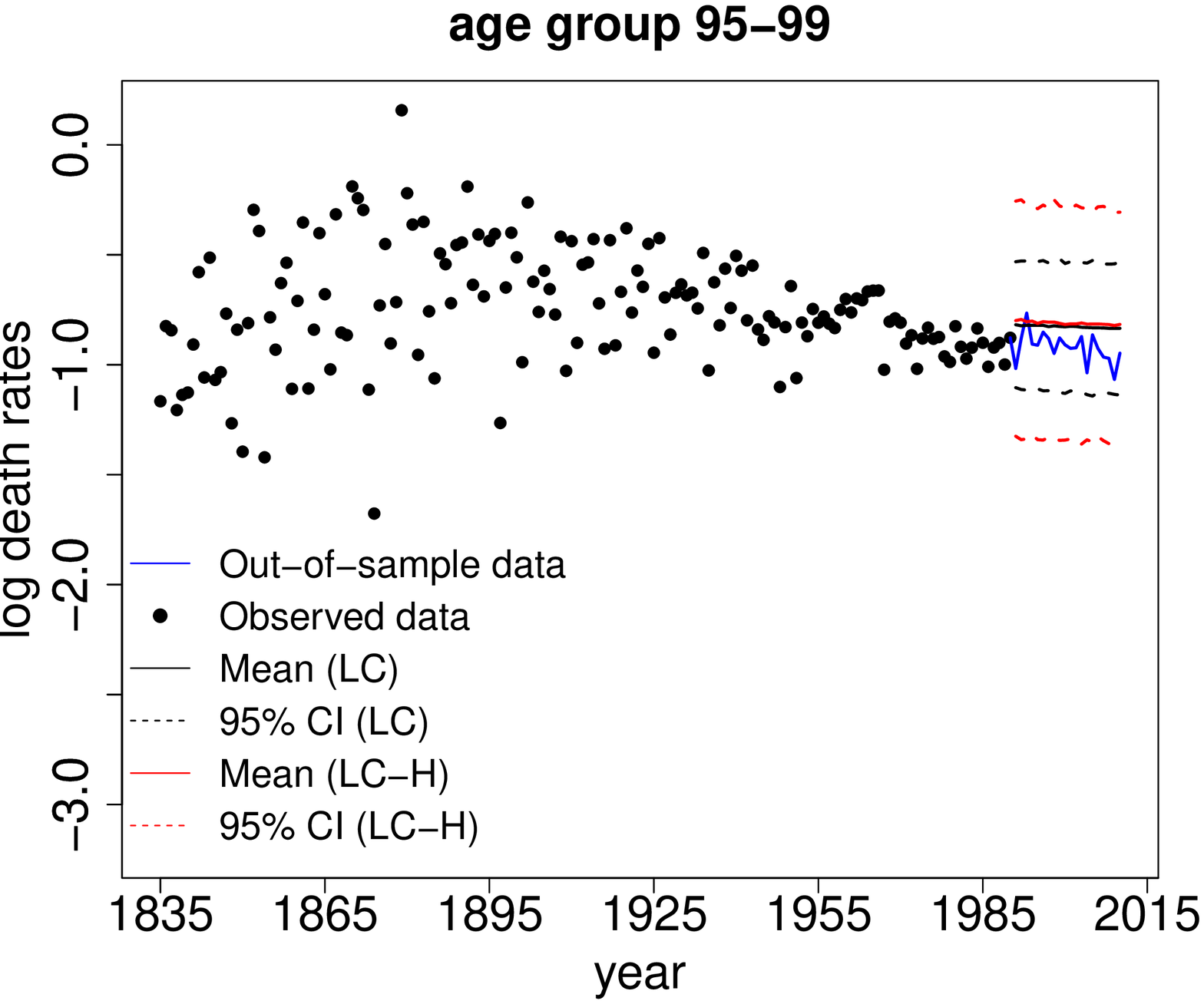}\includegraphics[width=5.5cm, height=5cm]{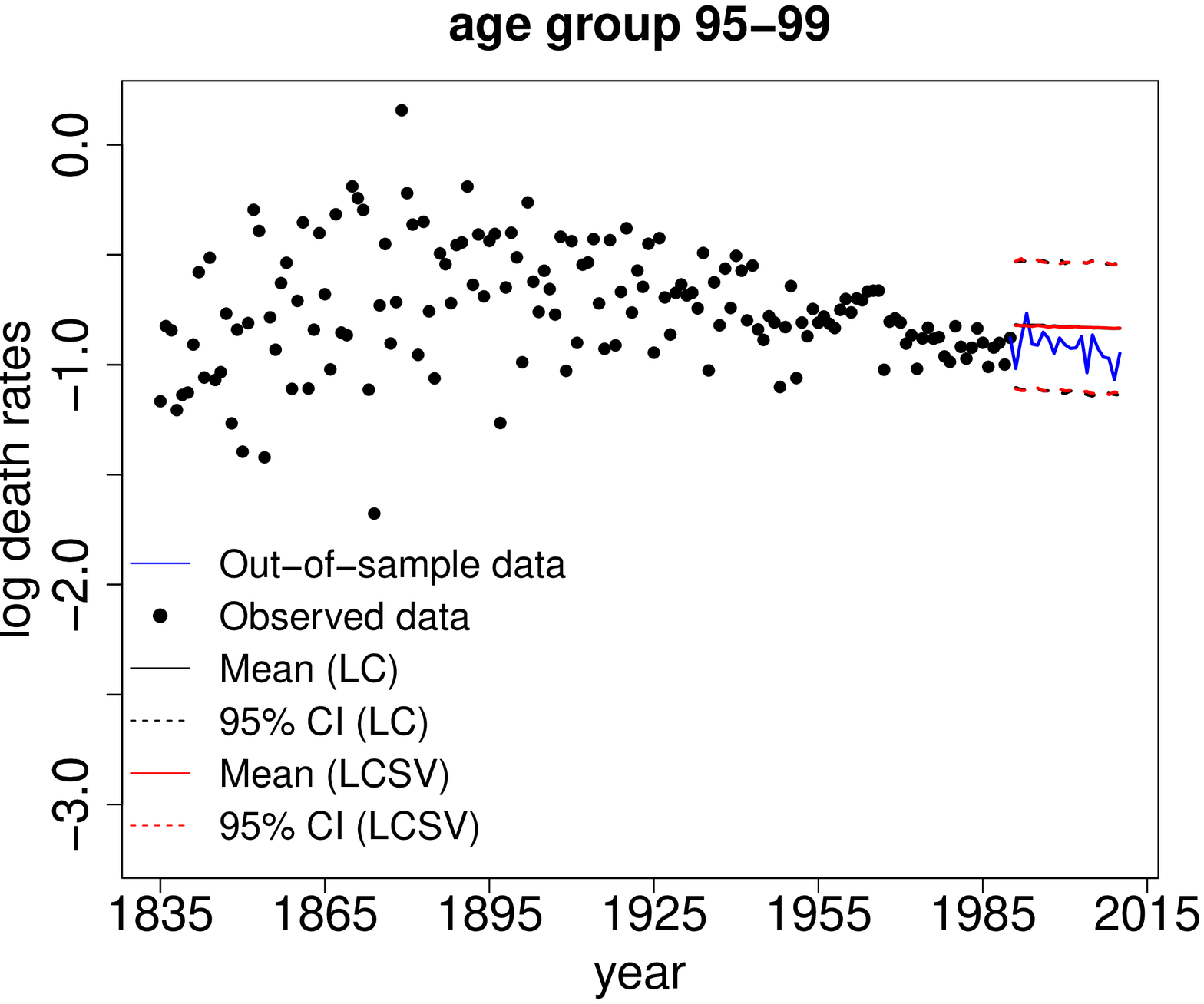}\includegraphics[width=5.5cm, height=5cm]{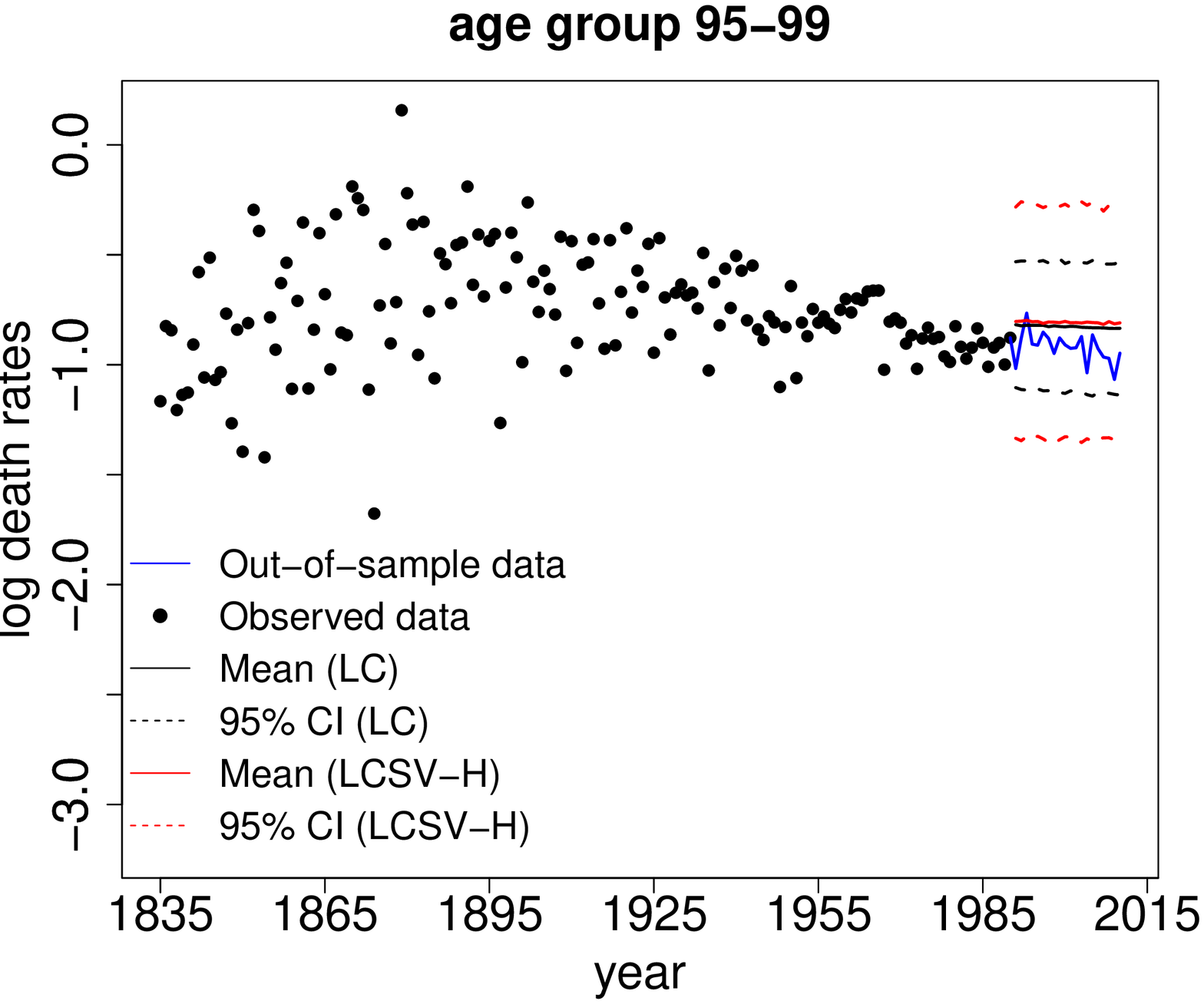}
\caption{\small{20-year out-of-sample forecasted log death rates for Danish male populationunder (left column) LC-H model, (middle column) LCSV model and (right column) LCSV-H model in comparison with LC model. Calibration period: 1835-1990.}}
\label{fig:forecastDeathRates18351990}
\end{center}
\end{figure}

To illustrate further the forecasting property of the LCSV model, we estimate the models for the period 1835-1990 and plot 20-year out-of-sample forecasted log death rates in Figure~\ref{fig:forecastDeathRates18351990}. It turns out the forecasting intervals predicted by the LCSV model tends to be narrower than the LC model, as the estimated $\sigma^2_\omega$ in the LC model is larger than the volatility level at the last estimation period for the LCSV model in this case. Note that the forecasted distributions produced by the LC-H model are biased compared to the benchmark LC model since the fitted rates at the last estimation period, that is year 1990, are different for the LC and LC-H model. This feature is known as jump-off error (\cite{LeeMi01}). One may remove this jump-off bias by forcing the forecasted death rates to start at the actual rates instead of the fitted rates (\cite{Bell97} and \cite{ShangBoHy11}). In this paper we do not perform this procedure, however.

\begin{figure}[h]
\begin{center}
\includegraphics[width=5.5cm, height=5cm]{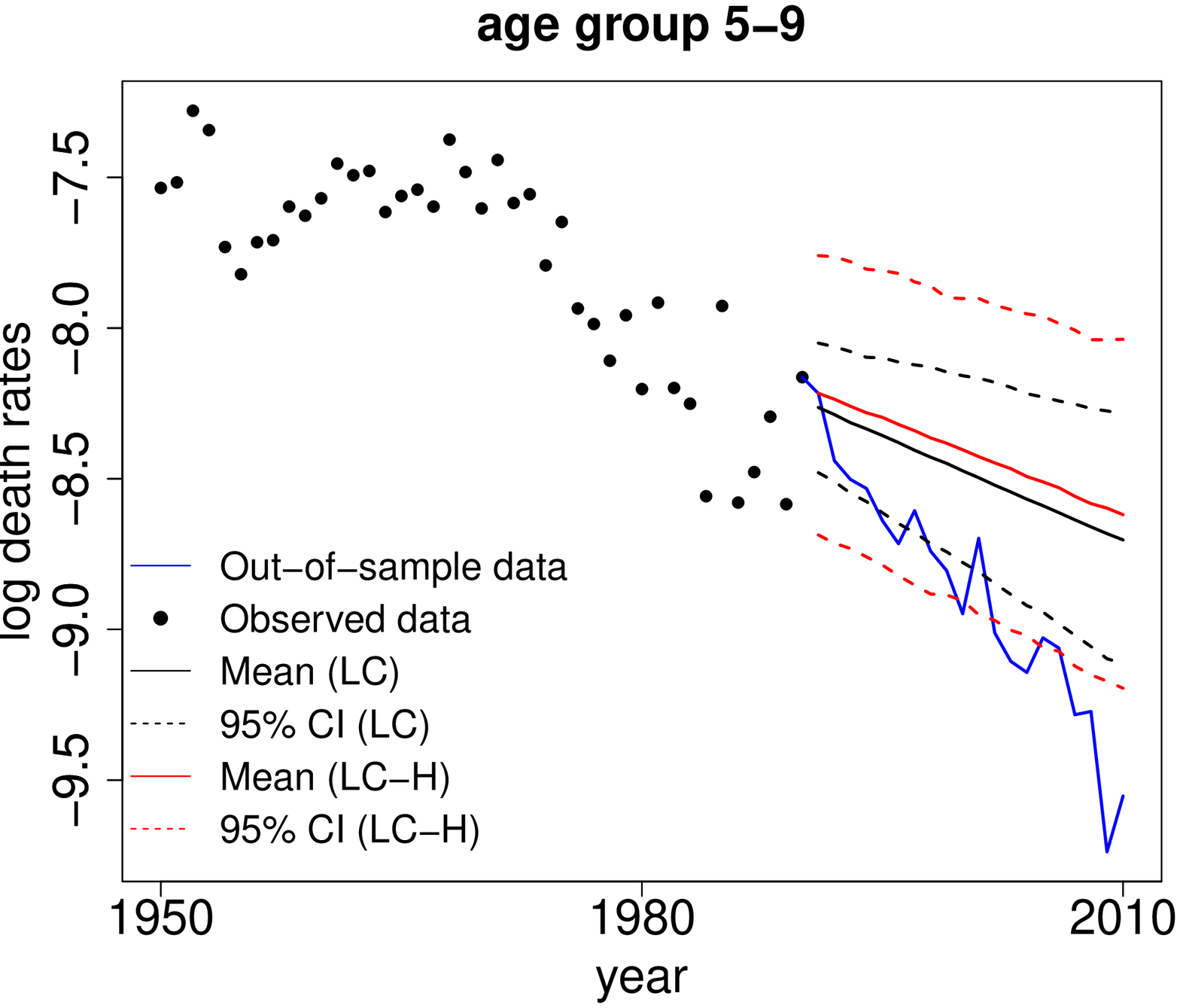}\includegraphics[width=5.5cm, height=5cm]{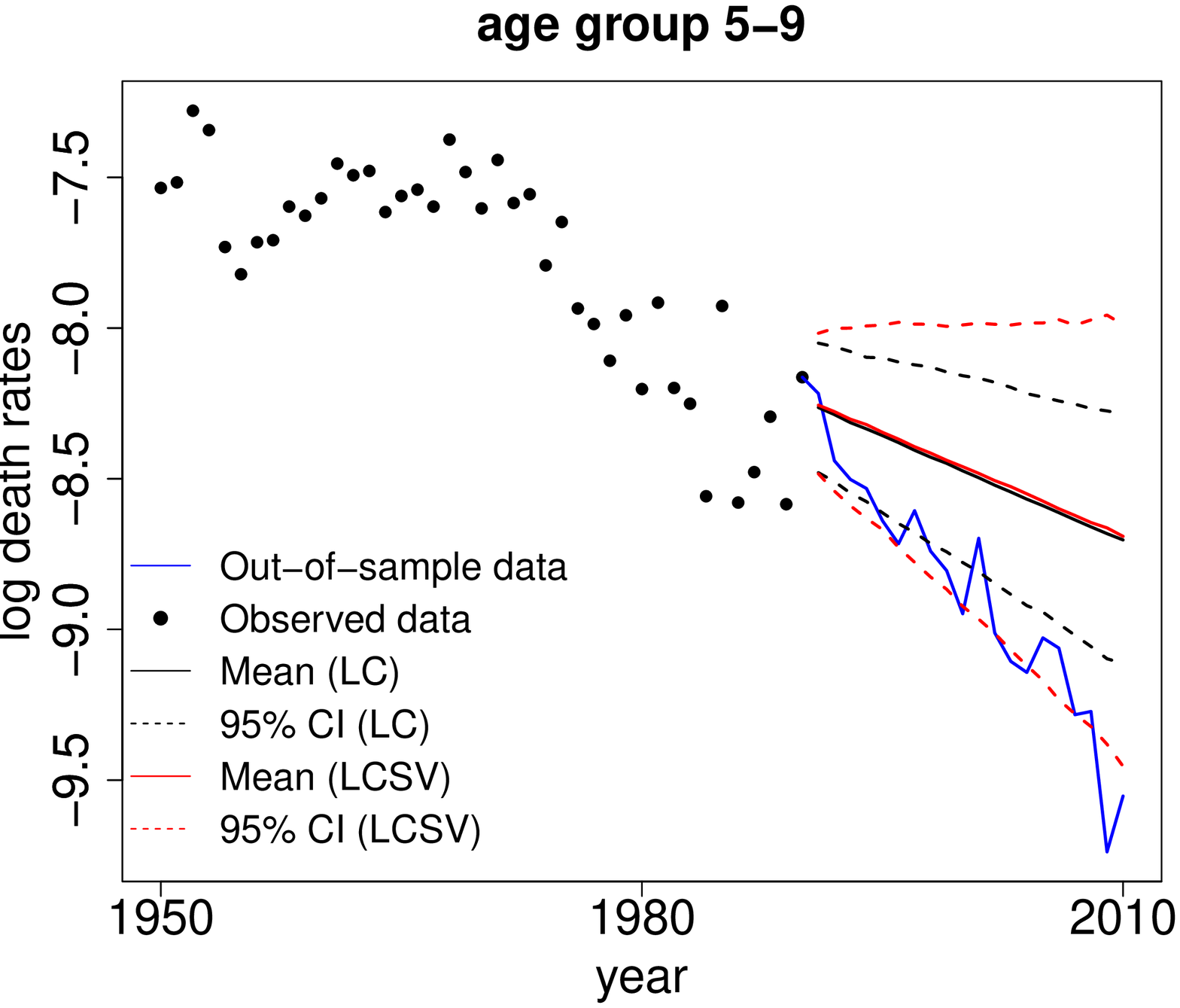}\includegraphics[width=5.5cm, height=5cm]{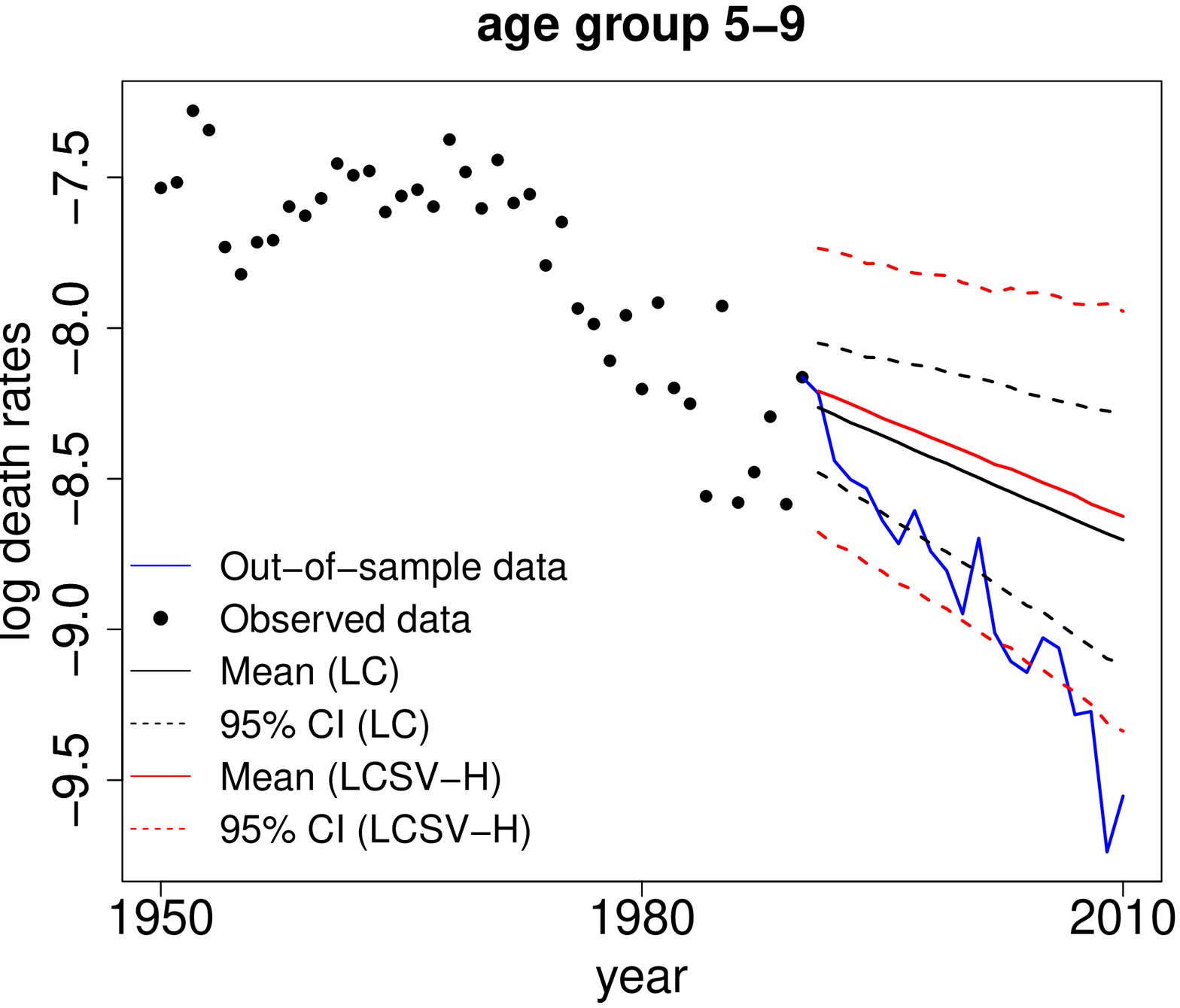}
\includegraphics[width=5.5cm, height=5cm]{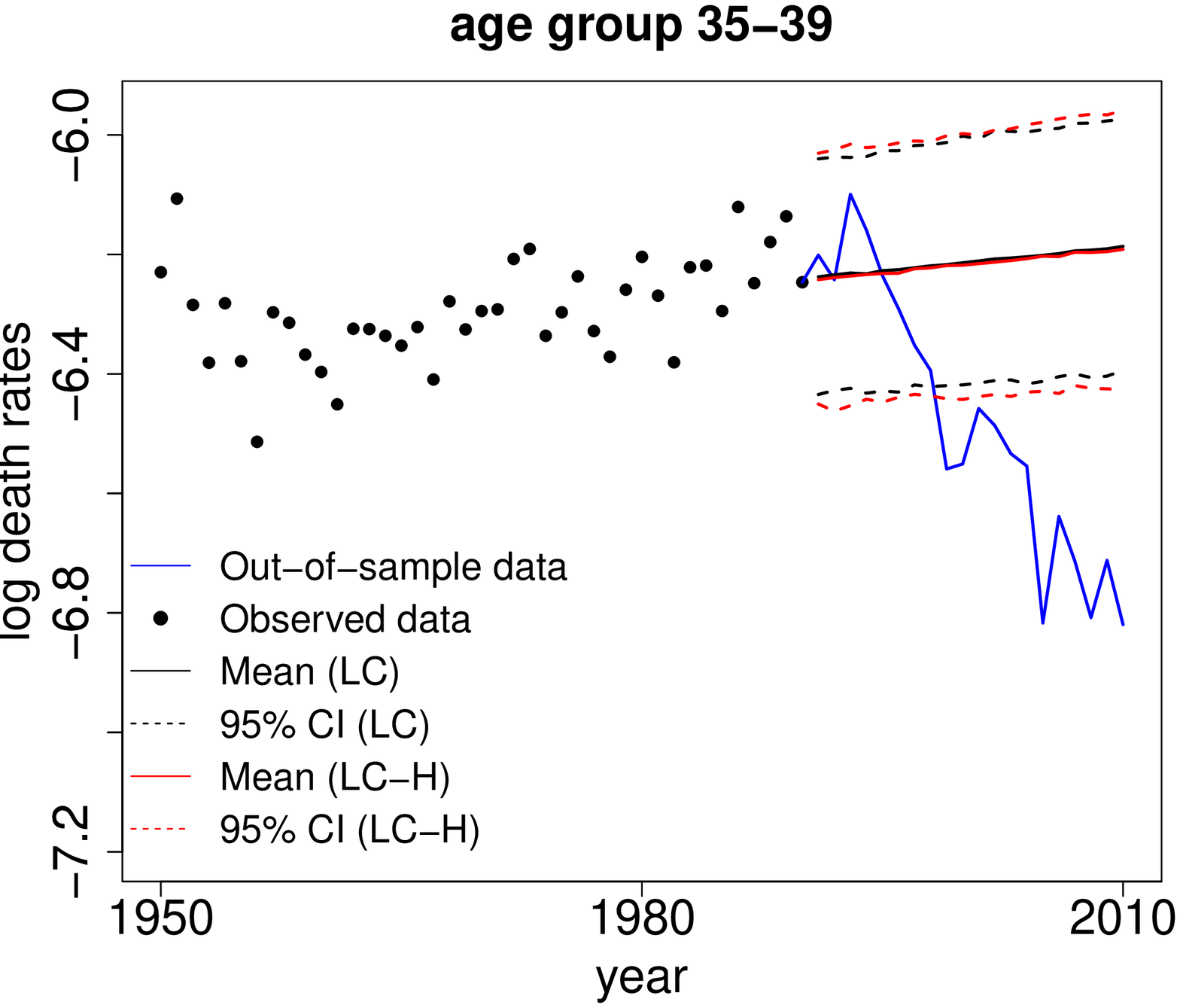}\includegraphics[width=5.5cm, height=5cm]{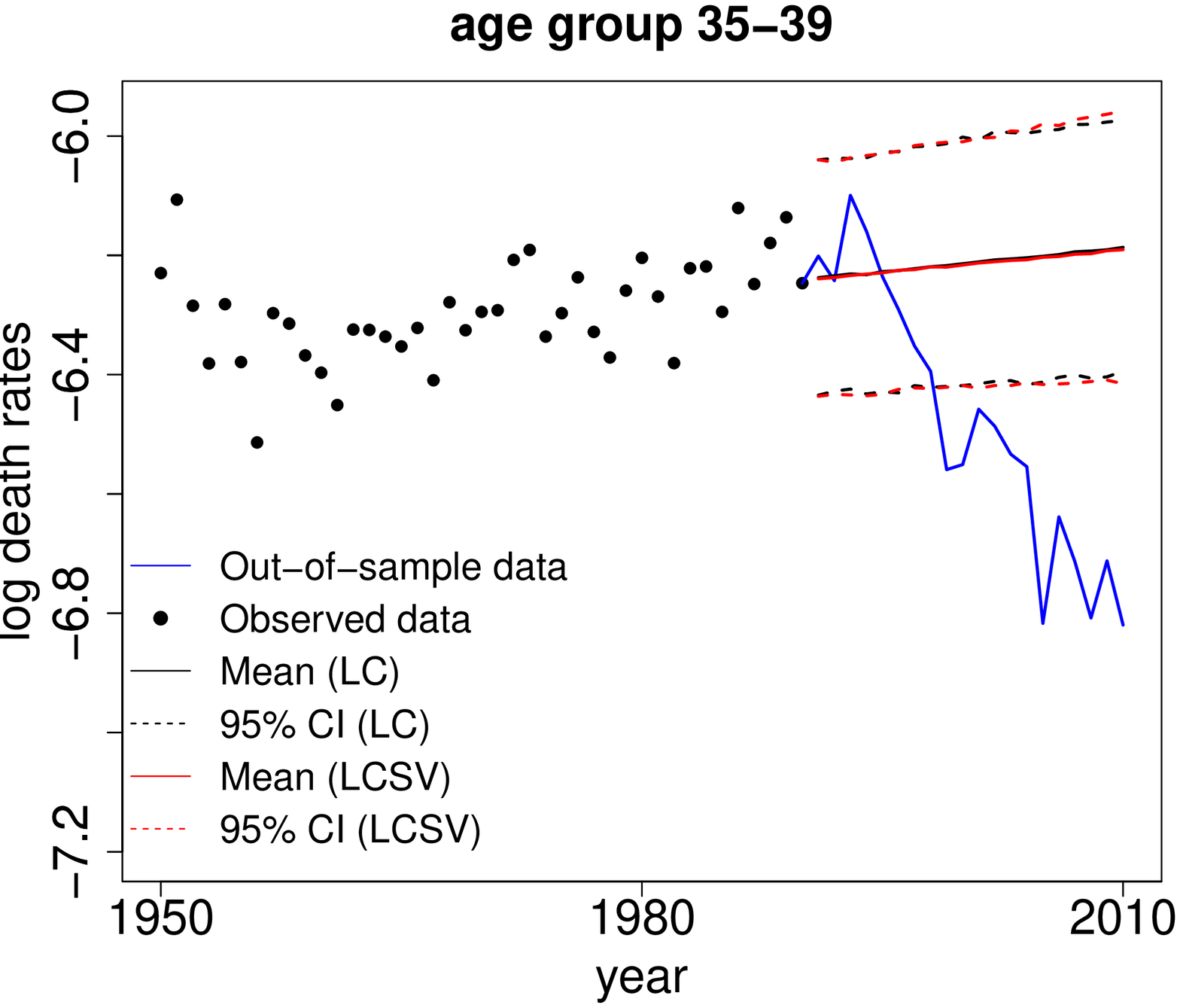}\includegraphics[width=5.5cm, height=5cm]{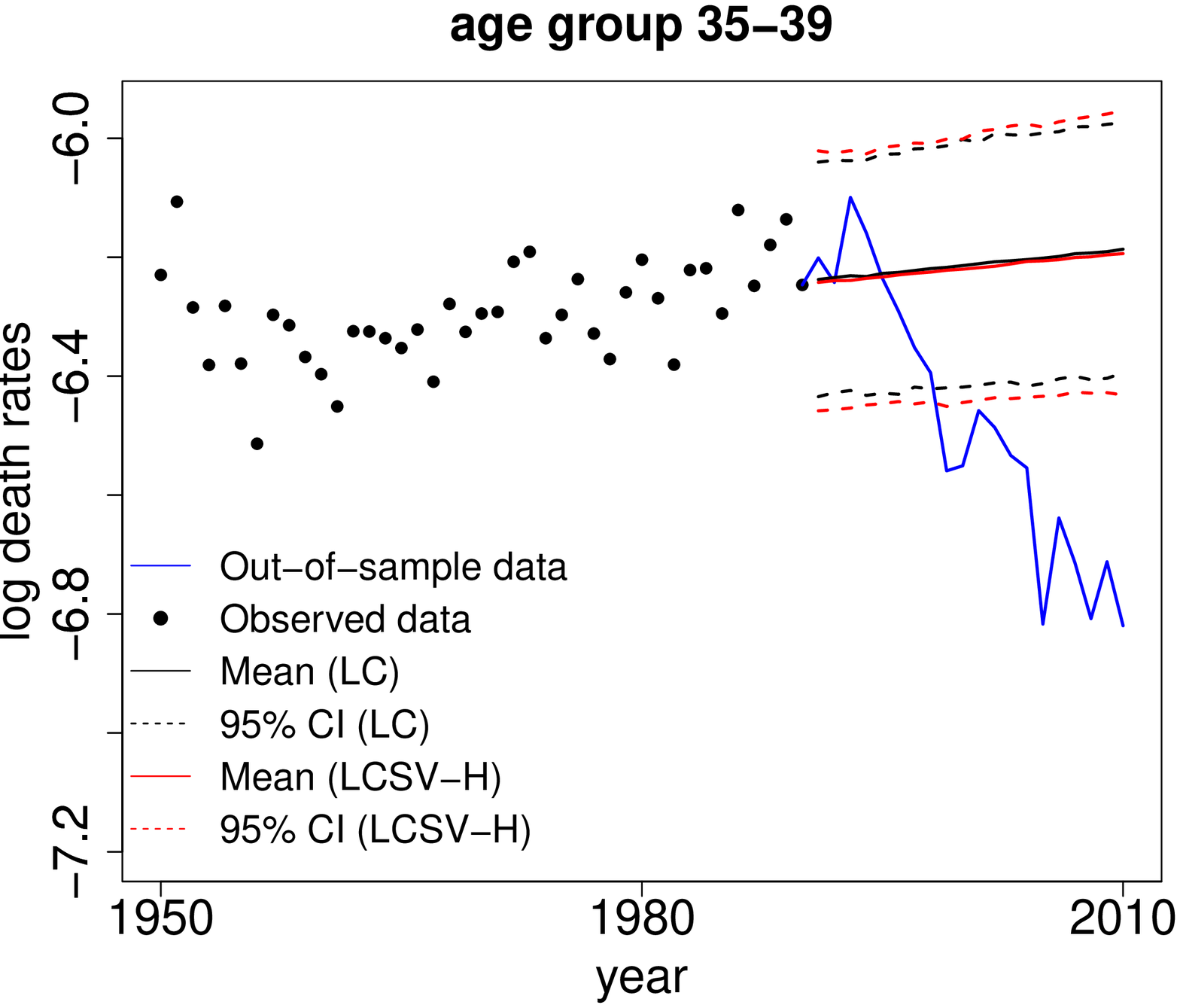}
\includegraphics[width=5.5cm, height=5cm]{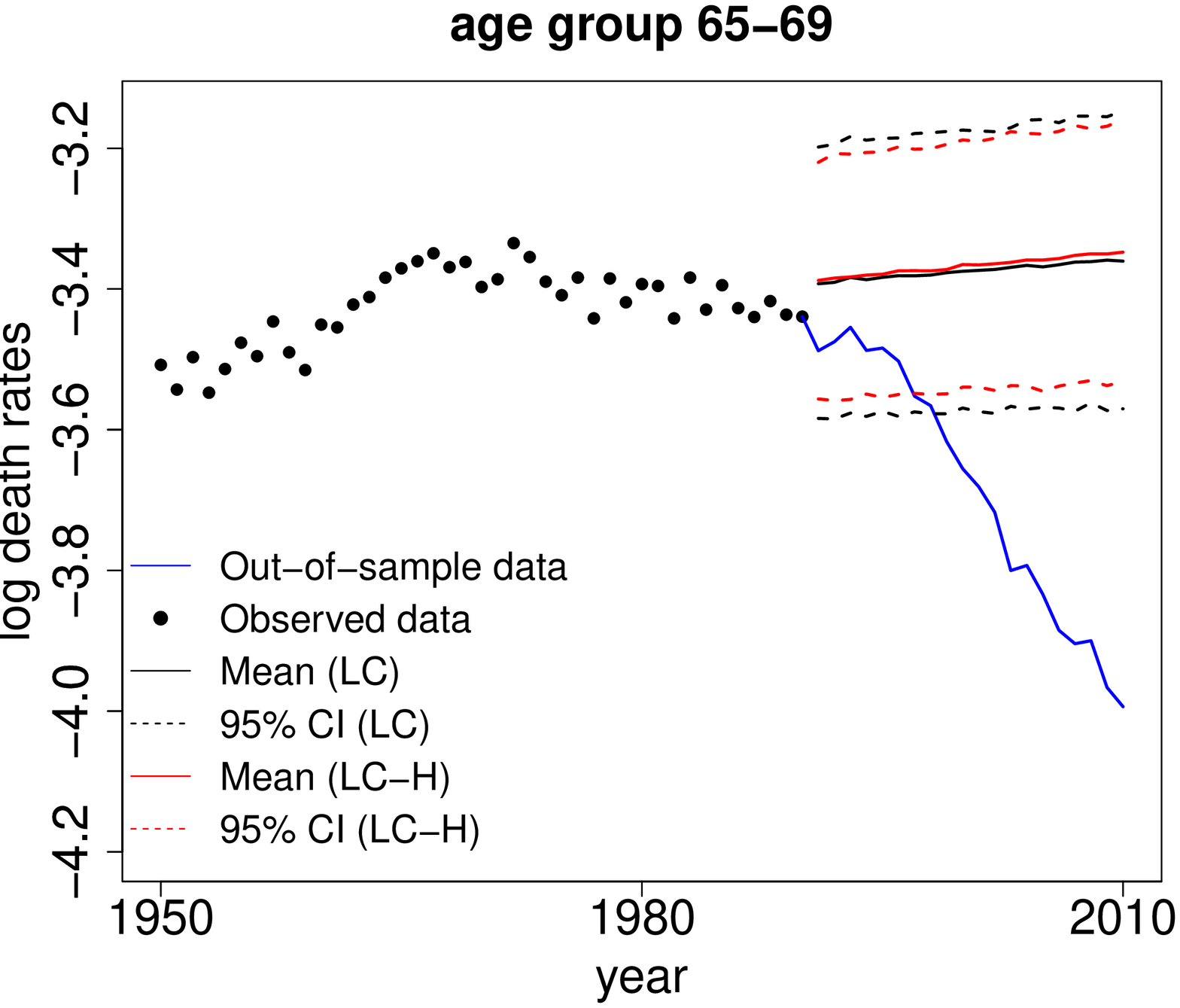}\includegraphics[width=5.5cm, height=5cm]{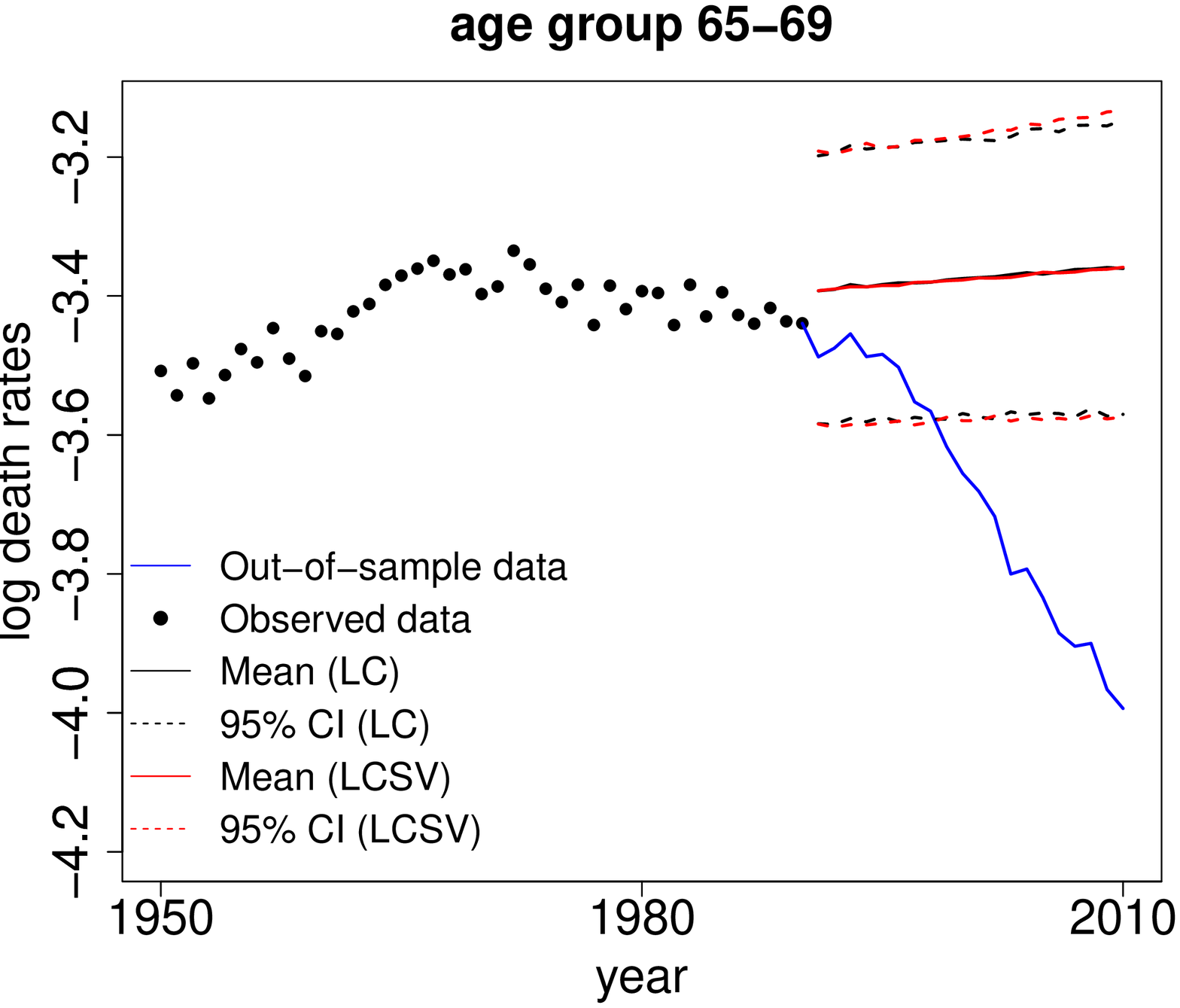}\includegraphics[width=5.5cm, height=5cm]{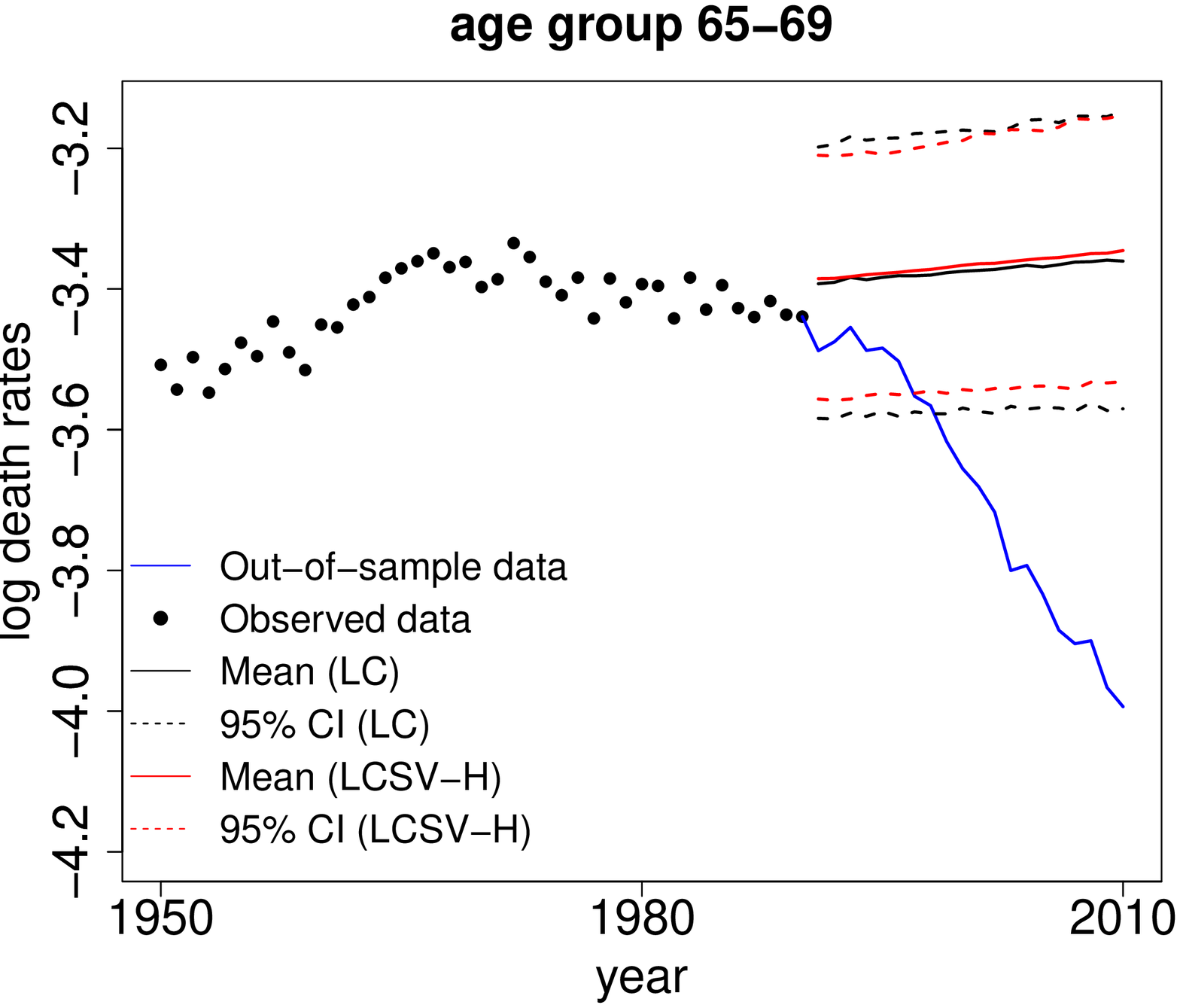}
\includegraphics[width=5.5cm, height=5cm]{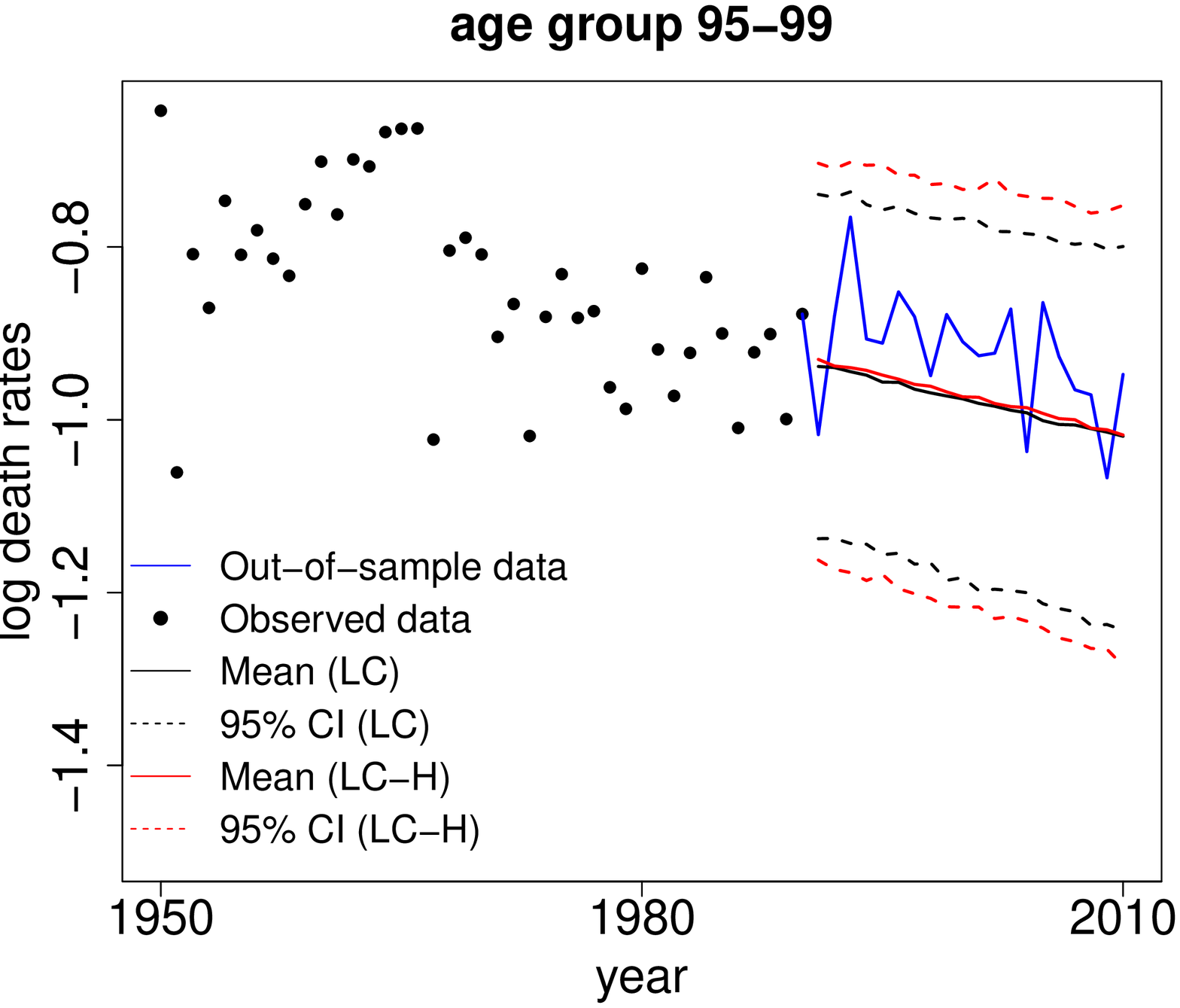}\includegraphics[width=5.5cm, height=5cm]{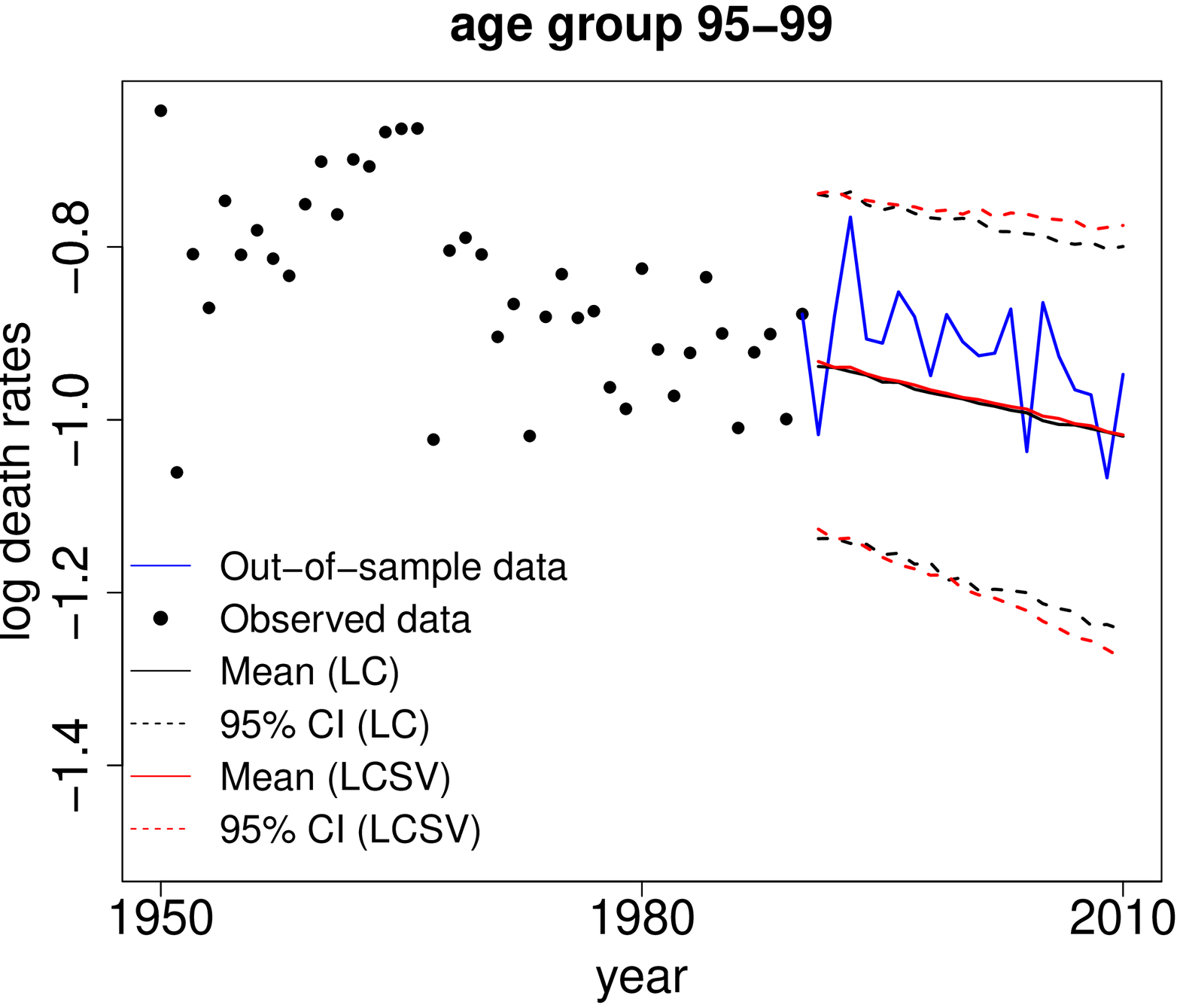}\includegraphics[width=5.5cm, height=5cm]{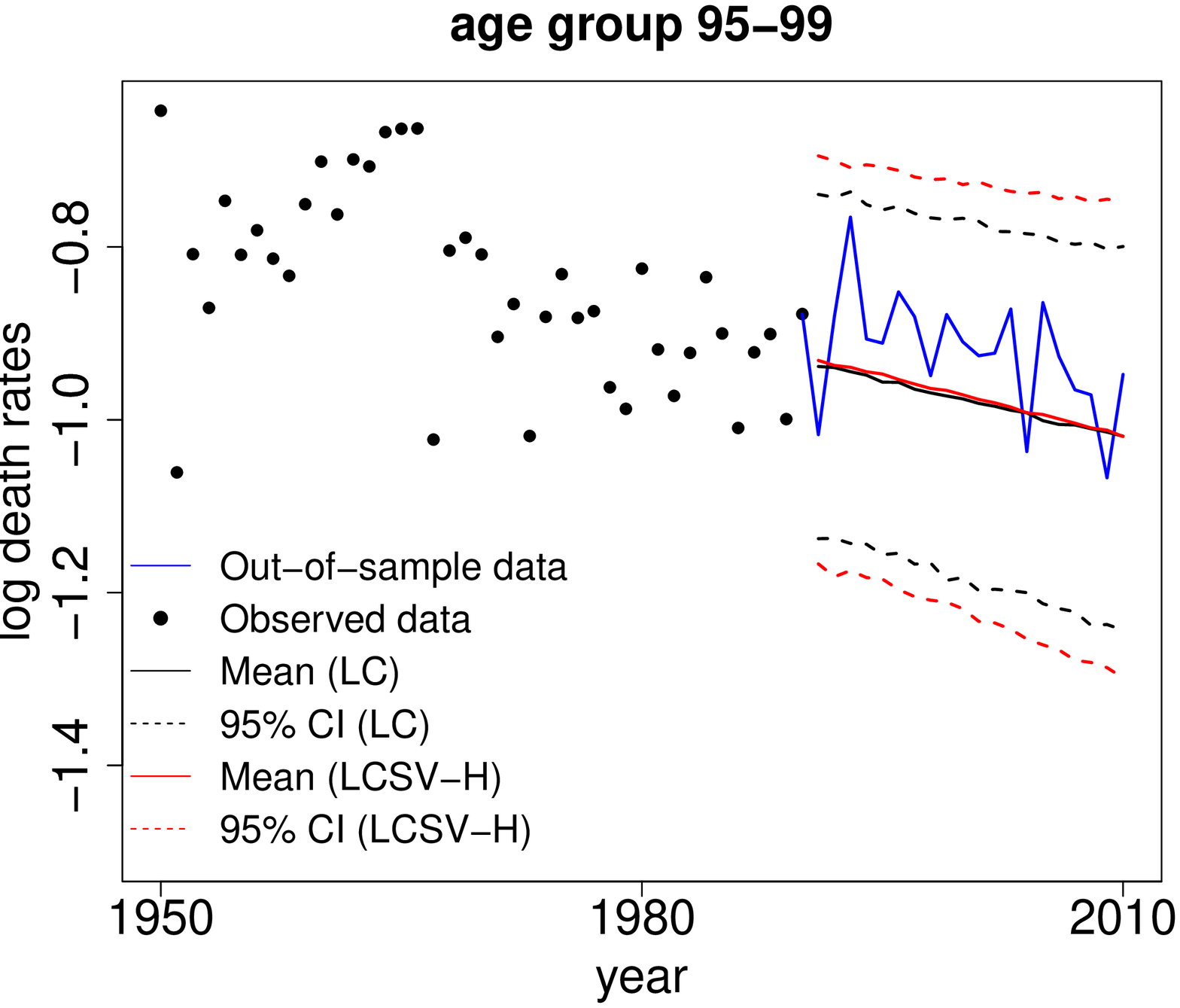}
\caption{\small{20-year out-of-sample forecasted log death rates for Danish male population under (left column) LC-H model, (middle column) LCSV model and (right column) LCSV-H model in comparison with LC model. Calibration period: 1950-1990.}}
\label{fig:forecastDeathRates19501990}
\end{center}
\end{figure}

Figure~\ref{fig:forecastDeathRates19501990} shows the forecasting distributions of log death rates where we assume a shorter calibration period from 1950 to 1990. For all the models, the estimated $\beta_x$ for all age groups, except for age groups 0, 1-4 and 5-9, are very close to zero. It is in fact expected since there is no clear downward trend in the observed mortality data besides the first few age groups, during the period 1950-1990. Therefore there is only small difference between the forecasting distributions produced by the LC model and the LCSV model, except for young age groups. Note that there is a clear change of downward trend for some of the middle age groups for the Danish male mortality data as shown in Figure~\ref{fig:forecastDeathRates19501990}. It results in the out-of-sample data falling out of the lower bound of the $95\%$ credible intervals and its consequence for the forecasting of life expectancy will be discussed in Section \ref{sec:lifeExp} and generally in Section \ref{sec:linearTrend}.

By comparing the forecast performance using in-sample data from 1835-1990 and from 1950-1990 displayed in Figure~\ref{fig:forecastDeathRates18351990} and \ref{fig:forecastDeathRates19501990}, we expose the influence that leaving out important historical events, that may affect the mortality rates markedly in a population, can have on the ability to accurately model trend and volatility structures in population dynamics. In particular we observe that one must be cautious as forecast performance can degrade markedly when important historical events are excluded from the sample as the forecast using data from 1835-1990 has clearly outperformed the forecast using only shorter calibration data from 1950-1990.

\subsubsection{Life expectancy}\label{sec:lifeExp}
Using the samples of the forecasted log death rates $y^{(\ell)}_{x,t}=\ln{\hat{m}}^{(\ell)}_{x,t}$, where $\ell=1,\dots,L$ and $L$ is the number of MCMC samples, we can obtain the so-called period life expectancy at different ages by constructing an abridged life table, since we use age group data, as follows (\cite{KoissiShHo06}, \cite{YusufMaSw}). We consider age group $x\in$ \{0, 1-4, 5-9, $\dots$, 95-99\} and $\tilde{x}$ is defined as the initial age of age group $x$, that is $\tilde{x} \in \{0, 1, 5, \dots, 90, 95\}$. Define $n_{\tilde{x}}$ as the length of the interval of age group $x$ (corresponds to $\tilde{x}$) and hence we have $n_0=1, n_1=4, n_5=5, \dots, n_{95}=5$. We then calculate the (crude) death probability\footnote{The death probability is ``crude" in the sense that the crude death rate is used for the calculation. For a discussion of crude and true death probabilities, see \cite{Dowdetal10}.} that a person aged $\tilde{x}$ in year $t$  will die in the next $n_{\tilde{x}}$ years as
\begin{equation}\label{eqn:nqx}
    {}_{n_{\tilde{x}}}\hat{q}^{(\ell)}_{\tilde{x},t} = \frac{n_{\tilde{x}}\, \hat{m}^{(\ell)}_{x,t}}{1+n_{\tilde{x}}(1-a(\tilde{x},n_{\tilde{x}}))\hat{m}^{(\ell)}_{x,t}},
\end{equation}
where $a(\tilde{x},n_{\tilde{x}})$ is the average fraction of the $n_{\tilde{x}}$ years lived by the people who is initially aged $\tilde{x}$ in that interval. Using the assumption that deaths are distributed uniformly in the interval, we set $a(\tilde{x},n_{\tilde{x}}) = 0.5$ for every $\tilde{x}$.\footnote{The Human Mortality Database provides abridged life tables with specific values for $a(\tilde{x},n_{\tilde{x}})$ in different period. For ease of comparison we use the typical assumption that $a(\tilde{x},n_{\tilde{x}}) = 0.5$. For the special case when age, instead of age group is considered (that is $\tilde{x}=x$), the one-year death probability $q_{x,t}$ in year $t$ is defined as the ratio of death counts and the population at the beginning of the year. Assuming half of the deaths occurred during the first half of the year, then we have
$q_{x,t}:=D_{x,t}/(E_{x,t}+0.5\,D_{x,t})=\hat{m}_{x,t}/(1+0.5\,\hat{m}_{x,t})$ where $E_{x,t}$ is the population at the middle of the year. The $n_{\tilde{x}}$-year death probability \eqref{eqn:nqx} for the general case when age group is considered can be derived similarly.} The hypothetical number of people alive at age $\tilde{x}+n_{\tilde{x}}$, $l^{(\ell)}_{\tilde{x}+n_{\tilde{x}},t}$, is determined by $l^{(\ell)}_{\tilde{x}+n_{\tilde{x}},t}=l^{(\ell)}_{\tilde{x},t}\left(1-{}_{n_{\tilde{x}}}q^{(\ell)}_{\tilde{x},t}\right)$ where $l^{(\ell)}_{0,t}$ is assumed to be $100,000$. We can then calculate the number of deaths ${}_{n_{\tilde{x}}}d^{(\ell)}_{\tilde{x},t} = l^{(\ell)}_{\tilde{x},t}-l^{(\ell)}_{\tilde{x}+n_{\tilde{x}},t}$ and the person-years lived
${}_{n_{\tilde{x}}}L^{(\ell)}_{\tilde{x},t}=n_{\tilde{x}}\left(l^{(\ell)}_{\tilde{x}+n_{\tilde{x}},t}+a(\tilde{x},n_{\tilde{x}}) \times {}_{n_{\tilde{x}}}d^{(\ell)}_{\tilde{x},t}\right
)$. The total future lifetime of the $l^{(\ell)}_{\tilde{x},t}$ persons who attain age $\tilde{x}$ is $T^{(\ell)}_{\tilde{x},t}=\sum_{i\geq \tilde{x}} {}_{n_{\tilde{x}}}L^{(\ell)}_{i,t}$, where $i\in\{0, 1, 5, \dots, 90, 95\}$. Finally, a sample of the period life expectancy at age $\tilde{x}$ is obtained as
\begin{equation}
    e^{(\ell)}_{\tilde{x},t} = T^{(\ell)}_{\tilde{x},t}/l^{(\ell)}_{\tilde{x},t}
\end{equation}
and the distributions are obtained in different forecasting year $t=T+k$ where $k\geq 1$.

\begin{remark}[Period and cohort life expectancy]
Period life expectancy assumes there is no trend for future death rates (it is evaluated based on the age-specific death rates in a fixed year $t$) while cohort life expectancy assumes death rates following the lifetime of a cohort and hence it takes mortality trend into account. For example, to evaluate period life expectancy at age $65$ in year $t$, one needs $\{{}_5q_{65,t},{}_5q_{70,t},\dots,{}_5q_{95,t}\}$ while for cohort life expectancy, $\{{}_5q_{65,t},{}_5q_{70,t+5},\dots,{}_5q_{95,t+30}\}$ are used instead. However, the cohort life expectancy for people born in recent years cannot be evaluated using data alone since some of the death rates data are yet to be observed. As a result we focus on period life expectancy so that our forecasts can be compared with the observed data.\end{remark}

\begin{figure}[h]
\begin{center}
\includegraphics[width=5.5cm, height=5cm]{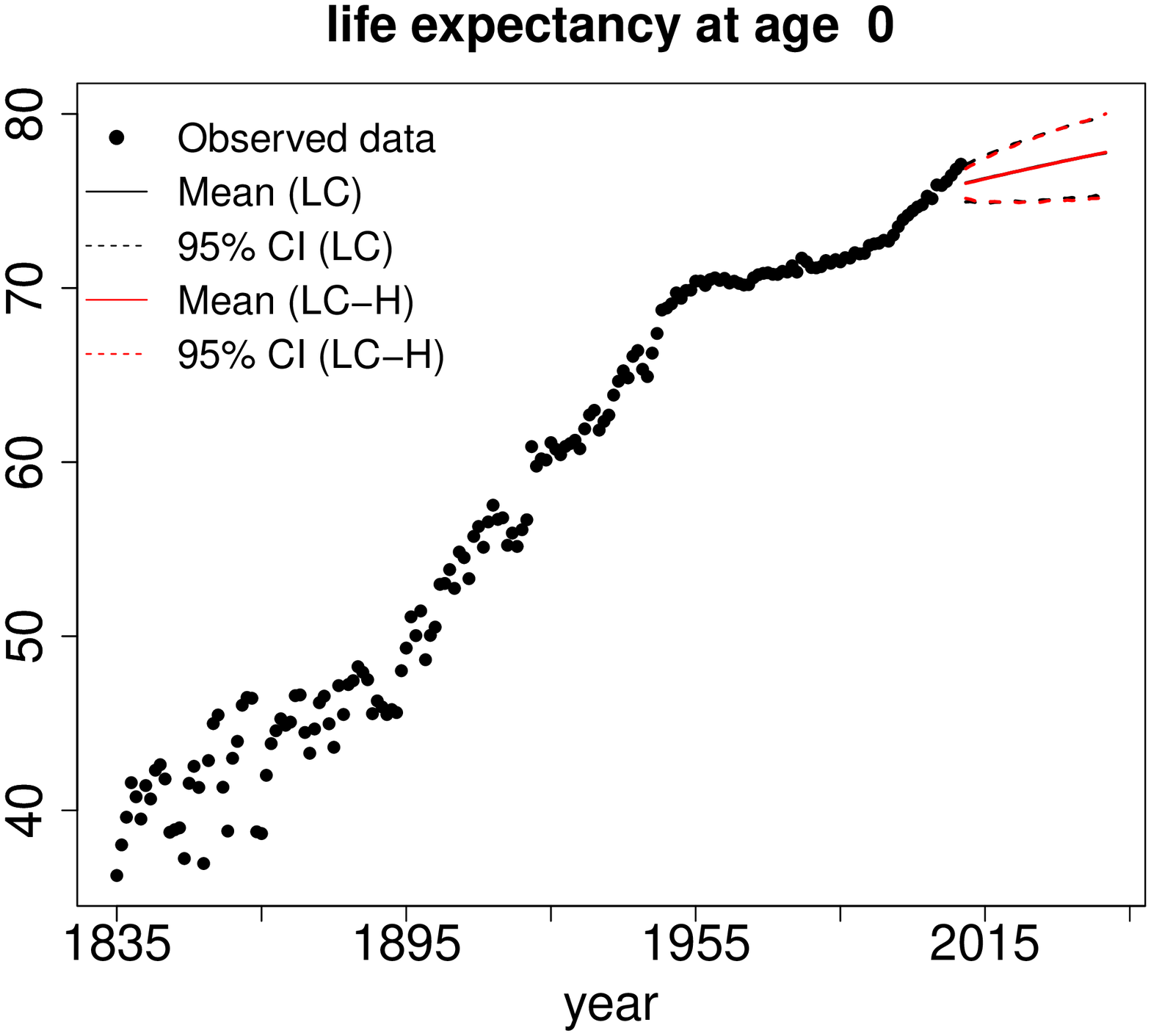}\includegraphics[width=5.5cm, height=5cm]{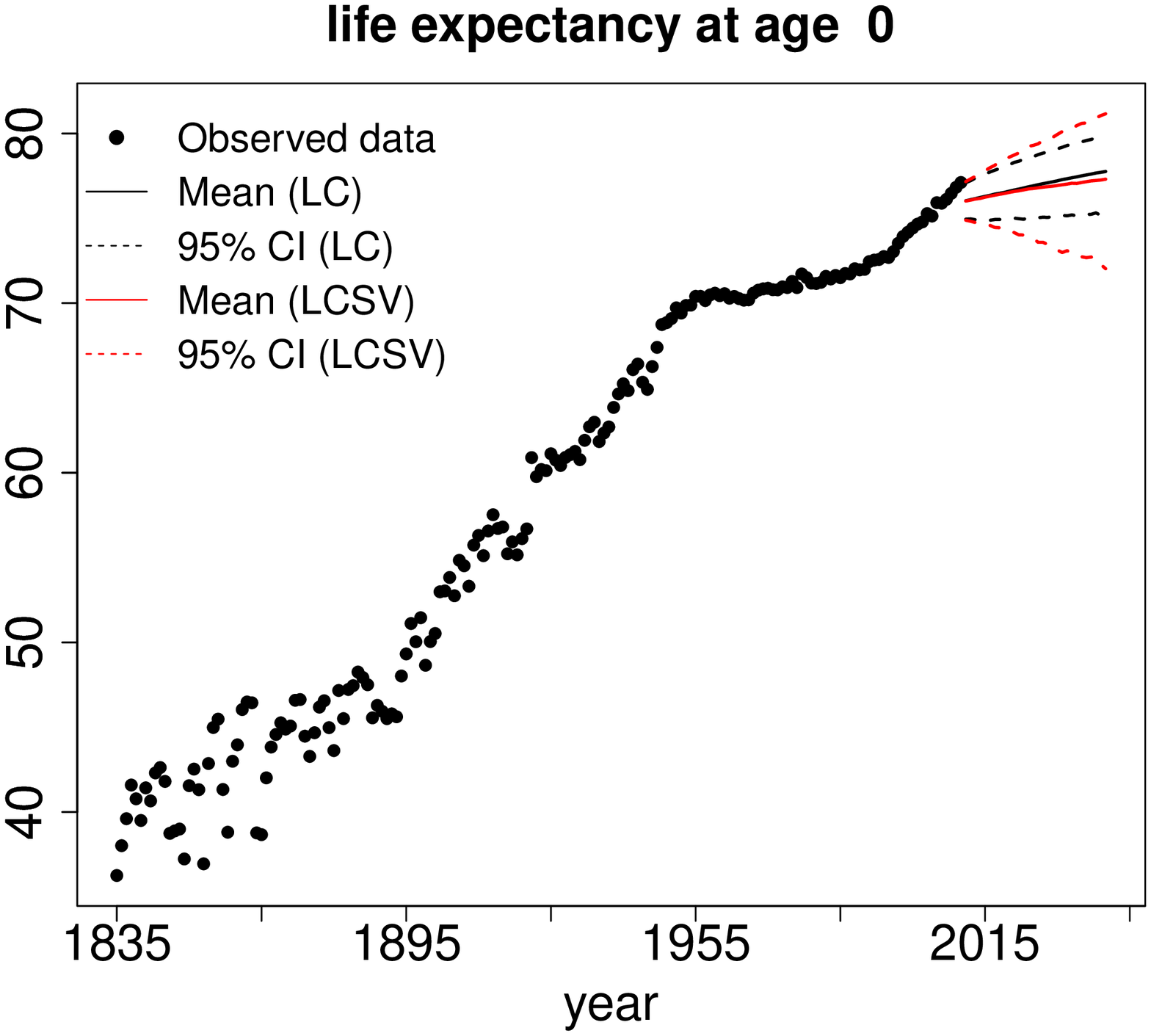}\includegraphics[width=5.5cm, height=5cm]{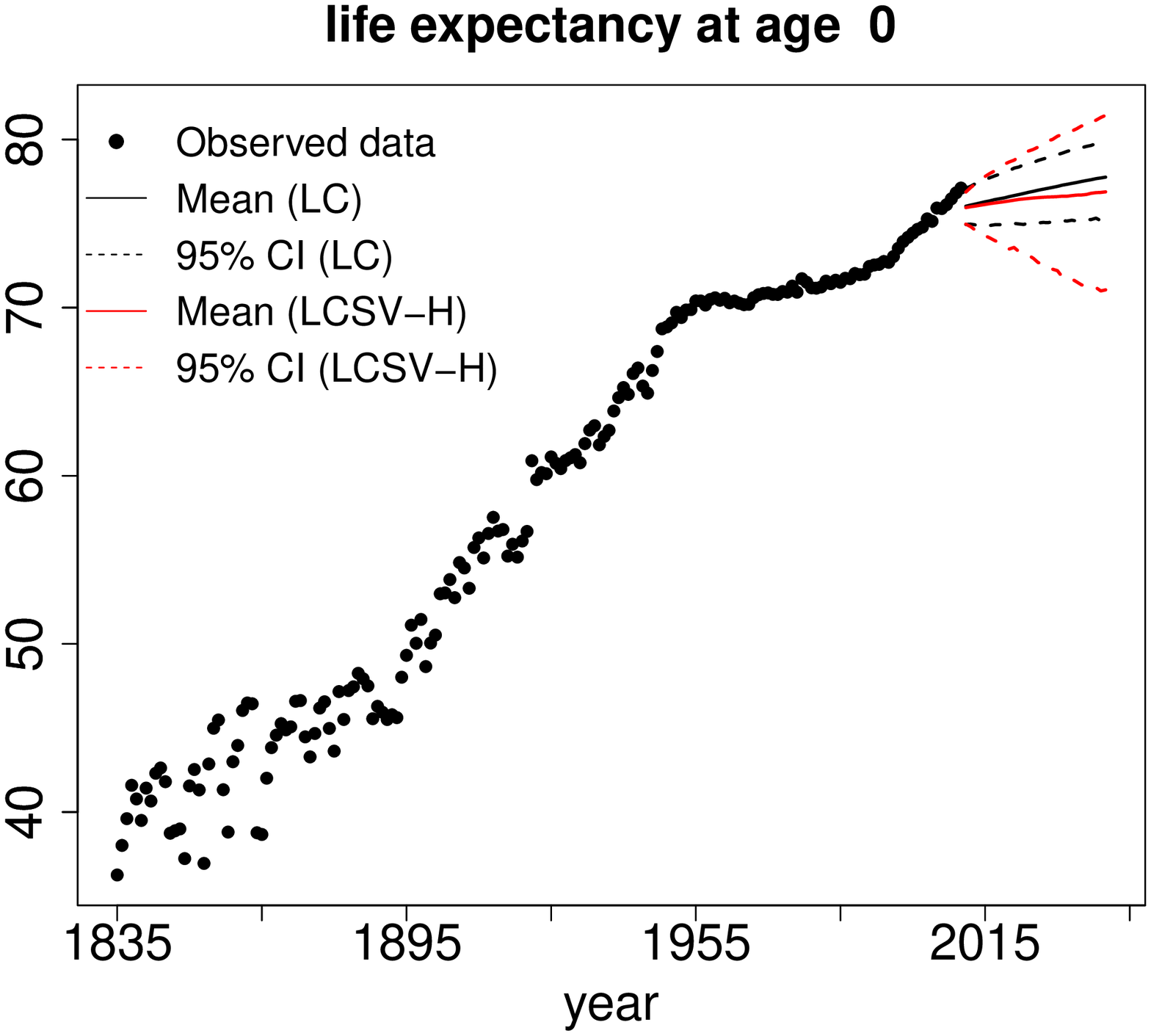}
\includegraphics[width=5.5cm, height=5cm]{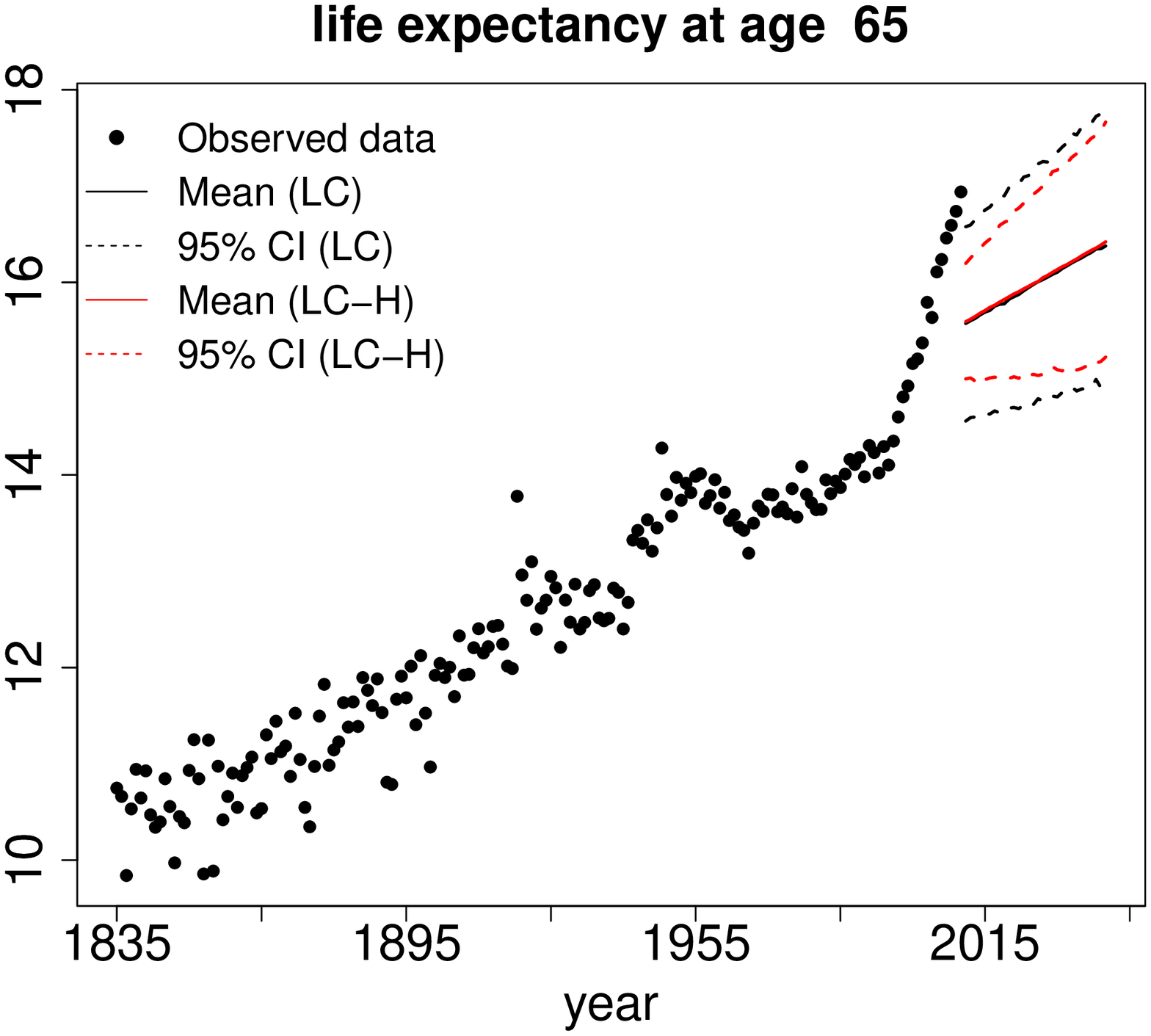}\includegraphics[width=5.5cm, height=5cm]{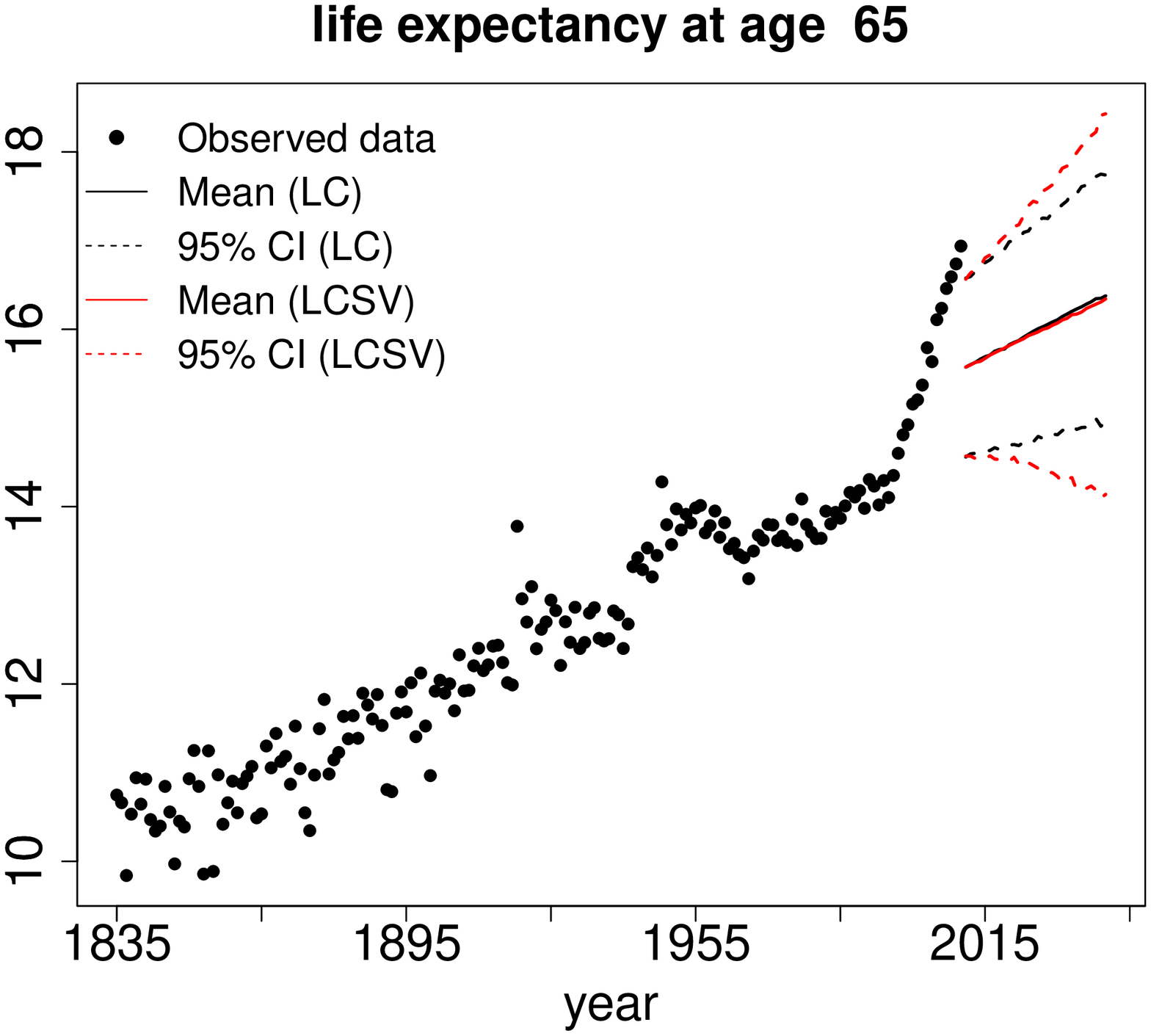}\includegraphics[width=5.5cm, height=5cm]{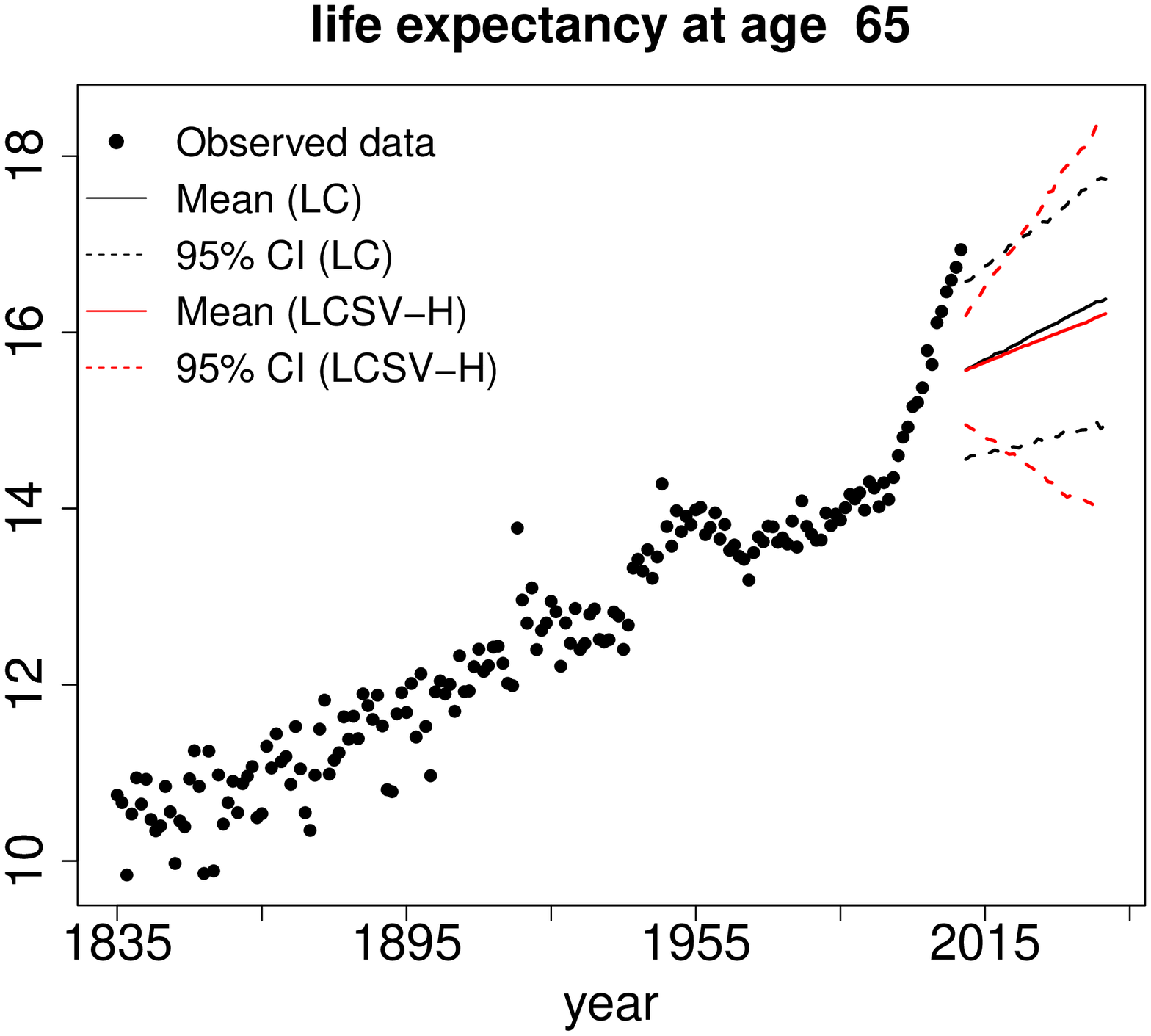}
\includegraphics[width=5.5cm, height=5cm]{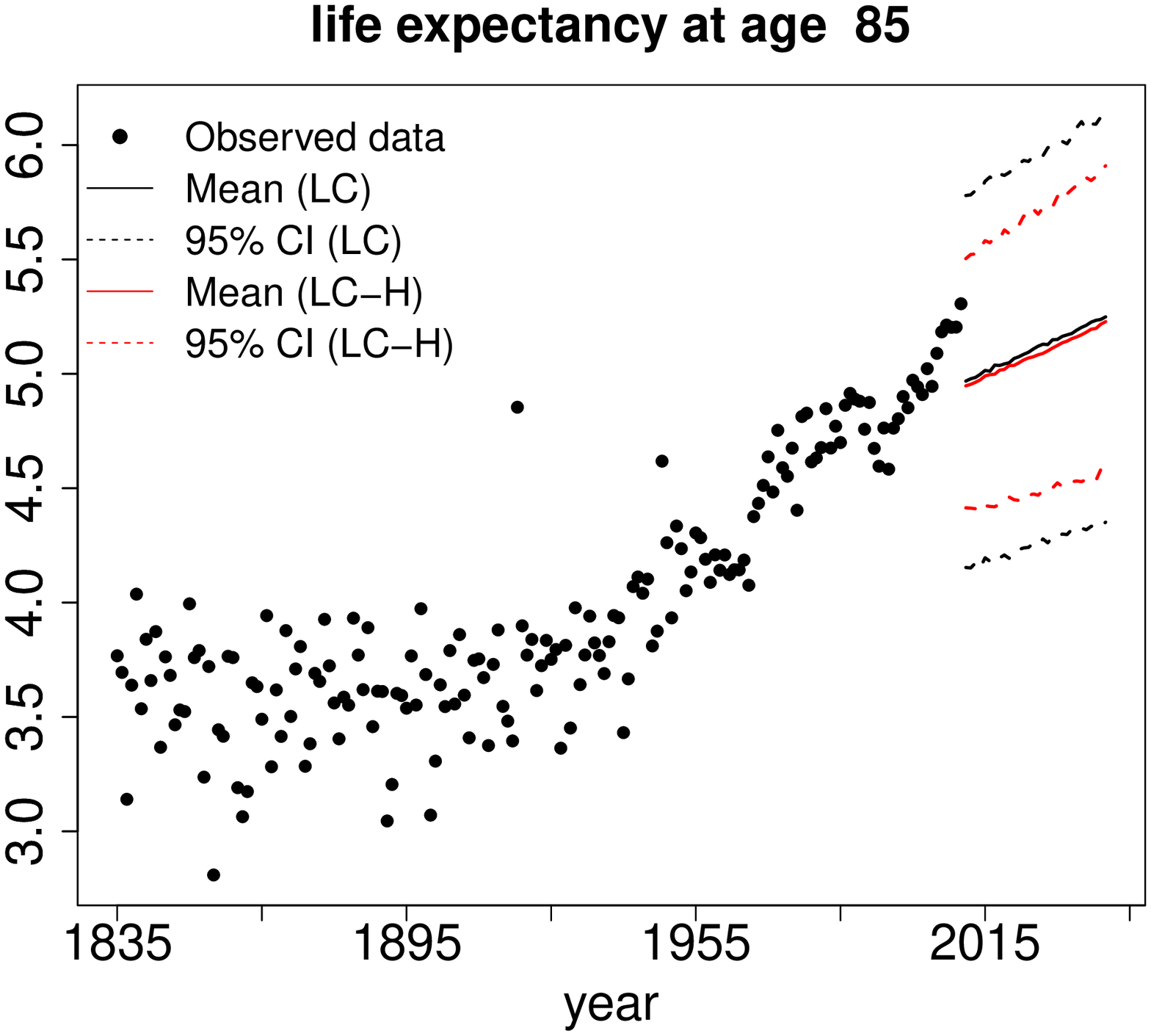}\includegraphics[width=5.5cm, height=5cm]{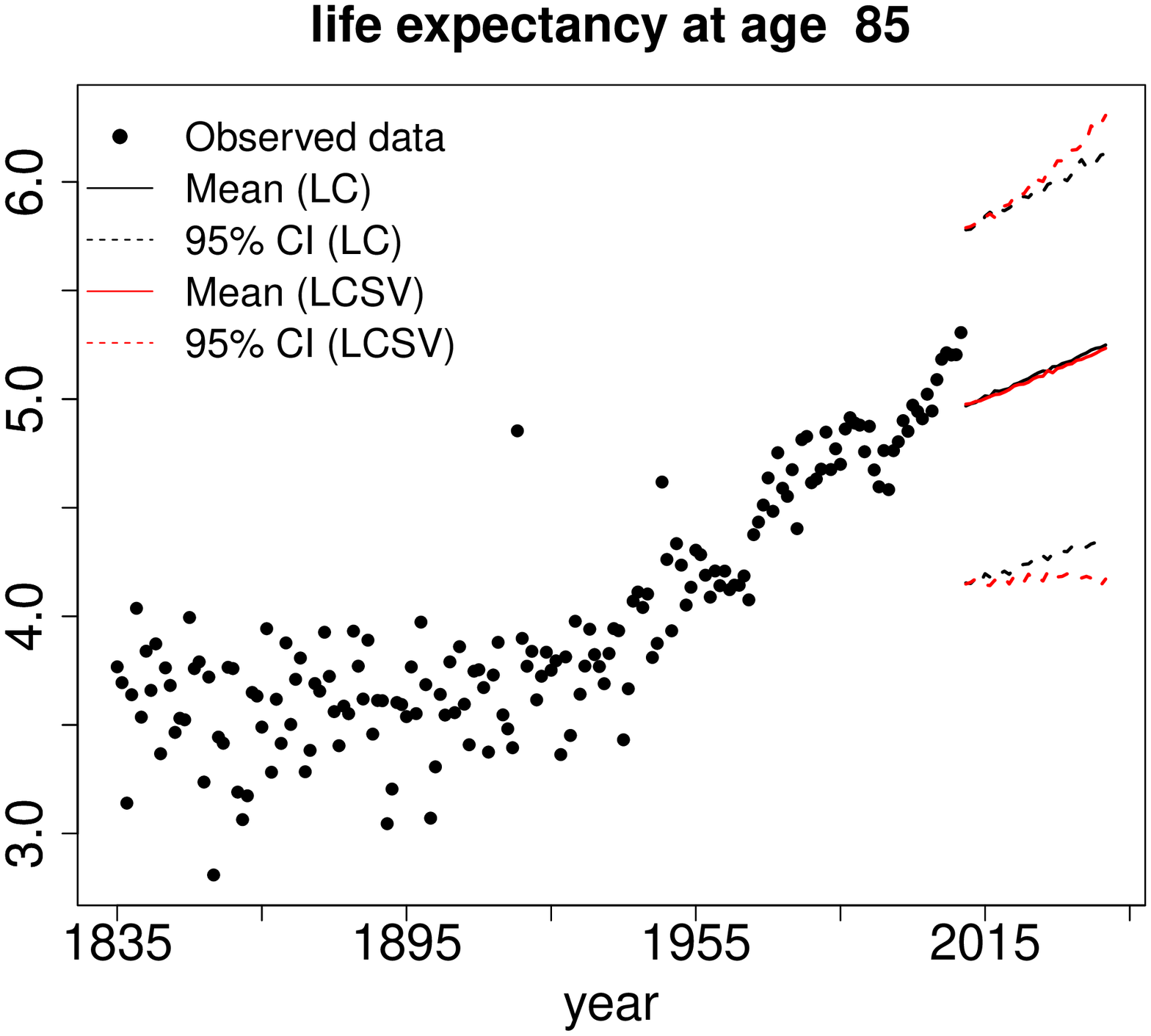}\includegraphics[width=5.5cm, height=5cm]{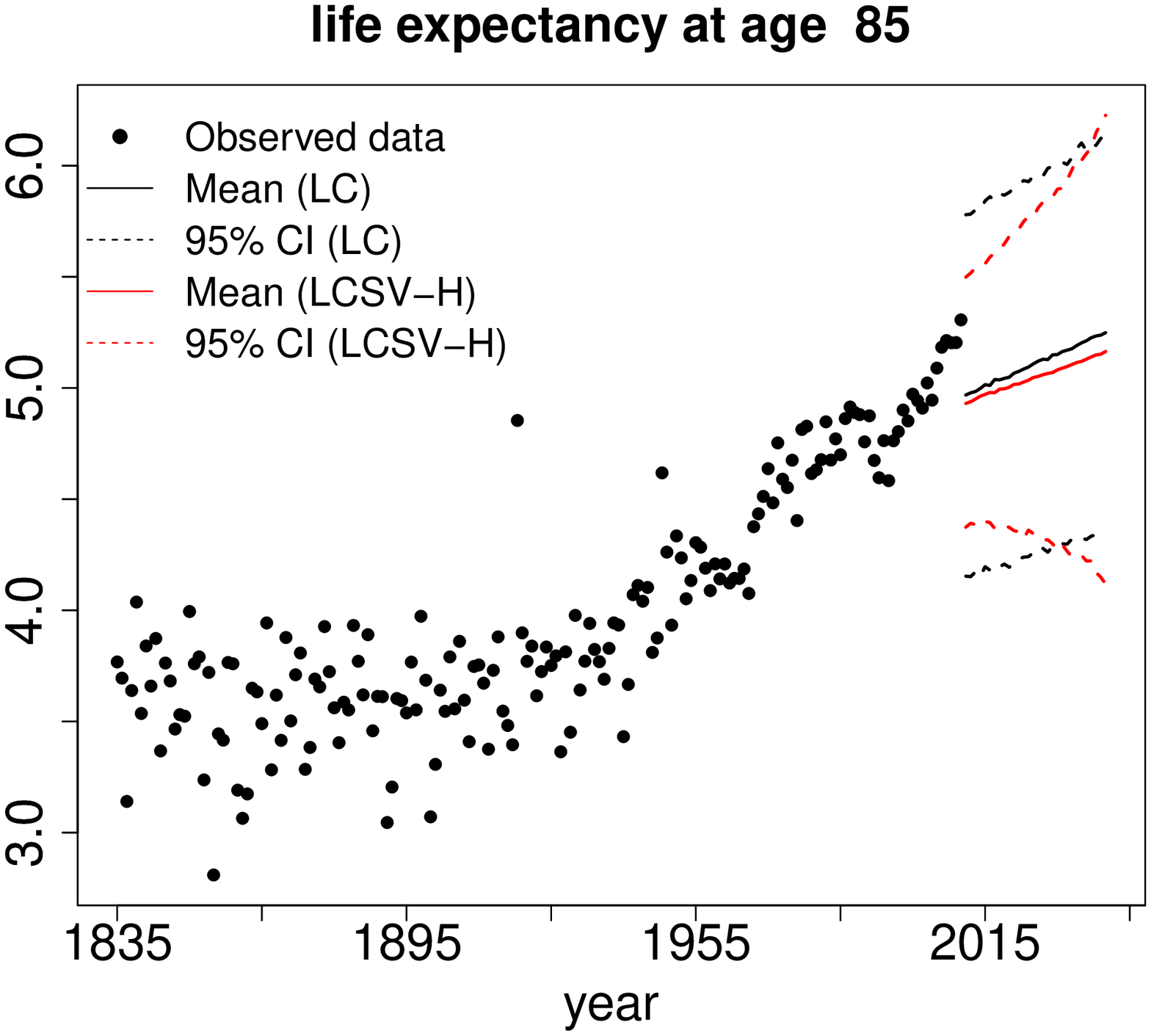}
\caption{\small{30-year forecasted life expectancy (2011-2041) at birth, age 65 and 85 for Danish male population under (left column) LC-H model, (middle column) LCSV model and (right column) LCSV-H model in comparison with LC model. Calibration period: 1835-2010.}}
\label{fig:forecastLifeExp18352010}
\end{center}
\end{figure}

Figure~\ref{fig:forecastLifeExp18352010} shows 30-year forecasted (period) life expectancy at birth, age 65 and age 85 for all the models estimated using data from 1835-2010. Interestingly, the forecasted life expectancy at birth is similar for the LC and LC-H model. It reflects the fact that forecast intervals of death rates produced by the LC-H model are wider for some age groups and narrower for others, compared to the LC model. These effects tend to cancel each other out as death rates are aggregated for all age groups to form the life expectancy at birth, resulting with a comparable life expectancy at birth distributions. This explanation does not apply to life expectancy at age 65 and 85, however, since only forecasted death rates for age groups larger than 65 and 85 are used to obtained the corresponding life expectancy distribution. As the forecast intervals of death rates generated by the LC-H model are narrower for old age groups compared with the LC model, the interval for the forecasted life expectancy at age 65 and 85 distribution produced by the LC-H model is observably narrower than the LC model.

The higher variability of the forecasted  death rates for the LCSV model translates to a wider forecast interval for life expectancy compared to the LC model. Similarly to the case of death rates forecasting, the LCSV-H model has both the features of the LC-H and LCSV model in terms of life expectancy prediction.

\begin{figure}[h]
\begin{center}
\includegraphics[width=5.5cm, height=5cm]{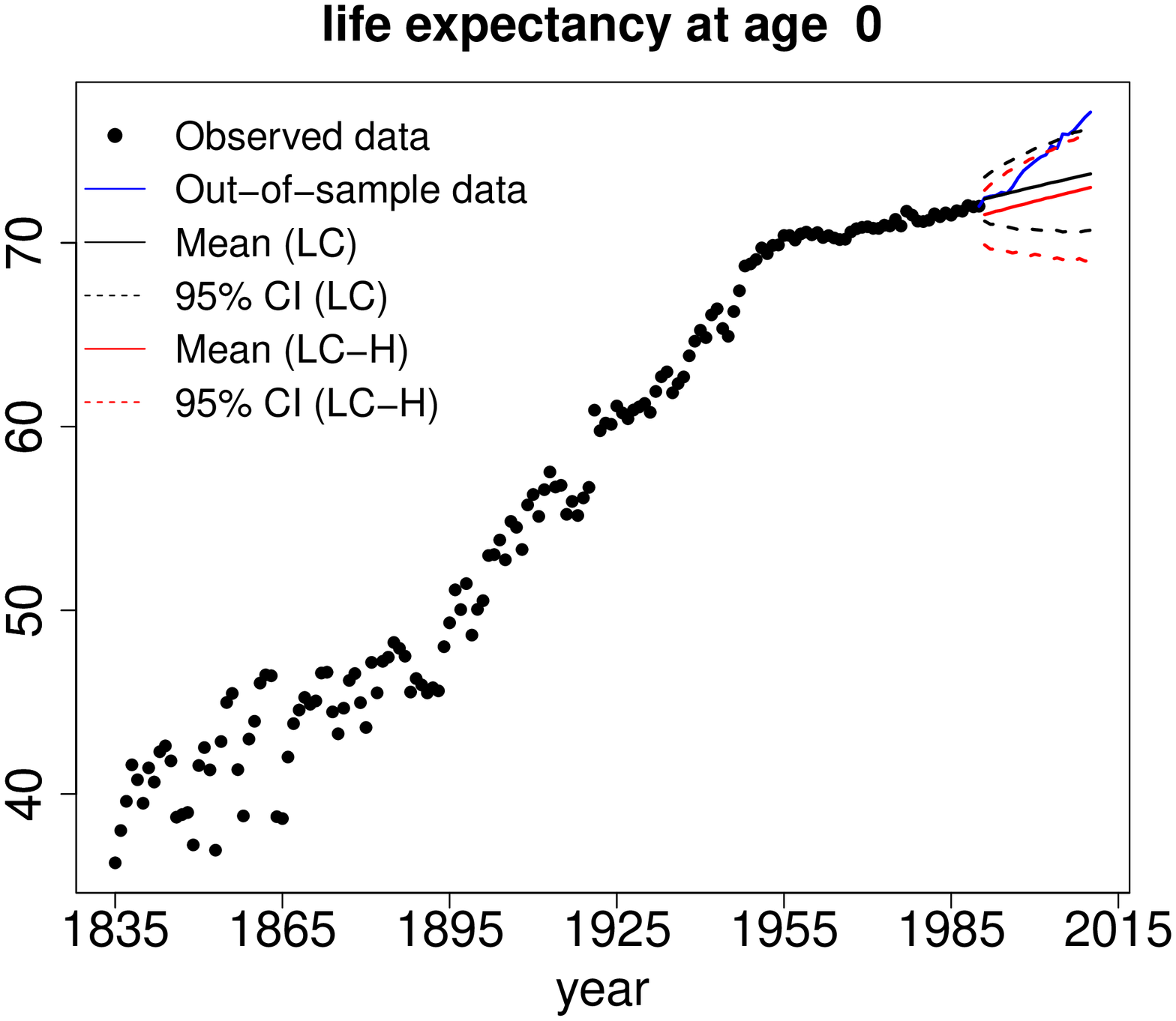}\includegraphics[width=5.5cm, height=5cm]{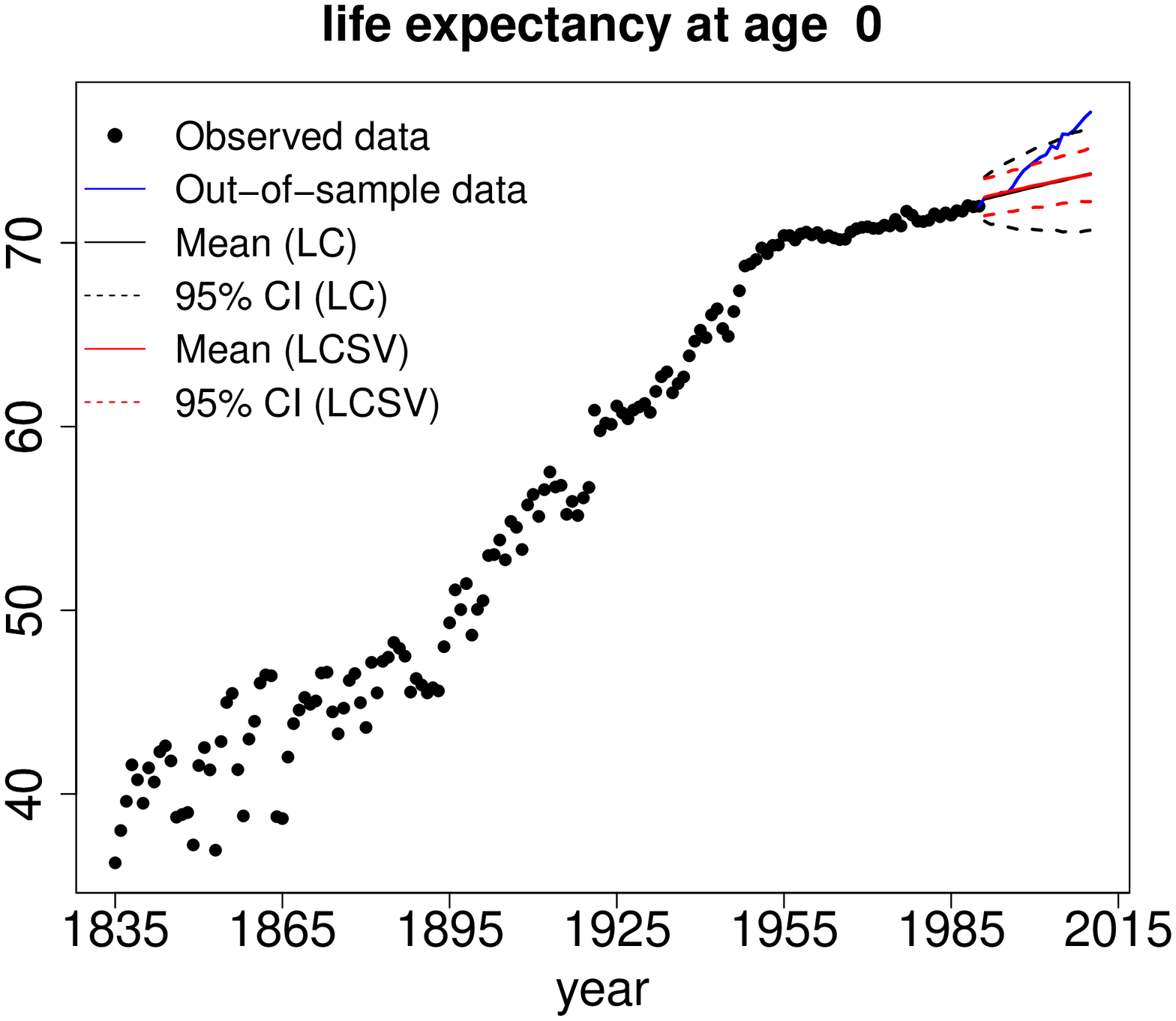}\includegraphics[width=5.5cm, height=5cm]{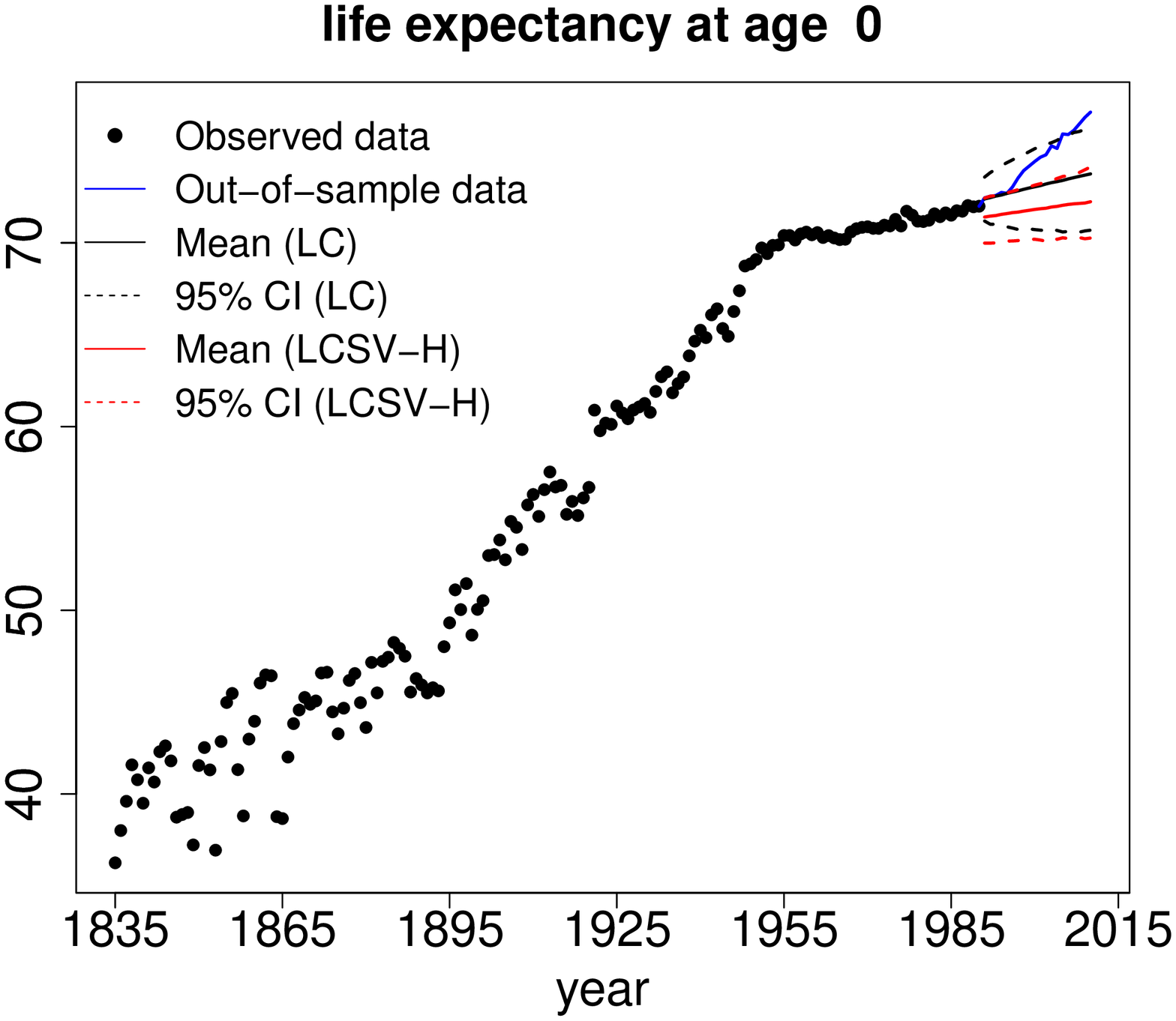}
\includegraphics[width=5.5cm, height=5cm]{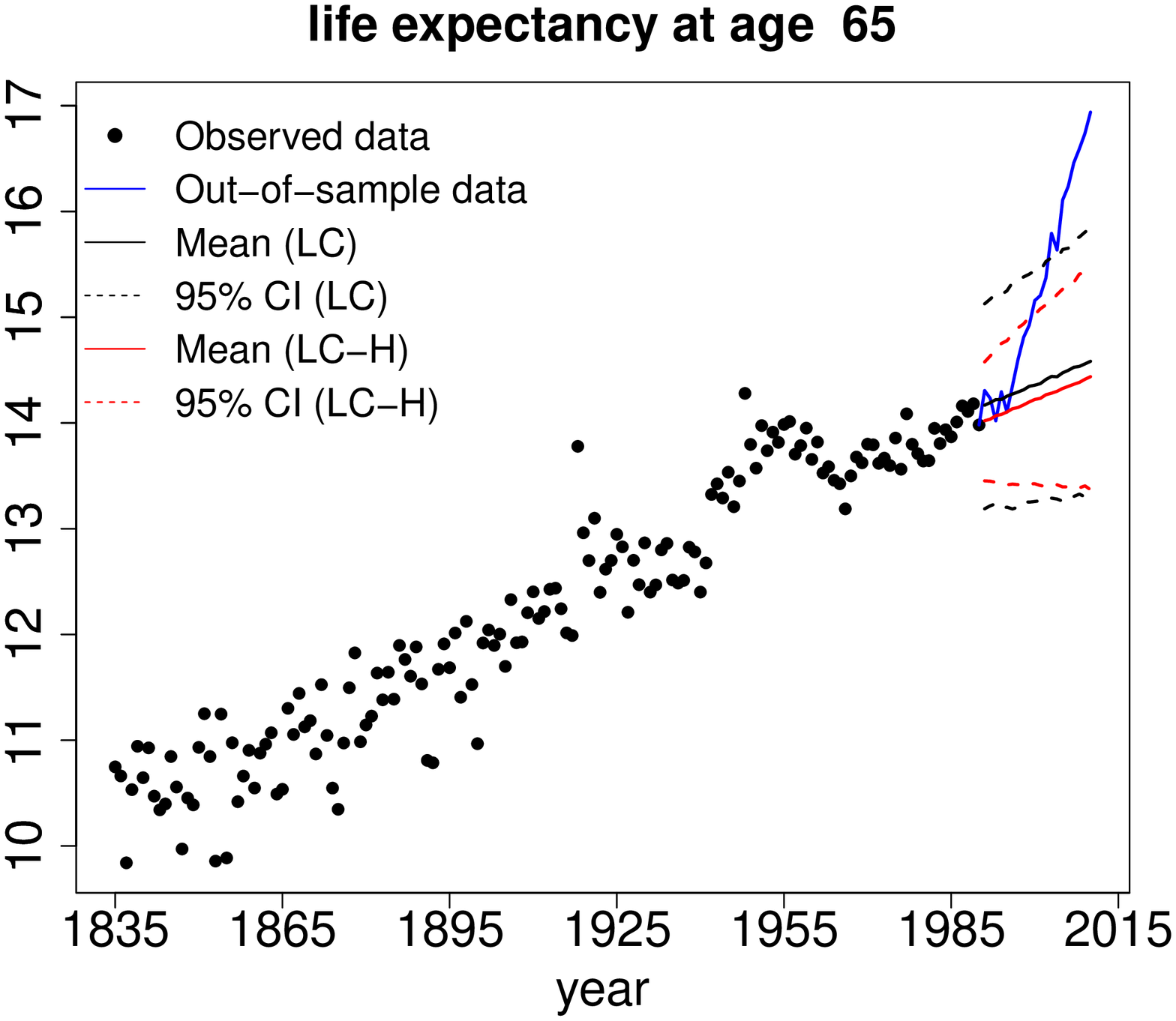}\includegraphics[width=5.5cm, height=5cm]{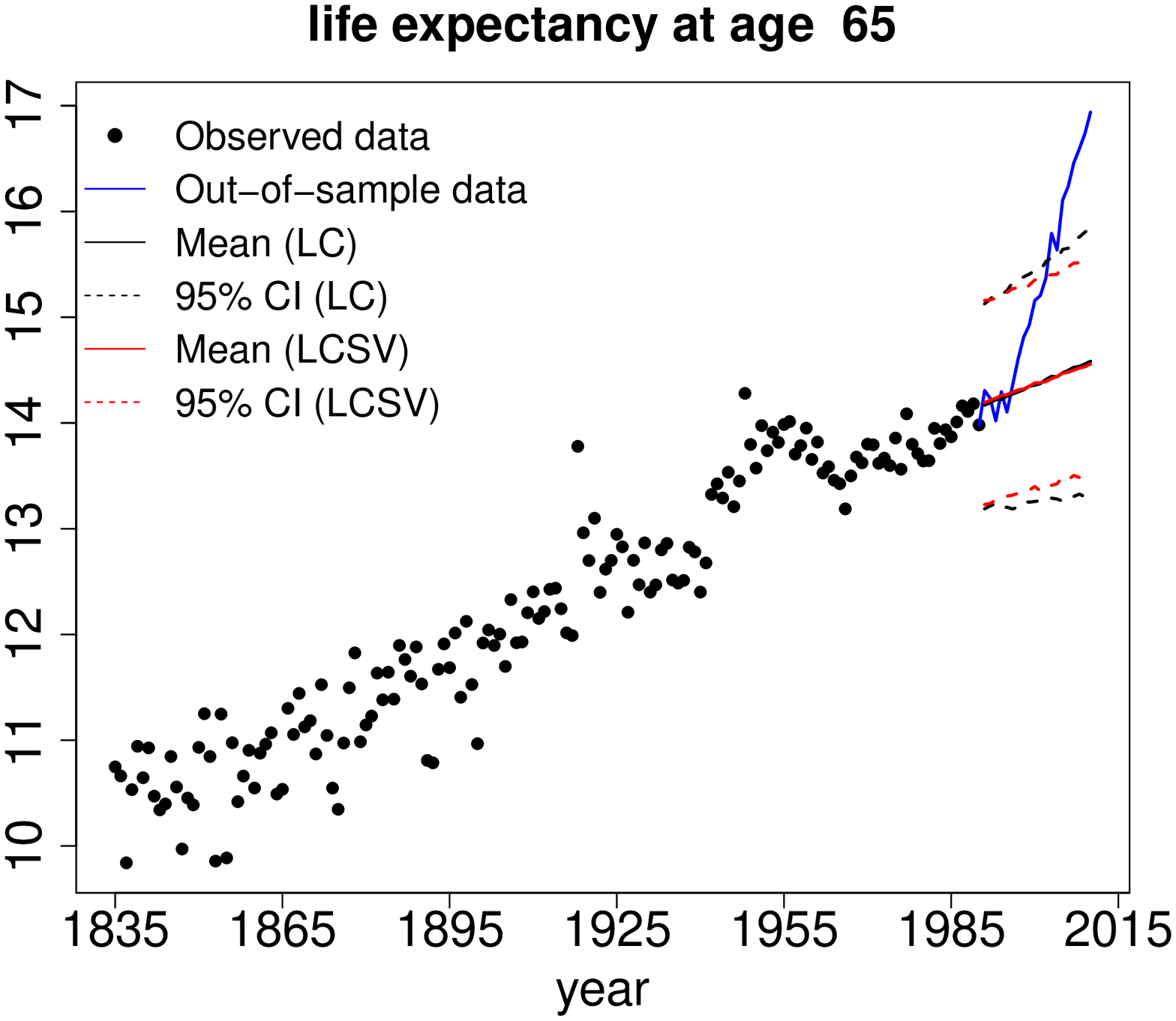}\includegraphics[width=5.5cm, height=5cm]{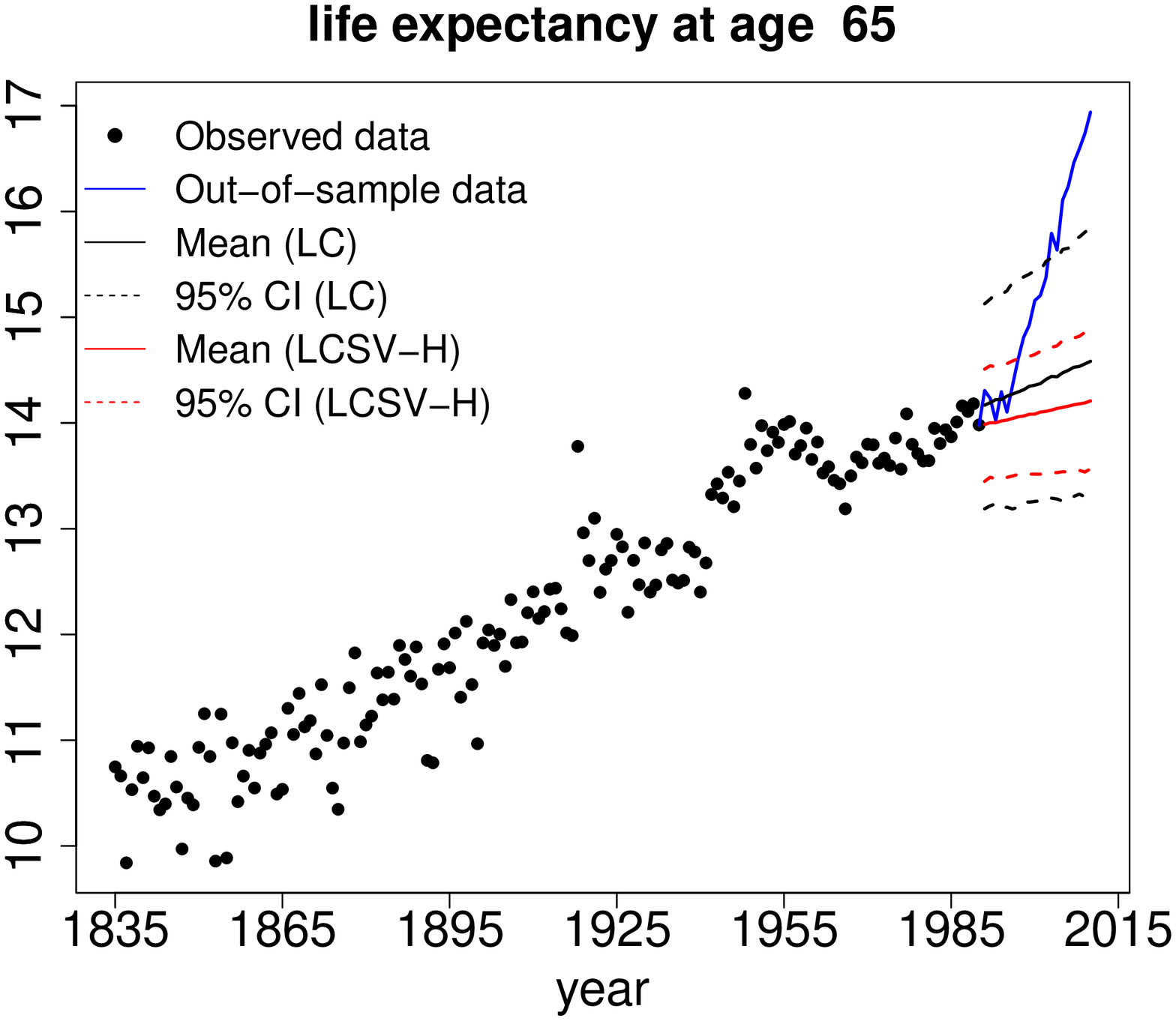}
\includegraphics[width=5.5cm, height=5cm]{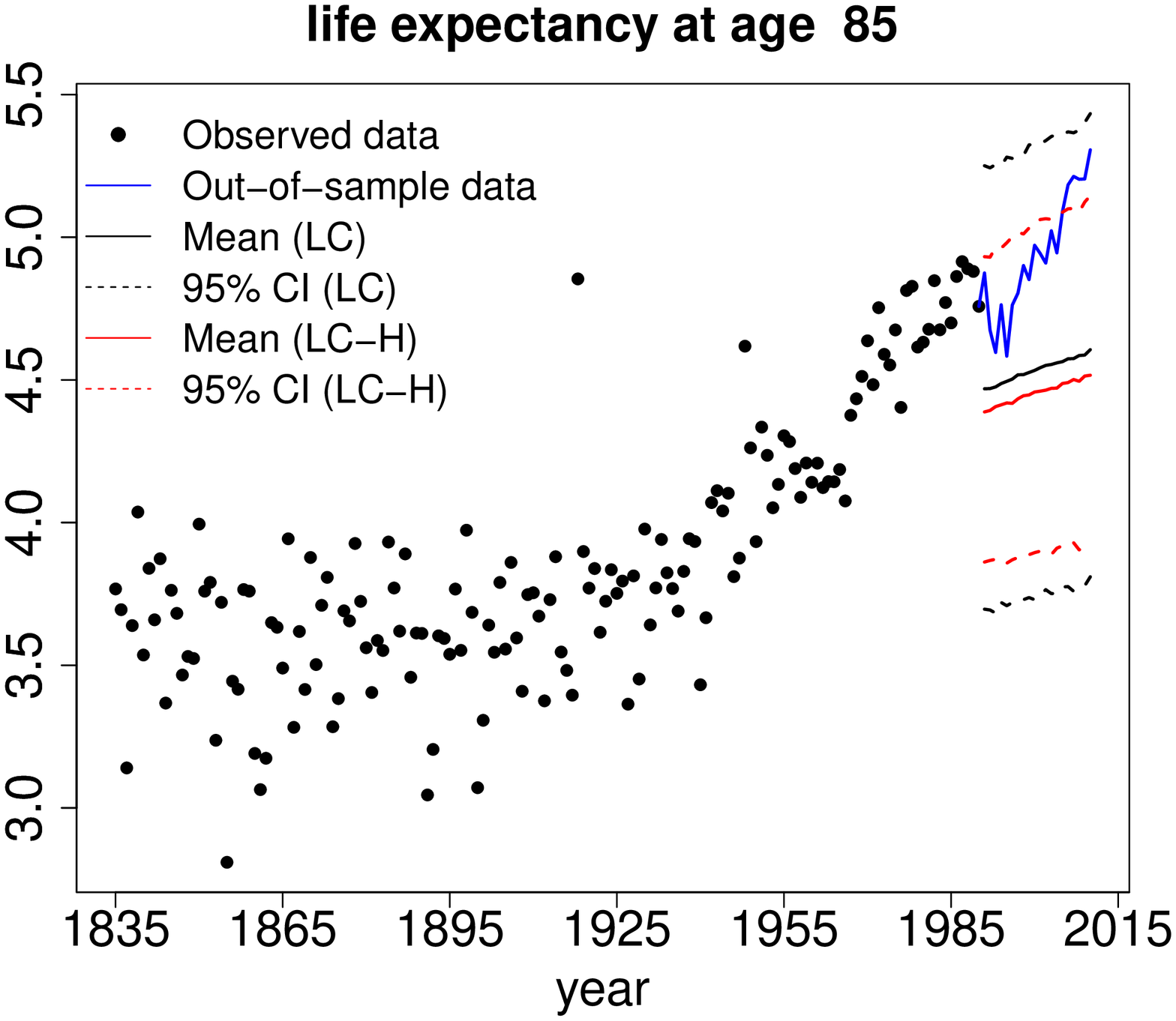}\includegraphics[width=5.5cm, height=5cm]{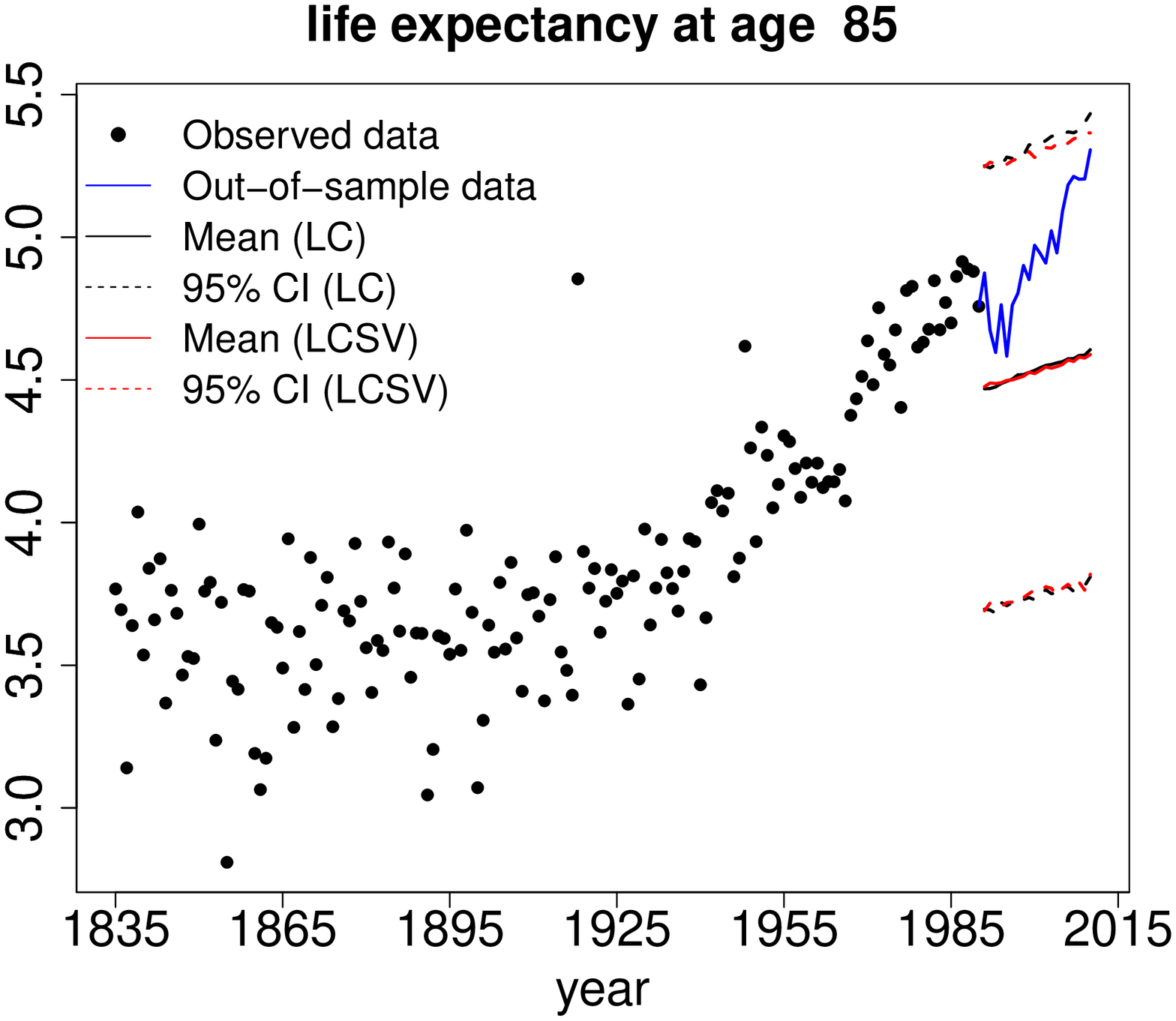}\includegraphics[width=5.5cm, height=5cm]{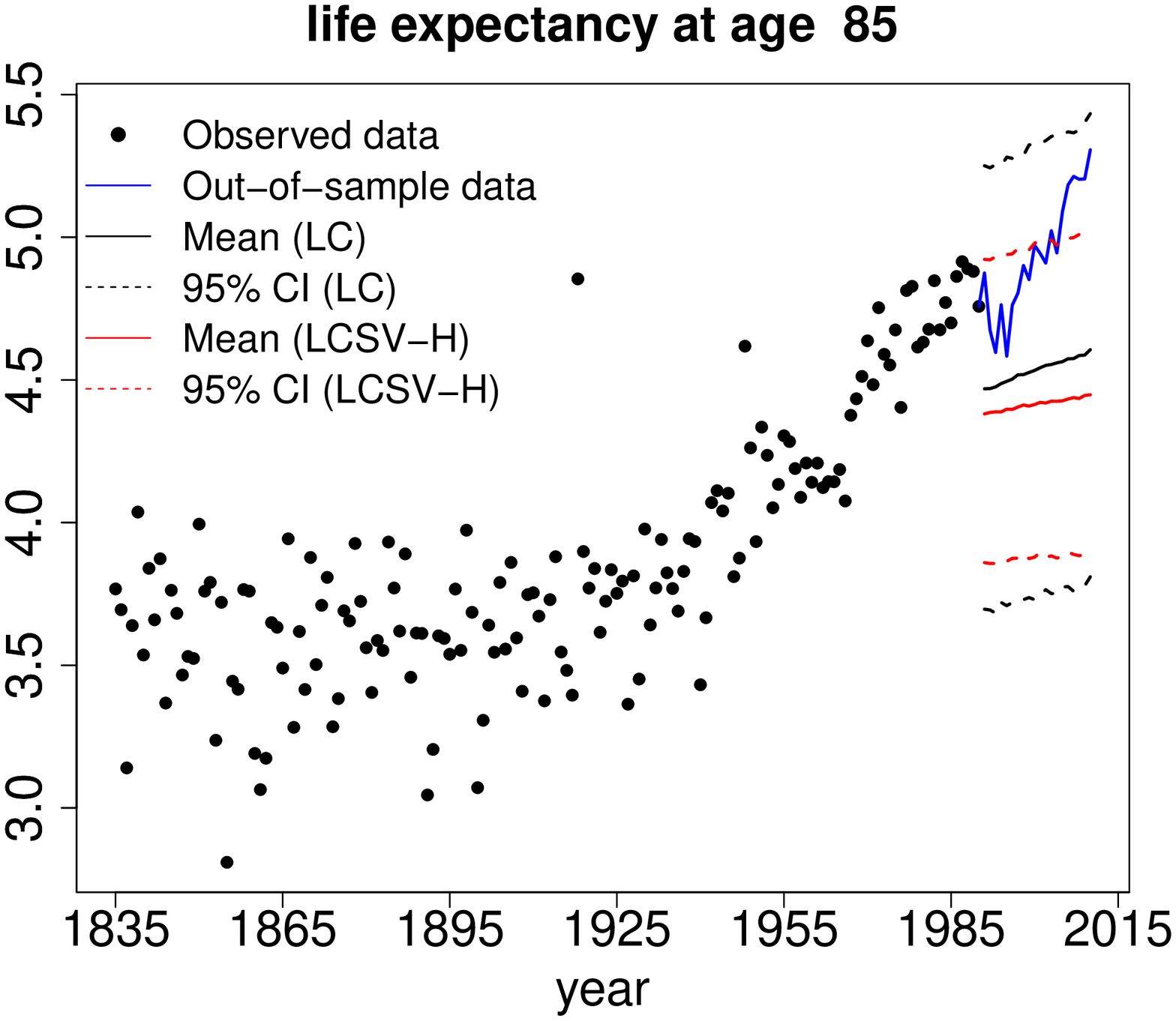}
\caption{\small{20-year out-of-sample forecasted life expectancy (1991-2010) at birth, age 65 and 85 for Danish male population under (left column) LC-H model, (middle column) LCSV model and (right column) LCSV-H model in comparison with LC model. Calibration period: 1835-1990.}}
\label{fig:forecastLifeExp18351980}
\end{center}
\end{figure}

\begin{figure}[h]
\begin{center}
\includegraphics[width=5.5cm, height=5cm]{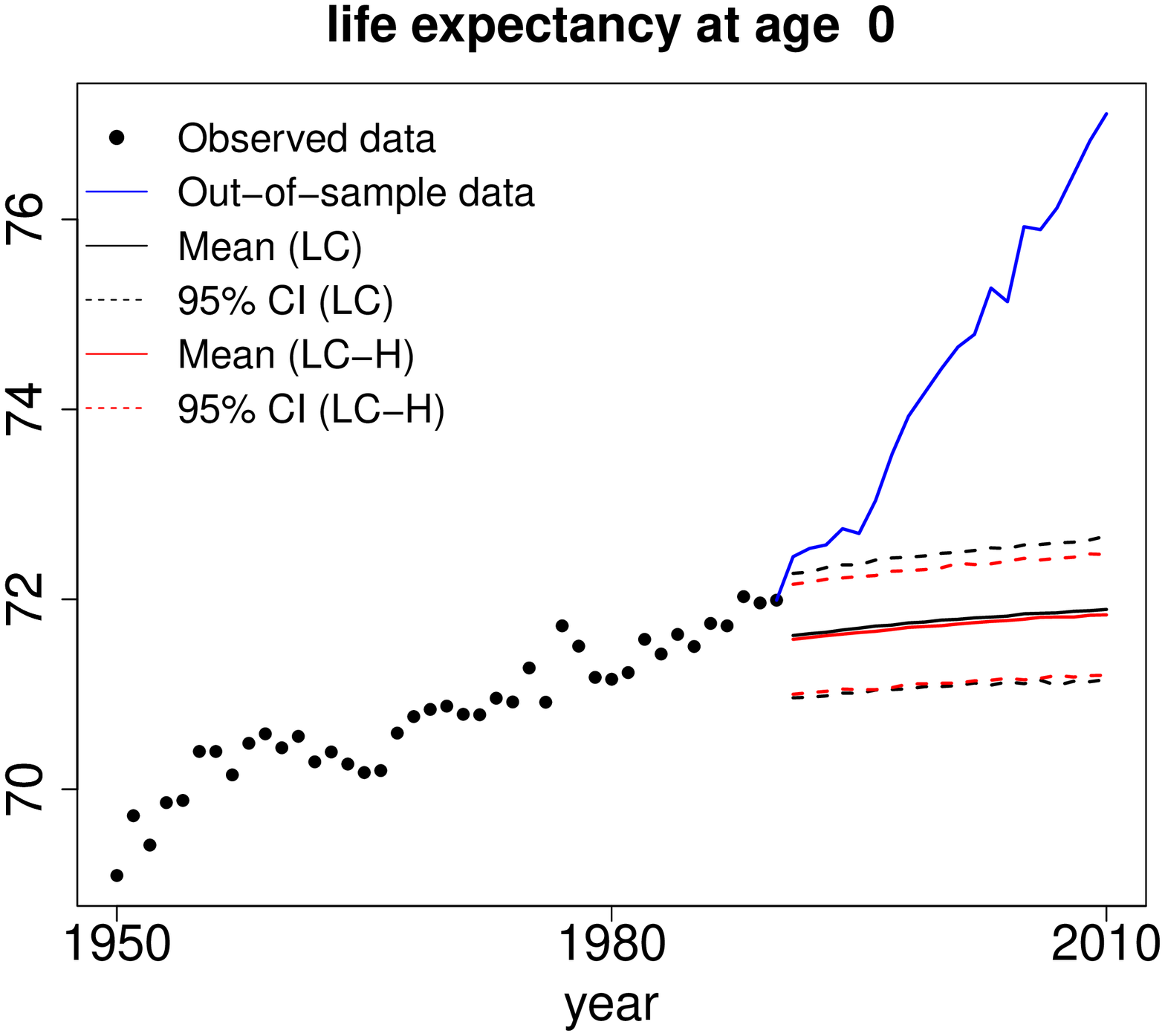}\includegraphics[width=5.5cm, height=5cm]{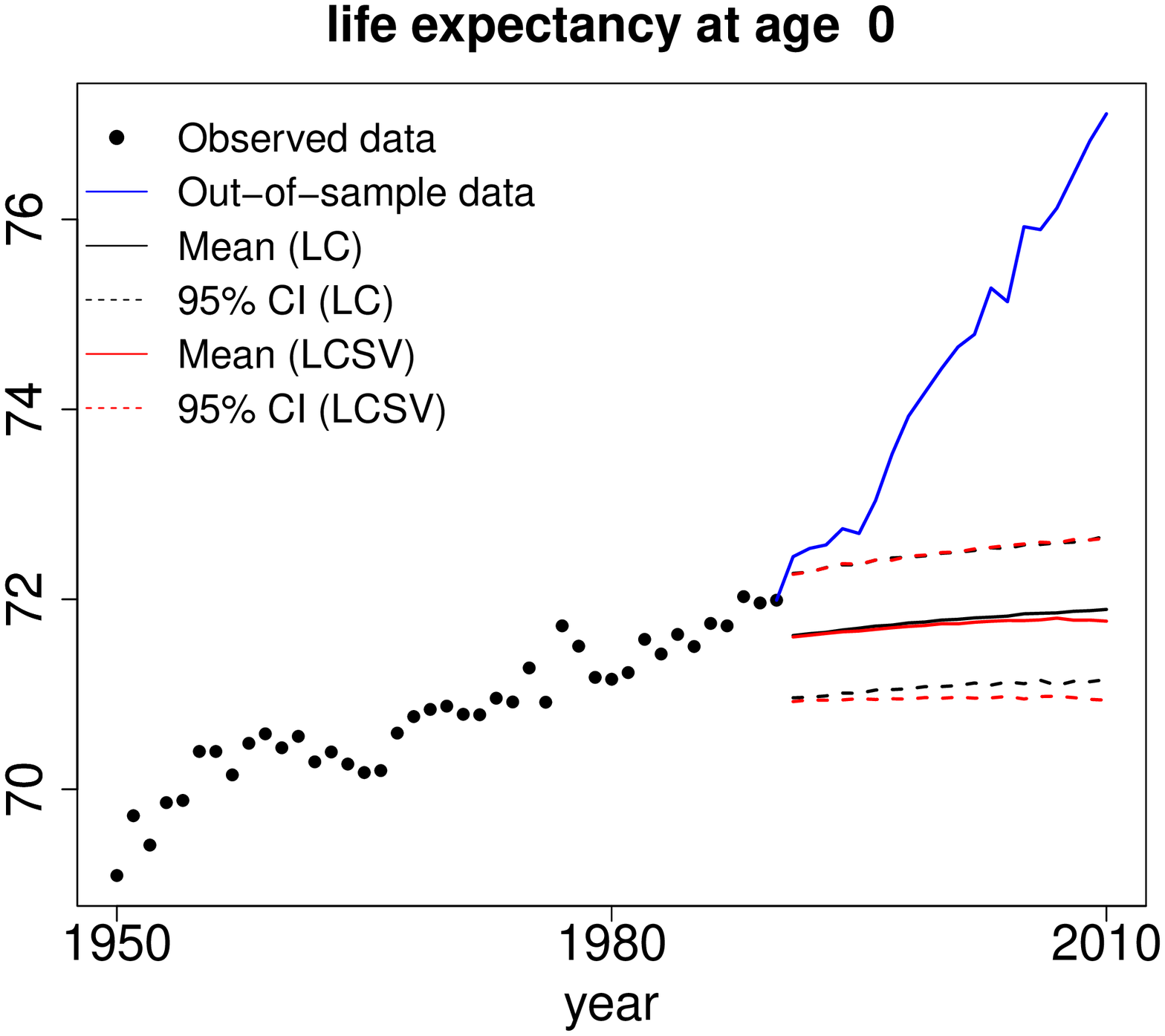}\includegraphics[width=5.5cm, height=5cm]{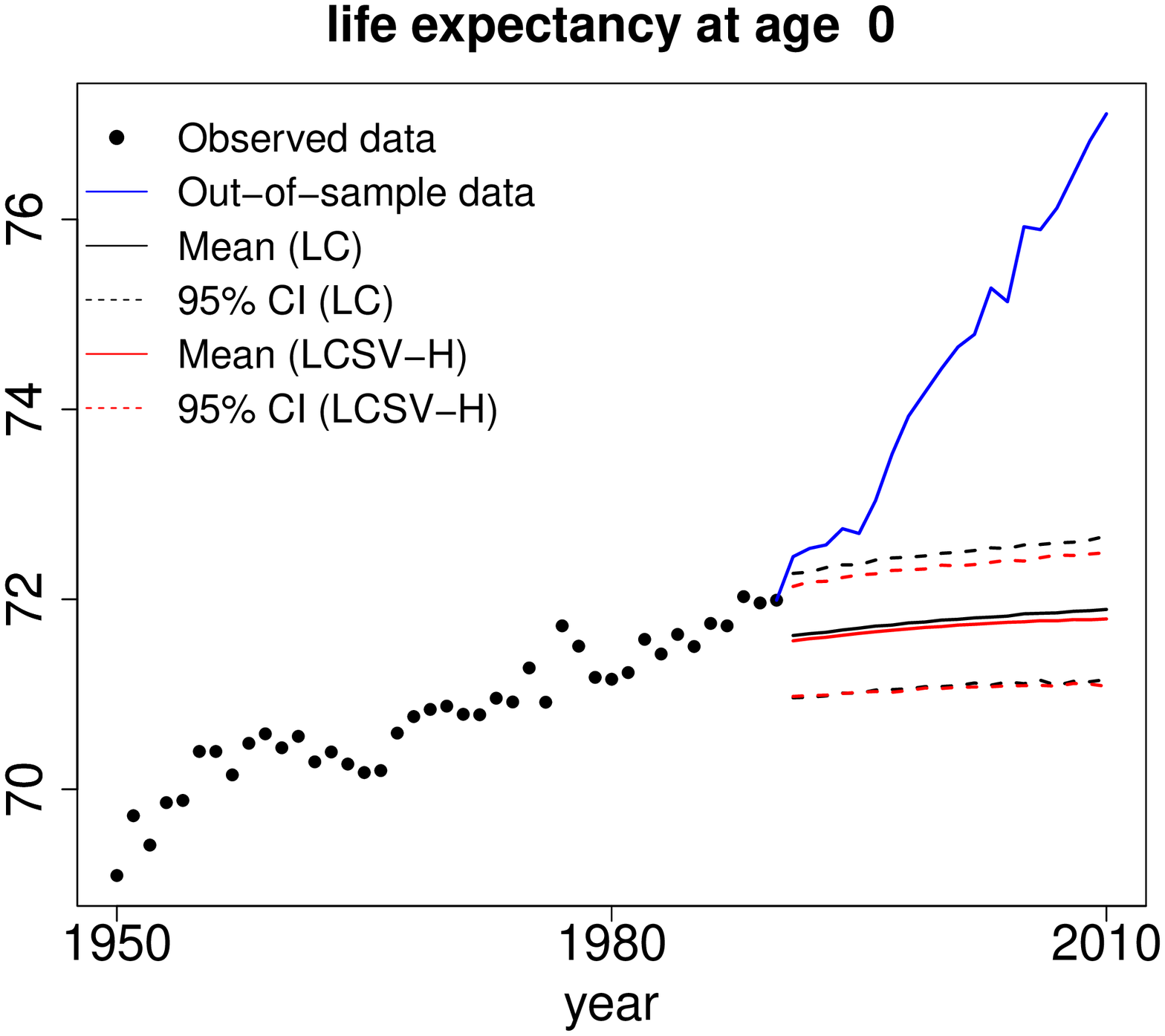}
\includegraphics[width=5.5cm, height=5cm]{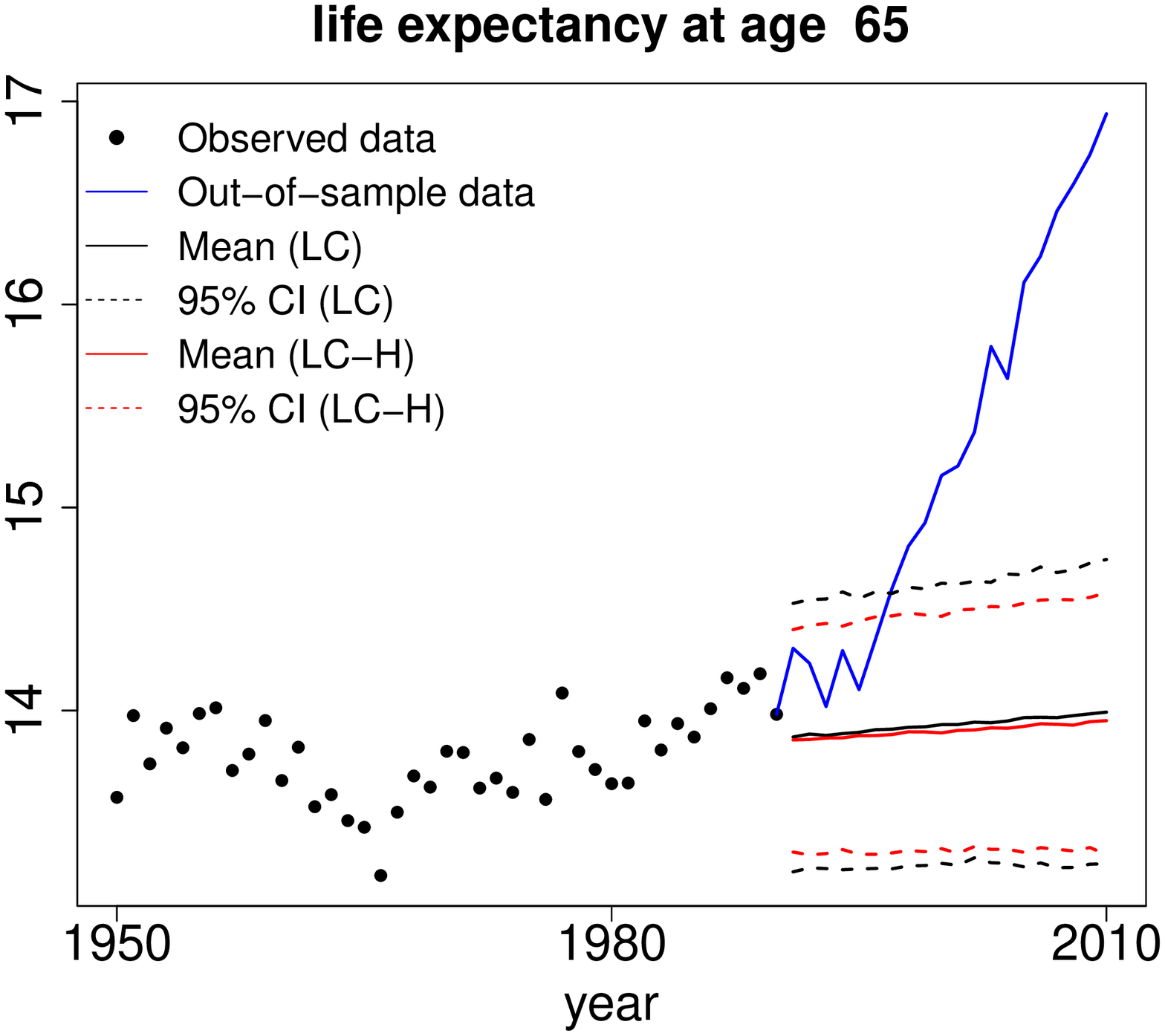}\includegraphics[width=5.5cm, height=5cm]{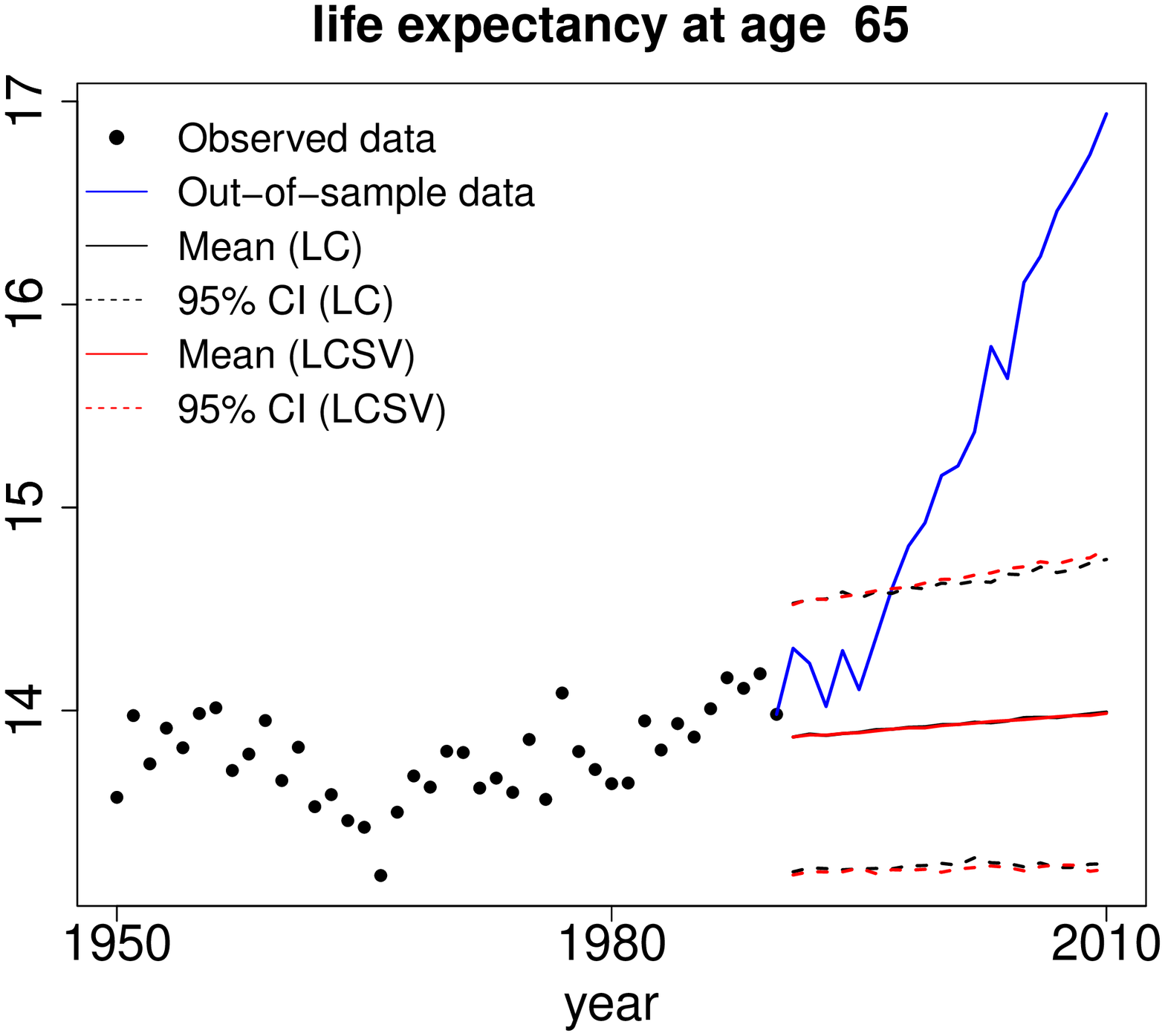}\includegraphics[width=5.5cm, height=5cm]{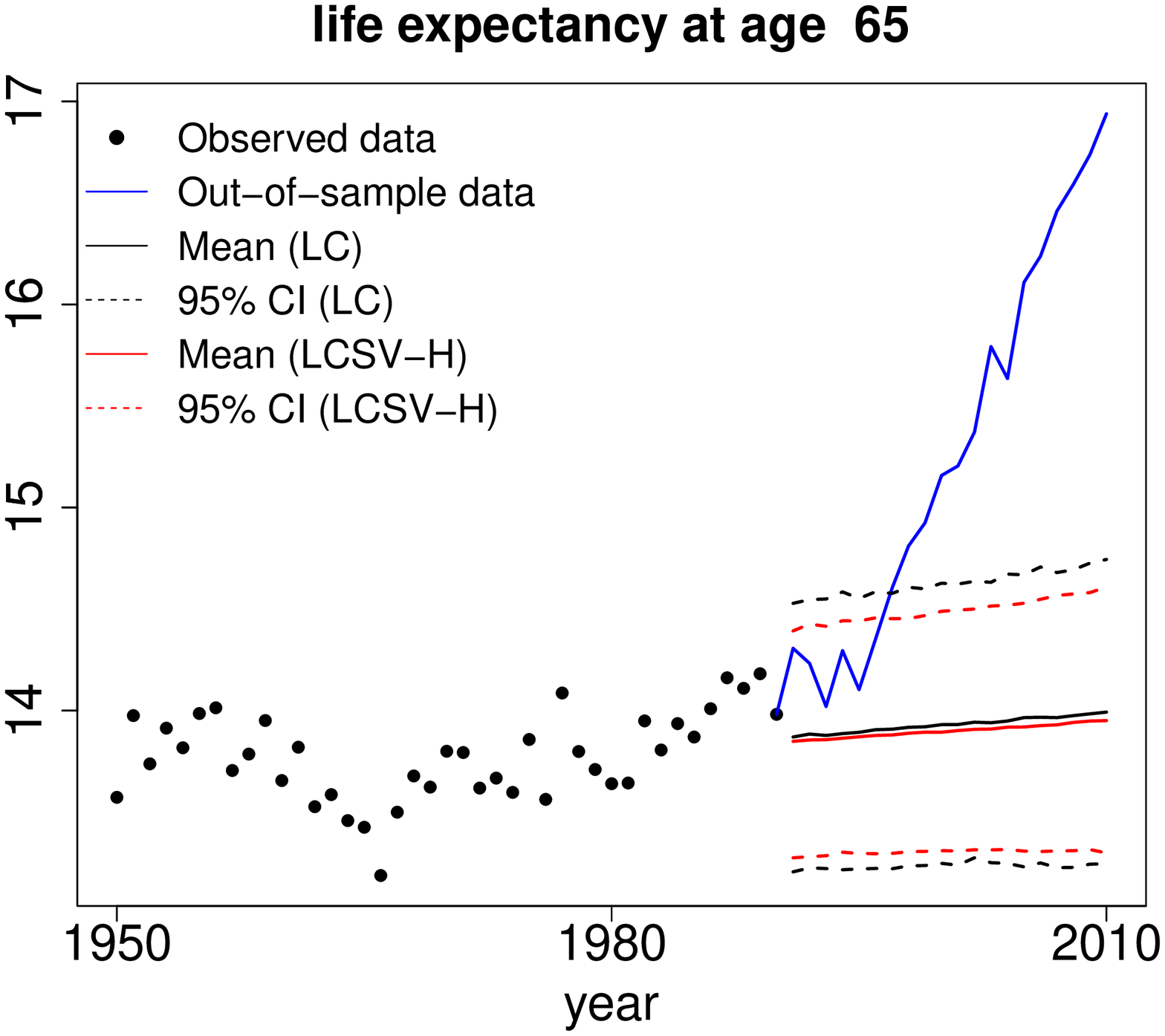}
\includegraphics[width=5.5cm, height=5cm]{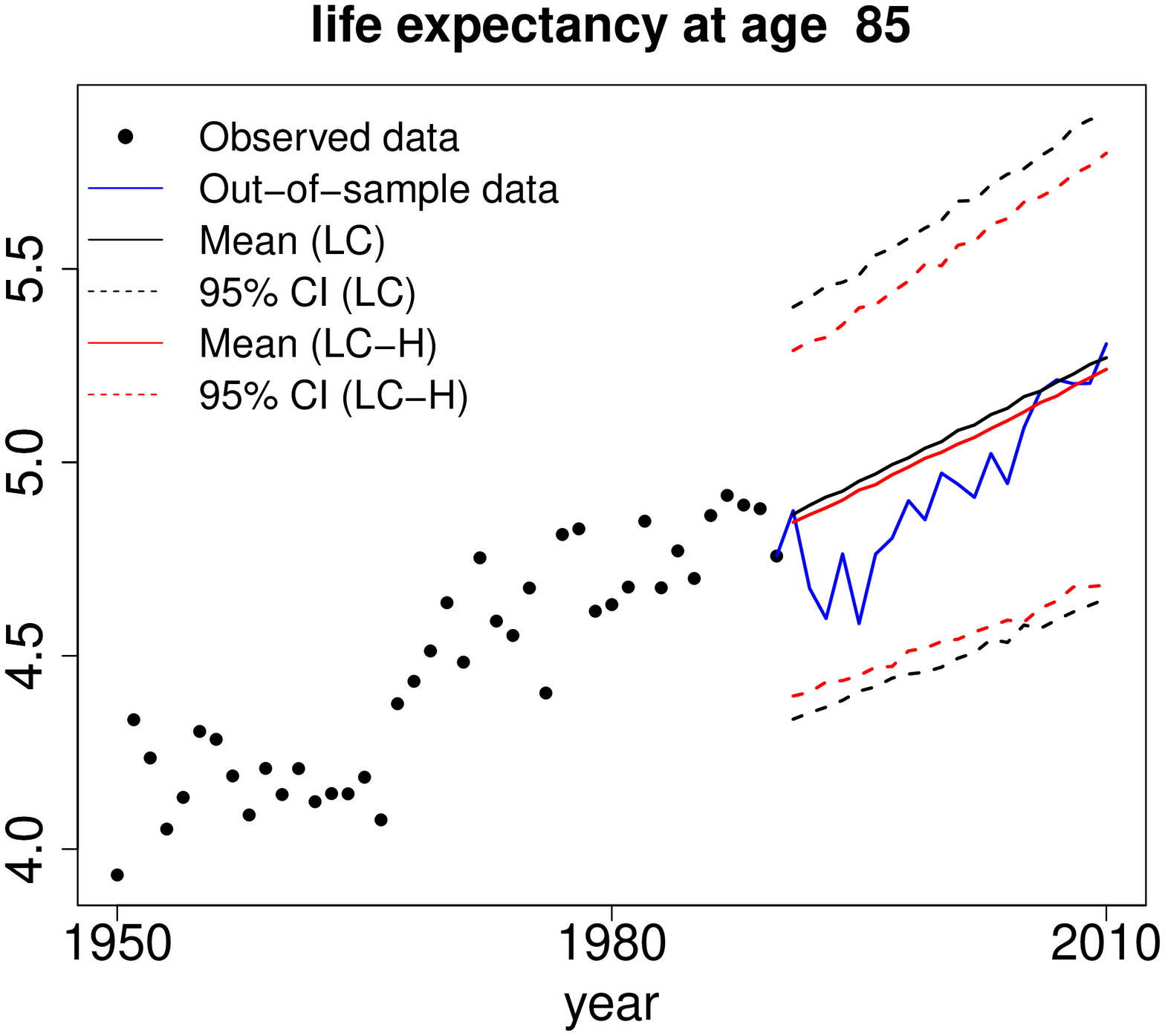}\includegraphics[width=5.5cm, height=5cm]{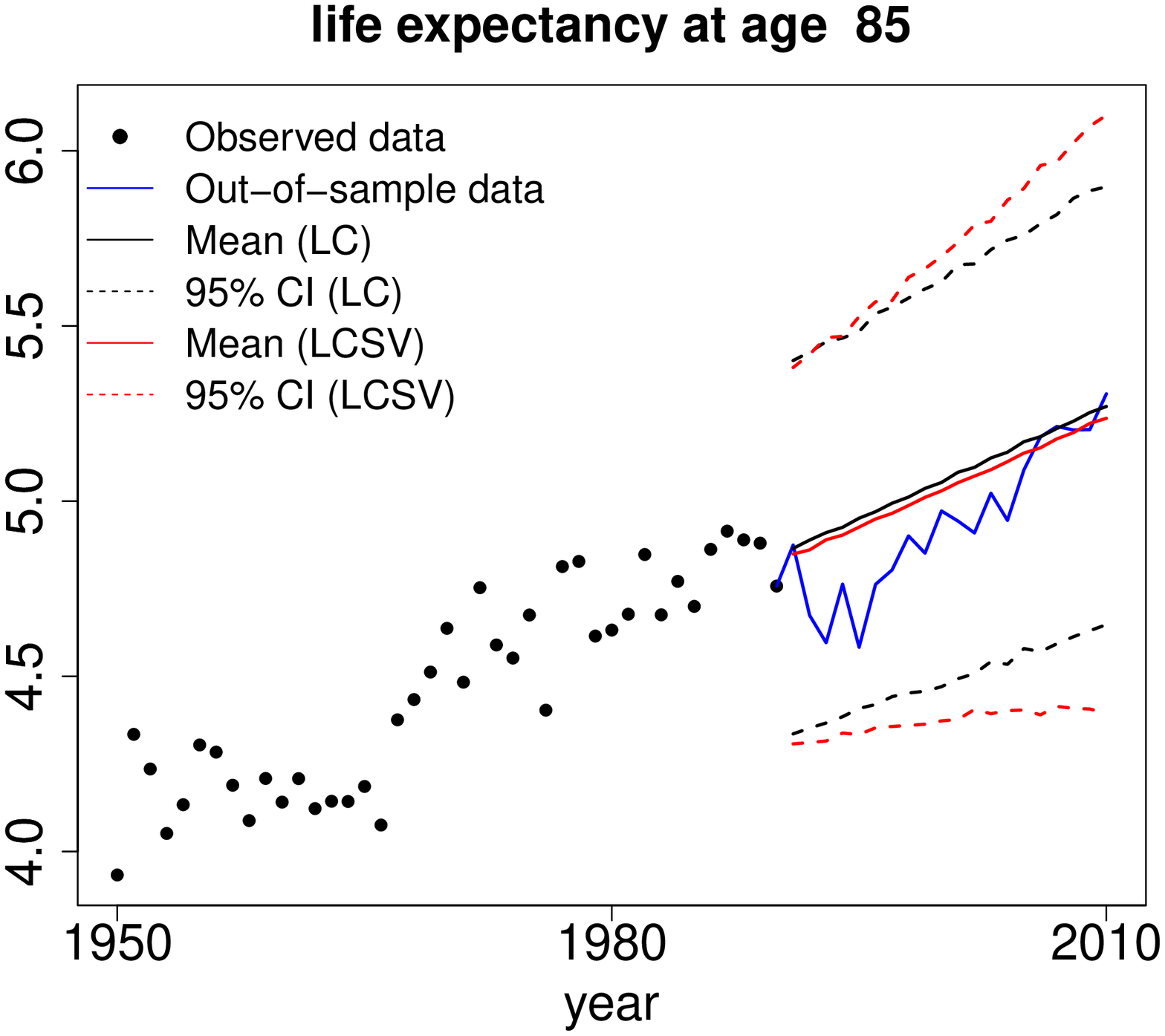}\includegraphics[width=5.5cm, height=5cm]{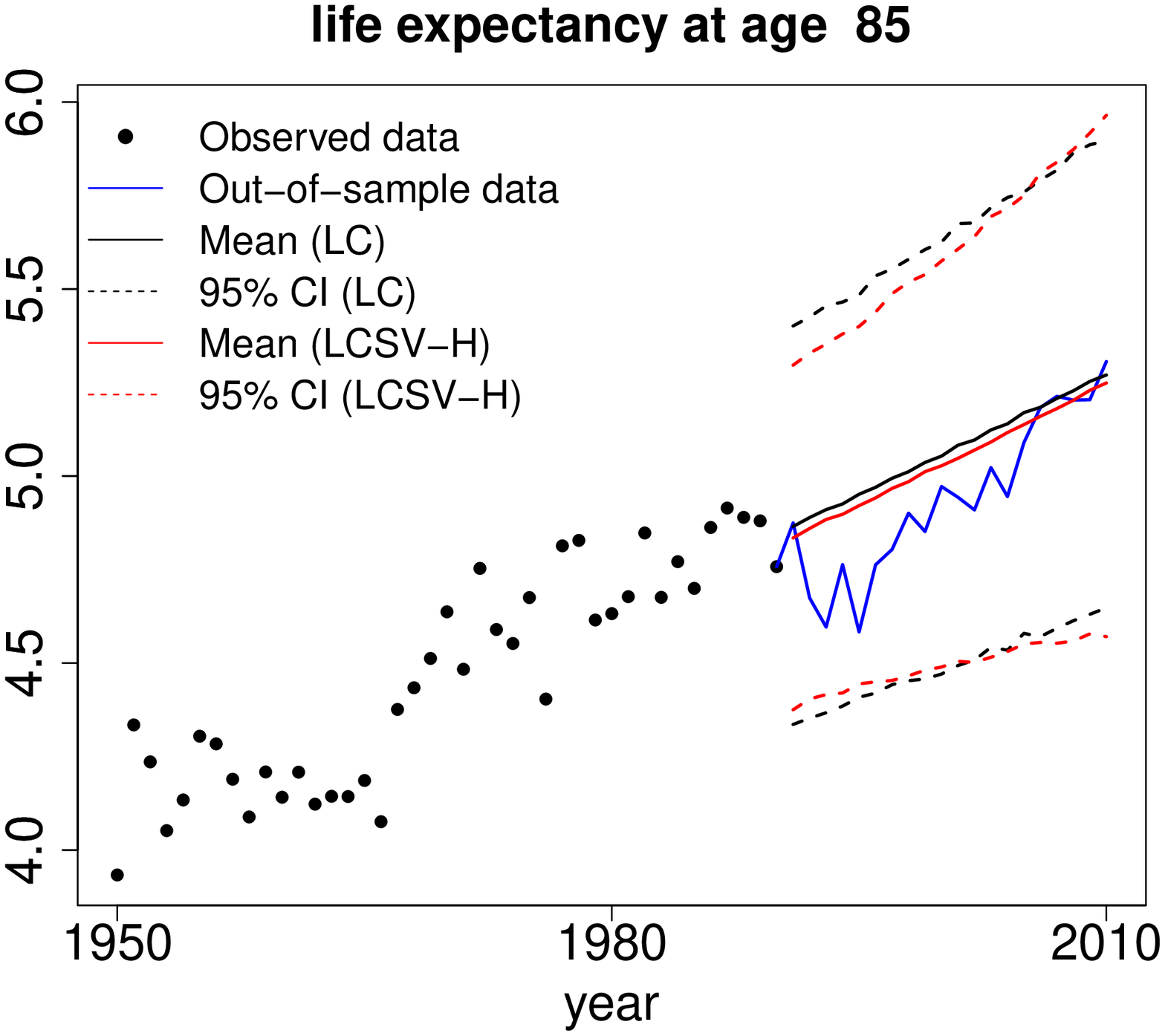}
\caption{\small{20-year out-of-sample forecasted life expectancy (1991-2010) at birth, age 65 and 85 for Danish male population under (left column) LC-H model, (middle column) LCSV model and (right column) LCSV-H model in comparison with LC model. Calibration period: 1950-1990.}}
\label{fig:forecastLifeExp19501980}
\end{center}
\end{figure}

As we use the fitted death rates instead of the observed death rates in the jump-off year (that is year 2010), there is a jump-off bias in the forecasted death rates. The forecasted life expectancy at age 65 is particularly sensitive to this jump off bias. It comes from a sudden decline of death rates for age groups larger than 65 beginning in year 1990, hence an increase of life expectancy at age 65 is observed. The jump-off bias is significantly smaller when the calibration period 1835-1990 is considered, see Figure~\ref{fig:forecastLifeExp18351980}. As expected, the forecasted distributions of life expectancy at birth and age 65 are similar for all the models estimated using mortality data from year 1950-1990 (Figure~\ref{fig:forecastLifeExp19501980}). Note that the $95\%$ credible intervals capture poorly the out-of-sample data in this case except for the life expectancy at age 85. It is a consequence of  the sudden change of significant downward trend for the death rates observed in the middle age groups of the Danish mortality data starting from around 1990, see Figure \ref{fig:forecastDeathRates19501990}, as well as the jump-off bias. We discuss about the linear trend assumption and jump-off bias in the next section.

\subsubsection{Linear trend assumption and jump-off bias}\label{sec:linearTrend}
In performing the forecasting of death rates and life expectancies, we use the models summarised in Table \ref{table:ModelSummary} where the LC-H, LCSV and LCSV-H models are variants of the Lee-Carter model in which a linear trend of the period effect is assumed. However, for the Danish male mortality data that we used, the overall trend is reasonably linear for the whole period 1835-2010, but the same may not be said for the shorter period 1950-2010 as Figure \ref{fig:forecastDeathRates18351990}-\ref{fig:forecastDeathRates19501990} indicate.

In particular, we observe in Figure \ref{fig:forecastDeathRates19501990} that there is a clear change of trend for the death rates of middle age groups. Such a change of trend is difficult, if not impossible, to predict in terms of timing and magnitude. We also perform the analysis on French male mortality data and found similar patterns. For forecasting purpose, one may therefore argue that expert opinion will be an important factor in predicting mortality. Even though any change of mortality trend in the short run cannot be predicted with reasonable accuracy using data alone, it can be \textit{detected} if the instantaneous volatility of mortality is quantified. For example, using the LCSV-H model, the log-volatility $\gamma$ is quantified and we observe from Figure \ref{fig:DENLCSVH18352010} that $\gamma$ started to increase around 1990 after several decades of declining. The change of volatility level not only affect the prediction intervals as discussed in previous sections, but also indicates that the change of mortality is heightened and one should be cautious whether a change of trend is taking place.

We also find that jump-off bias, discussed in Section \ref{sec:deathrates}-\ref{sec:lifeExp}, is an important factor in predicting deaths rates and life expectancies. We note that it is straight forward to remove the jump-off bias by adjusting \eqref{eqn:forecastLCsample}-\eqref{eqn:forecastLCSVsample} so that the actual death rates, instead of the fitted death rates, are used in the beginning of the forecasting period. We do not provide the corresponding plots in the paper but one can envisage the results simply by shifting the forecasted distribution so that the forecasted mean is attached to the (in-sample) data at the end of the estimation year. Removing the jump-off bias will have a significant impact on the accuracy of mortality forecasting especially when data exhibit clear trending.

%%%%%%%%%%%%%%%%%%%%%%%%%%%%%%%%%%%%%%%% Conlcusion %%%%%%%%%%%%%%%%%%%%%%%%%%%%%%%%%%%%%%%%%%%%%%
\section{Concluding Remarks}
\label{sec:conclusion}
We developed and presented a comprehensive state-space framework for stochastic mortality modelling. The state-space approach has two key advantages. First, it puts modelling, estimation and forecasting of mortality in a unified framework in contrast to common practice in this area. Second, the methodology permits realistic and sophisticated mortality models to be estimated and forecasted, which could be difficult to handled using other approaches.

We show that many of the popular mortality models exist in the literature can be cast in state-space form. We then suggest several classes of mortality models that can be classified as linear Gaussian state-space models and non-linear / non-Gaussian state-space models. Our proposals are not exhaustive but aim to illustrate the flexibility of the methodology. In particular we incorporate heteroscedasticity and stochastic volatility in mortality modelling, as an examination of mortality data suggests that volatility of death rates is not constant in the age and time dimension over a long time period. Moreover, we propose an alternative identification constraint for the Lee-Carter type modelling, which is tailored for the state-space approach.

Frequestist state-space inference for stochastic mortality models is carried out and explained based on the gradient and Hessian of the marginalized likelihood developed recently in statistics literature. We also utilise a modern approach to Bayesian inference for state-space modelling hinged on the PMCMC framework. In particular we develop a sampler using a combination of Rao-Blackwellized Kalman filter and particle filter for the latent state process full posterior conditionals, combined with Gibbs sampling steps for the static model parameters to estimate a stochastic volatility model for mortality proposed in this paper.

Using mortality data of Danish male population, we assess the extended models based on deviance conditional criterion. It is found that incorporating heteroscedasticity is a crucial improvement factor in model fitting, while model complexity is accounted for. The incorporation of stochastic volatility clearly enhances model performance for fitting of long term mortality time series. Estimation results for long calibration period support the assumption of stochastic volatility. We show that forecasting can be carried out straightforwardly in state-space framework under a Bayeisan setting. We examine the forecasting properties of the models using different calibration periods. The inclusion of heteroscedasticity and stochastic volatility substantially affects prediction intervals of death rate and life expectancy distributions. The linear trend assumption commonly found in mortality modelling and jump-off bias are discussed in light of the Danish mortality data.

State-space framework provides attractive features that are of importance to mortality modelling. The methods and results developed and shown in the paper will have significant implications for longevity risk management in actuarial applications which is a topic of future research.

\section*{Acknowledgements}
This research was supported by the CSIRO-Monash Superannuation Research
Cluster, a collaboration among CSIRO, Monash University, Griffith
University, the University of Western Australia, the University of
Warwick, and stakeholders of the retirement system in the interest
of better outcomes for all. This research was also partially supported under the
Australian Research Council's Discovery Projects funding scheme (project number:
DP160103489).

\begin{appendices}
\section{Differentiation Matrices in Gradient-based Estimation}\label{appA:Matrices}
For the LC-H model, the parameter vector is denoted by $\boldsymbol{\psi} =(\alpha_{x_2:x_p},\beta_{x_2:x_p},\sigma^2_{\varepsilon,x_1:x_p},\theta,\sigma^2_\omega)$ with dimension $n = 3p$ where $p$ is the number of age group considered. We are required to evaluate $\frac{\partial \boldsymbol{\alpha}}{\partial \boldsymbol{\psi}_i}$, $\frac{\partial \boldsymbol{\beta}}{\partial \boldsymbol{\psi}_i}$, $\frac{\partial \boldsymbol{\Sigma}}{\partial \boldsymbol{\psi}_i}$, $\frac{\partial \boldsymbol{\theta}}{\partial \boldsymbol{\psi}_i}$ and $\frac{\partial \boldsymbol{\sigma^2_\omega}}{\partial \boldsymbol{\psi}_i}$ in the gradient-based estimation (Section \ref{subsec:MLE}). Define
\begin{align*}
    \boldsymbol{\psi}_{\alpha} &:= \psi_{1:p-1} = \alpha_{x_2:x_p} (:=\boldsymbol{\alpha}_{-{x_1}}), \quad
    \boldsymbol{\psi}_{\beta} := \psi_{p:2p-2} = \beta_{x_2:x_p} (:=\boldsymbol{\beta}_{-{x_1}}) \\
    \boldsymbol{\psi}_{\sigma^2_\varepsilon} &:= \psi_{2p-1:3p-2} = \sigma^2_{\varepsilon, x_1:x_p}, \quad
    \psi_{\theta} := \psi_{3p-1} = \theta, \quad
    \psi_{\sigma^2_\omega} := \psi_{3p} = \sigma^2_\omega.
\end{align*}
Then we have
\begin{align*}
    \frac{\partial (\boldsymbol{\alpha}_{-{x_1}})_j}{\partial (\boldsymbol{\psi}_\alpha)_i} &= \delta_{ij} =
    \frac{\partial (\boldsymbol{\beta}_{-{x_1}})_j}{\partial (\boldsymbol{\psi}_\beta)_i}, \quad i,j=1,\dots,p-1 \\
    \frac{\partial (\Sigma)_{jj}}{\partial (\boldsymbol{\psi}_{\sigma^2_\varepsilon})_i} &= \delta_{ij}, \quad i,j=1,\dots,p \\
    \frac{\partial \theta}{\partial \psi_\theta} &= 1 = \frac{\partial \sigma^2_\omega}{\partial \psi_{\sigma^2_\omega}},
    %\frac{\partial \sigma^2_\omega}{\partial \psi_{\sigma^2_\omega}} &= 1 \\
\end{align*}
where $\delta_{ij} = 1$ if $j=i$ and zero otherwise; $\Sigma$ is a diagonal matrix with diagonal $\sigma^2_{\varepsilon, x_1:x_p}$. Note that $\frac{\partial (\boldsymbol{\alpha})_{1}}{\partial (\boldsymbol{\psi}_\alpha)_i} = \frac{\partial (\boldsymbol{\beta})_1}{\partial (\boldsymbol{\psi}_\beta)_i} = 0$ for $i=1,\dots,p-1$, where $\boldsymbol{\alpha}=\alpha_{x_1:x_p}$ and $\boldsymbol{\beta}=\beta_{x_1:x_p}$.

\section{A Review of SMC Method}
\label{app:SMC}
SMC, also known as particle filtering, can be viewed as a generalisation of Kalman filtering in state-space modelling context. The method is based on importance sampling and it has become an essential sampling-based tool in many domains (\cite{DoucetFrGo01}). In the following we give a brief review of the method using the LCSV model, \eqref{eqn:LCSVb}-\eqref{eqn:LCSVc}, as an example to derive a basic particle filtering algorithm. Our target density is the joint posterior distribution of the states for stochastic volatility:
\begin{equation}
    \pi(\gamma_{1:t}|\kappa_{0:t})
\end{equation}
where the parameters of the model are assumed to be known and is suppressed here for ease of notation. To apply importance sampling, we first calculate
\begin{align}
    \pi(\gamma_{1:t}|\kappa_{0:t}) &=
    \frac{\pi(\kappa_t|\gamma_{1:t},\kappa_{0:t-1})\pi(\gamma_{1:t}|\kappa_{0:t-1})}{\pi(\kappa_t|\kappa_{0:t-1})} \notag
    \\
    &=
    \frac{\pi(\kappa_t|\gamma_{1:t},\kappa_{0:t-1})\pi(\gamma_t|\gamma_{1:t-1},\kappa_{0:t-1})}{\pi(\kappa_t|\kappa_{0:t-1})}\pi(\gamma_{1:t-1}|\kappa_{0:t-1})
     \notag %\label{eqn:pnpn0}
    \\
    &=
    \frac{\pi(\kappa_t|\gamma_t, \kappa_{t-1})\pi(\gamma_t|\gamma_{t-1})}{\pi(\kappa_t|\kappa_{0:t-1})}\pi(\gamma_{1:t-1}|\kappa_{0:t-1}).
    \label{eqn:pnpn1}
\end{align}
The importance density is assumed to satisfy
\begin{equation}\label{eqn:g0n}
    g_{1:t}(\gamma_{1:t}|\kappa_{0:t}):=
    g_t(\gamma_t|\gamma_{1:t-1},\kappa_{0:t})g_{1:t-1}(\gamma_{1:t-1}|\kappa_{0:t-1})
\end{equation}
and the importance weight is given by
\begin{align}
    \tilde{w}_t &=
    \frac{\pi(\kappa_t|\gamma_t,\kappa_{t-1})\pi(\gamma_t|\gamma_{t-1})}{\pi(\kappa_t|\kappa_{t-1})g_t(\gamma_t|\gamma_{1:t-1},\kappa_{0:t})}
        \frac{\pi(\gamma_{1:t-1}|\kappa_{0:t-1})}{g_{0:t-1}(\gamma_{1:t-1}|\kappa_{0:t-1})}
        \notag
    \\
    &\propto
    \frac{\pi(\kappa_t|\gamma_t,\kappa_{t-1})\pi(\gamma_t|\gamma_{t-1})}{g_t(\gamma_t|\gamma_{1:t-1},\kappa_{0:t})}\, \tilde{w}_{t-1}
    \notag
    \\
    &:= \hat{w}_t \,\tilde{w}_{t-1}, \label{eqn:SISweight}
\end{align}
where $\hat{w}_t$ is called the incremental importance weight. The normalised importance weights are then obtained as $w^{(i)}_t := \tilde{w}^{(i)}_t/\sum^N_{j=1}\tilde{w}^{(j)}_t$. To summarise, suppose we have $N$ particle paths $(\gamma^{(i)}_{1:t-1},w^{(i)}_{t-1})_{i=1}^{N}$ to approximate the density $\pi(\gamma_{1:t-1}|\kappa_{0:t-1})$ at time $t-1$. Then, from \eqref{eqn:g0n}, the $i$-th particle path at time $t$ is given by $\gamma^{(i)}_{1:t}=(\gamma^{(i)}_{1:t-1},\gamma^{(i)}_t)$ where $\gamma^{(i)}_t$ is sampled from $g_t(\gamma_t|\gamma^{(i)}_{1:t-1},\kappa_{0:t})$. The target density $\pi(\gamma_{1:t}|\kappa_{0:t})$ is approximated by $(\gamma^{(i)}_{1:t},w^{(i)}_{t})_{i=1}^{N}$ where the normalised weight $w^{(i)}_{t}$ is obtained from \eqref{eqn:SISweight} and normalisation is carried out.

The problem of degeneracy, that is a majority of the particle paths may have negligible weight, can be handled by resampling. Specifically, we define the so-called effective sample size
\begin{equation}
    N_{eff} := \left(\sum^N_{i=1}(w^{(i)}_t)^2\right)^{-1}.
\end{equation}
At each time $t$, if $N_{eff}$ is smaller than some threshold (for example $80\%$ of $N$) then we draw $N$ samples (denoted by $N(i),i=1,\dots,N$) from a multinomial distribution with probability weights $w^{(i)}_t$, $i=1,\dots,N$, and replace the particle paths $\gamma^{(i)}_{1:t}$ by $\gamma^{(N(i))}_{1:t}$, and set $w^{(i)}_t=1/N$. The resampling step allows to keep the particle paths in proportion to their weights and tend to discard those that have negligible weights.
\end{appendices}

%%%%%%%%%%%%%%%%%%%%%%%%%%%%% BIBLIOGRAPHY %%%%%%%%%%%%%%%%%%%%%%%%%%%%%%%%%%%%%%%%%%%%%%%%%%%%%%%%%%%%%%%%%%%%%%%%%%%%%%%

%\clearpage
\bibliographystyle{elsart-harv}
\footnotesize{\bibliography{mcf}}

\end{document}